\documentclass[twocolumn,epjc3]{svjour3}  
\smartqed  
\RequirePackage{color,graphicx,url}

\journalname{Eur. Phys. J. A}

\usepackage{lmodern}
\usepackage[T1]{fontenc}
\usepackage{listings}

\usepackage{amsfonts,amsmath,amssymb,bm,xspace}
\usepackage{soul}
\usepackage{cite}

\graphicspath{{./figs/}}

\usepackage{hyperref}                

\usepackage{academicons} 
\newcommand{\orcid}[1]{\href{https://orcid.org/#1}{\textcolor[HTML]{A6CE39}{\aiOrcid}}}
\usepackage{xcolor}
\usepackage{hyperref}
\usepackage{listings}
\definecolor{SLpurple}{RGB}{155, 48, 255}

\newcommand{\bat}{\Bigr\rvert}

\newcommand{\sat}{\mathrm{sat}}
\newcommand{\sym}{\mathrm{sym}}

\newcommand{\rma}{\mathrm{r.m.}}
\newcommand{\ffg}{\mathrm{FFG}}
\newcommand{\nr}{\mathrm{NR}}
\newcommand{\rl}{\mathrm{R}}
\newcommand{\ur}{\mathrm{UR}}
\newcommand{\tov}{\mathrm{TOV}}

\newcommand{\NM}{\mathrm{NM}}
\newcommand{\SM}{\mathrm{SM}}

\newcommand{\nuc}{\mathrm{nuc}}
\newcommand{\lep}{\mathrm{lep}}
\newcommand{\el}{\mathrm{e}}
\newcommand{\tot}{\mathrm{tot}}

\newcommand{\intn}{\mathrm{int}}

\renewcommand{\max}{\mathrm{max}}
\newcommand{\obs}{\mathrm{obs}}
\newcommand{\chirp}{\mathrm{chirp}}

\usepackage{graphicx}

\hypersetup{%
pdfsubject=Paper,
pdfkeywords={nuclear physics} {MBPT} {chiral EFT} {asymmetric nuclear matter}
unicode=true,
breaklinks=true,
colorlinks   = true,
linkcolor = blue,
citecolor = blue,
menucolor = blue,
urlcolor = blue
}

\begin{document}

\sloppy
\title{The \texttt{nucleardatapy} toolkit for simple access to experimental nuclear data, astrophysical observations, and theoretical predictions}

\author{
J\'er\^ome Margueron\thanksref{addr1}\href{https://orcid.org/0000-0001-8743-3092}{\includegraphics[scale=0.3]{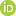}}
\and
Christian Drischler\thanksref{addr2,addr9}\href{https://orcid.org/0000-0003-1534-6285}{\includegraphics[scale=0.3]{figs/orcid.png}}
\and
Mariana Dutra\thanksref{addr3}\href{https://orcid.org/0000-0001-7501-0404}{\includegraphics[scale=0.3]{figs/orcid.png}}
\and
Stefano~Gandolfi\thanksref{addr4}\href{https://orcid.org/0000-0003-2656-6355}{\includegraphics[scale=0.3]{figs/orcid.png}} 
\and
Alexandros Gezerlis\thanksref{addr5}\href{https://orcid.org/0000-0003-2232-2484}{\includegraphics[scale=0.3]{figs/orcid.png}}
\and
Guilherme Grams\thanksref{addr6}\href{https://orcid.org/0000-0002-8635-383X}{\includegraphics[scale=0.3]{figs/orcid.png}}
\and
S\'ebastien~Guillot\thanksref{addr7,addr8}\href{https://orcid.org/???}{\includegraphics[scale=0.3]{figs/orcid.png}}
\and
Rohit Kumar\thanksref{addr9}\href{https://orcid.org/0000-0003-0786-2562}{\includegraphics[scale=0.3]{figs/orcid.png}}
\and
Sudhanva Lalit\thanksref{addr9}\href{https://orcid.org/0000-0001-7758-492X}{\includegraphics[scale=0.3]{figs/orcid.png}} 
\and
Odilon~Louren\c{c}o\thanksref{addr3}\href{https://orcid.org/0000-0002-0935-8565}{\includegraphics[scale=0.3]{figs/orcid.png}}
\and
Rahul Somasundaram\thanksref{addr4}\href{https://orcid.org/0000-0003-0427-3893}{\includegraphics[scale=0.3]{figs/orcid.png}} 
\and
Ingo Tews\thanksref{addr4}\href{https://orcid.org/0000-0003-2656-6355}{\includegraphics[scale=0.3]{figs/orcid.png}} 
\and
Isaac~Vida\~na\thanksref{addr10}\href{https://orcid.org/0000-0001-7930-9112}{\includegraphics[scale=0.3]{figs/orcid.png}} 
}

\institute{International Research Laboratory on Nuclear Physics and Astrophysics, Michigan State University and CNRS, East Lansing, MI 48824, USA \label{addr1}
\and
Institute of Nuclear and Particle Physics (INPP), Ohio University, Athens, OH 45701, USA \label{addr2}
\and
Facility for Rare Isotope Beams, Michigan State University, East Lansing, MI 48824, USA\label{addr9}
\and
Departamento de F\'isica e Laborat\'orio de Computa\c c\~ao Cient\'ifica Avan\c cada e Modelamento (Lab-CCAM), Instituto Tecnol\'ogico de Aeron\'autica, DCTA, 12228-900, S\~ao Jos\'e dos Campos, SP, Brazil \label{addr3}
\and
Theoretical Division, Los Alamos National Laboratory, Los Alamos, New Mexico 87545, USA\label{addr4}
\and
Department of Physics, University of Guelph, Guelph, Ontario N1G 2W1, Canada\label{addr5}
\and
Institut für Physik und Astronomie, Universität Potsdam, Haus 28, Karl-Liebknecht-Str. 24/25, 14476, Potsdam, Germany\label{addr6}
\and
IRAP, CNRS, 9 avenue du Colonel Roche, BP 44346, F-31028 Toulouse Cedex 4, France\label{addr7}
\and
Université de Toulouse, CNES, UPS-OMP, F-31028 Toulouse, France.
\label{addr8}
\and
Istituto Nazionale di Fisica Nucleare, Sezione di Catania, Dipartimento di Fisica e Astronomia ``Ettore Majorana'', Universit\`a di Catania, Via Santa Sofia 64, I-95123 Catania, Italy\label{addr10}
}

\maketitle


\date{Draft version: \today, Received: date / Accepted: date}

\abstract{Systematic comparisons across theoretical predictions for the properties of dense matter, nuclear physics data, and astrophysical observations (also called meta-analyses) are performed. Existing predictions for symmetric nuclear and neutron matter properties are considered, and they are shown in this paper as an illustration of the present knowledge. Asymmetric matter is constructed assuming the isospin asymmetry quadratic approximation. It is employed to predict the pressure at twice saturation energy-density based only on nuclear-physics constraints, and we find it compatible with the one from the gravitational-wave community. To make our meta-analysis transparent, updated in the future, and to publicly share our results, the Python toolkit \texttt{nucleardatapy} is described and released here. Hence, this paper accompanies \texttt{nucleardatapy}, which simplifies access to nuclear-physics data, including theoretical calculations, experimental measurements, and astrophysical observations. This Python toolkit is designed to easily provide data for: i) predictions for uniform matter (from microscopic or phenomenological approaches); ii) correlation among nuclear properties induced by experimental and theoretical constraints; iii) measurements for finite nuclei (nuclear chart, charge radii, neutron skins or nuclear incompressibilities, etc.) and hypernuclei (single particle energies); and iv) astrophysical observations. This toolkit provides data in a unified format for easy comparison and provides new meta-analysis tools. It will be continuously developed, and we expect contributions from the community in our endeavor.}



\noindent

{\bf Program Summary and Specifications}\\


\begin{small}

\noindent

{Program title:} \texttt{nucleardatapy}, version 1.0\\

{Licensing provisions:} CC BY-NC-ND 4.0\\

{Programming language:} Python.\\

{Repository:} \href{https://github.com/jeromemargueron/nucleardatapy}{nucleardatapy} on GitHub (public), see also Ref~\cite{nuda:rep}. \\


{Documentation:} \href{https://nucleardatapy.readthedocs.io/en/latest}{direct link}, see also Ref.~\cite{NudaDocumentation2025}.\\

{Tutorials:} \href{https://jeromemargueron.github.io/nucleardatapy/landing.html}{direct link}, see also Ref.~\cite{NudaTutorials2025}.\\

{Description of problem:} Nuclear physics and observational data for dense nuclear matter are dispersed across various sources. They may be stored in different formats, requiring the reconstruction of quantities, which might differ from one author to another. Meta-analyses are limited by access to a large number of predictions.\\


{Method of solution:} The \texttt{nucleardatapy} toolkit simplifies access to nuclear-physics data by collecting them in a single repository and providing the community with a simple Python toolkit, containing reconstructed quantities in a unified and transparent way. It also makes it easy to manipulate data further as well as to perform meta-analyses.\\


{Additional comments:} The Python toolkit employed to create the figures presented in this paper is open-source, see Sec~\ref{sec:toolkit} for details on the installation and use of the toolkit, and more information can also be available from the GitHub repository, see Ref~\cite{nuda:rep}. Documentation and tutorials are provided as well. Possible issues as well as further extensions can be suggested from the GitHub Issue option, or alternatively, could be discussed directly with the authors of this paper.\\


\end{small}




\section{Introduction}
\label{sec:intro}

Dense nuclear matter is a system instrumental for the understanding of compact stars, such as neutron stars, and high-energy astrophysical phenomena they are involved in, such as binary neutron star mergers.
However, it is difficult to directly measure its properties or to isolate its components from a nuclear experiment~\cite{Book:BohrMottelson:1969,Book:Ring:Schuck:1980} or an astrophysical observation~\cite{Book:Haensel:Potekhin:Yakovlev:2007,Book:Rezzolla:2018}. In finite nuclei, for instance, dense matter alone does not define their properties, but finite-size contributions, which also reflect the properties of the nuclear interaction, such as its finite range and its momentum-dependence, need to be considered. In neutron stars, it contributes together with leptons to the equation of state (EoS). 

Phase transitions in the crust, where forming non-uniform clusters is energetically favorable, or in the core, where new degrees of freedom are expected above a given density, have also to be considered~\cite{Book:Glendenning:2000}. It is, therefore, impossible to isolate dense nuclear matter from the other effects in the system under study. Instead, in a first step it is more appropriate to compare the global modeling of the system of interest (nuclei, hypernuclei, or neutron stars) with existing data. In the second step, theory is then employed to disentangle the contribution of dense nuclear matter from others.

For such a program, it has become common to perform Bayesian analyses, coupled to Markov-Chain Monte-Carlo sampling of the model parameter space, to link nuclear matter properties with experimental and/or astrophysical data and estimate uncertainties~\cite{JDobaczewski:2014,LNeufcourt:2018}. For instance, collections of scientific libraries are publicly available and can be employed for model parameter exploration~\cite{O2SCL,BAND,DPhillips:2021}. In such analyses, the constraints from fundamental approaches as well as from nuclear experiments are crucial to assess the quality of the models. These constraints are, therefore, commonly employed, but they are not systematically provided in a form that is easy to manipulate. In addition, experimental results and theoretical calculations are often available in different formats and references, making comparisons difficult. Since data plays a crucial role in assessing the quality of the models, it is important to furnish a source of data, checked by authors and by the community, which is easily accessible and improved by user feedback. The \texttt{nucleardatapy} toolkit is aimed to be such a community-driven tool. Hence, feedback from users is encouraged and will be fully considered. The motivation for  \texttt{nucleardatapy} is to facilitate and simplify the sharing of data in the nuclear and nuclear-astrophysics communities.

The \texttt{nucleardatapy} toolkit collects data that can be useful to calibrate low-energy nuclear models, such as energy-density functionals (EDFs). These EDF\footnote{A list of all abbreviations is provided in~\ref{app:acronyms}.} models are expected to reproduce at least nuclear binding energies and charge radii, which are present in the \texttt{nucleardatapy} toolkit, but one could also be interested in comparing model predictions, for example, for the isoscalar giant monopole resonance (ISGMR), the correlations between $E_\sym$ and $L_\sym$, as well as between $K_\sat$ and $Q_\sat$, and many others (empirical parameters such as $E_\sym$, $L_\sym$, $K_\sat$, and $Q_\sat$ are defined in Sec.~\ref{sec:mat:nep}). 
Additionally, microscopic ab initio and phenomenological calculations of uniform matter are provided by \texttt{nucleardatapy}. Finally, astrophysical data related to neutron star properties are provided in the toolkit. They are complementary to the new repository for astrophysical observations CompARE~\cite{CompARE:2023}.

The \texttt{nucleardatapy} toolkit also simplifies meta-analyses, by collecting and providing results from different approaches in a single format. The present study suggests several of these meta-analyses, and one that is particularly interesting is the prediction for the pressure at twice saturation energy-density and its comparison with the results from the gravitational-wave community. Furthermore, the \texttt{nucleardatapy} toolkit constructs the EoS for the ground state of uniform nuclear matter at $\beta$-equilibrium for the models predicting nuclear properties in symmetric and neutron matter, and assuming a simple approximation to describe asymmetric matter, see Sec.~\ref{sec:eos} for more details. A larger sample of EoSs, including finite temperature and non-uniform matter in the crust, is provided by the CompOSE repository~\cite{CompOSE:2022}, or the stellar collapse website~\cite{StellarCollapse}.

The present paper is divided into the following sections: we begin with the installation of the toolkit in Sec.~\ref{sec:toolkit}. In Sec.~\ref{sec:unif} we first review the constraints from microscopic and phenomenological approaches for symmetric and neutron matter (in some cases, asymmetric matter is also provided), e.g., the energy per particle, pairing gaps, empirical parameters, Landau parameters as well as the experimental constraints from heavy-ion collisions. We then show correlation diagrams for $K_\sat$-$Q_\sat$ and $E_\sym$-$L_\sym$ in Sec.~\ref{sec:corr}. We review the constraints from finite nuclei in Sec.~\ref{sec:nuc}, e.g., nuclear masses, binding energies, two-neutron separation energies, odd-even mass staggering, charge radii, and neutron skins. The constraints from hypernuclei are reviewed in Sec.~\ref{sec:hypernuclei} for single and double $\Lambda$ and single $\Xi^-$ hypernuclei. We address the neutron star crust and the related \texttt{crust} module in Sec.~\ref{sec:crust}, asymmetric matter at beta-equilibrium in Sec.~\ref{sec:eos}, and astrophysical data in Sec.~\ref{sec:astro}. Our conclusions are presented in Sec.~\ref{sec:conclusions}.

\section{Installation notes for the toolkit.}
\label{sec:toolkit}

The toolkit is written in Python, and once installed, the library is designed to be imported from everywhere, including Google colab.
To install the toolkit, launch the following command from a terminal on your computer (or on Google Colab):
\begin{lstlisting}[language=bash]
$ pip install nucleardatapy
\end{lstlisting}
The toolkit is available for direct download on the pypi website. Once installed, one can access this package from any Python script. If you are using a mobile device, details on installing and using the \texttt{nucleardatapy} toolkit are given in~\ref{app:ipad}. 
In all cases, one must import the toolkit in the usual way in Python:
\begin{lstlisting}[language=Python]
import nucleardatapy as nuda
\end{lstlisting}
We also import the \texttt{numpy} library as:
\begin{lstlisting}[language=Python]
import numpy as np
\end{lstlisting}
From now on, we shall call the \texttt{nucleardatapy} toolkit as \texttt{nuda}, and \texttt{numpy} as \texttt{np} in short.

To check that the toolkit is installed correctly and is accessible, enter into python and write at the prompt:
\begin{lstlisting}[language=Python]
import nucleardatapy as nuda
nuda.hello()
\end{lstlisting}
You should get \texttt{hello world!} in return for your hug!

The list of all functions and global variables available in the toolkit can be printed using:
\begin{lstlisting}[language=Python]
print(dir(nuda))
\end{lstlisting}
You can also have a detailed view of the routines with:
\begin{lstlisting}[language=Python]
print(help(nuda))
\end{lstlisting}
or
\begin{lstlisting}[language=Python]
print(nuda.__dict__)
\end{lstlisting}
as well as
\begin{lstlisting}[language=Python]
print(nuda.__dict__.keys())
\end{lstlisting}
In the following, we detail the functions available for use in the \texttt{nuda} toolkit and discuss the results. 

The \texttt{nuda} toolkit is divided into several modules: \texttt{astro}, \texttt{corr}, \texttt{crust}, \texttt{eos}, \texttt{fig}, \texttt{hnuc}, \texttt{matter}, and \texttt{nuc}. In addition, there are a few common files where global definitions are fixed: \texttt{cst.py} for fixing the constants, \texttt{env.py} for fixing environment flags (such as \texttt{verb}), \texttt{param.py} for provided parameters, for instance, the absolute paths to data files.

As an example, one can instantiate the variable \texttt{mass} with the neutron star mass measured by radio-astronomy in the following way:
\begin{lstlisting}[language=Python]
mass = nuda.astro.setupMasses()
\end{lstlisting}
where the default pulsar is PSR J1614–2230. All the properties available for the variable \texttt{mass} are accessible via:
\begin{lstlisting}[language=Python]
print(mass.__dict__)
\end{lstlisting}
More details about the \texttt{nuda.astro} class \texttt{nuda.astro.setupMasses()} are given in Sec.~\ref{sec:astro:masses}.

The public GitHub repository~\cite{nuda:rep} contains installation instructions as well as links to the documentation~\cite{NudaDocumentation2025} and tutorials~\cite{NudaTutorials2025}. Detailed tutorials explaining the usage of the \texttt{nuda} toolkit are given in the documentation. Tutorials are distributed in several chapters, each dedicated to providing examples of usage for a given module. For instance, a tutorial providing massive neutron star masses can be found in chapter 8: Astrophysical Data. In addition, the figures shown in this paper can be reproduced using the tutorial and are available from the toolkit repository in the folder \texttt{nucleardatapy\_sample}. The present paper provides details on how to employ the toolkit, for which it provides an introduction, while further details are given in the documentation~\cite{NudaDocumentation2025} and tutorials~\cite{NudaTutorials2025}.

We would also like to remind the users of the \texttt{nuda} toolkit to systematically provide citations to the original reference of the employed data. In the toolkit, all data are provided with their original reference, so when using these data in a scientific paper, references to data should be provided explicitly. To facilitate the quoting of the original references, we also provide a set of \texttt{.bib} files in the folder \texttt{biblio} on the GitHub repository~\cite{nuda:rep}. These files are the ones employed for this paper. They are sorted chronologically. One can also cite this toolkit referring to the present paper, in addition to citing the original publication and/or data.

\section{Theoretical predictions for the ground state of uniform nuclear matter: the \texttt{matter} module.}
\label{sec:unif}

Consider a system composed of neutrons and protons, whose rest mass contribution to the energy is additive:
\begin{equation}
E_{r.m.} \equiv N_n m_n c^2 + N_p m_p c^2 \, ,
\end{equation}
where $N_n$ ($N_p$) is the number of neutrons (protons) and $m_n$ ($m_p$) is the neutron (proton) mass. Introduce the nucleon mass $m_N$ defined as $m_N=(m_n+m_p)/2$. The rest mass contribution to the energy per particle ($e_{r.m.}$) reads,
\begin{equation}
e_{r.m.}\equiv \frac {E_{r.m.}} {N_\nuc} = x_n m_n c^2 + x_p m_p c^2 \, ,
\end{equation}
where the baryon number is $N_\nuc=N_n+N_p$, the neutron (proton) fraction is $x_n=N_n/N_\nuc$ ($x_p=N_p/N_\nuc$).

In uniform matter, neutron and proton densities are defined as $n_n=N_n/V$ and $n_p=N_p/V$, where $V$ is a mesoscopic volume containing $N_\nuc$ nucleons. The total nucleon density is $n_\nuc=n_n+n_p$ and the isospin parameter $\delta$ is defined as $\delta=(n_n-n_p)/n_\nuc$. We have,
\begin{equation}   
\delta = 
     \begin{cases}
       0 &\quad\text{symmetric matter (SM)}, \\
       1 &\quad\text{neutron matter (NM)}.\\ 
     \end{cases}
\end{equation}
The total energy $E_\tot$ is composed of the rest mass energy $E_{\rma}$ and the internal energy $E_\intn$ as,
\begin{equation}
E_\tot \equiv  E_{\rma} + E_\intn \, .
\label{eq:etot}
\end{equation}
Eq.~\eqref {eq:etot} is fixed by the requirement that non-relativistic approaches match relativistic ones at low momentum.

In the following, we consider predictions for the nucleon ground state, i.e., at zero temperature.

\subsection{Non-relativistic nucleonic free Fermi gas}
\label{sec:unif:ffgnr}

The nucleonic free Fermi gas (FFG) is a uniform nuclear quantum system composed of fermions in their ground state with no interactions. In this section, we consider only non-relativistic (NR) FFG systems, for which the NR single-particle internal energy $e_q^{\nr\ffg}(k)$ is ($q=n$, $p$), 
\begin{equation}
e_q^{\nr\ffg}(k) = \frac{\hbar^2 k^2}{2m_q} \, .
\end{equation}
The Fermi energy is defined as the internal energy of the last occupied state for $k=k_{F_q}$,
\begin{equation}
e_{F_q}^{\nr\ffg} \equiv \frac{\hbar^2 k_{F_q}^2}{2m_q} \equiv \mu_q -m_q c^2\, ,
\end{equation}
where $\mu_q$ is the ground state chemical potential. The internal energy per nucleon is obtained by adding the neutron and proton contributions as,
\begin{equation}
e_{\intn}^{\nr\ffg}(k_{Fn},k_{Fp}) \equiv \frac{3}{5}\left(\frac{\hbar^2 k_{Fn}^2}{2m_n} + \frac{\hbar^2 k_{Fp}^2}{2m_p} \right) \, ,
\label{eq:matter:ffg:enp}
\end{equation}
where the Fermi momentum is $k_{F_q}=(3\pi^2 n_q)^{1/3}$, assuming spin-saturated systems (equal number of spin up and down). The nucleonic Fermi momentum is defined as $k_{F_\nuc}=(3\pi^2 n_\nuc/2)^{1/3}$, and it corresponds to the momentum of the last occupied state in SM. Note that in SM, we have $k_{F_n}=k_{F_p}=k_{F_\nuc}$.
Introducing the density $n_\nuc$ and the isospin asymmetry $\delta$ in Eq.~\eqref{eq:matter:ffg:enp}, we obtain:
\begin{equation}
e_{\intn}^{\nr\ffg} = e_{\intn,\sat}^{\ffg} \left(\frac{n_\nuc}{n_\sat}\right)^{2/3}\frac{(1+\delta)^{5/3} + (1-\delta)^{5/3}}{2} \, ,
\label{eq:matter:ffg:e}
\end{equation}
where the energy per particle at saturation density in SM is
\begin{equation}
e_{\intn,\sat}^{\ffg} \equiv \frac{3}{5}\frac{\hbar^2}{2m_N} \left(\frac{3\pi^2}{2}n_\sat\right)^{2/3} \approx 22\hbox{ MeV}\, ,
\end{equation}
for $n_\sat\approx 0.16$~fm$^{-3}$, $m_N c^2\approx 939$~MeV and $\hbar c\approx 197$~MeV~fm.
Note that Eq.~\eqref{eq:matter:ffg:e} is the kinetic energy where the neutron and proton masses are fixed to be identical ($m_n\approx m_p\approx m_N$). This approximation is accurate and simplifies equations, but it is not necessary.
The internal energy density reads,
\begin{equation}
\epsilon_{\intn}^{\nr\ffg}(n_\nuc,\delta) = n_\nuc \, e_{\intn}^{\nr\ffg}(n_\nuc,\delta) \, ,
\end{equation}
the total energy per nucleon is
\begin{equation}
e(n_\nuc,\delta) = e_{r.m.}(n_\nuc,\delta)+e_{\intn}(n_\nuc,\delta) \, ,
\end{equation}
and the energy-density $\epsilon=e\, n_\nuc$ is defined as
\begin{equation}
\epsilon(n_\nuc,\delta) = \epsilon_{r.m.}(n_\nuc,\delta)+\epsilon_{\intn}(n_\nuc,\delta) \, .
\end{equation}
Note that the energy density $\epsilon$ and the mass density $\rho$ are related in the following way:
\begin{equation}
\epsilon(n_\nuc,\delta) = \rho(n_\nuc,\delta) c^2 \, ,
\end{equation}
where $c$ is the speed of light.

The symmetry energy is defined as the difference between NM and SM,
\begin{equation}
e_\sym(n_\nuc) \equiv e_{\intn}(n_\nuc,\delta=1) - e_{\intn}(n_\nuc,\delta=0) \, . \label{eq:ffg:esym}
\end{equation}
This definition applies to the NR FFG,
\begin{equation}
e_\sym^{\nr\ffg}(n_\nuc) = E_{\intn,\sat}^{\ffg} \left(\frac{n_\nuc}{n_\sat}\right)^{2/3}\left( 2^{2/3}-1\right)\, .
\end{equation}
The symmetry energy can also be expressed as a series expansion, as
\[e_\sym(n_\nuc)=e_{\sym,2}(n_\nuc)+e_{\sym,4}(n_\nuc)+\dots,\] where the quadratic and quartic contributions are defined as
\begin{eqnarray}
e_{\sym,2}(n_\nuc) &=& \frac 1 2 \frac{\partial^2 e(n_\nuc,\delta)}{\partial \delta^2} \, , \, \\
e_{\sym,4}(n_\nuc) &=& \frac 1 {24} \frac{\partial^4 e(n_\nuc,\delta)}{\partial \delta^4} \, .
\end{eqnarray}

For the NR FFG, we have 
\begin{eqnarray}
e_{\sym,2}^{\nr\ffg}(n_\nuc) &=& \frac{10}{18} E_{\intn,\sat}^{\ffg} \left(\frac{n_\nuc}{n_\sat}\right)^{2/3} \, , \, \\
e_{\sym,4}^{\nr\ffg}(n_\nuc) &=& \frac 5 {243} E_{\intn,\sat}^{\ffg} \left(\frac{n_\nuc}{n_\sat}\right)^{2/3} \, .
\end{eqnarray}

The nucleon pressure is defined as
\begin{equation}
p^{\nr\ffg}(n_\nuc,\delta)\equiv n_\nuc^2\, \frac{\partial e^{\nr\ffg}}{\partial n_\nuc} = \frac 2 3 \epsilon_{\intn}^{\nr\ffg} \, ,
\label{eq:pre}
\end{equation}
which means that the pressure does not depend on the rest mass energy of the particles (in this case).
It can also be calculated using
\begin{equation}
p^{\nr\ffg}(n_\nuc,\delta)=\frac{n_\nuc\, k_{F_\nuc}}{3} \frac{\partial e^{\nr\ffg}}{\partial k_{F_\nuc}} \, .
\end{equation}

The enthalpy per nucleon is defined as 
\begin{equation}
h(n_\nuc,\delta) \equiv \frac{d \rho(n_\nuc,\delta) c^2}{d n_\nuc} = e(n_\nuc,\delta) + \frac{p(n_\nuc,\delta)}{n_\nuc} \, ,
\label{eq:chempot}
\end{equation}
which is dominated, as lowest density, by the rest mass energy.

The derivative of the pressure is
\begin{equation}
\frac{\partial p^{\nr\ffg}}{\partial n_\nuc} = \frac{10}{9} e^{\nr\ffg}_\intn(n_\nuc,\delta) \, .
\end{equation}

The sound speed $c_s$ is defined as:
\begin{equation}
\left(c_{s}(n_\nuc,\delta)/c\right)^2 \equiv \frac{\partial p}{\partial \epsilon}
= \frac{1}{h(n_\nuc,\delta)} \frac{\partial p}{\partial n_\nuc} \, ,
\end{equation}
for fixed isospin asymmetry $\delta$. Note that at low density, when the rest mass energy dominates the total energy, we have
\begin{equation}
(c_s/c)^2 \xrightarrow[n \to 0]{} \frac{10}{9} \frac{e_\intn}{e_\rma} \propto n_\nuc^{2/3} \, .
\end{equation}

\begin{table}[tb]
\begin{center}
\caption{Functions from \texttt{nucleardatapy} toolkit associated with free Fermi gas quantities. The physics property is given in the first row, the corresponding method is in the second row, and finally the third row provides the associated attribute of the object \texttt{ffg} instantiated via the class \texttt{matter.setupFFGNuc}. The input variables can be scalar or \texttt{numpy} float array. We have: \texttt{den\_nuc} for $n_\nuc$ and \texttt{delta} for $\delta$. Note that both non-relativistic ($\nr$) expressions and relativistic ones are provided. }
\label{table:FFG}
\tabcolsep=0.12cm
\def\arraystretch{1.5}
\begin{tabular}{lll}
\hline\noalign{\smallskip}
& methods & instantiation of \texttt{ffg} \\
& & \texttt{ffg=nuda.matter.setupFFGNuc}\\
\hline\noalign{\smallskip}
$k_{F_\nuc}$ & kf\_nuc (den) & ffg.kf\_nuc \\ 
$k_{F_n}$ & kf\_n (den\_n) & ffg.kf\_n \\ 
$k_{F_p}$ & - & ffg.kf\_p \\ 
$n_\nuc$ & den(kf) & ffg.den\\
$n_n$ & den\_n(kf\_n) & ffg.den\_n\\
$n_p$ & - & ffg.den\_p\\
$e_{F_n}$ & eF\_n(kf\_n) & ffg.eF\_n\\
$e_{F_n}^\nr$ & eF\_n\_nr(kf\_n) & ffg.eF\_n\_nr\\
$e_{F_p}$ & - & ffg.eF\_p\\
$e_{F_p}^\nr$ & - & ffg.eF\_p\_nr\\
$e_\intn^{\nr\ffg}$ & effg\_nr(kf\_n) & ffg.e2a\_int\_nr\\
$\epsilon^{FFG}$ & - & ffg.eps\\
$\epsilon_\intn^{FFG}$ & - & ffg.eps\_int\\
$e_\sym^{\nr\ffg}$ & - & ffg.esym\\
$e_{\sym,2}^{\nr\ffg}$ & - & ffg.esym2\\
$e_{\sym,4}^{\nr\ffg}$ & - & ffg.esym4\\
$p^{\nr\ffg}$ & - & ffg.pre\_nr\\
$p^{\ffg}$ & - & ffg.pre\\
$(c_s/c)^2$ & - & ffg.cs2\\
\noalign{\smallskip}\hline
\end{tabular}
\end{center}
\end{table}

These quantities are given in \texttt{nuda} toolkit, and the correspondence between these quantities and the \texttt{nuda} toolkit functions is given in table~\ref{table:FFG}. For instance, to obtain the FFG energies for a set of \texttt{kf\_n}, one shall write the following lines in Python:
\begin{lstlisting}[language=Python]
kf_n = np.linspace(0.01,0.2,20)
print(`FFG energy:',nuda.matter.effg_nr(kf_n))
\end{lstlisting}
Alternatively, one can use the \texttt{matter.setupFFGNuc} class in the following way (for SM):
\begin{lstlisting}[language=Python]
den = np.linspace(0.01,0.3,20)
delta = np.zeros( den.size )
ffg = nuda.matter.setupFFGNuc( den, delta)
ffg.print_outputs()
\end{lstlisting}
This option, in fact, provides more than just the energy. Most common attributes are the following ones: \texttt{ffg.den\_nuc} for $n_\nuc$, \texttt{ffg.e2a\_nuc} for $e^{\ffg}$, \texttt{ffg.e2a\_nuc\_int} for $e^{\ffg}_\intn$, \texttt{ffg.eps\_nuc} for $\epsilon^{\ffg}$, \texttt{ffg.pre\_nuc} for $p^\ffg$, \texttt{ffg.cs2\_nuc} for $(c_s^\ffg/c)^2$... For more attributes, see table~\ref{table:FFG}, while the full list of all attributes associated with the object \texttt{ffg} can be obtained via:
\begin{lstlisting}[language=Python]
print(ffg.__dir__)
\end{lstlisting}

\begin{figure*}[t]
\centering
\includegraphics[scale=0.9]{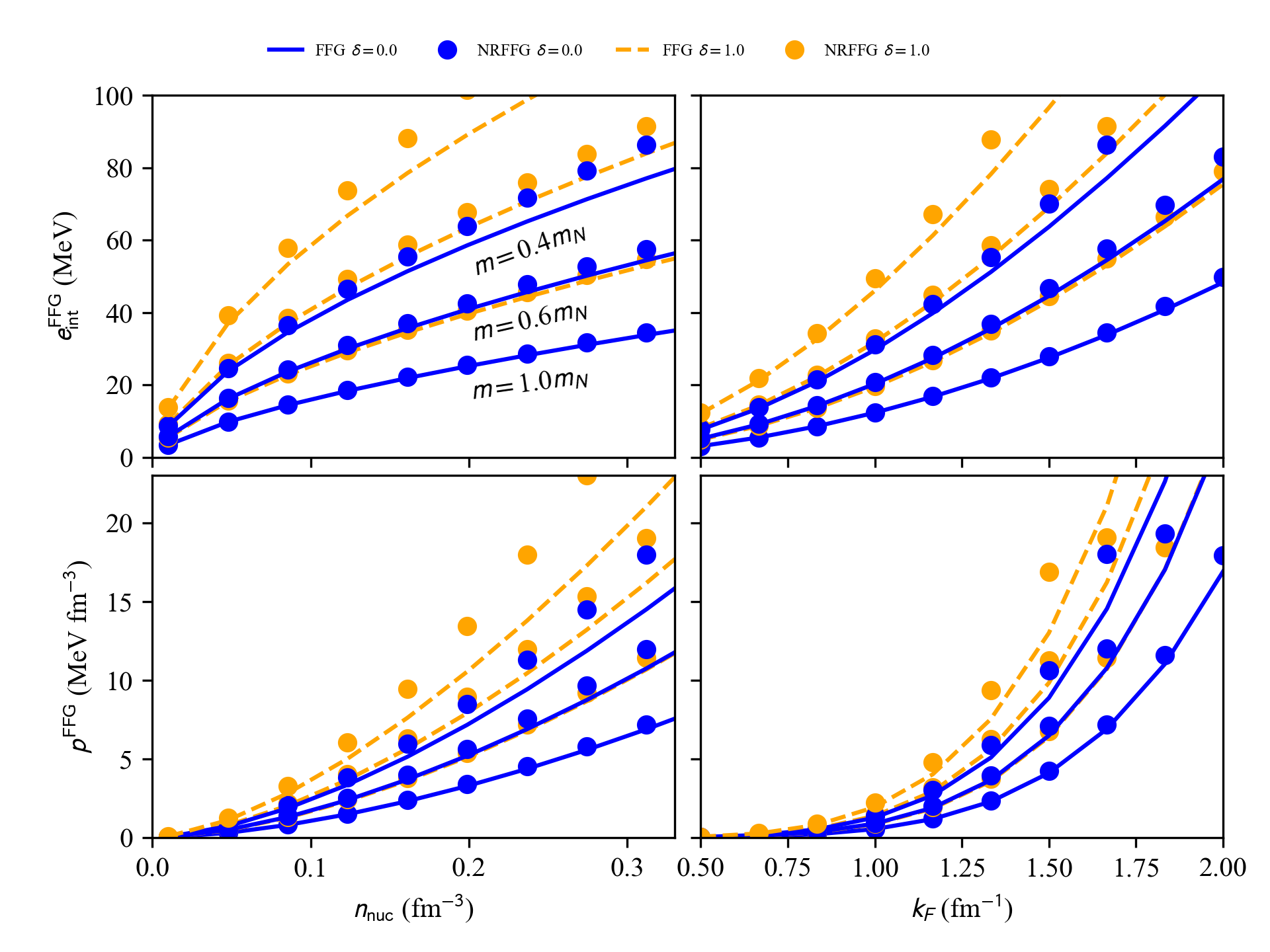}
\caption{FFG energy in NM ($\delta=1$) and in SM ($\delta=0$) (top) and FFG pressure in NM and SM (bottom) as a function of the density $n_\nuc$ (left) and the Fermi momentum $k_{F}$ (right). Lines (Symbols) show the relativistic (non-relativistic) FFG results. The nucleon mass is fixed to three constant values: $m_N$, $0.4m_N$, and $0.6m_N$. This figure is generated with \texttt{matter\_setupFFGNuc\_plot.py}.}
\label{fig:ffg}
\end{figure*}

\subsection{Relativistic nucleonic free Fermi gas}
\label{sec:unif:ffgrl}

The relativistic expression for the free Fermi gas energy density is
\begin{equation}
\epsilon^{\rl\ffg}(n) = C\Bigg[ 2 p_F E_F^3 - m^2 p_F E_F -m^4\log \frac{p_F+E_F}{m} \Bigg] \, ,
\label{eq:matter:erffg}
\end{equation}
where $n$ is the particle density, $E_F=\sqrt{m^2+p_F^2}$, $p_F=\hbar c \, k_F$ the Fermi impulsion, $C=g/(16\pi^2)$, $g$ the degeneracy ($g=4$ in SM).
The non-relativistic limit is recovered for low momenta ($p_F$, $k_F \rightarrow 0$), or equivalently low density, as
\begin{equation}
\epsilon^{\rl\ffg}(n) \xrightarrow[n \to 0]{} \left( m c^2+ \frac 3 {10} \frac{p_F^2}{m c^2} \right) \, n \, ,
\end{equation}
and the ultra-relativistic limit ($p_F\gg mc^2$) gives:
\begin{equation}
e^{\ur\ffg} = p_F = \hbar c \left( \frac{6 \pi^2}{g} n\right)^{1/3}\,.
\end{equation}

The pressure is
\begin{equation}
p^{\rl\ffg}(n) = C^\prime\Bigg[ 2p_F E_F^3 - 5m^2 p_F E_F +3m^4\log \frac{p_F+E_F}{m} \Bigg] \, ,
\end{equation}
where $C^\prime=C/3$, and for the ultra-relativistic limit:
\begin{equation}
p^{\ur\ffg}=\frac{1}{3} \epsilon^{\ur\ffg} \, .
\end{equation}

The enthalpy per particle is
\begin{equation}
h^{\rl\ffg}(n) = \frac{8C}{3 n}\Bigg[ p_F E_F^3 - m^2 p_F E_F \Bigg] \, .
\end{equation}
The derivative of the pressure is
\begin{equation}
\frac{\partial p^{\rl\ffg}}{\partial n} = \frac 1 3 \frac{p_F^2}{E_F} \, .
\end{equation}

The sound speed $c_s$ is:
\begin{equation}
\left(c_{s}(n)/c\right)^2 = \frac{1}{h(n)} \frac{\partial p}{\partial n} \, ,
\end{equation}
and for the ultra-relativistic limit,
\begin{equation}
(c_s^{\ur\ffg}/c)^2 = 1/3 \, ,
\end{equation}
which is also called the conformal limit for the sound speed.

\begin{figure}[t]
\centering
\includegraphics[scale=0.52]{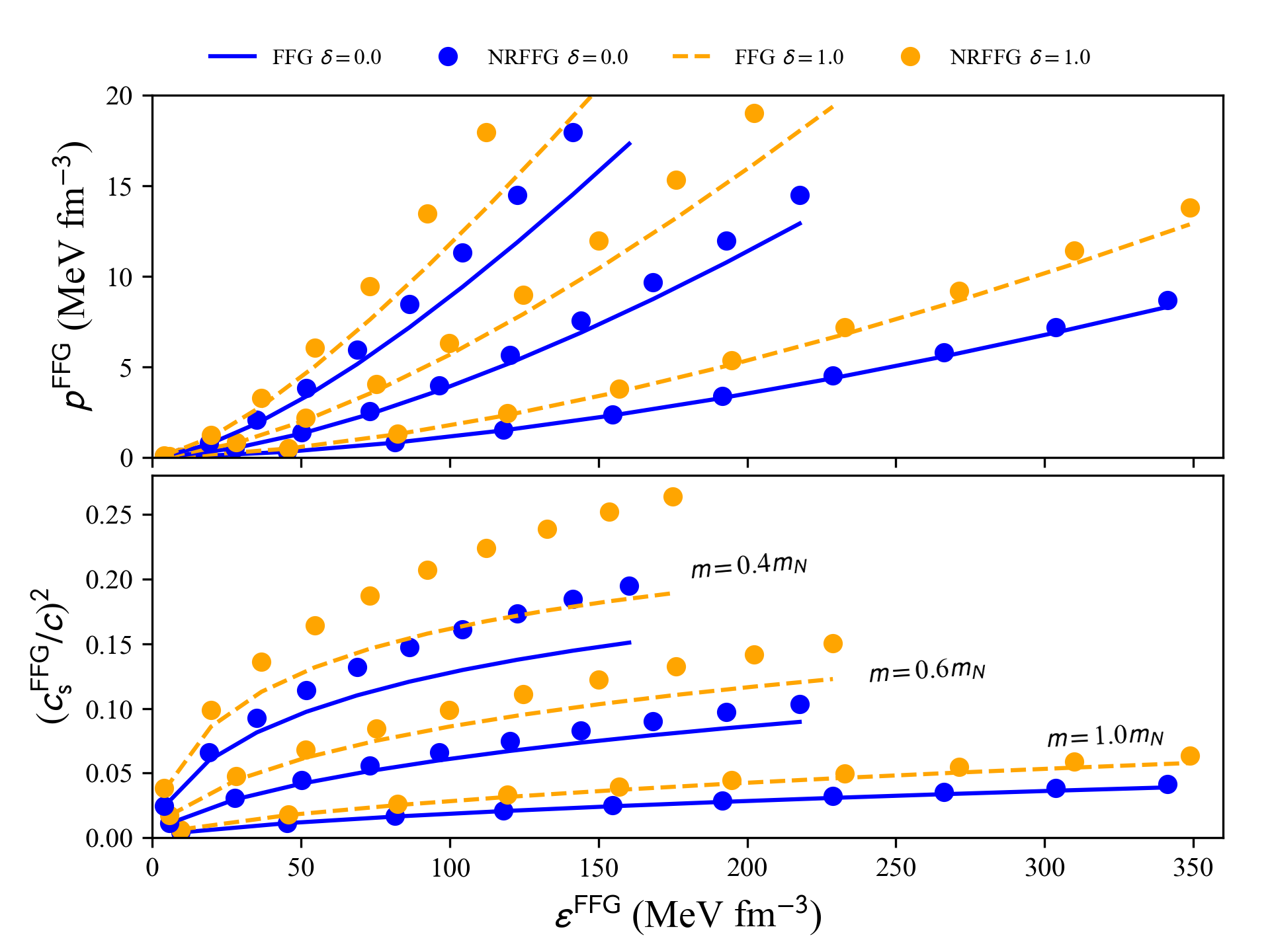}
\caption{FFG EoS: pressure $p$ (top) and sound speed $(c_s/c)^2$ (bottom) for SM (solid blue lines) and NM (dashed yellow lines) function of the energy density $\epsilon$. Circles (lines) show the non-relativistic (relativistic) FFG results. This figure is generated with \texttt{matter\_setupFFGNuc\_plot.py}.}
\label{fig:ffg:EOS}
\end{figure}

For nuclear matter, we have
\begin{equation}
e_\nuc=\frac{E_\nuc}{A}=x_n \frac{E_n}{N} + x_p \frac{E_p}{Z}=x_n \frac{\epsilon_n(n_n)}{n_n} + x_p \frac{\epsilon_p(n_p)}{n_p} \, ,
\end{equation}
where $x_i=n_i/n_\nuc$ ($i=n$, $p$), $n_\nuc=n_n+n_p$ and the energy densities $\epsilon_q$ are obtained from Eq.~\eqref{eq:matter:erffg}.

The object \texttt{ffg} can be instantiated to the class \texttt{matter.setupFFGNuc} in the following way (for SM in this example):
\begin{lstlisting}[language=Python]
den = np.linspace(0.01,0.3,20)
delta = np.zeros( den.size )
ffg = nuda.matter.setupFFGNuc( den, delta)
ffg.print_outputs()
\end{lstlisting}

The FFG energy and pressure in SM (solid blue) and NM (dashed yellow) are shown in Fig.~\ref{fig:ffg} as a function of the density $n_\nuc$ (left) and Fermi momentum $k_{F}$ (right). A comparison of the non-relativistic results (circles) with the relativistic ones (lines) is shown. We note that in the case of nucleons, the relativistic corrections to the FFG are small for densities below $2n_\sat$ and for $m=m_N$.
The impact of the relativistic corrections increases as the nucleon mass $m$ decreases. We show in Fig.~\ref{fig:ffg} situations where the mass $m$ is reduced to $0.6m_N$ and $0.4m_N$, which correspond to the current expectation for the effective mass at $\approx n_\sat$ and $\approx 2n_\sat$. Note, however, that the effect of the density dependence of the effective mass (rearrangement contribution) is not considered here. As the mass $m$ is reduced, the difference between the non-relativistic and the relativistic quantities is increased, and these differences are larger in NM compared to SM.

The EoS is shown in Fig.~\ref{fig:ffg:EOS}, with the pressure $p$ function of the energy density $\rho c^2$ (top panel) and the sound speed squared $(c_s/c)^2$ function of the energy density $\epsilon$. Similarly to Fig.~\ref{fig:ffg}, we show relativistic and non-relativistic quantities, in SM and NM, and for different choices for the nucleon mass. We observe that the sound speed increases as the mass $m$ decreases: it reaches 0.07 (0.15) for $0.6m_N$ ($0.4m_N$) at saturation density.

\subsection{Relativistic free Fermi gas for leptons}
\label{sec:unif:ffgrl:lep}

\begin{figure}[t]
\centering
\includegraphics[scale=0.52]{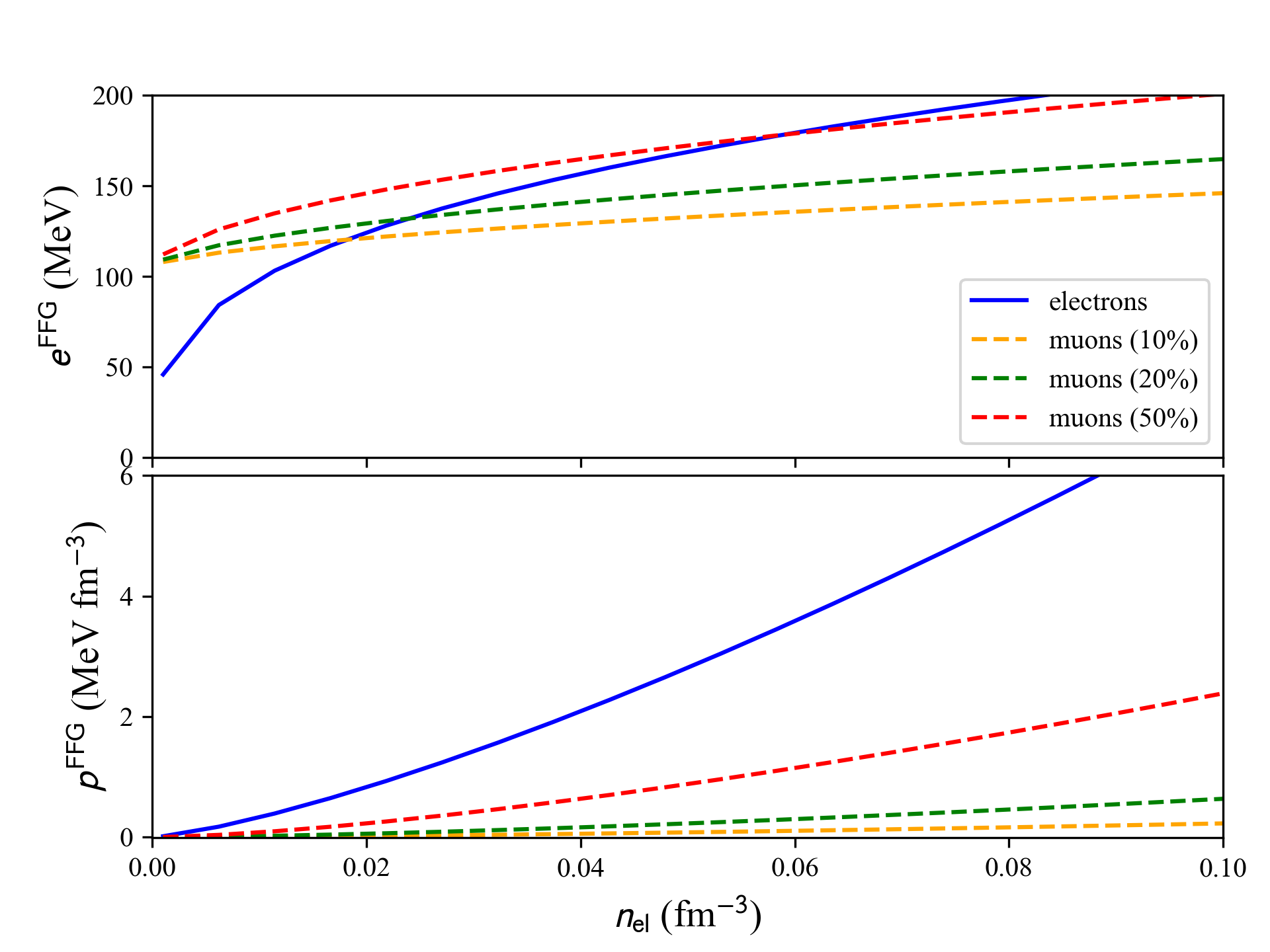}
\caption{Leptonic FFG: energy per particle for electrons (solid) and muons (dashed) (top) and pressure (bottom). We have considered three scenarios for muons: 10\% of the electron density (yellow), 20\% (green) and 50\% (red). This figure is generated with \texttt{matter\_setupFFGLep\_plot.py}.}
\label{fig:ffg:lep}
\end{figure}

Electrons and muons are contributing to the leptonic energy in neutron stars. The mass of the $\tau$ is too high to contribute in neutron stars. We have
\begin{equation}
\epsilon_\lep = \frac{E_\el+E_\mu}{V} = \epsilon_\el + \epsilon_\mu \, ,
\end{equation}
and 
\begin{equation}
e_\lep=\frac{E_\lep}{N_\lep}=\tilde{x}_\el \frac{E_\el}{N_\el} + \tilde{x}_\mu \frac{E_\mu}{N_\mu}=\tilde{x}_\el \frac{\epsilon_\el}{n_\el} + \tilde{x}_\mu \frac{\epsilon_\mu}{n_\mu} \, ,
\end{equation}
where $N_\lep=N_\el+N_\mu$, $\tilde{x}_i=N_i/N_\lep=n_i/n_\lep$ ($i=\el$, $\mu$).
Electrons and muons are present in the core of cold neutron stars (in fact, muons are present for densities $n\gtrapprox n_\sat$). In the following example, the terms $\epsilon_\el$ and $\epsilon_\mu$ are properties of the object \texttt{lep} instantiated from the class \texttt{nuda.matter.setupFFGLep} in the following way:
\begin{lstlisting}[language=Python]
den_el = np.linspace(0.01,0.3,num=20)
den_mu = 0.2 * den_el
lep = nuda.matter.setupFFGLep( den_el, den_mu)
lep.print_outputs()
\end{lstlisting}
where \texttt{den\_el} and \texttt{den\_mu} are the number densities (in fm$^{-3}$) associated with the electrons and the muons. The electron (muon) energy density $\epsilon_\el$ ($\epsilon_\mu$) is given as an attribute of the object \texttt{lep} as \texttt{lep.eps\_el} (\texttt{lep.eps\_mu}).

The relativistic leptonic FFG is shown in Fig.~\ref{fig:ffg:lep}. Solid lines stand for the electron contribution while dashed lines for the muon one for $n_\mu=0.1 n_\text{el}$ (yellow), $n_\mu=0.2 n_\text{el}$ (green), $n_\mu=0.5 n_\text{el}$ (red). At low density, the rest mass contribution to the energy dominate, while the rest mass contribution decreases as the density increases. This effect is more pronounced for the electrons and less for the muons. For the pressure, there is no rest mass contribution and the electron pressure is always larger than the muon pressure.

\subsection{Microscopic models}
\label{sec:unif:micro}

In this section, we list the microscopic models available in the \texttt{nuda} toolkit. These microscopic predictions are grouped into several many-body approaches. The following instruction provides the list of many-body groups:
\begin{lstlisting}[language=Python]
mbs, mbs_lower = nuda.matter.micro_mbs()
print(`mbs available:',mbs)
\end{lstlisting}
It provides the following list: [ `VAR', `AFDMC', `BHF2', `BHF23', `QMC', `MBPT', `NLEFT' ], which can be completed if necessary. The following instruction provides a list of models inside a given group defined by the variable \texttt{mb} in the class \texttt{nuda.matter.micro\_models\_mb()}:
\begin{lstlisting}[language=Python]
mbody = `VAR'
models, models_lower=nuda.matter.micro_models_mb( mb=mbody )
print(`for mb:',mbody,`models:',models)
\end{lstlisting}

Instead of a single many-body group, a list of groups can be provided to obtain the associated list of model predictions, as in this example:
\begin{lstlisting}[language=Python]
mbodies = [ `VAR', `QMC' ]
models, models_lower = nuda.matter.micro_models_mbs( mbs = mbodies )
print(`for mbs:',mbodies,`models:',models)
\end{lstlisting}

The complete list of available microscopic models is given with the following instructions:
\begin{lstlisting}[language=Python]
models, models_lower = nuda.matter.micro_models( )
print(`All micro models:',models)
\end{lstlisting}

The class \texttt{matter.setupMicro(model)} provides the following attributes: \texttt{nm\_e2a}, \texttt{sm\_e2a} (energy per particle in SM and NM), 
\texttt{nm\_e2a\_int}, \texttt{sm\_e2a\_int} (internal energy per particle in SM and NM), \texttt{nm\_eps}, \texttt{sm\_eps} (energy density in SM and NM) and the pressure \texttt{nm\_pre} and \texttt{sm\_pre}, related to the cubic-spline derivative of the energy per particle. The chemical potentials are obtained, similarly, from a cubic-spline of the energy per nucleon
and are encoded in the properties: \texttt{sm\_chempot}, \texttt{nm\_chempot}.
If provided by the authors, the pairing gap \texttt{nm\_gap} and \texttt{sm\_gap} are also given by the toolkit, otherwise, these properties are defined as \texttt{None}.

The call for the results of a microscopic calculation for a given \texttt{model}, here the FP EoS, can be done in the following way:
\begin{lstlisting}[language=Python]
micro = nuda.matter.setupMicro( model = `1981-VAR-AM-FP' )
micro.print_outputs()
\end{lstlisting}

In the following, the variable \texttt{model} will be varied on all the EoS available in the \texttt{nuda} toolkit. We adopt the following convention \texttt{model}=`YYYY-X-Y-Z', where:
\begin{itemize}
\item `YYYY' stands for the year associated with the publication of the results; 
\item `X' for the many-body technique, e.g., VAR, AFDMC, QMC, etc.;
\item `Y' is `NM' for neutron matter only, and `AM' for SM and NM; 
\item `Z' is optional: it is the name of the EoS when it exists, e.g., FP, APR, or of the nuclear interaction, e.g., Av18.
\end{itemize}

For this example, output quantities are attributes of the object \texttt{micro}, for instance: \texttt{micro.nm\_e2a} (\texttt{micro.nm\_e2a\_int}) and \texttt{micro.sm\_e2a}  (\texttt{micro.sm\_e2a\_int}) arrays for the nucleon (internal) energy per particle in NM and SM with obvious notations\footnote{New microscopic models have recently been added to the toolkit: many-body perturbation theory 'MBPT', self-consistent Green's functions 'SCGF', and coupled-cluster 'CC' from Ref.~\cite{Marino:2024,Marino:2025, Marino:2025:Zenodo} in 2024 and self-consistent Green's functions 'SCGF' From Ref.~\cite{Rios:2020} in 2020.}.

\subsubsection{Variational approaches}

\begin{figure*}[t]
\centering
\includegraphics[scale=0.9]{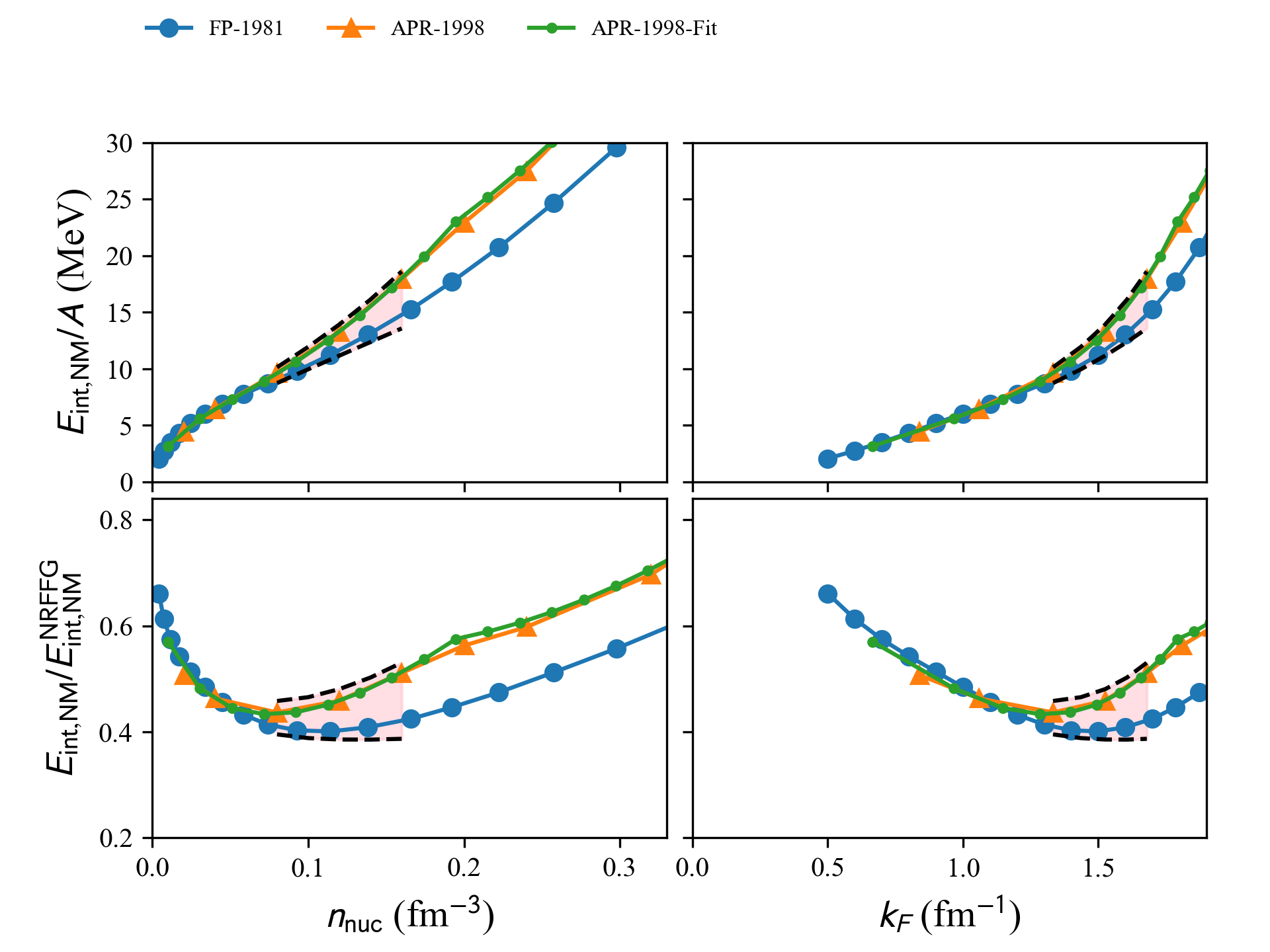}
\caption{Internal energy per nucleon in neutron matter (NM) $E_{\NM}^\intn$ (top) and $E_{\NM}^\intn$ over the non-relativistic free Fermi gas energy (bottom) as a function of the density (left) and the Fermi momentum (right) for the variational models (FP and APR) available in \texttt{nuda} toolkit. The reference band detailed in Sec.~\ref{sec:unif:band} is also shown (pink area). We have applied here and in the next figures a selection of the models: all those passing through the reference band are shown in solid lines, the ones not passing through it in dashed lines. Since all models shown in this figure pass through the reference band, they are all shown with solid lines. Figure generated with \texttt{matter\_setupMicro\_plot.py}.}
\label{fig:micro:var:e2a}
\end{figure*}

In \texttt{nuda} toolkit, two variational EoS are provided: FP~\cite{BFriedman:1981} and APR~\cite{APR:1998}. They are described below. Note that in the case of APR EoS, the authors of the original paper provide a fit in addition to their results. In \texttt{nuda} toolkit, we therefore provide their fit, based on continuous functions, which allows us to perform accurate first and second-order derivations.

A typical call for a given \texttt{model}, see hereafter, is:
\begin{lstlisting}[language=Python]
micro = nuda.matter.setupMicro( model=`1998-VAR-AM-APR' )
micro.print_outputs()
\end{lstlisting}
where the object \texttt{micro} contains a large set of properties, some of them listed by the command \texttt{micro.print\_outputs()}. The entire list of properties are obtained in the following way: 
\begin{lstlisting}[language=Python]
print(micro.__dict__)
\end{lstlisting}

We now list the options available for the variable \texttt{model}. Note that the code will stop if the variable \texttt{model} is misspelled.\\

\noindent
\texttt{model}=`1981-VAR-AM-FP':\\
The Friedman-Pandharipande (FP) EoS is obtained from a variational hypernetted-chain approach based on $v_{14}$ two-nucleon interaction and complemented by TNI three-nucleon interaction. The data provided by \texttt{nuda} are extracted from Ref.~\cite{BFriedman:1981}, more specifically from the second column of Table 1 for SM and Table 2 for NM.\\

\noindent
\texttt{model}=`1998-VAR-AM-APR':\\
The Akmal-Pandharipande-Ravenhall (APR) EoS is obtained from a variational chain summation calculation for SM and NM based on Argonne $v_{18}$ two-nucleon interaction and complemented with three-nucleon interaction and boost corrections (leading order relativistic correction). The data provided by \texttt{nuda} are extracted from Ref.~\cite{APR:1998}, more specifically from the last column (corrected) of Table VI for SM and the last column of Table VII for NM.\\

\begin{figure*}[t]
\centering
\includegraphics[scale=0.9]{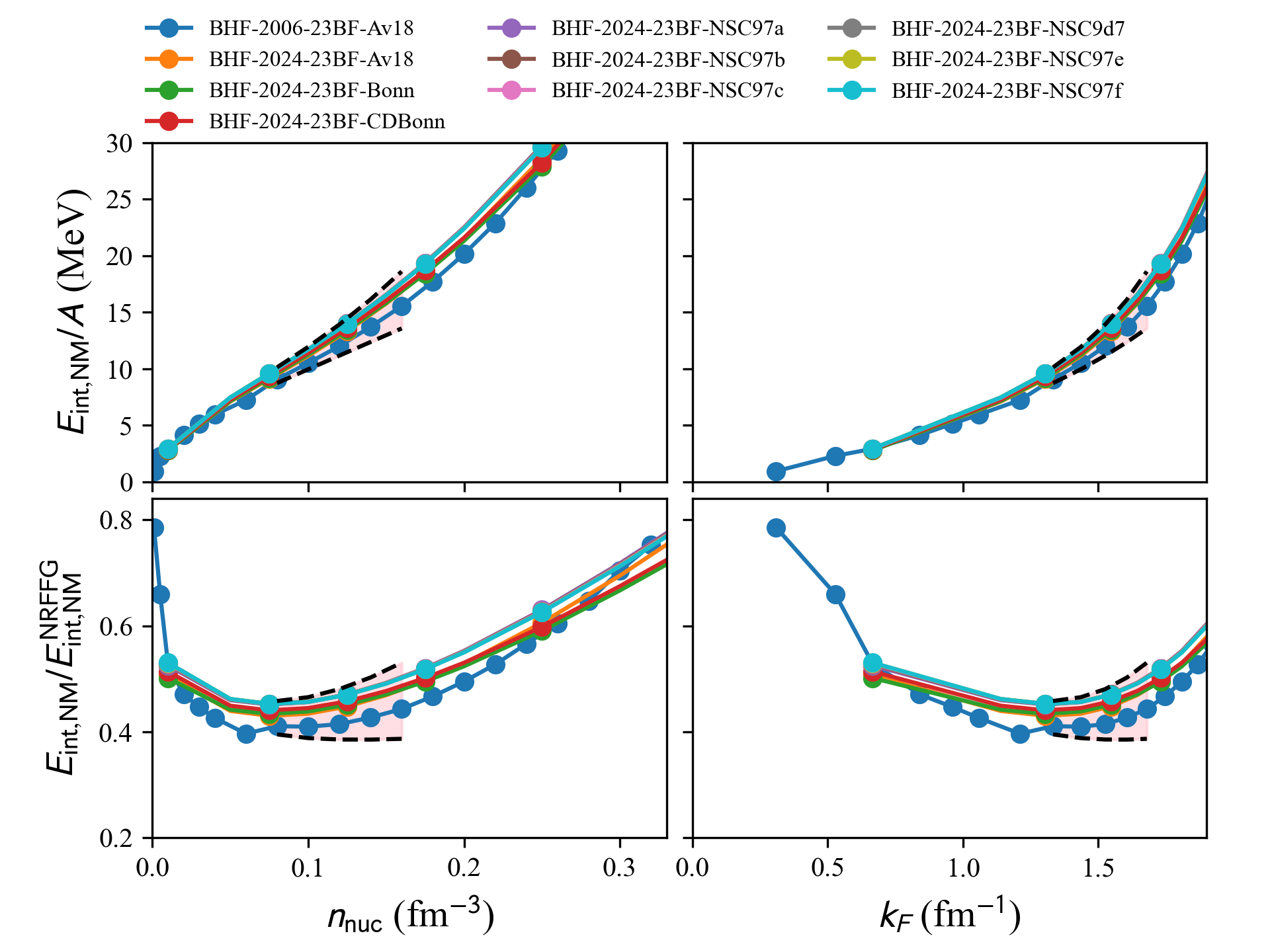}
\caption{Same as Fig.~\ref{fig:micro:var:e2a} for BHF23 models with 2+3BF available in \texttt{nuda} toolkit. Figure generated with \texttt{matter\_setupMicro\_plot.py}.}
\label{fig:micro:bhf:e2a}
\end{figure*}

\noindent
\texttt{model}=`1998-VAR-AM-APR-fit':\\
The suggested fit of the APR EoS in Ref.~\cite{APR:1998} is provided for a set of densities and isospin asymmetries given in the arrays \texttt{var1} (densities) and \texttt{var2} (isospin asymmetries). The call of the class \texttt{nuda.matter.setupMicro() is slightly modified as:}
\begin{lstlisting}[language=Python]
den = np.linspace(0.01,0.4,20)
asy = 0.0
micro = nuda.matter.setupMicro( model=`1998-VAR-AM-APR-fit', var1=den, var2=asy )
micro.print_outputs()
\end{lstlisting}
Note that \texttt{var1} can be an array of density, while \texttt{var2} is a scalar fixing the isospin asymmetry.

We show in Fig.~\ref{fig:micro:var:e2a} the variational models FP and APR. The symbols represent the data provided in the tables from Refs.~\cite{BFriedman:1981,APR:1998} while the solid line is the fit provided in Ref.~\cite{APR:1998}. The top panels show the total NM energy and the bottom panel shows the energy divided by the NRFFG, so $E_{\NM}/E_{\nr\ffg}$ highlights the contribution of the potential energy given by the models.
The band represents the reference uncertainties presented in Sec.~\ref{sec:unif:band}. Interestingly, the FP and APR EoS are located inside the reference band, which is solely defined by different many-body approaches based on similar $\chi$EFT interactions~\cite{ITews:2018,CDrischler:2021}.

\subsubsection{Brueckner-Hartree-Fock}

There are several calculations based on the Brueckner-Hartree-Fock (BHF) approach and employing several two- and three-body interactions. In \texttt{nuda} toolkit, we provide a large number of calculations originating from a few papers. For instance, BHF calculations where a large set of nuclear interactions involving two- and three-nucleons have been performed in Ref.~\cite{IVidana:2024} for two-nucleon interactions only, for two- and three-nucleon phenomenological interactions, and for two- and three-nucleon microscopic interactions. The toolkit provides the results for all these calculations and also for other calculations. 

A typical call for a given \texttt{model}, see hereafter, is:
\begin{lstlisting}[language=Python]
micro = nuda.matter.setupMicro( model=`2006-BHF-AM' )
micro.print_outputs()
\end{lstlisting}
Each calculation can be obtained by fixing the variable \texttt{model} to one of the following values:\\

\noindent
\texttt{model}=`2006-BHF-AM':\\
Calculation based on an extended version of the Brueckner-Hartree-Fock (BHF) approach and presented in Ref.~\cite{LGCao:2006}.\\

\noindent
Only two-nucleon interaction:\\
\texttt{model}=`2024-BHF-AM-2BF-Av18'.\\
\texttt{model}=`2024-BHF-AM-2BF-BONN'.\\
\texttt{model}=`2024-BHF-AM-2BF-CDBONN'.\\
\texttt{model}=`2024-BHF-AM-2BF-NSC97a'.\\
\texttt{model}=`2024-BHF-AM-2BF-NSC97b'.\\
\texttt{model}=`2024-BHF-AM-2BF-NSC97c'.\\
\texttt{model}=`2024-BHF-AM-2BF-NSC97d'.\\
\texttt{model}=`2024-BHF-AM-2BF-NSC97e'.\\
\texttt{model}=`2024-BHF-AM-2BF-NSC97f'.\\

\noindent
Two- and three-nucleon phenomenological interaction:\\
\texttt{model}=`2024-BHF-AM-23BF-Av18'.\\
\texttt{model}=`2024-BHF-AM-23BF-BONN'.\\
\texttt{model}=`2024-BHF-AM-23BF-CDBONN'.\\
\texttt{model}=`2024-BHF-AM-23BF-NSC97a'.\\
\texttt{model}=`2024-BHF-AM-23BF-NSC97b'.\\
\texttt{model}=`2024-BHF-AM-23BF-NSC97c'.\\
\texttt{model}=`2024-BHF-AM-23BF-NSC97d'.\\
\texttt{model}=`2024-BHF-AM-23BF-NSC97e'.\\
\texttt{model}=`2024-BHF-AM-23BF-NSC97f'.\\

\noindent
Two- and three-nucleon microscopic interaction:\\
\texttt{model}=`2024-BHF-AM-23BFmicro-Av18'.\\
\texttt{model}=`2024-BHF-AM-23BFmicro-BONNB'.\\
\texttt{model}=`2024-BHF-AM-23BFmicro-NSC93'.\\

We show in Fig.~\ref{fig:micro:bhf:e2a} a set of BHF models with 2+3BF interactions. Calculations with only 2BF are provided in the toolkit, as detailed hereinbefore, but they are not interesting to show in Fig.~\ref{fig:micro:bhf:e2a} because they do not saturate. The band represents the reference uncertainties presented in Sec.~\ref{sec:unif:band}. Again, it is interesting to note that BHF23 calculations are compatible with the reference band.

\subsubsection{Auxiliary Field Diffusion Monte-Carlo}

\begin{figure*}[t]
\centering
\includegraphics[scale=0.9]{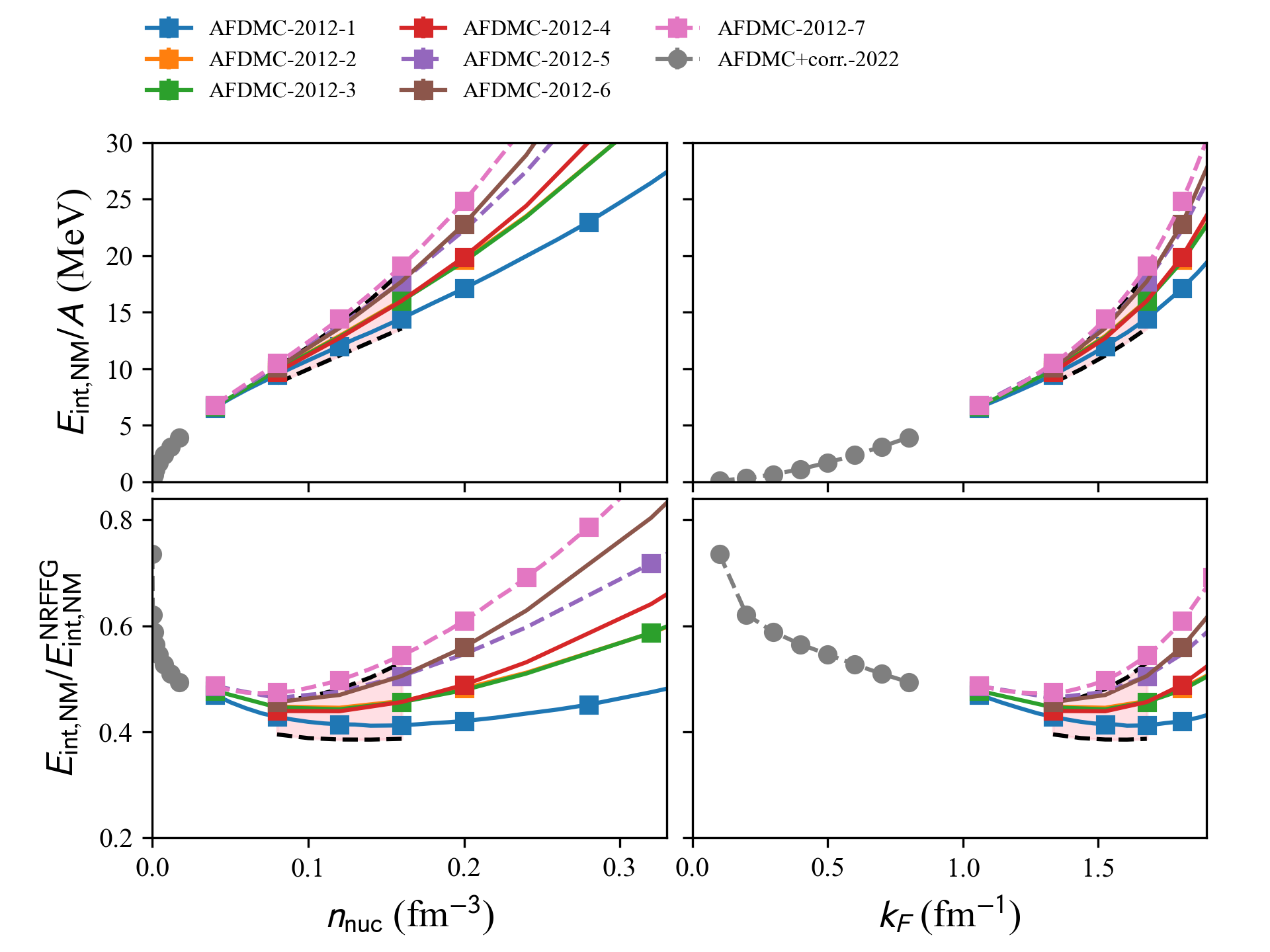}
\caption{Same as Fig.~\ref{fig:micro:var:e2a} for AFDMC models with 3BF available in the \texttt{nucleardatapy} toolkit. Note that the models in dashed lines are the ones that do not pass through the reference band. Figure generated with \texttt{matter\_setupMicro\_plot.py}.}
\label{fig:micro:afdmc:e2a}
\end{figure*}

A set of Auxiliary Field Diffusion Monte Carlo (AFDMC) is provided in the \texttt{nuda} toolkit. All these EoS are calculated in NM, even the fit provided in `2012-AFDMC-NM-FIT-n'.

A typical call for a given \texttt{model}, see hereafter, is:
\begin{lstlisting}[language=Python]
micro = nuda.matter.setupMicro( model=`2012-AFDMC-NM-RES-5' )
micro.print_outputs()
\end{lstlisting}
Each calculation can be obtained by fixing the variable \texttt{model} to one of the following values:\\




\noindent
\texttt{model}=`2012-AFDMC-NM-RES-n' (with n=1,7):\\
AFDMC calculations with AV8$^\prime$ two-nucleon interaction and seven models for the three-nucleon interaction have been performed in Ref.~\cite{SGandolfi:2012}. The index $n$ runs over the different three-nucleon interactions considered: none ($n=1$), $V_{2\pi}^{PW}+V_{\mu=150}^R$ ($n=2$), $V_{2\pi}^{PW}+V_{\mu=300}^R$ ($n=3$), $V_{3\pi}+V_R$ ($n=4$),
$V_{2\pi}^{PW}+V_{\mu=150}^R$ ($n=5$), $V_{3\pi}+V_R$ ($n=6$), UIX ($n=7$).\\

\noindent
\texttt{model}=`2012-AFDMC-NM-FIT-n' (with n=1,7):\\
Fit of AFDMC calculations with AV8$^\prime$ two-nucleon interaction and seven models for the three-nucleon interaction have been performed in Ref.~\cite{SGandolfi:2012}. \\

\noindent
\texttt{model}=`2022-AFDMC-NM':\\
The AFDMC approach is employed to calculate the binding energy and the $^1$S$_0$ pairing gap in NM using realistic nuclear Hamiltonians that include two- and three-body interactions in Ref.~\cite{SGandolfi:2022}. The trial state is properly optimized to capture the essential pairing correlations and the results are extrapolated from the finite box to the thermodynamic limit using the symmetry-restored projected Bardeen-Cooper-Schrieffer (PBCS) theory. The pairing gap shows a modest suppression with respect to the mean-field BCS values, see Fig.~\ref{fig:micro:gap:1s0}, except around the peak of the pairing gap for $k_F$ around 0.8~fm$^{-1}$. \\

We show in Fig.~\ref{fig:micro:afdmc:e2a} a set of AFDMC models. The band represents the reference uncertainties presented in Sec.~\ref{sec:unif:band}. Most of AFDMC calculation are located inside the reference band, except `AFDMC-2012-7', which is more repulsive than the upper bound of the reference band and `AFDMC+corr.-2022', which is only provided for densities below the densities of the reference band.

\subsubsection{Quantum Monte Carlo}

\begin{figure*}[t]
\centering
\includegraphics[scale=0.9]{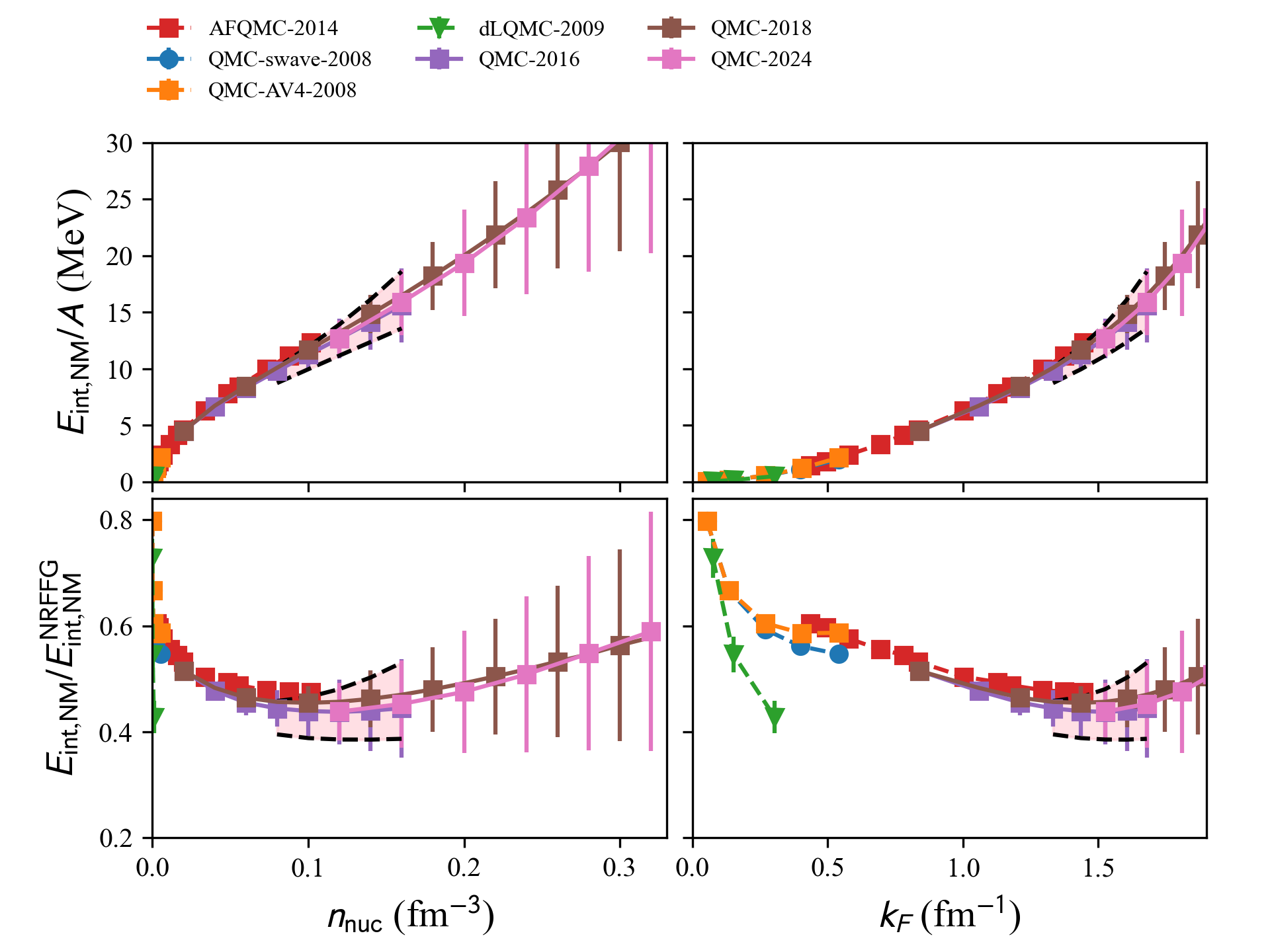}
\caption{Same as Fig.~\ref{fig:micro:var:e2a} for QMC models available in the \texttt{nucleardatapy} toolkit. Figure generated with \texttt{matter\_setupMicro\_plot.py}.}
\label{fig:micro:qmc:e2a}
\end{figure*}

Several Quantum Monte Carlo (QMC) calculations based on phenomenological 2BF as well as on $\chi$EFT nuclear interactions are available from the \texttt{nuda} toolkit. 

A typical call for a given \texttt{model}, see hereafter, is:
\begin{lstlisting}[language=Python]
micro = nuda.matter.setupMicro( model=`2016-QMC-NM' )
micro.print_outputs()
\end{lstlisting}
Each calculation can be obtained by fixing the variable \texttt{model} to one of the following values:\\

\noindent
\texttt{model}=`2008-QMC-NM-swave',\\
Quantum Monte-Carlo methods (variational and Green's function) are employed to study dilute neutron matter using $s$ wave terms of the AV18 interaction~\cite{RBWiringa:1995} in Ref.~\cite{AGezerlis:2008}. Calculations are performed for $k_{F_n}=0.05$~fm$^{-1}$ up to 0.55~fm$^{-1}$.\\

\noindent
\texttt{model}=`2009-DLQMC-NM':\\
This is the determinantal lattice QMC performed in Ref.~\cite{TAbe:2009}.\\

\noindent
\texttt{model}=`2010-QMC-NM-AV4':\\
Quantum Monte-Carlo methods (variational and Green's function) are employed to study dilute neutron matter using $s$ and $p$-wave terms, where the $s$-wave term is the same as in Ref.~\cite{AGezerlis:2008} and the $p$-wave term is determined by the AV4 interaction~\cite{RBWiringa:2002} in Ref.~\cite{AGezerlis:2010}. Calculations are performed for $k_{F_n}=0.05$~fm$^{-1}$ up to 0.55~fm$^{-1}$.\\

\noindent
\texttt{model}=`2014-AFQMC-NM':\\
NM variational Monte Carlo calculation using chiral nuclear forces at N3LO (N2LO) for the two-nucleon (three-nucleon) nuclear force in~\cite{GWlazlowski:2014}.\\

\noindent
\texttt{model}=`2016-QMC-NM':\\
QMC calculations of NM with chiral three-body forces at N2LO are reported in~\cite{ITews:2016}. 
Calculations are performed for densities  $n_{\rm nuc}=0.02$~fm$^{-3}$ up to 0.16~fm$^{-3}$.\\

\noindent
\texttt{model}=`2018-QMC-NM':\\
QMC calculations of NM with N2LO chiral Hamiltonians, including 3N interactions with only the two-pion exchange, and 3N interactions containing the two-pion exchange plus shorter-range contact terms with two different spin-isospin operators. Calculations~\cite{ITews:2018} are performed for densities $n_{\rm nuc}=0.04$~fm$^{-3}$ up to 0.32~fm$^{-3}$.\\

\noindent
\texttt{model}=`2024-QMC-NM':\\
Ref.~\cite{ITews:2024} reported QMC calculations of NM with large-cutoff (400 MeV - 700 MeV) derived within N2LO chiral EFT, adjusted to nucleon-nucleon scattering phase shifts, the triton binding energy, as well as the triton beta-decay half-life.
Calculations are performed for densities  $n_{\rm nuc}=0.08$~fm$^{-3}$ up to 0.2~fm$^{-3}$. \\

We show in Fig.~\ref{fig:micro:qmc:e2a} a set of QMC models. The band represents the reference uncertainties presented in Sec.~\ref{sec:unif:band}. All the centroid predictions from QMC calculations are located inside the reference band, except `AFQMC-2014', which is more repulsive than the upper boundary of the reference band, especially at low density.

\subsubsection{Many-Body Perturbation Theory}

\begin{figure*}[t]
\centering
\includegraphics[scale=0.9]{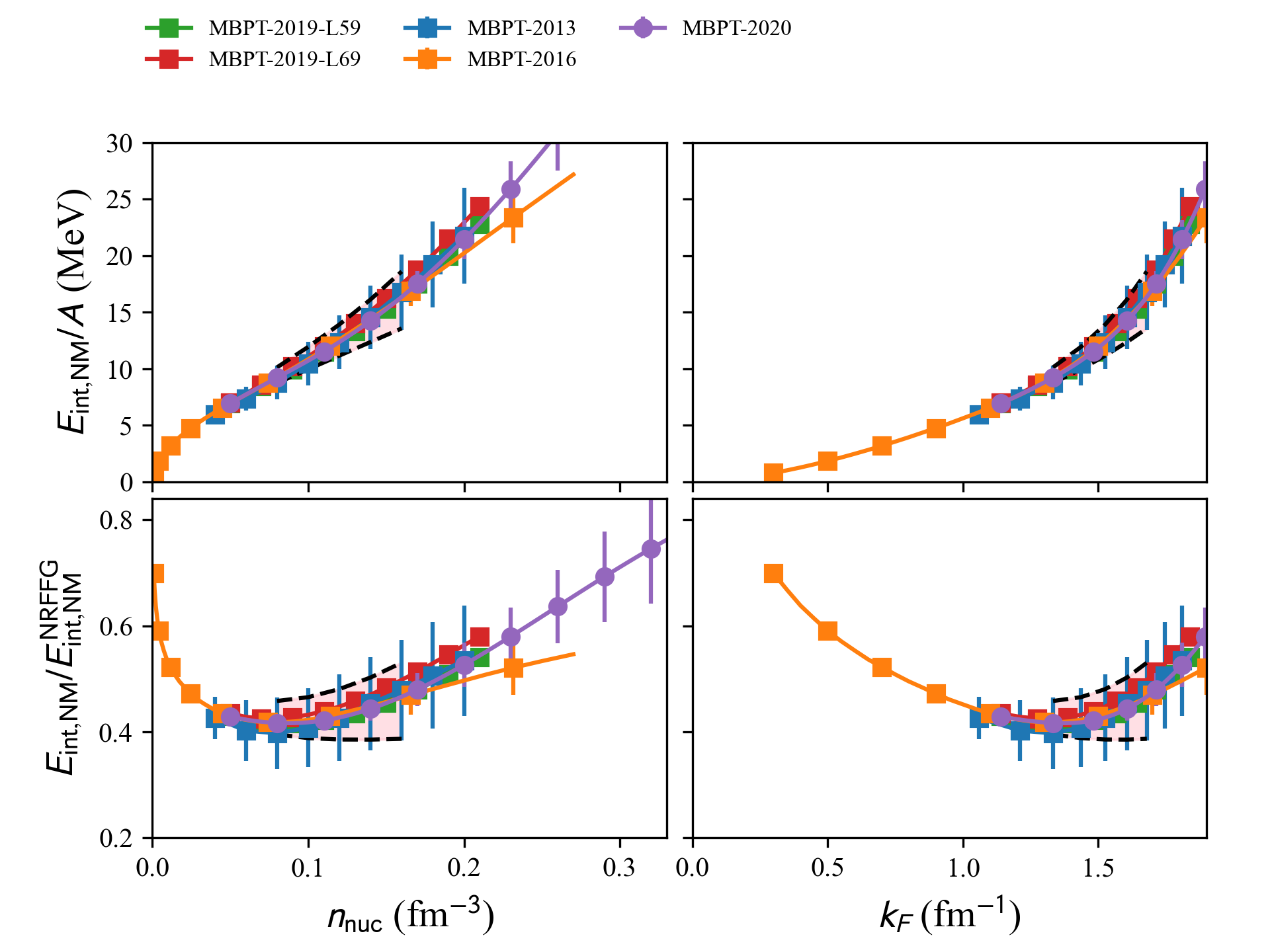}
\caption{Same as Fig.~\ref{fig:micro:var:e2a} for MBPT models available in the \texttt{nucleardatapy} toolkit. Figure generated with \texttt{matter\_setupMicro\_plot.py}.}
\label{fig:micro:mbpt:e2a}
\end{figure*}

Several Many-Body Perturbation Theory (MBPT) calculations with N2LO and N3LO $\chi$EFT interactions are available in the \texttt{nuda} toolkit.

A typical call for a given \texttt{model}, see hereafter, is:
\begin{lstlisting}[language=Python]
micro = nuda.matter.setupMicro( model=`2016-MBPT-AM' )
micro.print_outputs()
\end{lstlisting}
Each calculation can be obtained by fixing the variable \texttt{model} to one of the following values:\\


\noindent
\texttt{model}=`2013-MBPT-NM':\\
MBPT calculations based on the two-nucleon and three-nucleon interaction at N3LO (next-to-next-to-next-to-leading order) have been performed in NM in~\cite{ITews:2013}. The three-nucleon interactions at N3LO were considered only at the Hartree-Fock level.\\

\noindent
\texttt{model}=`2016-MBPT-AM':\\
Explicit calculations of isospin-asymmetric nuclear matter based on chiral two-nucleon interactions at N3LO and three-nucleon interactions at N2LO calculated in second-order MBPT are reported in Ref.~\cite{CDrischler:2016}.

\noindent
Results for isospin asymmetric matter are also provided in Ref.~\cite{CDrischler:2016}. They can be obtained from the \texttt{nuda} toolkit requesting the following properties:
\texttt{micro.am\_e2a\_int\_av[i]} and \texttt{micro.am\_e2a\_int\_err[i]} for the average and uncertainty band for the internal energy in asymmetric matter, where $\delta=i/10$ and $i=0$, $10$. The nucleon density in asymmetric matter can be obtained in the following way: \texttt{micro.am\_den[i]}. The neutron and proton fractions can be obtained as: \texttt{micro.am\_xn[i]} and \texttt{micro.am\_xp[i]}. Detail results from each of the hamiltonians H$_j$, see Ref.~\cite{CDrischler:2016} for more details, can be obtained from the following property:
\texttt{micro.am\_e2a\_int[i,j]}.\\

\noindent
\texttt{model}=`2019-MBPT-AM-DHSL59', `2019-MBPT-AM-DHSL69':\\
Reference~\cite{CDrischler:2019} presented a Monte Carlo framework for higher-order MBPT calculations. The results are presented for SM and NM based on chiral two-, three-, and four-nucleon interactions at N2LO and N3LO, respectively. DHSL59 calculations represent $L_{\sym} = 59$ MeV while DHSL69 has $L_{\sym} = 69$ MeV.
The calculations were performed for densities  $n_{\rm nuc}=0.05$~fm$^{-3}$ up to 0.21~fm$^{-3}$.\\

\noindent
\texttt{model}=`2020-MBPT-AM':\\
References~\cite{CDrischler:2020a,CDrischler:2020b} perform a statistical uncertainty quantification of NM based on the MBPT calculations in Ref.~\cite{CDrischler:2019} using chiral EFT interactions up to N3LO. The model applies Bayesian machine learning with Gaussian processes to quantify correlated truncation errors due to truncating the $\chi$EFT at a finite order (so-called EFT truncation errors). Predictions for the energy per particle, pressure, and speed of sound of neutron matter up to twice nuclear saturation density, along with constraints on the nuclear symmetry energy and its derivative, are provided. The authors (i.e., the BUQEYE collaboration) made their source codes publicly available~\cite{BUQEYEsoftware}.\\

We show in Fig.~\ref{fig:micro:mbpt:e2a} a set of MBPT calculations. The band represents the reference uncertainties presented in Sec.~\ref{sec:unif:band}.
Again, it is interesting that MBPT calculations are compatible with the reference band.

\subsubsection{Nuclear Lattice Effective Field Theory}

\begin{figure*}[t]
\centering
\includegraphics[scale=0.9]{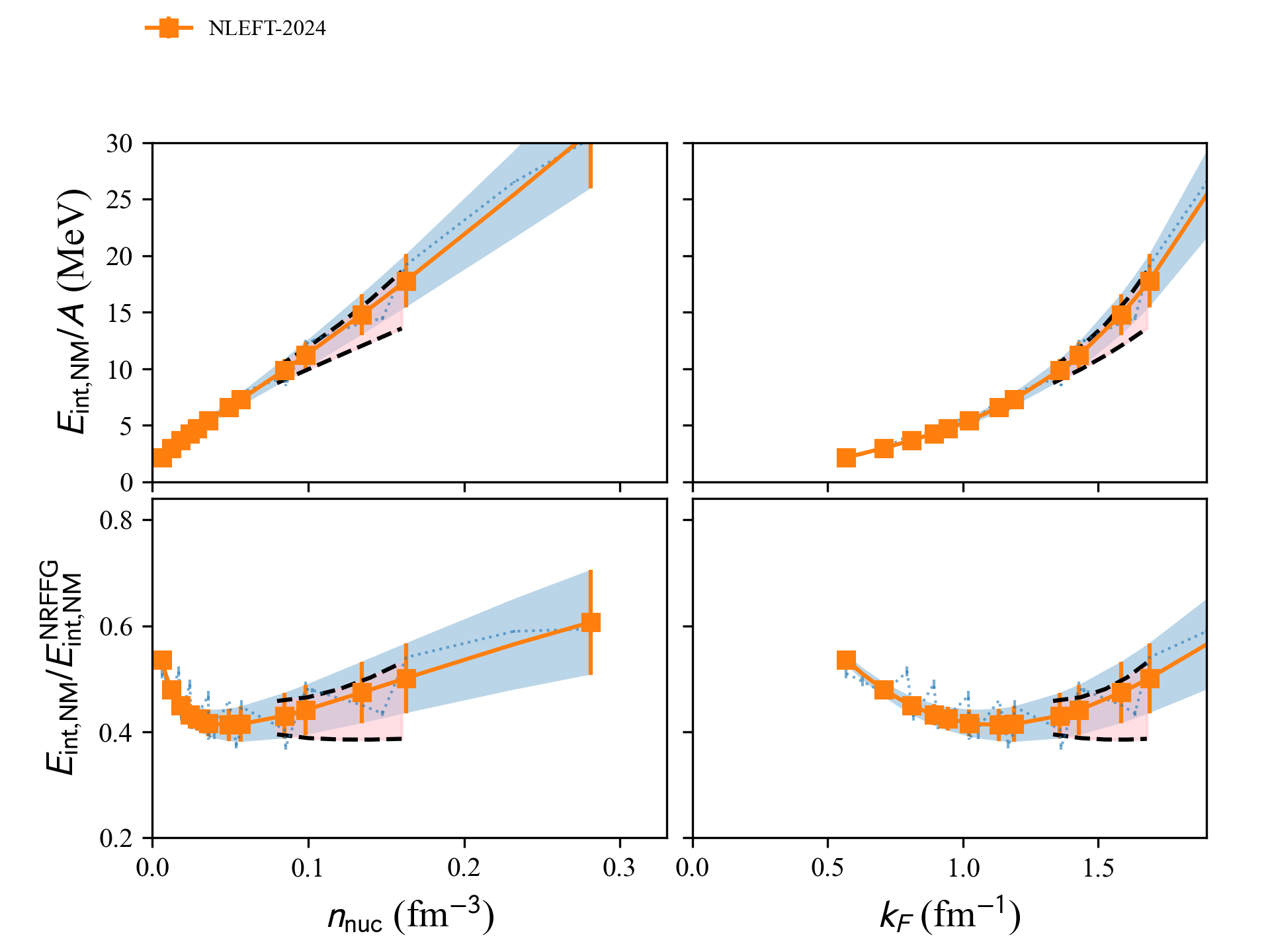}
\caption{Same as Fig.~\ref{fig:micro:var:e2a} for NLEFT models available in the \texttt{nucleardatapy} toolkit. Figure generated with \texttt{matter\_setupMicro\_plot.py}.}
\label{fig:micro:nleft:e2a}
\end{figure*}

The complex problem of systems where the interactions may be complicated is addressed using wavefunction matching, which transforms the interaction between particles into an easily computable interaction. A recent calculation has been done in Ref.~\cite{SElhatisari:2024} for NM and SM using Nuclear Lattice Effective Field Theory (NLEFT). These calculations are performed in different box sizes, and the finite-size effects have to be removed. To do so, we adjust a function $f(k_F)$ which depends only on $k_F$ and which represents the difference between the energy per particle $E/A$ and the non-relativistic Fermi gas prediction:
\begin{equation}
E_\text{NLEFT}(k_F) \approx f(k_F) \, E^{\nr\ffg}(k_F) ,
\label{eq:NLEFT}
\end{equation}
and we consider the following polynomial expansion for the function $f$,
\begin{equation}
f(k_F) = a + b k_F + c k_F^2 + d k_F^3\, ,
\label{eq:NLEFT:fit}
\end{equation}
where the parameters $a$, $b$, $c$, and $d$ are obtained by fitting the NLEFT prediction for the energy per particle in SM and NM from ref.~\cite{SElhatisari:2024}, see Table~\ref{table:NLEFT}.
We impose $a=1.0$ to get the low-density limit given by the FFG.

\begin{table}[t]
\begin{center}
\caption{Parameter of the function \eqref{eq:NLEFT:fit} fitted on the NLEFT predictions for the energy per particle in SM and NM from Ref.~\cite{SElhatisari:2024} and considering $a=1$, to get the FFG limit as the density goes to zero.}
\label{table:NLEFT}
\tabcolsep=0.1cm
\def\arraystretch{1.5}
\begin{tabular}{lccc}
\hline\noalign{\smallskip}
parameters & $b$ & $c$ & $d$ \\
matter &  (fm) & (fm$^2$) & (fm$^3$) \\
\hline
SM & $-9.171\pm0.505$ & $10.01\pm1.40$ & $-3.070\pm0.239$\\
NM & $-1.188\pm0.014$ & $0.719\pm0.022$ & $-0.113\pm0.002$\\
\hline
\end{tabular}
\end{center}
\end{table}

A typical call for a given \texttt{model}, see hereafter, is:
\begin{lstlisting}[language=Python]
micro = nuda.matter.setupMicro( model=`2024-NLEFT-AM' )
micro.print_outputs()
\end{lstlisting}
The present toolkit provides results for only one NLEFT calculation. It is, however, ready for more results by specifying them in the variable \texttt{model}. Here is an example:\\

\noindent
\texttt{model}=`2024-NLEFT-AM':\\
Consider the predictions for SM and NM from Ref.~\cite{SElhatisari:2024} and perform the fit defined by Eq.~\eqref{eq:NLEFT}. The parameters of the function $f(k_F)$ defined in Eq.~\eqref{eq:NLEFT:fit} are those given in Table~\ref{table:NLEFT} for SM and NM.\\

We show in Fig.~\ref{fig:micro:nleft:e2a} results from NLEFT given in Ref.~\cite{SElhatisari:2024}. The blue area represents the result of the fit to NLEFT calculations considering the uncertainties. The band represents the reference uncertainties presented in Sec.~\ref{sec:unif:band}. Here also, there is a nice overlap between the NLEFT results and the reference band.

The bare data extracted from the tables given in Ref.~\cite{SElhatisari:2024} are stored in the properties \texttt{micro.nm\_e2a\_int\_data} and \texttt{micro.sm\_e2a\_int\_data}, they are shown in blue dotted line in Fig.~\ref{fig:micro:nleft:e2a}, while the results of the fit are stored in  \texttt{micro.nm\_e2a\_int} and \texttt{micro.sm\_e2a\_int} (with errors in \texttt{micro.nm\_e2a\_err} and \texttt{micro.sm\_e2a\_err}) and are shown in solid line in Fig.~\ref{fig:micro:nleft:e2a}.

\subsection{Uncertainties in the model predictions}\label{sec:unif:band}

To compare different results from different models, it is necessary to know the uncertainties associated with these models. Since these uncertainties are not always estimated, we suggest a function that provides generic average uncertainties that could be associated to a given model. These uncertainties are extracted from recent QMC and MBPT approaches, where they are accurately estimated, as we detail hereafter.

There are different sources of uncertainties in the model prediction, which are addressed in different ways by different authors. The first source of uncertainties is related to the Hamiltonian itself, 
which is mainly due to the truncation of the operator basis of nuclear interactions and is partially connected to the statistical uncertainties in the fits to data. 
These uncertainties are estimated by some authors, especially in the most recent publications, see Figs.~\ref{fig:micro:qmc:e2a} and \ref{fig:micro:mbpt:e2a} for instance. 

\begin{figure}[t]
\centering
\includegraphics[scale=0.52]{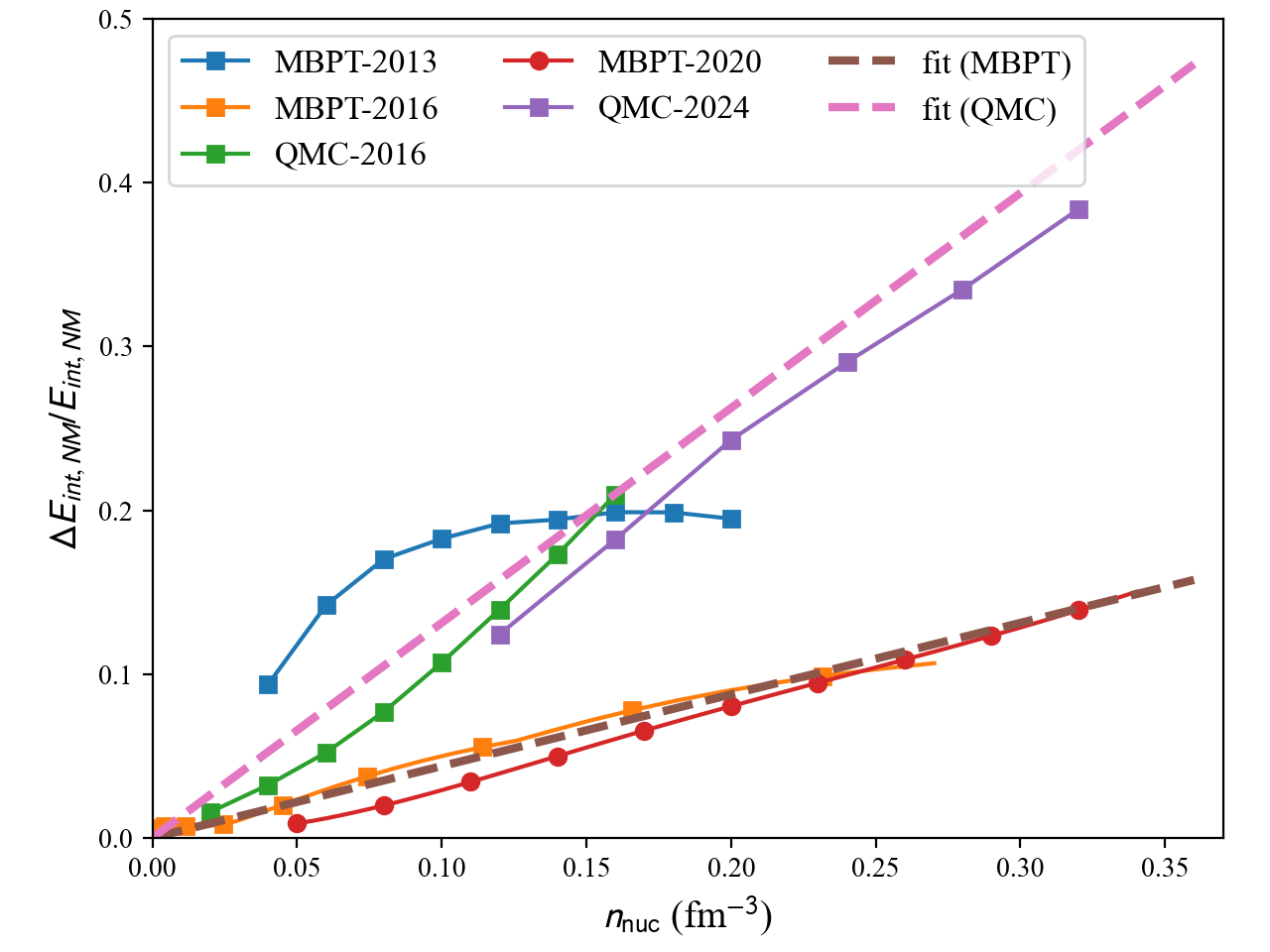}
\caption{Relative uncertainties in NM estimated by different calculations: MBPT-2013~\cite{ITews:2013}, MBPT-2016~\cite{CDrischler:2016}, QMC-2016~\cite{ITews:2016}, MBPT-2020~\cite{CDrischler:2020a,CDrischler:2019}, QMC-2024~\cite{ITews:2024}. The dashed lines represent fits to MBPT, except MBPT-2013, and QMC results, see the text for more details. Figure generated with \texttt{matter\_setupMicroErr\_plot.py}.}
\label{fig:micro:err}
\end{figure}

We show in Fig.~\ref{fig:micro:err}, the relative uncertainties in NM, $\Delta E_{NM}/E_{NM}$, as a function of the density for various calculations: MBPT-2013~\cite{ITews:2013}, MBPT-2016~\cite{CDrischler:2016}, QMC-2016~\cite{ITews:2016}, MBPT-2020~\cite{CDrischler:2020a,CDrischler:2019}, QMC-2024~\cite{ITews:2024}. Simple fits of the relative uncertainty,
\begin{equation}
\frac{\Delta E_{NM}}{E_{NM}}(n_\nuc,\hbox{MBPT}) = 0.07 \frac{n_\nuc}{n_\sat}\, ,
\label{fig:micro:fit:MBPT}
\end{equation}
for MBPT, except MBPT-2013, and
\begin{equation}
\frac{\Delta E_{NM}}{E_{NM}}(n_\nuc,\hbox{QMC}) = 0.21 \frac{n_\nuc}{n_\sat}\, ,
\label{fig:micro:fit:QMC}
\end{equation}
for QMC with $n_\sat=0.16$~fm$^{-3}$~\cite{JMargueron:2018a,CDrischler:2024ebw} are also shown (dashed lines) in Fig.~\ref{fig:micro:err}. There are two groups of uncertainties, the larger one originating from QMC calculations and the smaller one from MBPT calculations, except MBPT-2013. We have defined these two possibilities in the function defining the uncertainties in \texttt{nuda} toolkit. Here is an example of the use of the function \texttt{uncertainty\_stat()} defined in the toolkit:
\begin{lstlisting}[language=Python]
den = np.linspace(0.01,0.2,10)
err = nuda.matter.uncertainty_stat( den , err=`MBPT' )
print(`err:',err)
\end{lstlisting}
where the variable \texttt{err} can be \texttt{`MBPT'} (by default) or \texttt{`QMC'}.

\begin{figure}[t]
\centering
\includegraphics[scale=0.52]{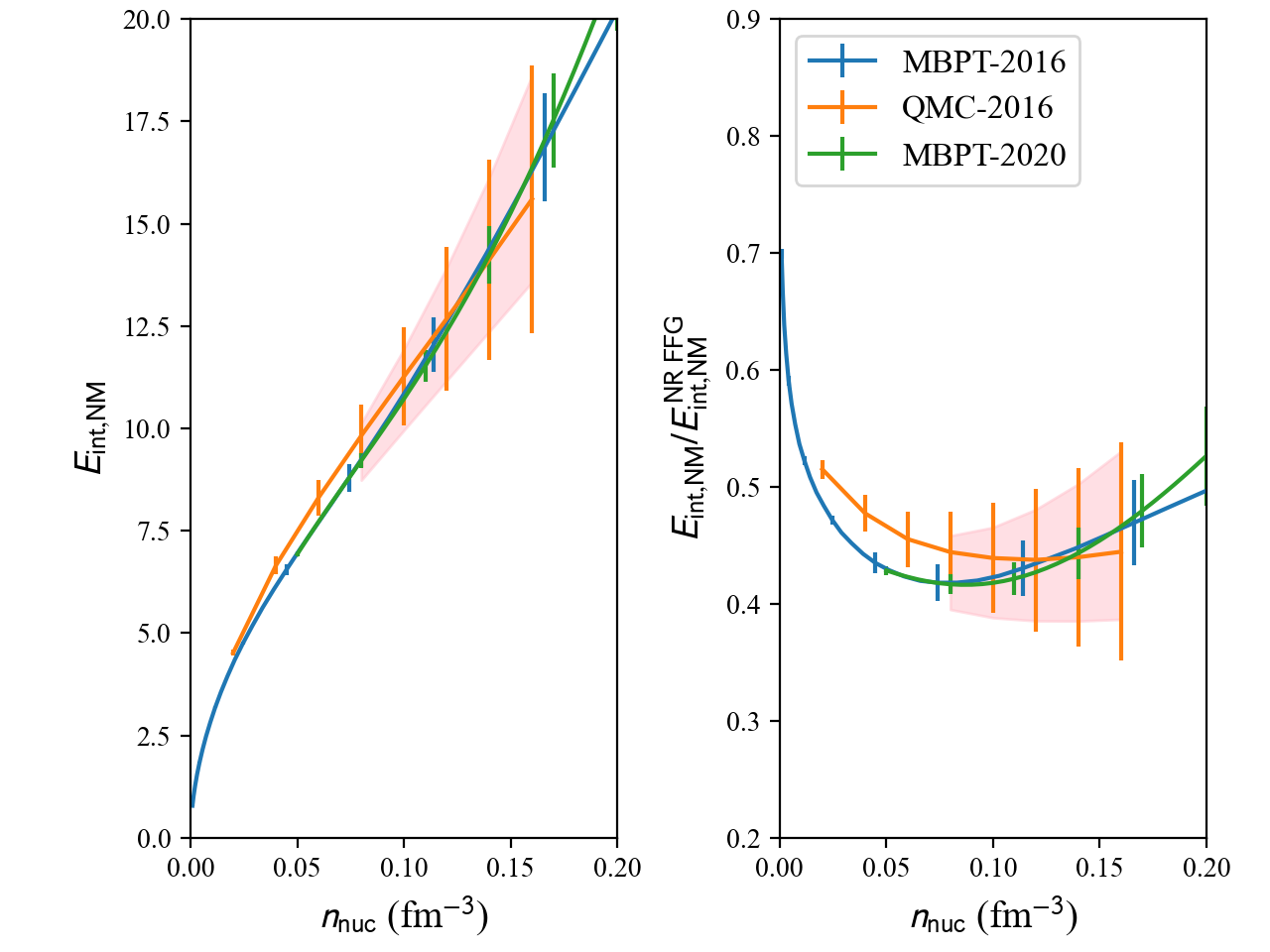}
\caption{Uncertainty reference band in NM obtained from the analysis of different predictions: MBPT-2016~\cite{CDrischler:2016}, QMC-2016~\cite{ITews:2016} and MBPT-2020~\cite{CDrischler:2020a, CDrischler:2019}. Figure generated with \texttt{matter\_setupMicro\_band\_plot.py}.}
\label{fig:micro:band:NM}
\end{figure}

The uncertainty originating from the many-body model is much more difficult to estimate. It has been suggested, for instance, to compare the results from different many-body techniques, while using the same interaction~\cite{Mbaldo:2012}. Since some of the many-body techniques require simplified interactions, this approach imposes that all interactions align with the simplest one. The authors of Ref.~\cite{Mbaldo:2012} confirm, however, that the tensor and spin-orbit components of the two-nucleon force and their in-medium treatment are responsible for most of the observed discrepancies among these approaches. Similar recent studies are using more realistic nuclear interactions, such as $AV_{18}$~\cite{MPiarulli:2020,ALovato:2022}. All these comparisons are very interesting, but they do not provide a simple estimation of the systematic uncertainty. In the following, we propose a method that provides results quantitatively similar to those obtained in Refs.~\cite{Mbaldo:2012,MPiarulli:2020,ALovato:2022}.

Let us now explain how we obtain uncertainties in NM, SM, and for $e_\sym$. First, we must select the models from which the uncertainty will be estimated. The users can make their own selection from different many-body techniques, such as AFDMC, QMC, BHF2, BHF23, AFQMC, and MBPT. Then the dispersion among the selected models is estimated from a Gaussian distribution, 
\begin{equation}
\frac{1}{\sqrt{2\pi}\sigma} \exp \left(-\frac{(e-e_\text{cent})^2}{2\sigma^2} \right) \, ,
\end{equation}
associated to each of the model, considering the centroid $e_\text{cent}$ and the $\sigma$ obtained by each model. For the case where $\sigma$ has not been estimated, we adopt the fit~\eqref{fig:micro:fit:MBPT} with the uncertainty from MBPT. The global centroids and uncertainties are then extracted from the probability distribution built upon the sum of the individual Gaussian distributions. 

\begin{figure}[t]
\centering
\includegraphics[scale=0.52]{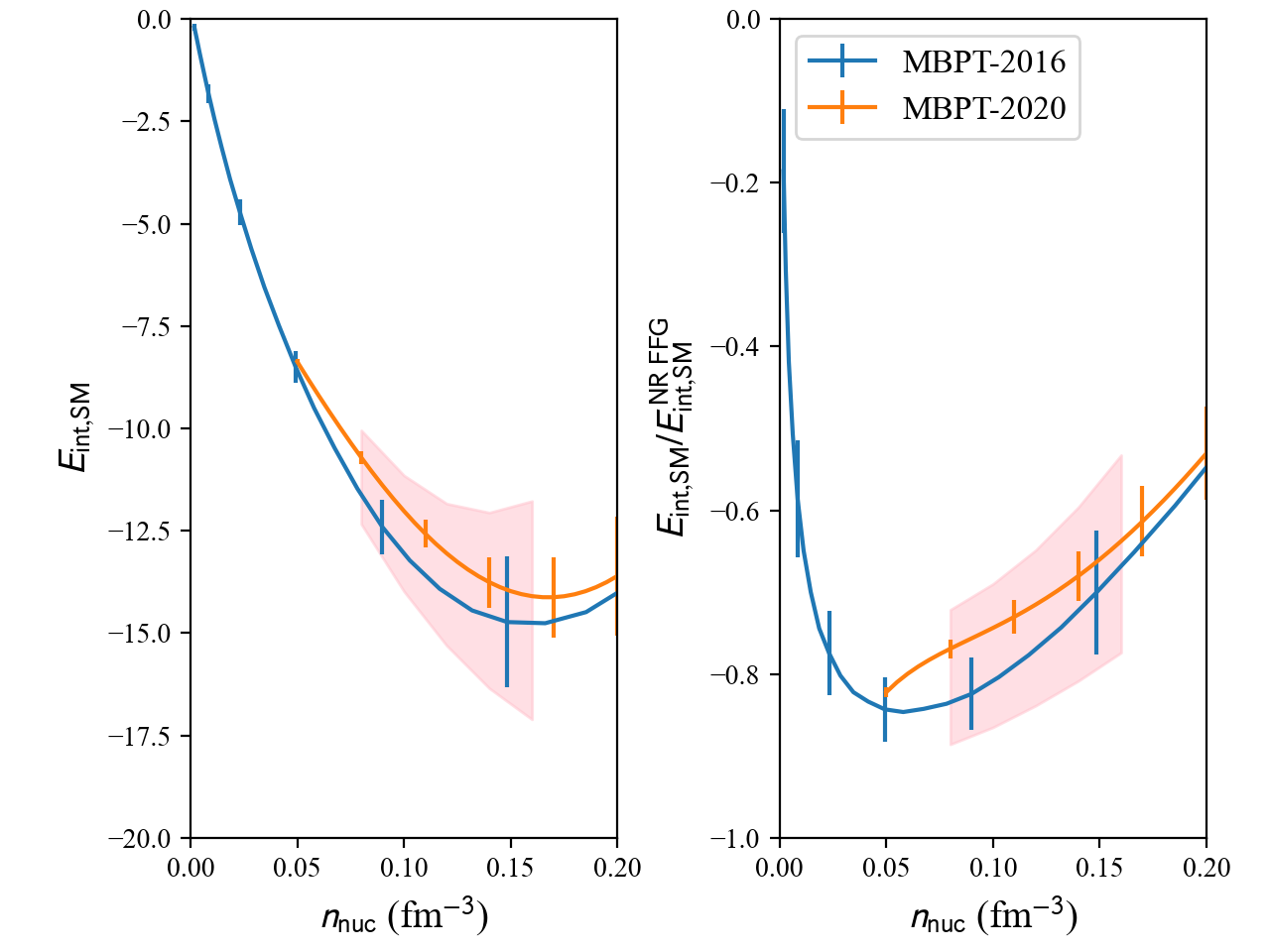}
\caption{Uncertainty reference band in internal energy for SM obtained from the analysis of different predictions: MBPT-2016~\cite{CDrischler:2016} and MBPT-2020~\cite{CDrischler:2020a,CDrischler:2019}. Figure generated with \texttt{matter\_setupMicro\_band\_plot.py}.}
\label{fig:micro:band:SM}
\end{figure}

\begin{figure}[t]
\centering
\includegraphics[scale=0.52]{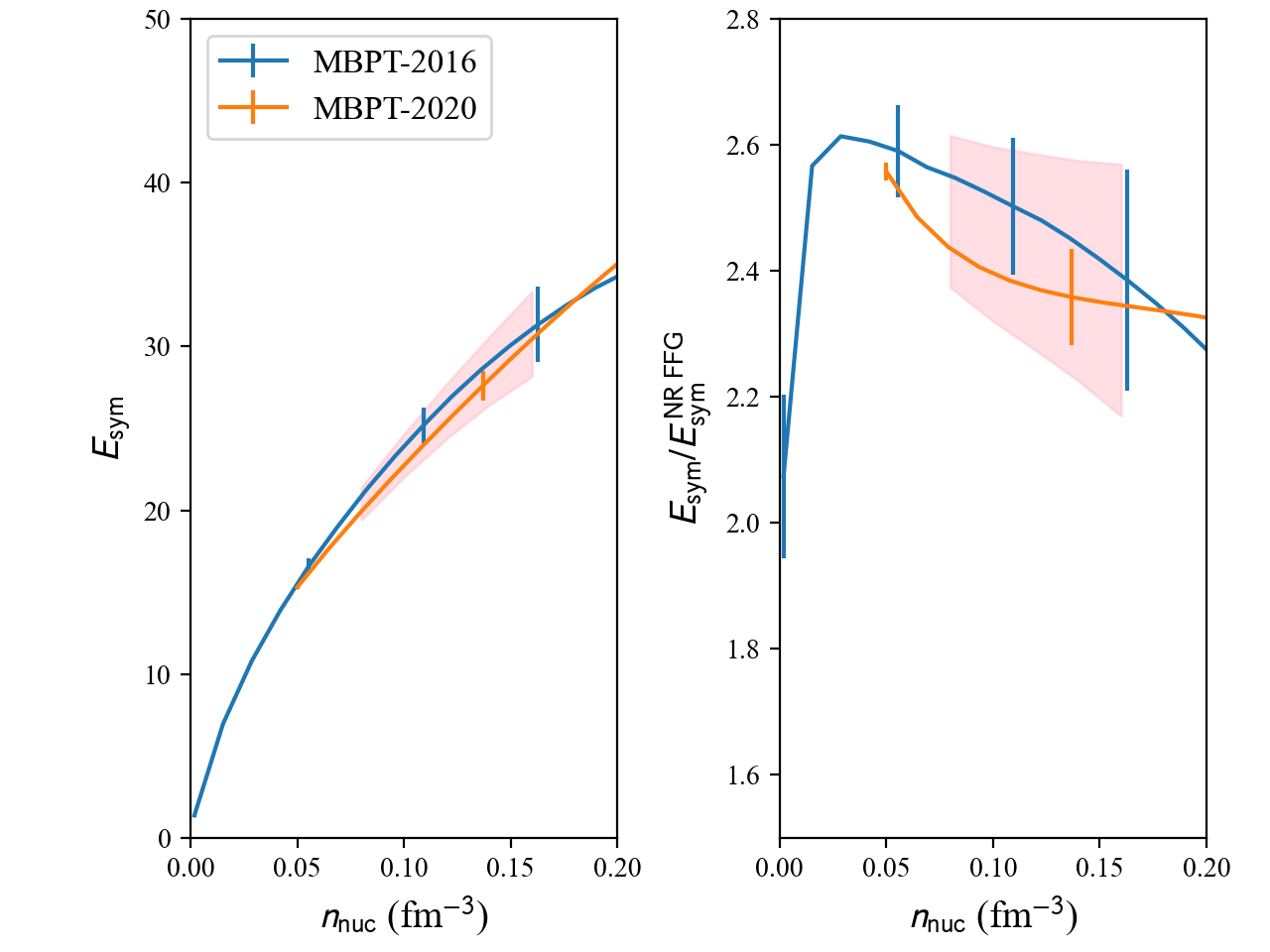}
\caption{Uncertainty reference band for the symmetry energy obtained from the analysis of different predictions: MBPT-2016~\cite{CDrischler:2016} and MBPT-2020~\cite{CDrischler:2020a,CDrischler:2019}. Figure generated with \texttt{matter\_setupMicro\_band\_plot.py}.}
\label{fig:micro:band:Esym}
\end{figure}

Results are shown in Fig.~\ref{fig:micro:band:NM} for NM reference band, Fig.~\ref{fig:micro:band:SM} for SM reference band, and Fig.~\ref{fig:micro:band:Esym} for the reference band representing the symmetry energy. Different microscopic calculations have been employed to define these bands. For instance, the reference band in SM does not include QMC-2016~\cite{ITews:2016}, since QMC-2016 only predicts NM. We employ MBPT-2016~\cite{CDrischler:2016} and MBPT-2020~\cite{CDrischler:2020a,CDrischler:2019} for the reference band in SM, and for the symmetry energy since NM and SM are provided in these calculations. 

The reference bands shown in Figs.~\ref{fig:micro:band:NM}, \ref{fig:micro:band:SM}, and \ref{fig:micro:band:Esym} are functions of the selected models. Choosing other models will produce different reference bands. However, we believe that the choice made here is reasonable and fairly represents the present uncertainties in the prediction for SM and NM. 

\begin{table}[t]
\begin{center}
\caption{Uncertainty reference band for the internal energy per particle in NM, in SM, and for $e_{\sym}$ obtained using MBPT-2016, MBPT-2020, and QMC-2016 (only for the band in NM) and MBPT-2016, MBPT-2020 for the band in SM and for Esym. See text for more details.}
\label{table:band}
\tabcolsep=0.3cm
\def\arraystretch{1.5}
\begin{tabular}{cccc}
\hline\noalign{\smallskip}
$n_\nuc$ & $e_{\NM}$ & $e_{\SM}$ & $e_{\sym}$ \\
(fm$^{-3}$) & (MeV) & (MeV) & (MeV) \\
\hline\noalign{\smallskip}
0.08 & 9.30$\pm$1.07  & -11.2$\pm$0.636 & 20.4$\pm$0.704 \\
0.10 & 10.9$\pm$1.37  & -12.6$\pm$0.786 & 23.3$\pm$0.938 \\
0.12 & 12.5$\pm$1.76  & -13.6$\pm$0.963 & 26.1$\pm$1.19 \\
0.14 & 14.4$\pm$2.24  & -14.2$\pm$1.19 & 28.5$\pm$1.48 \\
0.16 & 16.2$\pm$2.83  & -14.4$\pm$1.48 & 30.8$\pm$1.85 \\
\noalign{\smallskip}\hline
\end{tabular}
\end{center}
\end{table}

A practical way to define the reference band in the \texttt{nuda} toolkit is now provided. The class \texttt{nuda.matter.setupMicroBand()} provided by
\texttt{nuda} toolkit performs an average over various microscopic predictions, given as a list to the input \texttt{models}, and in the case defined by the variable \texttt{matter}. The following instructions: 
\begin{lstlisting}[language=Python]
band = nuda.matter.setupMicroBand( models  = [ `2016-QMC-NM', `2016-MBPT-AM', `2020-MBPT-AM' ], matter=`NM' )
band.print_outputs()
\end{lstlisting}
perform an averaging over the following models: `2016-QMC-NM', `2016-MBPT-AM', and `2020-MBPT-AM', and for NM. The input variable \texttt{models} contains the list of models defining the reference band and the variable \texttt{matter} can be set to `NM' for NM, `SM' for SM, and `Esym' for the symmetry energy. 

The class \texttt{nuda.matter.setupMicroBand} first analyzes the density range associated with each model and defines a new density range, avoiding extrapolation. This new interval is split into \texttt{nden} points (by default \texttt{nden=10}), where the reference band will be calculated. The class \texttt{nuda.matter.setupMicroBand} sums the Gaussian distribution associated with each of the input models and then computes total centroids and uncertainties, defined as standard deviation. The variable \texttt{ne} defines the number of points in the energy (vertical) direction (by default \texttt{ne=200}). The final result should be independent of the choice for \texttt{ne}. For the density range considered, it provides as output the centroids and uncertainties.

Here is an example where the variables \texttt{nden} and \texttt{ne} are explicitly defined:
\begin{lstlisting}[language=Python]
band = nuda.matter.setupMicroBand( models, matter, nden=5, ne=100 )
\end{lstlisting}

The density mesh can also be imposed by the user. For example, in the present study, we have considered the following density mesh, \texttt{den=[0.08, 0.10, 0.12, 0.14, 0.16]}, to obtain the results shown in Tab.~\ref{table:band}. In this case, however, since the density is imposed by the user, there is no check concerning extrapolation. It is the responsibility of the user to fix the density mesh in agreement with the density range of the models defining the reference band. Here is an example where the \texttt{den} is explicitly fixed:
\begin{lstlisting}[language=Python]
den = np.linspace(0.08,0.16,5)
band = nuda.matter.setupMicroBand( models, matter=`NM', den=den )
\end{lstlisting}

In addition, the energy boundaries (vertical) are fixed by default: \texttt{e2a\_min=-20}~MeV and  \texttt{e2a\_max=50}~MeV. This choice can also be modified by giving explicit values for the variables  \texttt{e2a\_min} and \texttt{e2a\_min}. Here is an example where these boundaries are explicitly fixed:
\begin{lstlisting}[language=Python]
band = nuda.matter.setupMicroBand( models, matter=`NM', e2a_min=-18.0, e2a_max=30.0 ).
\end{lstlisting}

Finally, the reference band can be increased or decreased by an arbitrary factor defined in \texttt{xfac}. By default, \texttt{xfac=1.0} in NM, \texttt{xfac=1.8} in SM, and \texttt{xfac=1.4} for the symmetry energy. The reason for introducing the factor \texttt{xfac} is to reconcile microscopic and phenomenological approaches in SM. Here is an example where the scaling factor \texttt{xfac} is explicitly fixed:
\begin{lstlisting}[language=Python]
band = nuda.matter.setupMicroBand( models, matter=`NM', xfac=1.2 ).
\end{lstlisting}

Output quantities are the following: 
\texttt{band.den} contains the densities, \texttt{band.e2a\_int} for the internal energy per nucleon, i.e., the centroids averaging over the different input models and corresponding to the densities given in \texttt{band.den}, and \texttt{band.e2a\_std} for the standard deviation among the input models.

\begin{figure*}[t]
\centering
\includegraphics[scale=0.9]{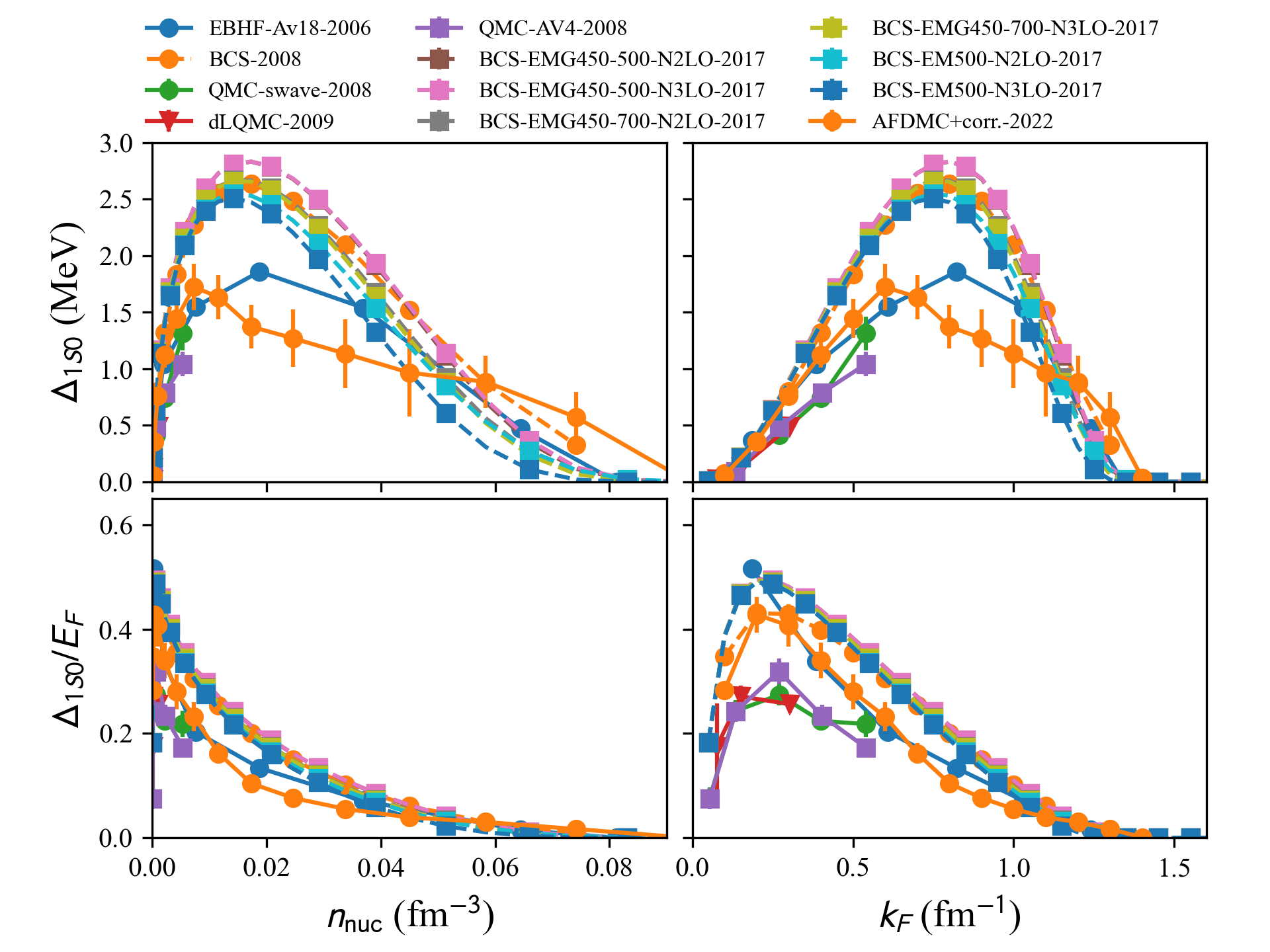}
\caption{$^1$S$_0$ pairing gap in neutron matter (NM) over the Fermi energy (top) and the pairing gap (bottom) as a function of the density (left) and the neutron Fermi momentum (right) for the complete list of model available in the \texttt{nuda} toolkit. Figure generated with \texttt{matter\_setupMicro\_gap\_plot.py}.}
\label{fig:micro:gap:1s0}
\end{figure*}

\subsection{Check models versus a reference band}
\label{sec:unif:check}

We now present another class to check models. It compares a given model and a reference band, and it replies depending on whether the model is inside the reference band or outside. To do so, the user should provide the model to check, define a reference band, and then call the class \texttt{nuda.matter.setupCheck}. It can be done as:
\begin{lstlisting}[language=Python]
# define the model to check:
model = `1998-VAR-AM-APR'
micro = nuda.matter.setupMicro( model=model )
# define the reference band:
den = np.array([0.08,0.1,0.12,0.14,0.16])
models = [ `2016-MBPT-AM', `2016-QMC-NM', `2020-MBPT-AM' ]
band = nuda.matter.setupMicroBand( models, den=den, matter=`NM' )
# check if eos=micro is inside the band:
check = nuda.matter.setupCheck( eos = micro, band=band )
\end{lstlisting}
The object \texttt{check} contains different properties: if the model passes inside (outside) the band, then the properties \texttt{check.isInside} (\texttt{check.isOutside}) is True (False), otherwise, it is False (True). This class has been employed, for instance, in Figs.~\ref{fig:micro:var:e2a}-\ref{fig:micro:nleft:e2a} to represent the results as solid lines (if \texttt{check.isInside} is True) or dashed lines (if \texttt{check.isInside} is False).

\subsection{Pairing gap in $^1$S$_0$ and $^3$PF$_2$ channels in NM}
\label{sec:unif:gap}

\begin{figure*}[t]
\centering
\includegraphics[scale=0.9]{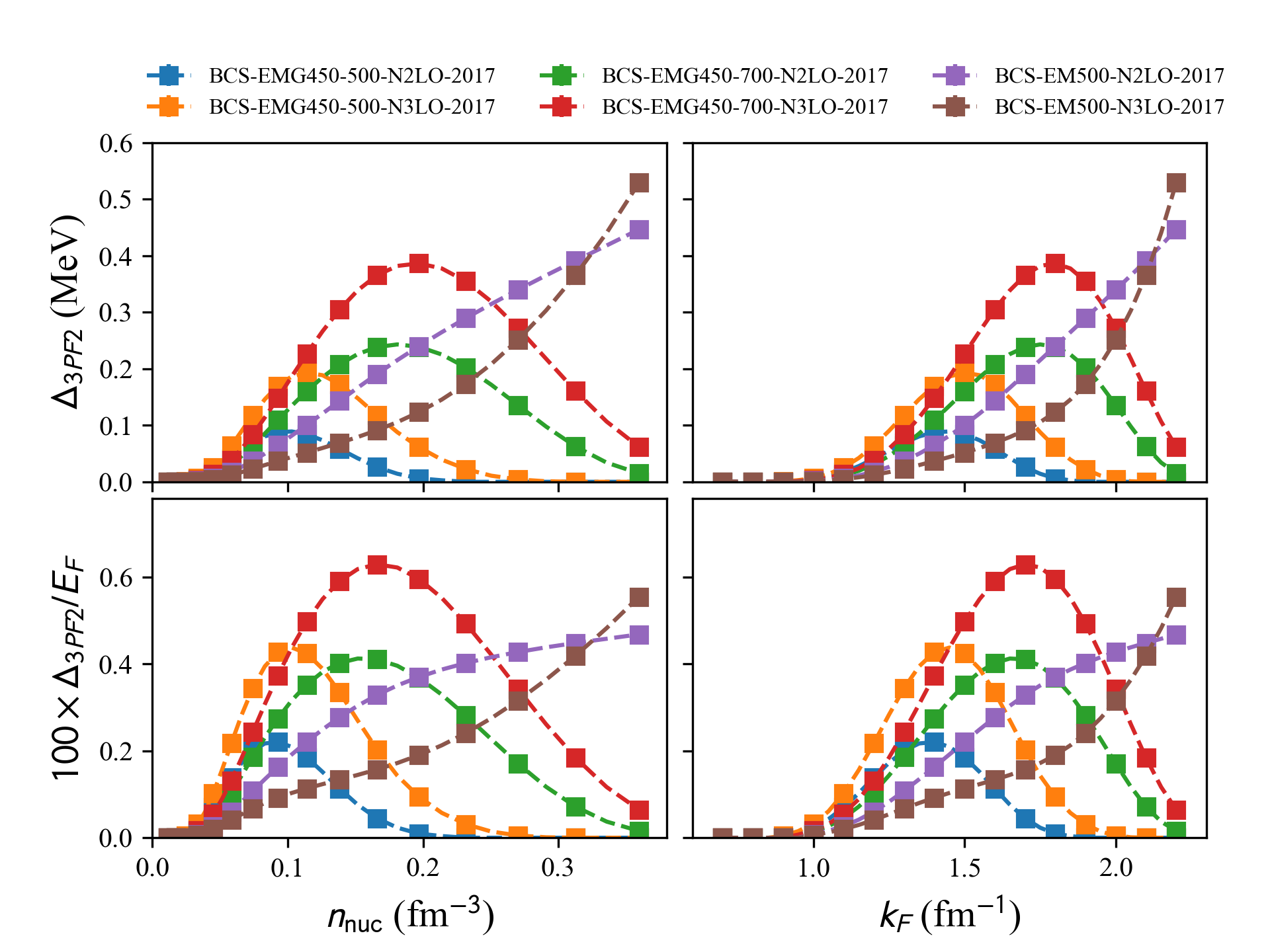}
\caption{$^3$PF$_2$ pairing gap in neutron matter (NM) over the Fermi energy (top) and the pairing gap (bottom) as a function of the density (left) and the neutron Fermi momentum (right) for the complete list of model available in the \texttt{nuda} toolkit. Figure generated with \texttt{matter\_setupMicro\_gap\_plot.py}.}
\label{fig:micro:gap:3pf2}
\end{figure*}

Pairing in nuclear systems results from the interplay between the bare nuclear force and medium polarizations, inducing correlations due to the strong bare nuclear interaction. Correlations beyond BCS are therefore non-negligible. Several many-body approaches have been employed to address the question of the strength of the pairing gap in neutron matter and symmetric matter. These many-body approaches are listed hereafter.

The \texttt{nuda} toolkit provides results for the pairing gap in $^1$S$_0$ and $^3$PF$_2$ channels in NM and SM. The complete list of available phenomenological models is given with the following instructions:
\begin{lstlisting}[language=Python]
print( nuda.matter.micro_gap_models( matter=`NM' ) )
\end{lstlisting}
where \texttt{matter} can be `NM' for predictions in NM, or `SM' for SM.
Once the variable \texttt{model} is chosen, the results can be obtained by calling the class \texttt{matter.setupMicroGap()} in the following way:
\begin{lstlisting}[language=Python]
gap = nuda.matter.setupMicroGap( model = `2006-BHF-NM' )
gap.print_outputs()
\end{lstlisting}
where the variable \texttt{model} can be chosen among the options listed hereafter.\\

\noindent
\texttt{model}=`2006-BHF-NM':\\
In this calculation, vertex and self-energy corrections are treated on the same footing.
A realistic two-body force (V18) is renormalized in the medium by considering RPA and dispersive corrections. Calculations are performed in NM, see Ref.~\cite{LGCao:2006b} for more details.\\

\noindent
\texttt{model}=`2006-BHF-SM':\\
The same many-body approach as \texttt{model}=`2006-BHF-NM' is performed here, except that SM is considered instead of NM, see Ref.~\cite{LGCao:2006b} for more details.\\

\noindent
\texttt{model}=`2008-BCS-NM':\\
BCS calculations based on the Argonne A8$^\prime$ two-nucleon interaction are performed in Ref.~\cite{AFabrocini:2008}. The data provided by \texttt{nuda} are extracted from Ref.~\cite{AFabrocini:2008}, more specifically from the uncorrelated BCS case given in Table 1 for the Fermi momentum $k_{F}$, the effective mass $m_n^*/m$ and the chemical potential $\mu_n$ in NM. \\

\noindent
\texttt{model}=`2008-QMC-NM-Swave':\\
Presented in Sec.~\ref{sec:unif:micro}.\\

\noindent
\texttt{model}=`2009-AFDMC-NM':\\
Presented in Sec.~\ref{sec:unif:micro}.\\

\noindent
\texttt{model}=`2008-QMC-NM-AV4':\\
Presented in Sec.~\ref{sec:unif:micro}.\\

\begin{figure*}[t]
\centering
\includegraphics[scale=0.9]{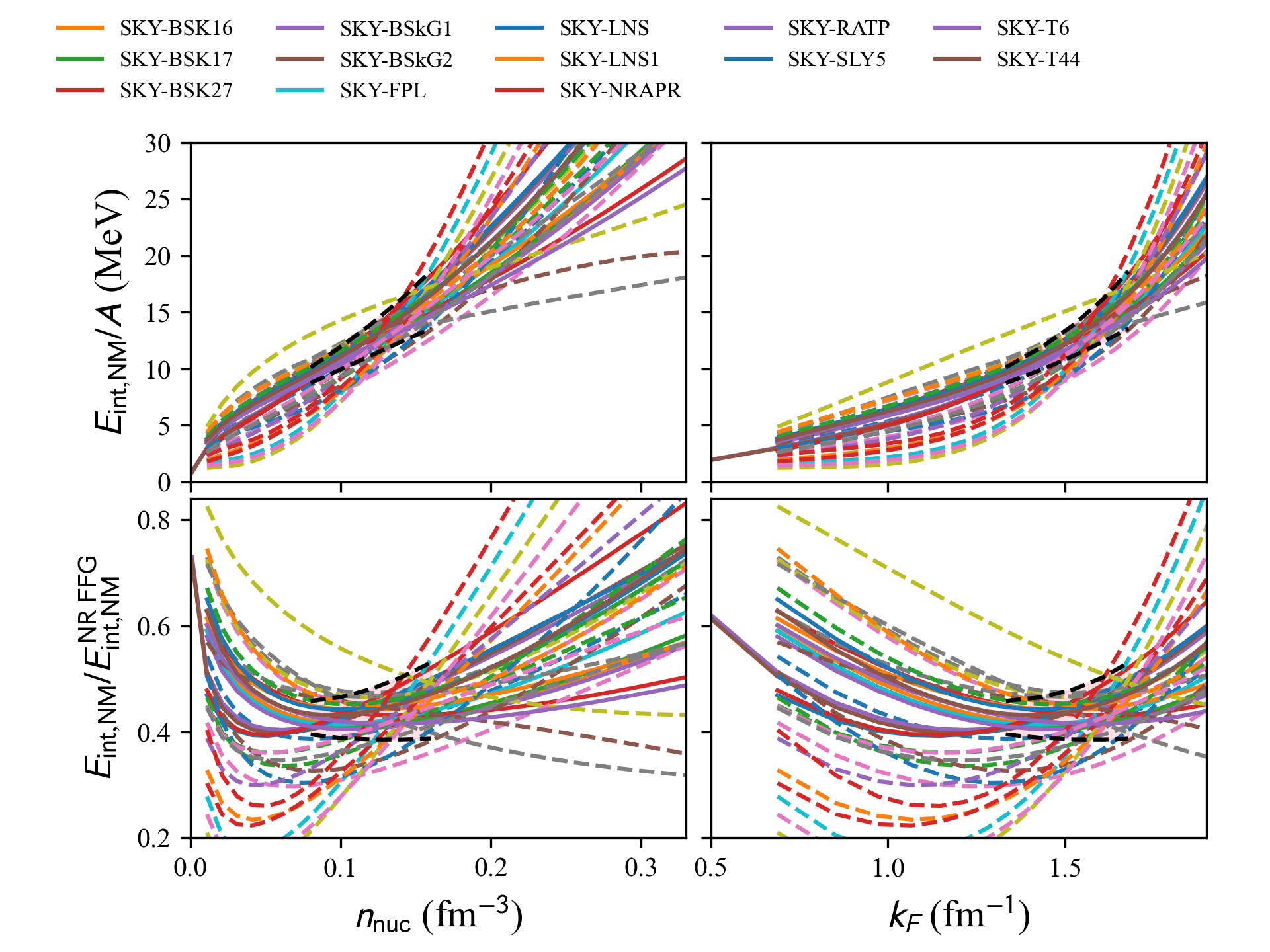}
\caption{Energy in neutron matter (NM) over the free Fermi gas energy (top) and the energy per particle (bottom) as a function of the density (left) and the neutron Fermi momentum (right) for the complete list of phenomenological models based on the standard Skyrme interaction available in the \texttt{nuda} toolkit. Note that the Skyrme models given in the legend are those which are compatible with the reference band (this prescription applies only to this figure).
Figure generated with \texttt{matter\_setupPheno\_plot.py}.}
\label{fig:pheno:sky:e2a:nm}
\end{figure*}

\begin{figure*}[t]
\centering
\includegraphics[scale=0.9]{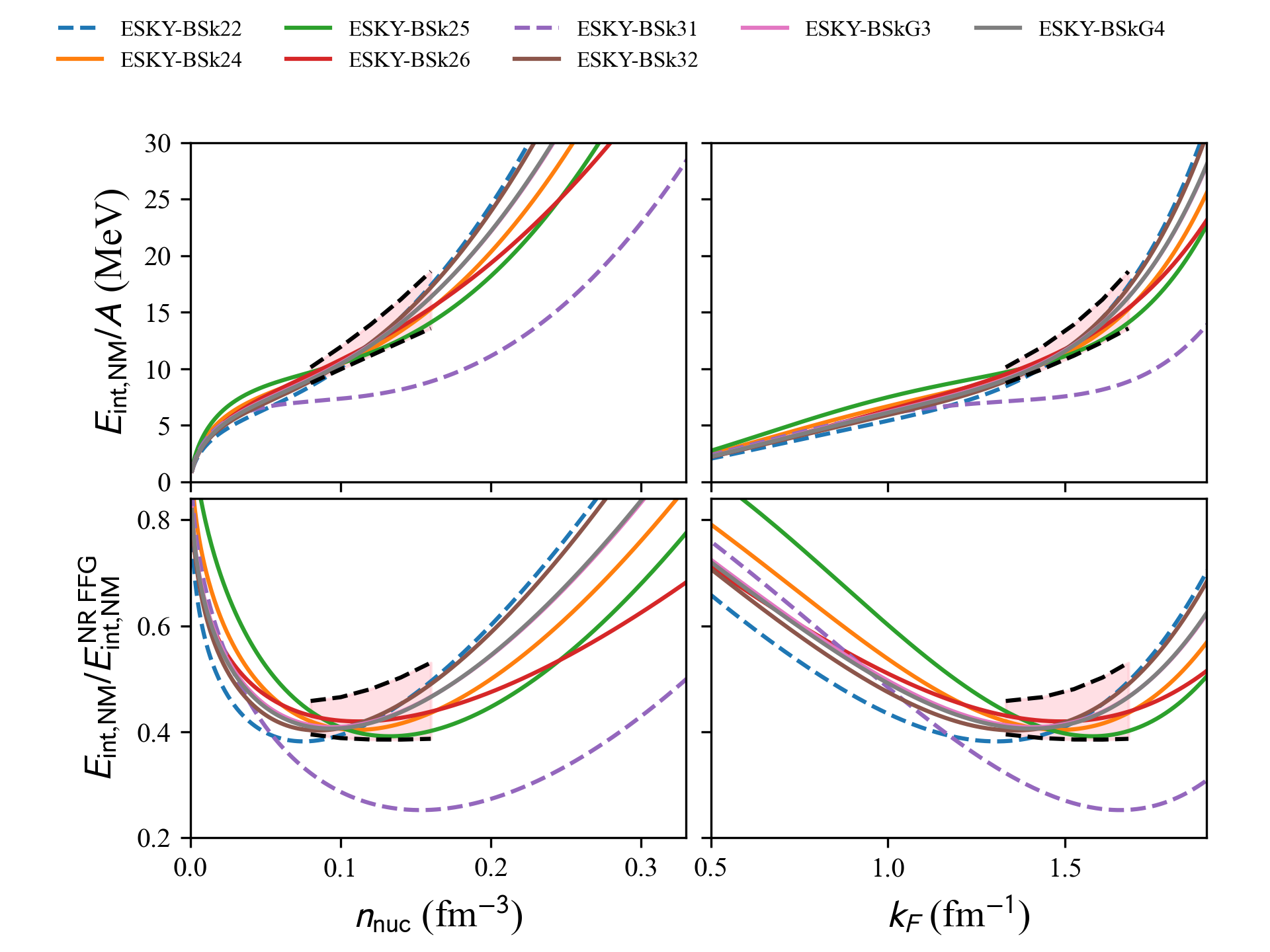}
\caption{Same as Fig.~\ref{fig:pheno:sky:e2a:nm} for the complete list of extended Skyrme EDF available in \texttt{nuda} toolkit. Figure generated with \texttt{matter\_setupPheno\_plot.py}.}
\label{fig:pheno:esky:e2a:nm}
\end{figure*}

\noindent
\texttt{model}=`2017-MBPT-NM-GAP-EMG-450-500-N2LO', `2017-MBPT-NM-GAP-EMG-450-500-N3LO', `2017-MBPT-NM-GAP-EMG-450-700-N2LO', `2017-MBPT-NM-GAP-EMG-450-700-N3LO', `2017-MBPT-NM-GAP-EM-500-N2LO', `2017-MBPT-NM-GAP-EM-500-N3LO.\\
Solutions of the BCS equation in the $^1$S$_0$ and $^3$PF$_2$ channels with various $\chi$EFT interactions given in Ref.~\cite{CDrischler:2017}.\\

\noindent
\texttt{model}=`2022-AFDMC-NM':\\
Presented in Sec.~\ref{sec:unif:micro}.\\

Results of various calculations are shown in Fig.~\ref{fig:micro:gap:1s0} for the pairing gap in the $^1$S$_0$ channel in NM. Among these results, the ones labeled as BCS (in dashed lines) are performed by solving the lowest-order BCS gap equation. They predict a pairing gap peaked at $2.75\pm 0.15$~MeV for $k_{F}\approx 0.8$~fm$^{-1}$. The other calculations (in solid lines) consider correlations beyond the bare BCS ones using different many-body techniques, see details here before. Interestingly, all calculations, including correlations beyond BCS, predict a reduction of the pairing gap.

Results of various calculations for the pairing gap in the $^3$PF$_2$ channel in NM are shown in Fig.~\ref{fig:micro:gap:3pf2}. All the calculations shown in Fig.~\ref{fig:micro:gap:3pf2} are performed at the BCS approximation using various $\chi$EFT nuclear interactions. The pairing gaps are small, in absolute values, but there is a strong model dependence of the results.

\subsection{Phenomenological models}
\label{sec:unif:pheno}

The complete list of available phenomenological models is given with the following instruction:
\begin{lstlisting}[language=Python]
print( nuda.matter.pheno_models( ) )
\end{lstlisting}

The \texttt{nuda} toolkit provides results for the following models: `Skyrme', `ESkyrme', `NLRH', `DDRH', and `DDRHF'.

The complete list of available phenomenological parametrizations for a given model is given with the following instruction:
\begin{lstlisting}[language=Python]
print( nuda.matter.pheno_params( model=`Skyrme' ) )
\end{lstlisting}
Once the variables \texttt{model} and \texttt{param} are chosen, the results can be obtained by calling the class \texttt{matter.setupPheno()} in the following way:
\begin{lstlisting}[language=Python]
pheno = nuda.matter.setupPheno( model=`Skyrme', param=`SLy5' )
pheno.print_outputs()
\end{lstlisting}
The object \texttt{pheno} contains a large set of properties, some of them listed by the command \texttt{pheno.print\_outputs()}. The entire list of properties is obtained in the following way: 
\begin{lstlisting}[language=Python]
print(pheno.__dict__)
\end{lstlisting}

Output quantities are, for instance: \texttt{pheno.nm\_e2a} (\texttt{pheno.nm\_e2a\_int}) and \texttt{pheno.sm\_e2a}  (\texttt{pheno.sm\_e2a\_int}) for the nucleon (internal) energy per particle in NM and SM.

\subsubsection{Skyrme EDFs}

Skyrme EDFs are obtained from the Skyrme contact nuclear interaction complemented with density-dependent terms. Reviews can be found in the following Refs.~\cite{Bender:2003,JRStone:2007}.\\

A typical call for a given \texttt{model} and parameter set \texttt{param}, see hereafter, is:
\begin{lstlisting}[language=Python]
micro = nuda.matter.setupPheno( model = `Skyrme', param=`SLy5' )
micro.print_outputs()
\end{lstlisting}
Each calculation can be obtained by fixing the variables \texttt{model} and \texttt{param} to one of the following values:\\

\noindent
\texttt{model}=`Skyrme'.\\
\texttt{param}= `BSK14'~\cite{SGoriely:2007}, `BSK16'~\cite{NChamel:2008}, `BSK17'~\cite{SGoriely:2009}, `BSK27'~\cite{SGoriely:2013_BSk27}, `BSkG1'~\cite{GScamps:2021}, `BSkG2'~\cite{WRyssens:2022},  `F-'~\cite{TLesinski:2006}, `F+'~\cite{TLesinski:2006}, `F0'~\cite{TLesinski:2006}, `FPL', `LNS'~\cite{LGCao:2006}, `LNS1'~\cite{DGambacurta:2011}, `LNS5'~\cite{DGambacurta:2011}, `NRAPR'~\cite{AWSteiner:2005}, `RATP'~\cite{MRayet:1982}, `SAMI'~\cite{XRocaMaza:2013}, `SGII'~\cite{NVanGiai:1981}, `SIII'~\cite{MBeiner:1975}, `SKGSIGMA' (Gs)~\cite{JFriedrich:1986}, `SKI2'~\cite{PGReinhard:1995}, `SKI4'~\cite{PGReinhard:1995}, `SKMP'~\cite{LBennour:1989}, `SKMS'~\cite{JBartel:1982}, `SKO'~\cite{PGReinhard:1999}, `SKOP'~\cite{PGReinhard:1999}, `SKP'~\cite{JDobaczewski:1984}, `SKRSIGMA' (Rs)~\cite{JFriedrich:1986}, `SKX'~\cite{ABrown:1998}, `Skz2'~\cite{JMargueron:2002}, `SLY4'~\cite{EChabanat:1998}, `SLY5'~\cite{EChabanat:1998}, `SLY230A'~\cite{EChabanat:1997}, `SLY230B'~\cite{EChabanat:1997}, `SV'~\cite{MBeiner:1975}, `T6'~\cite{FTondeur:1984}, `T44'~\cite{TLesinski:2007}, `UNEDF0'~\cite{MKortelainen:2010}, `UNEDF1'~\cite{MKortelainen:2012}.\\

\noindent
\texttt{model}=`ESkyrme'.\\
\texttt{param}= `BSk22'~\cite{SGoriely:2013}, `BSk24'~\cite{SGoriely:2013}, `BSk25'~\cite{SGoriely:2013}, `BSk26'~\cite{SGoriely:2013}, `BSk31'~\cite{SGoriely:2016}, `BSk32'~\cite{SGoriely:2016}, `BSkG3'~\cite{GGrams:2023},
`BSkG4'~\cite{GGrams:2025}.

\begin{figure*}[t]
\centering
\includegraphics[scale=0.9]{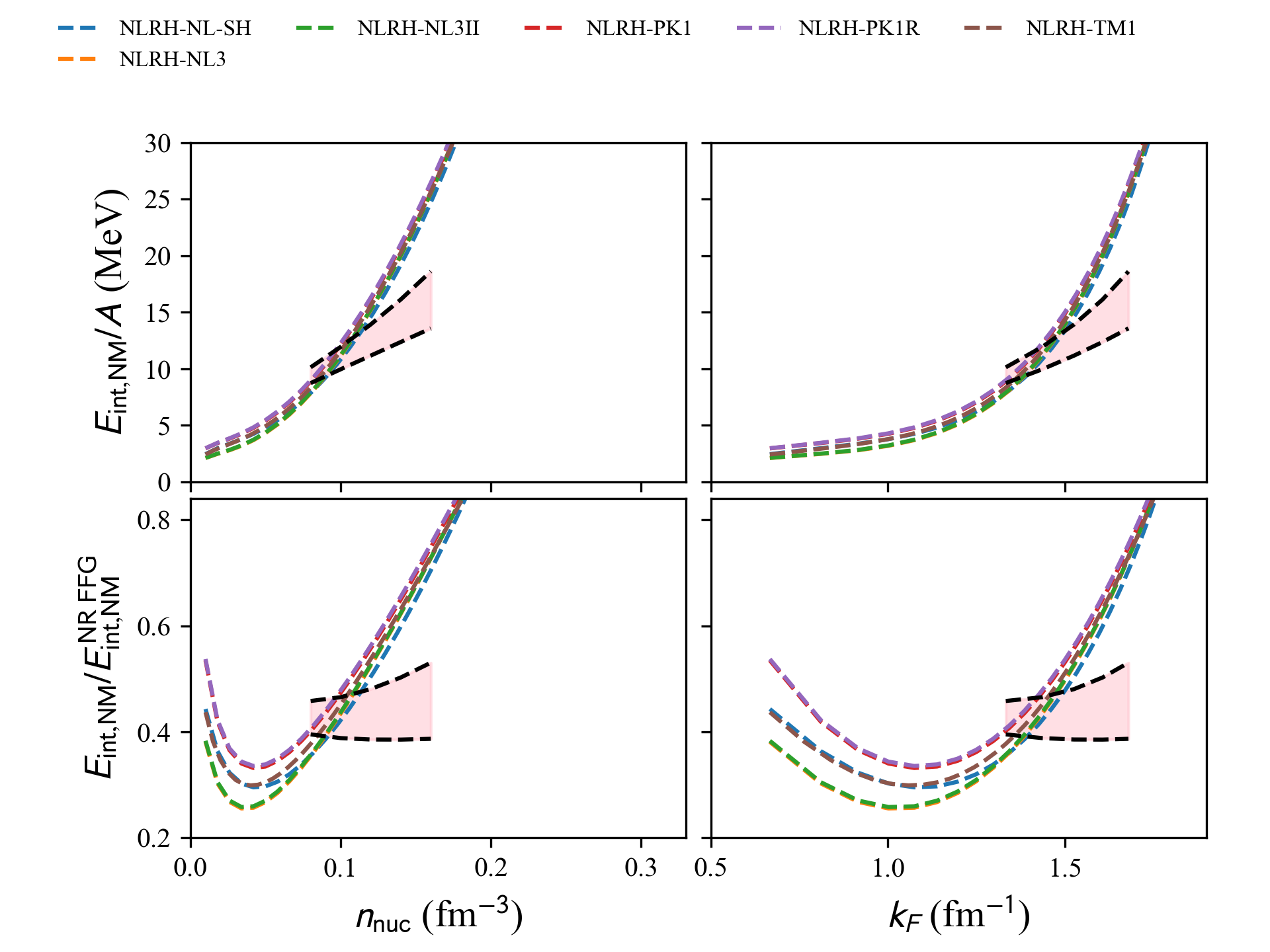}
\caption{Same as Fig.~\ref{fig:pheno:sky:e2a:nm} for the complete list of NLRH EDF available in \texttt{nuda} toolkit. Figure generated with \texttt{matter\_setupPheno\_plot.py}.}
\label{fig:pheno:nlrh:e2a:nm}
\end{figure*}

Results in NM are shown in fig.~\ref{fig:pheno:sky:e2a:nm} for Skyrme EDFs and \ref{fig:pheno:esky:e2a:nm} for extended Skyrme EDFs. For Skyrme EDFs, the legend only names the models passing the reference band, since there is a long list of models. For extended Skyrme EDFs, as well as for other models presented hereafter, the entire list of models available in the \texttt{nuda} toolkit is given in the legend of the figures.
 
\subsubsection{Non-linear RH (NLRH)}

The nonlinear relativistic Hartree approaches, often referred to as relativistic mean-field (RMF) approaches, are an extension of the original Walecka model~\cite{BDSerot:1997}, considering nonlinear meson-nucleon couplings. Nonlinear couplings were introduced initially to soften the EoS and reduce the incompressibility modulus at saturation density.
A typical call for a given \texttt{model} and parameter set \texttt{param}, see hereafter, is:
\begin{lstlisting}[language=Python]
micro = nuda.matter.setupPheno( model=`nlrh', param=`nl3' )
micro.print_outputs()
\end{lstlisting}
Each calculation can be obtained by fixing the variables \texttt{model} and \texttt{param} to one of the following values:\\

\noindent
\texttt{model}=`NLRH'.\\
\texttt{param}=`NL-SH'~\cite{MMSharma:1993}, `NL3'~\cite{GALalazissis:1997}, `NL3s'~\cite{GALalazissis:1997}, `PK1'~\cite{WLong:2004}, `PK1R'~\cite{WLong:2004}, `TM1'~\cite{KSumiyoshi:1995}.\\

Results in NM are shown in fig.~\ref{fig:pheno:nlrh:e2a:nm} for NLRH EDFs. The NLRH models available in the \texttt{nuda} toolkit are all represented, and it is interesting to note that they all disagree with the reference band previously discussed. Fig.~\ref{fig:pheno:nlrh:e2a:nm} shows that all NLRH are more repulsive than expectations from the NM reference band.

\subsubsection{Density-dependent RH (DDRH)}

\begin{figure*}[t]
\centering
\includegraphics[scale=0.9]{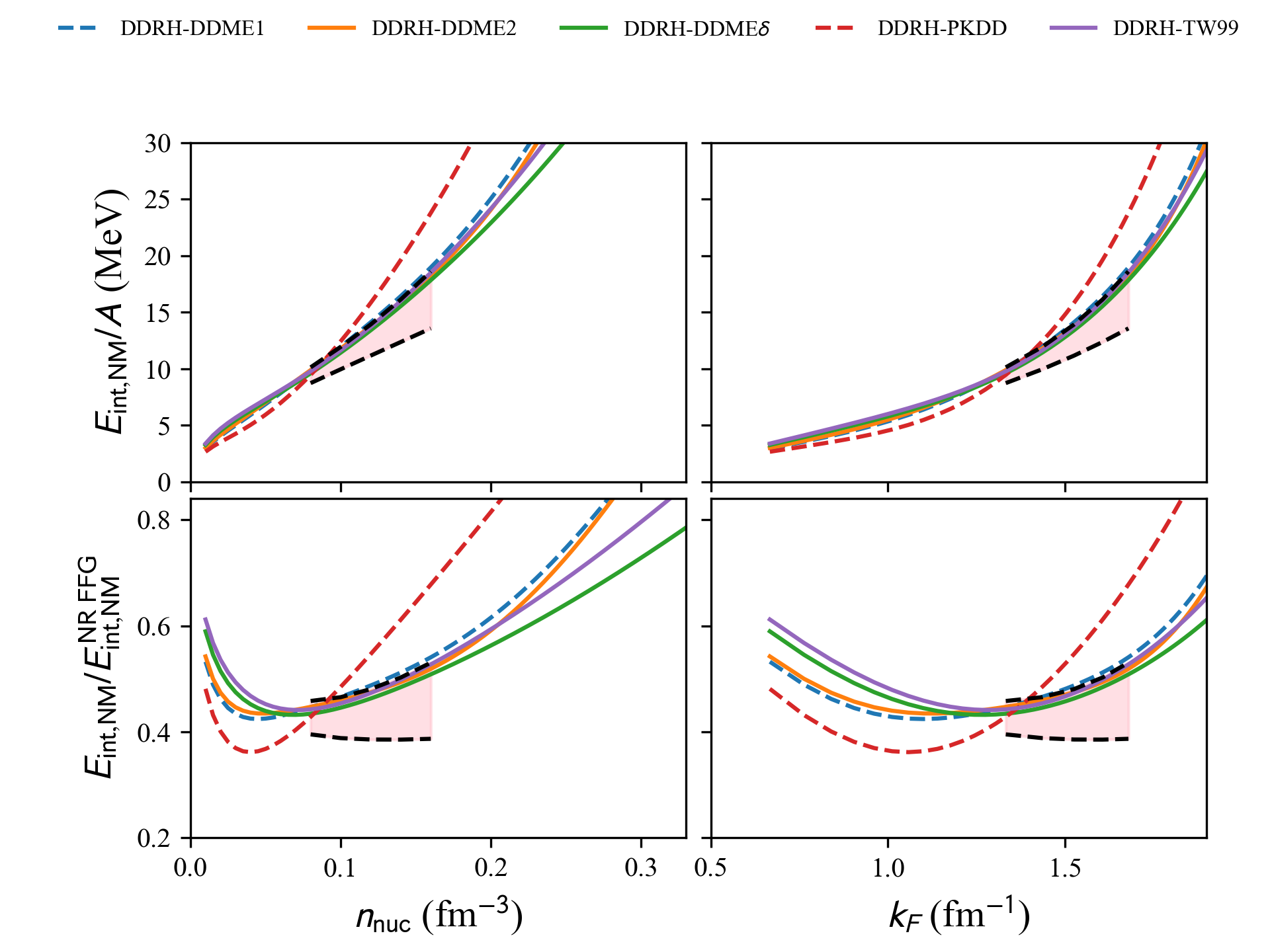}
\caption{Same as Fig.~\ref{fig:pheno:sky:e2a:nm} for the complete list of DDRH EDF available in \texttt{nuda} toolkit. Figure generated with \texttt{matter\_setupPheno\_plot.py}.}
\label{fig:pheno:ddrh:e2a:nm}
\end{figure*}

The density-dependent relativistic Hartree approach has emerged after the NLRH one. It consists of considering density-dependent coupling constants instead of the non-linear meson-nucleon coupling in the NLRH model. These density-dependent coupling constants are adjusted to mimic Dirac-Brueckner-Hartree-Fock results, and they produce softer EoS similar to the NLRH approach.

A typical call for a given \texttt{model} and parameter set \texttt{param}, see hereafter, is:
\begin{lstlisting}[language=Python]
micro = nuda.matter.setupPheno( model=`ddrh', param=`ddme2' )
micro.print_outputs()
\end{lstlisting}
Each calculation can be obtained by fixing the variables \texttt{model} and \texttt{param} to one of the following values:\\

\noindent
\texttt{model}=`DDRH'.\\
\texttt{param}=`DDME1'~\cite{TNiksic:2002}, `DDME2'~\cite{GALalazissis:2005}, `DDMEd' (DD-ME$\delta$)~\cite{XRocaMaza:2011}, `PKDD'~\cite{WLong:2004}, `TW99'~\cite{STypel:1999}.

Results in NM are shown in Fig.~\ref{fig:pheno:ddrh:e2a:nm} for DDRH EDFs. The relativistic DDRH models available in the \texttt{nuda} toolkit are all represented. The density-dependent coupling constant softens the energy per particle, and among the models plotted in Fig.~\ref{fig:pheno:ddrh:e2a:nm}, three are in agreement with the reference band previously discussed: TW99, DDME$\delta$, and DDME2. The model DDME1 is very close to the reference band, while the model PKDD is the most repulsive one.

\subsubsection{Density-dependent RHF (DDRHF)}

\begin{figure*}[t]
\centering
\includegraphics[scale=0.9]{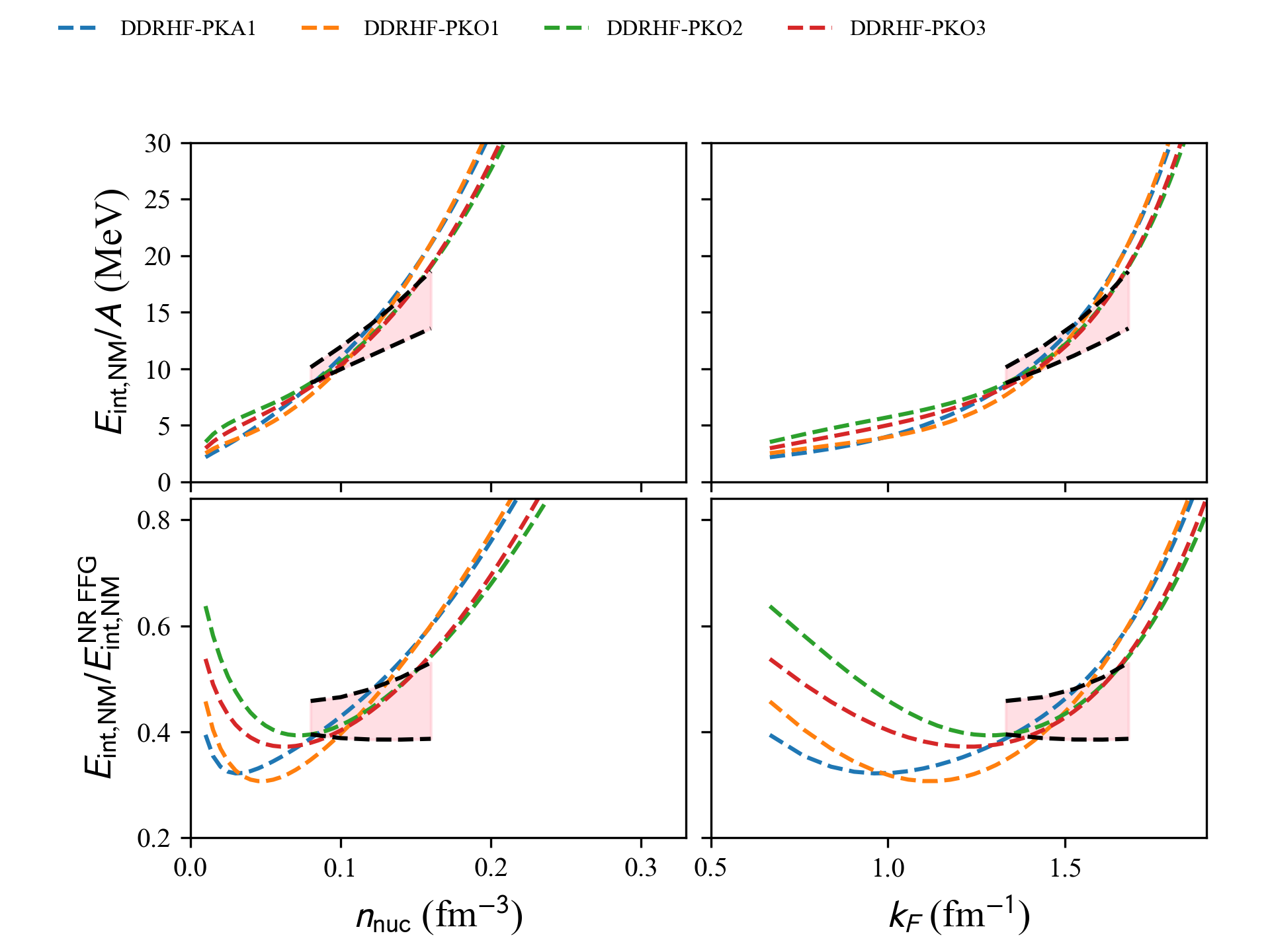}
\caption{Same as Fig.~\ref{fig:pheno:sky:e2a:nm} for the complete list of DDRHF EDF available in \texttt{nuda} toolkit. Figure generated with \texttt{matter\_setupPheno\_plot.py}.}
\label{fig:pheno:ddrhf:e2a:nm}
\end{figure*}

The density-dependent relativistic Hartree-Fock (DDRHF) approach adds the Fock term in the mean field to the DDRH model. By introducing the Fock term, it allows contributions from pions and rho-tensors in uniform matter.

A typical call for a given \texttt{model} and parameter set \texttt{param}, see hereafter, is:
\begin{lstlisting}[language=Python]
micro = nuda.matter.setupPheno( model = `ddrhf', param=`pka1' )
micro.print_outputs()
\end{lstlisting}
Each calculation can be obtained by fixing the variables \texttt{model} and \texttt{param} to one of the following values:\\

\noindent
\texttt{model}=`DDRHF'.\\
\texttt{param}=`PKO1'~\cite{WLong:2006}, `PKO2'~\cite{WLong:2008}, `PKO3'~\cite{WLong:2008}, `PKA1'~\cite{WLong:2007}.

Results in NM are shown in Fig.~\ref{fig:pheno:ddrhf:e2a:nm} for DDRH EDFs. The relativistic DDRH models available in the \texttt{nuda} toolkit are all represented. The models represented are not compatible with the reference band, but they are less repulsive in average than the previously represented NLRH models. The impact of the Fock term is to make the energy per particle more repulsive than DDRH models shown in Fig.~\ref{fig:pheno:ddrh:e2a:nm}. Among the models shown in Fig.~\ref{fig:pheno:ddrhf:e2a:nm}, PKO2 and PKO3 are the closest to the reference band.



\subsection{Energy per particle in NM and SM from a meta-analysis}

We now collect all microscopic and phenomenological models' predictions available in the toolkit to perform meta-analyses.

A typical call for given input variables, see above, is:
\begin{lstlisting}[language=Python]
micro = nuda.matter.setupMicro( model=`2024-NLEFT-AM' )
pheno = nuda.matter.setupPheno( model=`Skyrme', param=`SLy5' )
\end{lstlisting}
The energy per nucleon in SM and NM is defined as an attribute of the objects \texttt{micro} and \texttt{pheno}:
\begin{lstlisting}[language=Python]
print( `in NM' )
print( `E/A(micro):', micro.nm_e2a_int )
print( `E/A(pheno):', pheno.nm_e2a_int )
print( `in SM' )
print( `E/A(micro):', micro.sm_e2a_int )
print( `E/A(pheno):', pheno.sm_e2a_int )
\end{lstlisting}

\begin{figure}[t]
\centering
\includegraphics[scale=0.52]{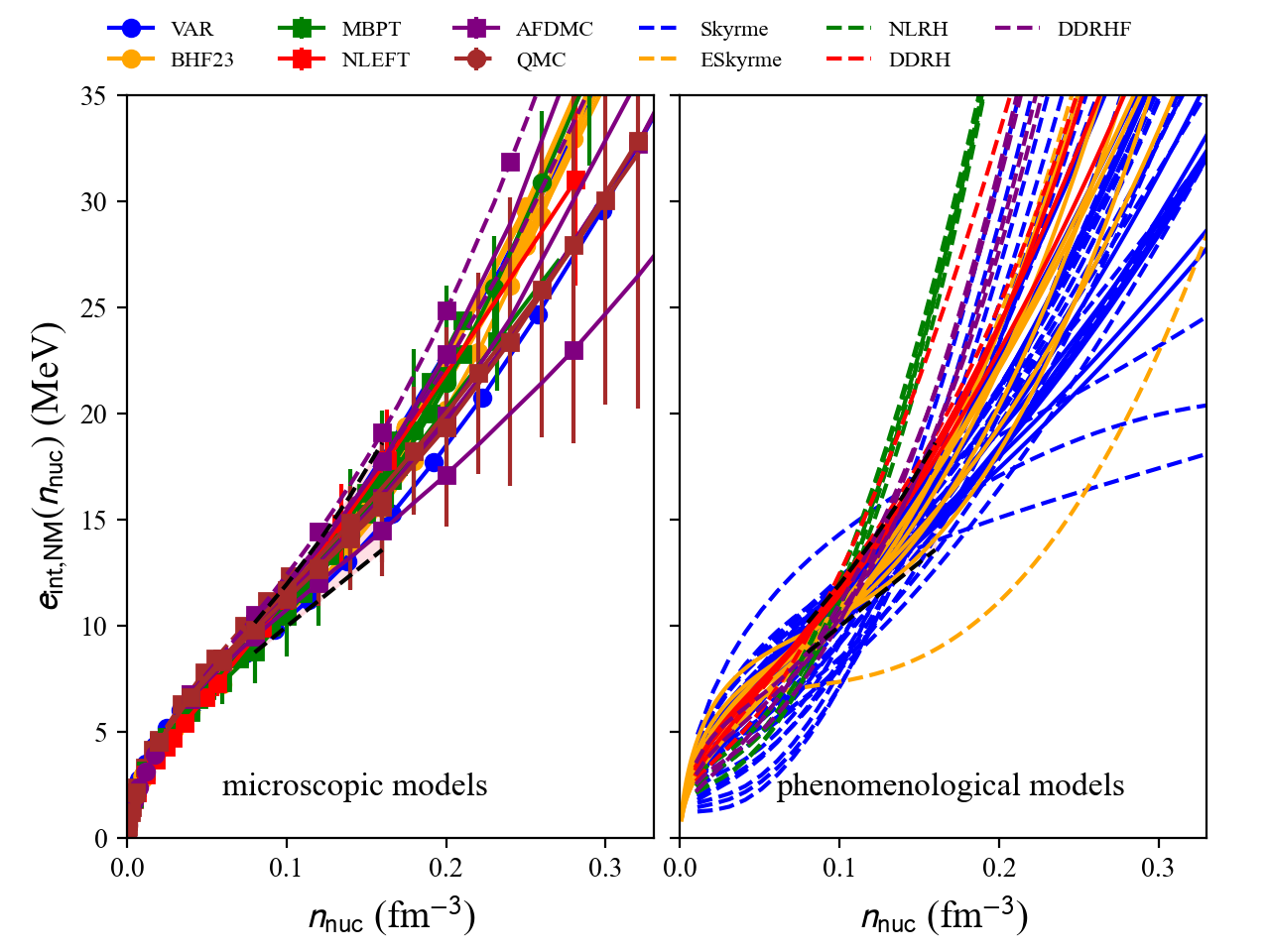}
\caption{Internal energy per nucleon in NM for the list of microscopic (left) and phenomenological (right) models available in the \texttt{nuda} toolkit. Figure generated with \texttt{matter\_all\_plot.py}.}
\label{fig:enm}
\end{figure}

In Fig.~\ref{fig:enm} the model predictions for the internal energy per nucleon in NM for microscopic approaches (left) and phenomenological ones (right) as a function of the nucleon density are compared. Models belonging to the same many-body approach (mb, micro) or model (model, pheno) are shown with the same color. We use the class \texttt{nuda.matter.setupCheck()} to distinguish between the model passing inside the reference band in NM (solid line) from the others being outside (dashed line): all the models drawn in Fig.~\ref{fig:enm} in solid line pass through the reference band in NM. Almost all microscopic models pass through the reference band in NM, while it selects a lot of phenomenological models.

\begin{figure}[t]
\centering
\includegraphics[scale=0.52]{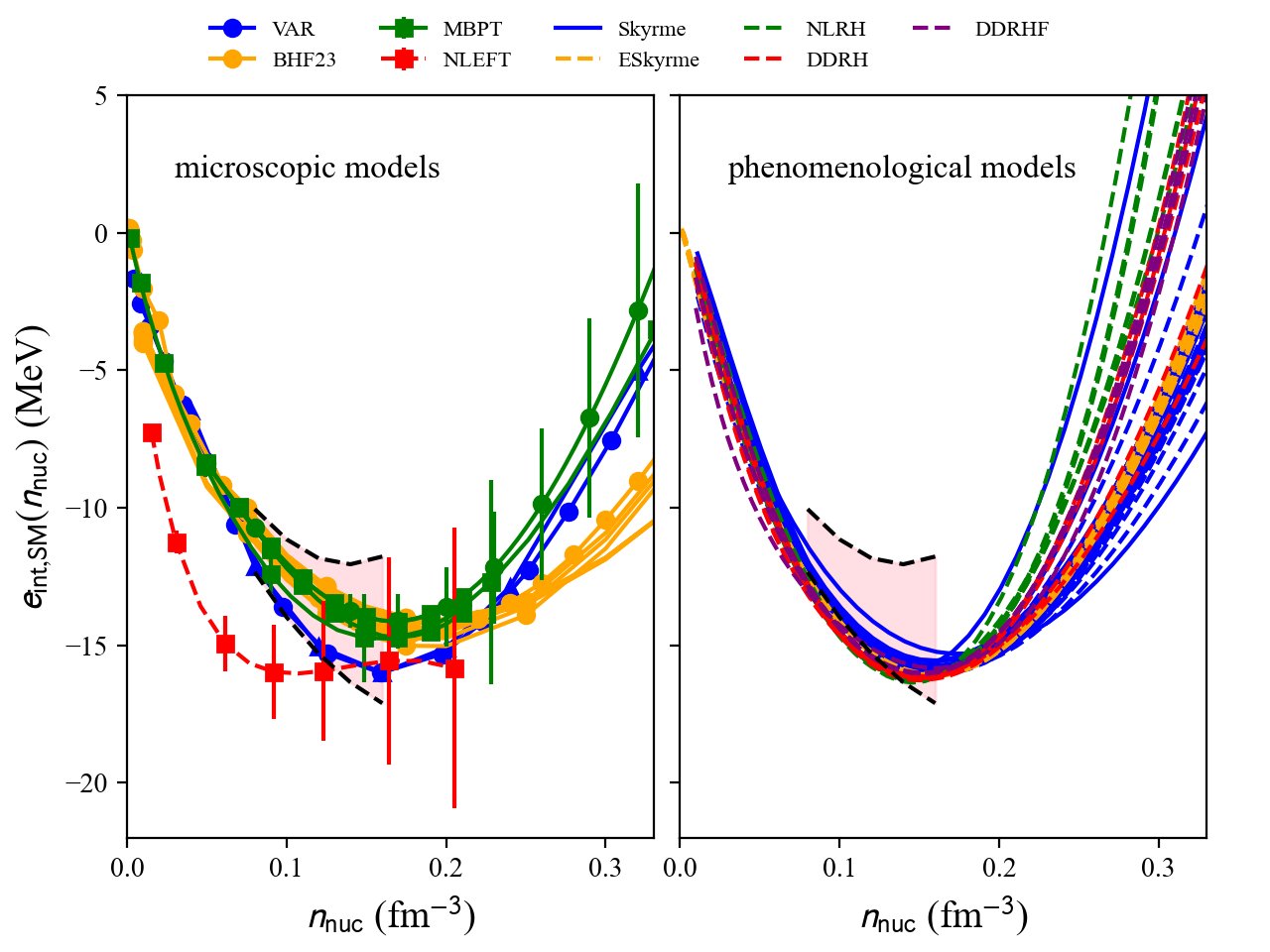}
\caption{Same as Fig.~\ref{fig:enm} for SM. Figure generated with \texttt{matter\_all\_plot.py}.}
\label{fig:esm}
\end{figure}

We now represent in Fig.~\ref{fig:esm} a comparison of the internal energy per nucleon in SM predicted by microscopic approaches (left) and phenomenological ones (right) as a function of the nucleon density. Models belonging to the same many-body approach (micro) or model (pheno) are shown with the same color. Similarly to the NM case, the solid lines in Fig.~\ref{fig:esm} refer to models passing inside the reference band, but now defined in SM. NLEFT predicts that the internal energy per nucleon in SM is more attractive at low density than the reference band. This is most probably due to the strong deuteron correlation in low-density symmetric matter. In NLEFT, there are indeed cluster correlations in the ground state that are present in the results. The dispersion between the phenomenological model is much reduced in SM, compared to NM. The reference band remains, however, quite selective, especially at low density.

\subsection{Symmetry energy in uniform matter}

The symmetry energy is defined as, see also Eq.~\eqref{eq:ffg:esym},
\begin{equation}
e_\sym(n_\nuc) = e_{\NM}(n_\nuc)-e_{\SM}(n_\nuc) \, .
\label{eq:esym}
\end{equation}
The calculation of the symmetry energy requires the energy per nucleon in SM and NM. The microscopic models providing results in NM and SM can be listed in the following way (for a given choice of \texttt{micro\_mb} inside the list provided by \texttt{micro\_mbs}):
\begin{lstlisting}[language=Python]
micro_mbs, micro_mbs_lower=nuda.matter.micro_esym_mbs()
micro_mb=`1998-VAR-AM-APR'
micro_models, micro_models_lower=nuda.matter.micro_esym_models_mb( mb=micro_mb )
\end{lstlisting}
Similarly, for phenomenological models:
\begin{lstlisting}[language=Python]
pheno_models, pheno_models_lower=nuda.matter.pheno_esym_models( )
pheno_model=`nlrh'
pheno_params, pheno_params_lower=nuda.matter.pheno_esym_params( model=pheno_model )
\end{lstlisting}

One should then select the variables \texttt{micro\_model} in the list \texttt{micro\_models}, and \texttt{pheno\_param} from \texttt{pheno\_params} to obtain the symmetry energy from the microscopic and phenomenological models as:
\begin{lstlisting}[language=Python]
microEsym=nuda.matter.setupMicroEsym( model=micro_model )
phenoEsym=nuda.matter.setupPhenoEsym( model=pheno_model, param=pheno_param )
\end{lstlisting}

The symmetry energy is defined as an attribute of the microscopic and phenomenological models: \texttt{microEsym.esym} and \texttt{phenoEsym.esym}. These quantities are calculated for a density mesh defined in \texttt{microEsym.den} and \texttt{phenoEsym.den}.

\begin{figure}[t]
\centering
\includegraphics[scale=0.52]{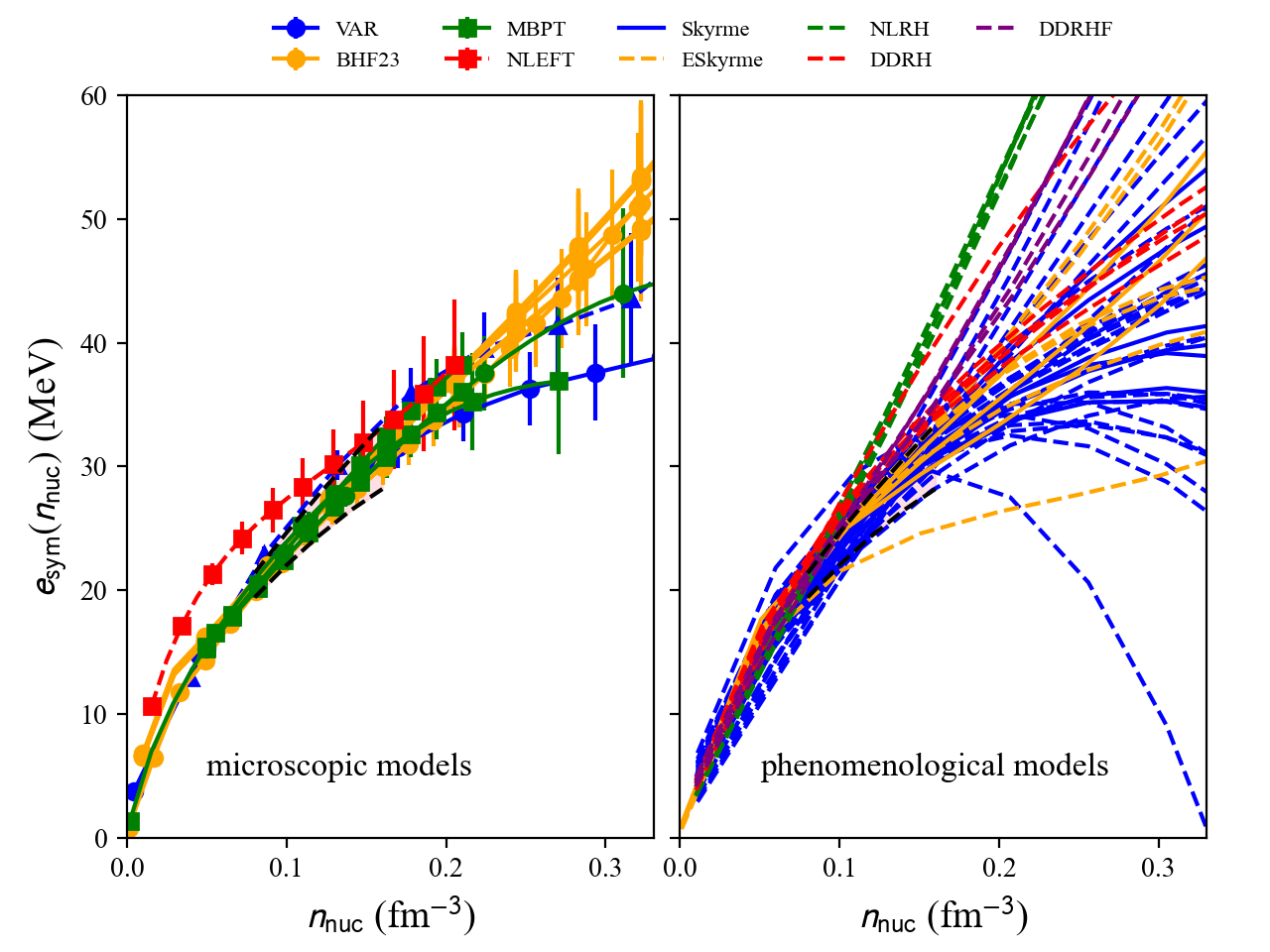}
\caption{Same as Fig.~\ref{fig:enm} for $e_\sym$. Figure generated with \texttt{matter\_all\_plot.py}.}
\label{fig:esym}
\end{figure}

The symmetry energy is shown in Fig.~\ref{fig:esym} for the microscopic models providing results in SM and NM. The models compatible with the reference band are shown as solid lines, while the others are shown as dashed lines. NLEFT predicts a large symmetry energy at low density, as expected from its prediction in SM. The prediction for the symmetry energy from NLEFT is outside the reference band. The other microscopic predictions are quite consistent, while the phenomenological predictions for the symmetry energy are quite dispersed above saturation density. This reflects the large dispersion among the phenomenological models in NM.

\subsection{Nucleon pressure in NM and SM}

The pressure measures the change in energy resulting from a modification of the volume, or density, of matter at constant number of particles. In uniform matter, the nucleon pressure is defined as:
\begin{equation}
p_\nuc(n_\nuc) = n_\nuc^2\frac{\partial e_\nuc(n_\nuc,\delta)}{\partial n_\nuc}\bat_{\delta} \, .
\end{equation}
The nucleon pressure is related to the first derivative of the energy per nucleon, discussed in previous sections. Pressure is a thermodynamic property of the system that counterbalances against compression, e.g., due to gravity, providing the stability of neutron stars. It is therefore important to calculate the pressure.

\begin{figure}[t]
\centering
\includegraphics[scale=0.52]{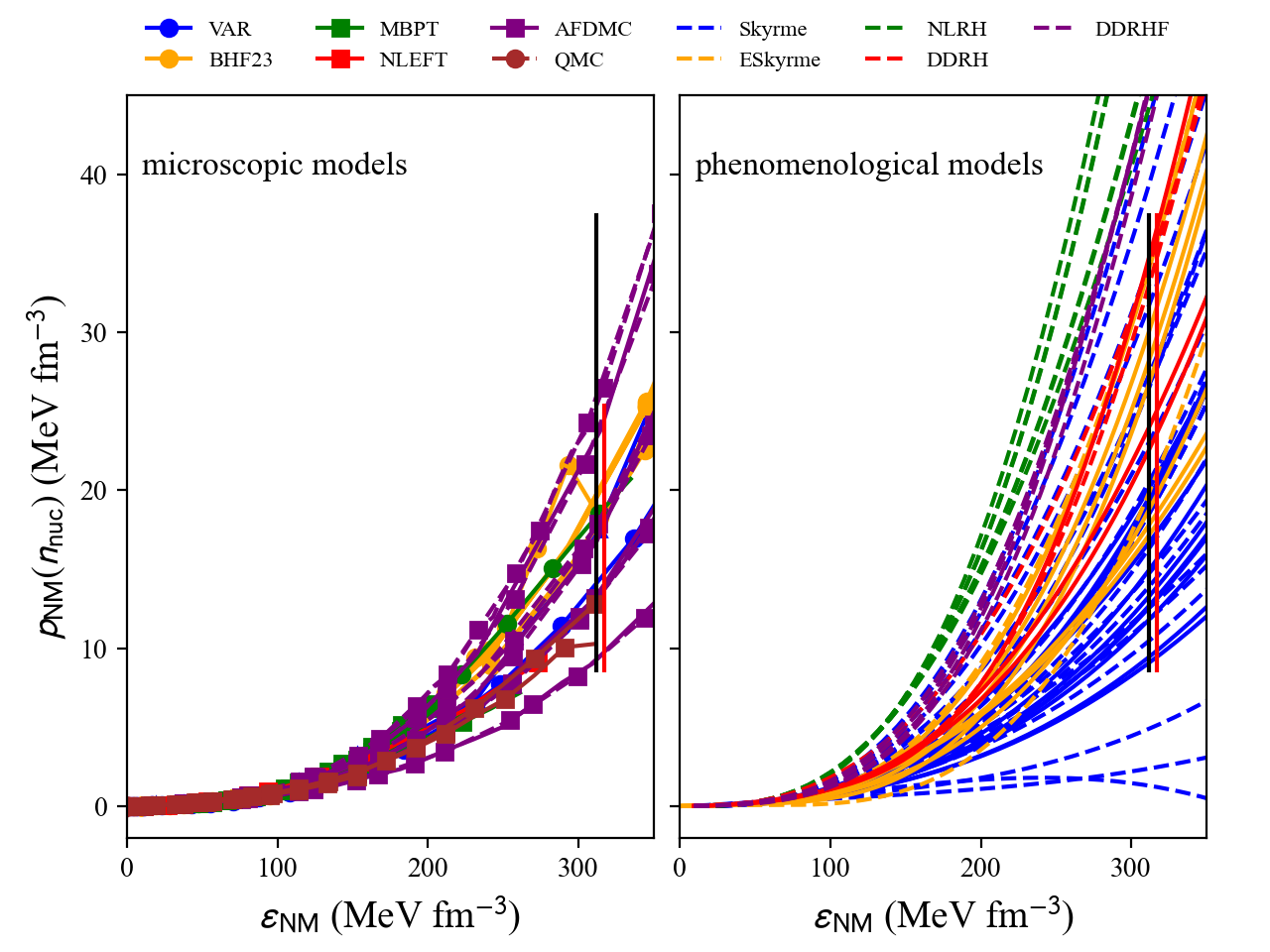}
\caption{Pressure in NM as a function of the energy density $\epsilon$ for the list of microscopic (left) and phenomenological (right) models available in the \texttt{nuda} toolkit. Figure generated with \texttt{matter\_all\_plot.py}.}
\label{fig:pnm}
\end{figure}

A typical call for given input variables, see above, is:
\begin{lstlisting}[language=Python]
micro = nuda.matter.setupMicro( model=`2024-NLEFT-AM' )
pheno=nuda.matter.setupPheno( model=`Skyrme', param=`SLy5' )
\end{lstlisting}
The nucleon pressure in SM and NM is defined as an attribute of the objects \texttt{micro} and \texttt{pheno}:
\begin{lstlisting}[language=Python]
print( `in NM' )
print( `p(micro):', micro.nm_pre )
print( `p(pheno):', pheno.nm_pre )
print( `in SM' )
print( `p(micro):', micro.sm_pre )
print( `p(pheno):', pheno.sm_pre )
\end{lstlisting}

Fig.~\ref{fig:pnm} shows a comparison of the nucleon pressure in NM predicted by microscopic approaches (left) and phenomenological ones (right) as a function of the nucleon energy density. Models belonging to the same many-body approach (micro) or model (pheno) are shown with the same color. Solid lines are associated with models that satisfy the reference band in NM, while dashed lines are for the models that disagree with the reference band in NM. We are interested in the prediction for the nucleon pressure at twice saturation energy-density ($\epsilon_\sat \approx 155$~MeV~fm$^{-3}$). The vertical bars represent the dispersion between the model predictions, for those in agreement with the reference band in NM. The red bars are different for microscopic and phenomenological models, while the black ones are identical for both cases and include the red ones. Note that the dispersion is larger for the phenomenological models compared to the microscopic ones. This was already observed for the energy per particle in NM, see Fig.~\ref{fig:enm} for instance.

\begin{figure}[t]
\centering
\includegraphics[scale=0.52]{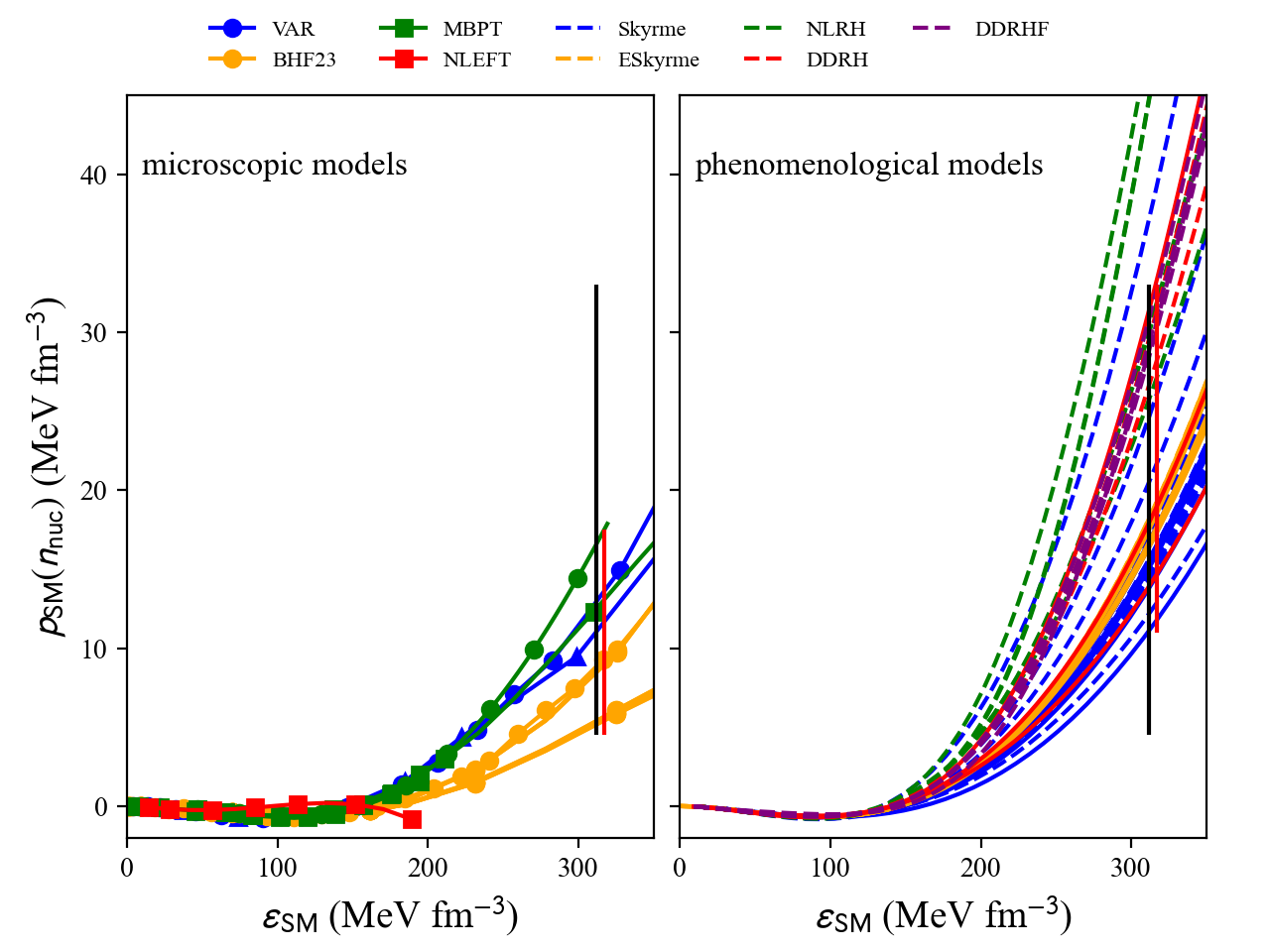}
\caption{Same as Fig.~\ref{fig:pnm} for SM. Figure generated with \texttt{matter\_all\_plot.py}.}
\label{fig:psm}
\end{figure}

\begin{figure*}[t]
\centering
\includegraphics[scale=0.9]{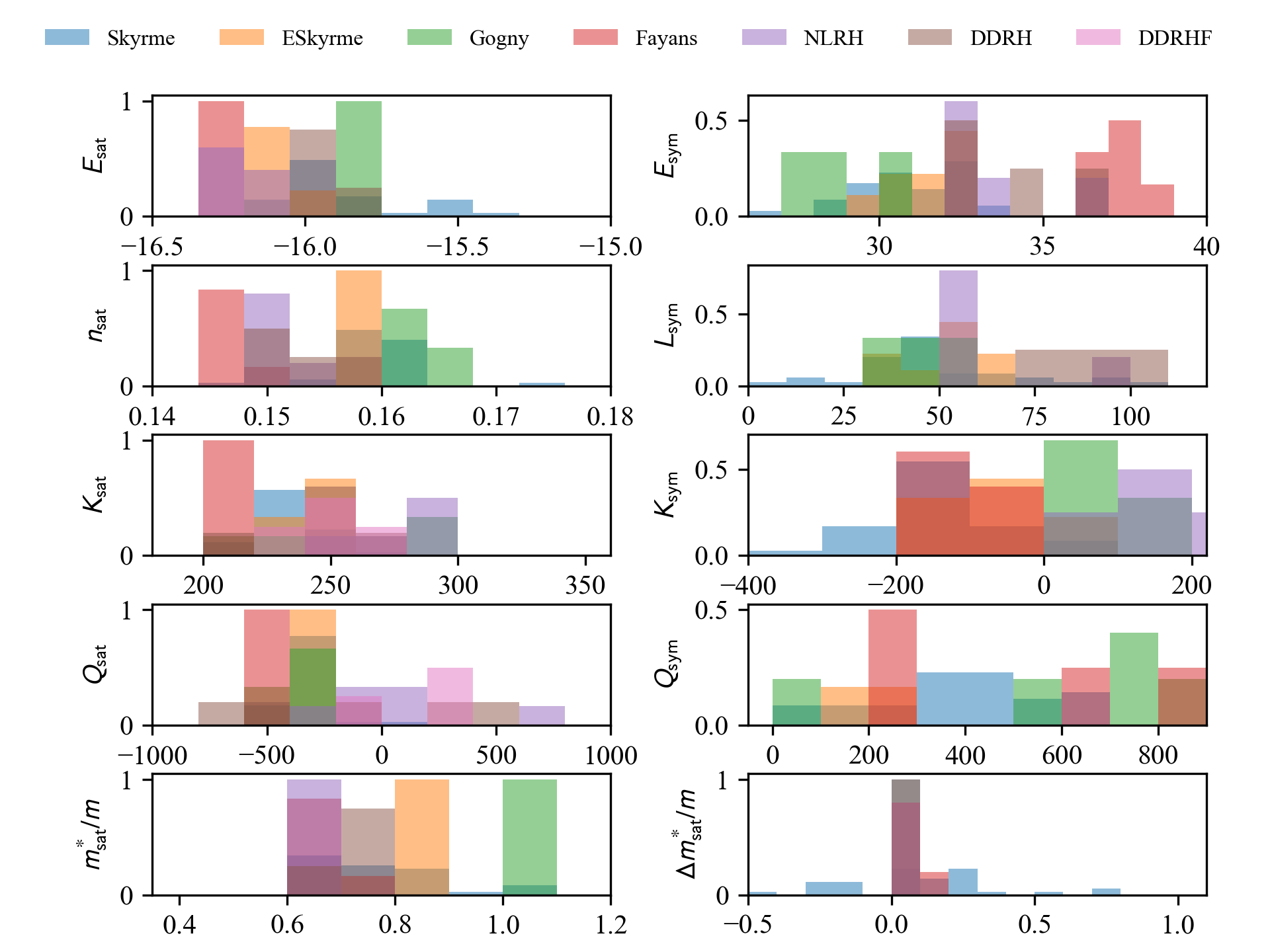}
\caption{Distribution of NEP for phenomenological models available in the \texttt{nuda} toolkit. Figure generated with \texttt{matter\_setupNEPDiststats\_plot.py}.}
\label{fig:nep}
\end{figure*}

Fig.~\ref{fig:psm} is similar to Fig.~\ref{fig:pnm}, except that it shows the pressure in SM instead of NM. The reference band in NM is still employed to separate the models in solid lines from the ones in dashed lines. While the model dispersion is larger for the phenomenological models compared to the microscopic ones (as in NM), the softest EoS predicted by microscopic models (BHF23) are not present in the phenomenological models encoded in the \texttt{nuda} toolkit. It could be due to the finite and low number of phenomenological models in the toolkit. It therefore illustrates that it is necessary to combine microscopic and phenomenological models for our meta-analysis. The vertical bars are constructed as explained in the discussion of Fig.~\ref{fig:pnm}.
The size of the bars in NM and SM is reported in Table \ref{table:pre}, discussed in Sec.~\ref{sec:eos:beta}.

\subsection{Nuclear empirical parameters}
\label{sec:mat:nep}

The nuclear empirical parameters (NEP) are defined as the derivatives for the energy per nucleon in SM, $e_\SM$, and the symmetry energy $e_\sym$ in the following way:
\begin{eqnarray}
P^N_{\sat} &=& \left(3n_\sat\right)^N \frac{\partial^N e_\SM(n_\nuc)}{\partial n_\nuc^N}\bat_{n_\sat} \, , \\
P^N_{\sym} &=& \left(3n_\sat\right)^N \frac{\partial^N e_\sym(n_\nuc)}{\partial n_\nuc^N}\bat_{n_\sat} \, ,
\end{eqnarray}
where $P=(E,L,K,Q,Z)$.
We have $L_\sat=0$, since the value of $n_\sat$ is defined as the density for which the pressure in SM is zero.

The complete list of models for which NEP is available is provided in the following way:
\begin{lstlisting}[language=Python]
print( nuda.matter.nep_models( ) )
\end{lstlisting}
For the moment, only phenomenological models are listed.

\begin{table*}[t]
\begin{center}
\caption{Centroids and standard deviations averaging over several models from the \texttt{nuda} toolkit. This table has been generated with \texttt{matter\_setupNEPstat\_script.py} and \texttt{matter\_setupNEPStats\_script.py}.}
\label{table:NEP}
\tabcolsep=0.24cm
\def\arraystretch{1.5}
\begin{tabular}{rrrrrrrrrrrrr}
\hline\noalign{\smallskip}
NEP & $E_{\sat}$ & $n_{\sat}$ & $K_{\sat}$ & $Q_{\sat}$ & $Z_{\sat}$ &  $E_{sym}$ & $L_{sym}$ & $K_{sym}$ & $Q_{sym}$ & $Z_{sym}$ &  $m^*_{sat}/m$ & $\Delta m^*_{sat}/m$ \\
 & MeV & fm$^{-3}n$ & MeV & MeV & MeV &  MeV & MeV & MeV & MeV & MeV &   &  \\
\hline\noalign{\smallskip}
\multicolumn{13}{l}{Skyrme (38): `BSK14', `BSK16', `BSK17', `BSK27', `BSkG1', `BSkG2',`F-', `F+', `F0', `FPL', `LNS', }\\
\multicolumn{13}{l}{`LNS1', `LNS5', `NRAPR', `RATP', `SAMI', `SGII', `SIII', `SKGSIGMA', `SKI2', `SKI4', `SKMP', `SKMS',} \\
\multicolumn{13}{l}{ `SKO', `SKOP', `SKP', `SKRSIGMA', `SKX', `Skz2', `SLY4', `SLY5', `SLY230A', `SLY230B', `SV', `T6',}\\
\multicolumn{13}{l}{ `T44', `UNEDF0', `UNEDF1'.}\\
 centroid & -15.89 & 0.159 &  238.0 & -350 & 1445 &  30.88 & 49.3 & -135 &  372 & -2180 & 0.77 &  0.126 \\
 std.dev. & 0.18 & 0.004 &  25.7 & 86 & 494 &  1.52 & 21.2 & 88 &  184 & 1044 & 0.14 &  0.300 \\
\multicolumn{13}{l}{GSkyrme (5): `SkK180', `SkK200', `SkK220', `SkK240', `SkKM'.}\\
 centroid & -15.83 & 0.156 &  212.0 & -936 & - &  30.00 & 16.2 & -368 &  195 & - & 1.10 &  1.876 \\
 std.dev. & 0.03 & 0.003 &  20.4 & 677 & - &  0.00 & 13.7 & 114 &  249 & - & 0.00 &  0.000 \\
\multicolumn{13}{l}{ESkyrme (8): `BSk22', `BSk24', `BSk25', `BSk26', `BSk31', `BSk32', `BSkG3', `BSkG4'.}\\
  centroid & -16.08 & 0.158 &  241.7 & -312 & - &  31.00 & 52.3 & -53 &  - & - & 0.83 &  0.000 \\
 std.dev. & 0.03 & 0.000 &  3.3 & 35 & - &  1.00 & 9.7 & 64 &  - & - & 0.03 &  0.000 \\
\multicolumn{13}{l}{NLRH (6): `NL-SH', `NL3', `NL3II', `PK1', `PK1R', `TM1'.}\\
centroid & -16.28 & 0.147 &  288.3 & 98 & 5734 &  37.40 & 116.2 & 72 &  24 & -3218 & 0.68 &  0.062 \\
std. dev. & 0.03 & 0.002 &  31.2 & 271 & 2920 &  0.80 & 3.8 & 26 &  129 & 940 & 0.01 &  0.002 \\
\multicolumn{13}{l}{DDRH (5): `DDME1', `DDME2', `DDMEd', `PKDD', `TW99'.}\\
centroid & -16.20 & 0.152 &  243.8 & -121 & 4248 &  33.48 & 61.1 & -102 &  578 & -4206 & 0.66 &  0.064 \\
std. dev. & 0.06 & 0.001 &  14.7 & 473 & 390 &  1.73 & 14.8 & 17 &  296 & 1970 & 0.02 &  0.033 \\
\multicolumn{13}{l}{DDRHF (4): `PKA1', `PKO1', `PKO2', `PKO3'.}\\
centroid & -15.97 & 0.154 &  248.1 & 389 & 5269 &  33.97 & 90.0 & 128 &  523 & -9955 & 0.73 &  0.015 \\
std. dev. & 0.08 & 0.004 &  11.6 & 350 & 838 &  1.37 & 11.1 & 51 &  237 & 4156 & 0.03 &  0.003 \\
\multicolumn{13}{l}{Total (74): `Skyrme', `GSkyrme', `ESkyrme', `Gogny', `Fayans', `NLRH', `DDRH', `DDRHF'.}\\
centroid & -15.97 & 0.157 &  242.1 & -298 & 2542 &  31.67 & 56.3 & -103 &  349 & -3141 & 0.79 &  0.248 \\
std. dev. & 0.20 & 0.005 &  29.0 & 369 & 2058 &  2.52 & 29.2 & 137 &  251 & 2670 & 0.15 &  0.553 \\
\noalign{\smallskip}\hline
\end{tabular}
\end{center}
\end{table*}

A typical call for a given \texttt{model} and parameter set \texttt{param}, see hereafter, is:
\begin{lstlisting}[language=Python]
nep = nuda.matter.setupNEP( model=`ddrhf', param=`pka1' )
\end{lstlisting}
For the choice of variables \texttt{model} and \texttt{param}, the NEP can be obtained in the following way:
\begin{lstlisting}[language=Python]
print( `NEP' )
print( `sat:', nep.Esat, nep.nsat, nep.Ksat, nep.Qsat, nep.Zsat )
print( `sym:', nep.Esym, nep.Lsym, nep.Ksym, nep.Qsym, nep.Zsym )
\end{lstlisting}

The statistical distribution of NEP for a given \texttt{model} can be obtained by calling the class \texttt{nuda.matter.setupNEPStat\_model()} in the following way:
\begin{lstlisting}[language=Python]
dist = nuda.matter.setupNEPStat_model( model=`Skyrme' )
dist.print_outputs()
mdist.print_latex( )
\end{lstlisting}
It is performed in the sample script \texttt{matter\_setupNEPStat\_script.py}.

The statistical distribution of NEP is shown in Fig.~\ref{fig:nep} for the \texttt{model} available in the \texttt{nuda} toolkit. The weights of the models have been adjusted to make this distribution independent of the number of parameter sets for each model. For low-order NEP, such as $E_\sat$, $n_\sat$, and $E_\sym$, the predictions of the phenomenological models are rather well grouped, even for $E_\sym$, for which the distribution is the broadest. Then comes $K_\sat$ and $L_\sym$, which are less well-known. The NEP $Q_\sat$, $K_\sym$, and $Q_\sym$ are yet totally unknown. Note, however, that for $Q_\sat$ the distribution for non-relativistic models is different from the distribution from relativistic ones: the NR models prefer a large and negative value for $Q_\sat$ while the relativistic models prefer $Q_\sat$ to be large and positive. Such a difference between relativistic and NR models has already been noted in Ref.~\cite{JMargueron:2018a}, and could also be at the origin of the different values for $K_\sat$ that these models prefer~\cite{EKhan:2012,EKhan:2013,JMargueron:2019}.

The statistical analysis summing over a set of models is performed in the sample script \texttt{matter\_setupNEPStats\_script.py}:
\begin{lstlisting}[language=Python]
dist = nuda.matter.setupNEPStat_models( models = ['Skyrme', `Gogny'] )
dist.print_latex( )
\end{lstlisting}
The input variable \texttt{models} contains a list of models for which we want to perform the statistical averaging. It can be a choice of a few models, as shown in the previous example, or it can be made over all models available in the \texttt{nuda} toolkit. It the latter case, the script is:
\begin{lstlisting}[language=Python]
models, models_lower=nuda.matter.nep_models()
dist = nuda.matter.setupNEPStat_models( models=models )
dist.print_latex( )
\end{lstlisting}

Results are shown in Table~\ref{table:NEP}, where the average of the NEP for each class of phenomenological model is provided, as well as a global average over all available models (meta-average). The NEPs $Q_\sat$ and $K_\sym$ are found to be very model-dependent. The centroid for $Q_\sat$ is negative for all non-relativistic models, and it is found to be positive for relativistic models, except `DDRH' ones, for which the distribution is compatible with zero. For $K_\sym$, the distribution of non-relativistic models is peaked at negative values, while relativistic models, except `DDRH' predict it to be positive.
Since there is a large number of non-relativistic Skyrme models in the \texttt{nuda} toolkit, the meta-analysis is influenced by their prediction, and the peaks for $Q_\sat$ and $K_\sym$ are negative. However, the standard deviation is large, reflecting the contribution of relativistic models for these parameters.

The Landau effective mass and the splitting of the Landau effective mass are introduced in the following Sec.~\ref{sec:ms}. For most of the models, the effective mass $m_\sat^*/m$ is lower than 1.0, except for the generalized Skyrme models (GSkyrme) for which it is fixed to be 1.1. The splitting of the effective mass is peaked at positive values, but its standard deviation reflects the fact that many models predict it to be negative.

\subsection{Landau effective mass}
\label{sec:ms}

\begin{figure}[t]
\centering
\includegraphics[scale=0.52]{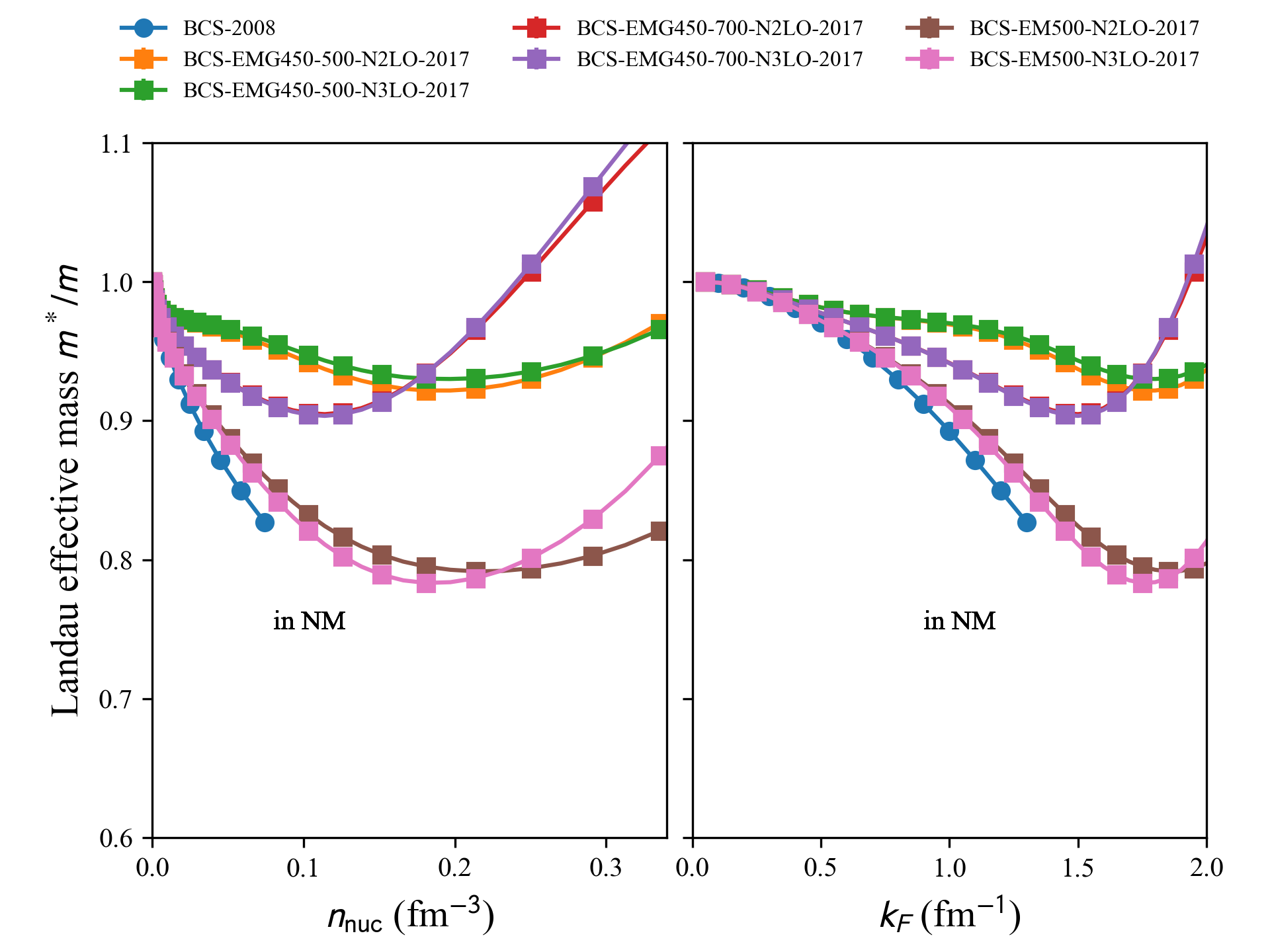}
\caption{Landau effective mass in NM predicted by the microscopic models available in the \texttt{nuda} toolkit. Figure generated with \texttt{matter\_setupMicro\_effmass\_plot.py}.}
\label{fig:effmass:nm}
\end{figure}

The \texttt{nuda} toolkit provides the effective mass calculated by the authors (when it is calculated). It is related to the group velocity as:
\begin{equation}
\frac{m_q}{m_q^*} = \frac{m_q}{\hbar^2} \left(\frac{1}{k}\frac{d e_q(k)}{d k}\right)\bat_{k=k_{F_q}} \, ,
\label{eq:effmass}
\end{equation}
where $e_q(k)$ is the single particle energy defined as
\begin{equation}
e_q(k) = \frac{\hbar^2 k^2}{2m_q}+\Sigma_q(k)+\Sigma_{q,0}\, ,
\end{equation}
where $\Sigma_q(k)$ is the momentum dependent self-energy. The quantity $m_q^*$ is also called the Landau effective mass.

The effective mass is often represented in terms of two quantities $\kappa_\sat$ ($=\kappa_s$) and $\kappa_\sym$, where $\kappa_\sym=\kappa_\sat-\kappa_v$, where $\kappa_v$ is the enhancement factor 
of the Thomas-Reiche-Kuhn sum rule~\cite{OBohigas:1979}.
We have:
\begin{equation}
\frac{m_q}{m^*_q(n_\nuc,\delta)} = 1 + \left( \kappa_\sat
+\tau_3 \kappa_\sym \delta \right) \frac{n_\nuc}{n_\sat} \, .
\end{equation}
We obtain in SM, $m_\sat^*/m=1/(1+\kappa_\sat)$, and the isospin splitting of the effective mass in NM:
\begin{equation}
\frac{\Delta m_\sat^*}{m} = \frac{m_n^*}{m_n} - \frac{m_p^*}{m_p} 
= -\frac{2\kappa_\sym}{(1+\kappa_\sat)^2-\kappa_\sym^2} \, .
\end{equation}

The complete list of available models is given with the following instructions:
\begin{lstlisting}[language=Python]
models, models_lower, models_all, models_all_lower=nuda.matter.micro_effmass_models( matter=`NM' )
print( models )
\end{lstlisting}

\begin{figure}[t]
\centering
\includegraphics[scale=0.52]{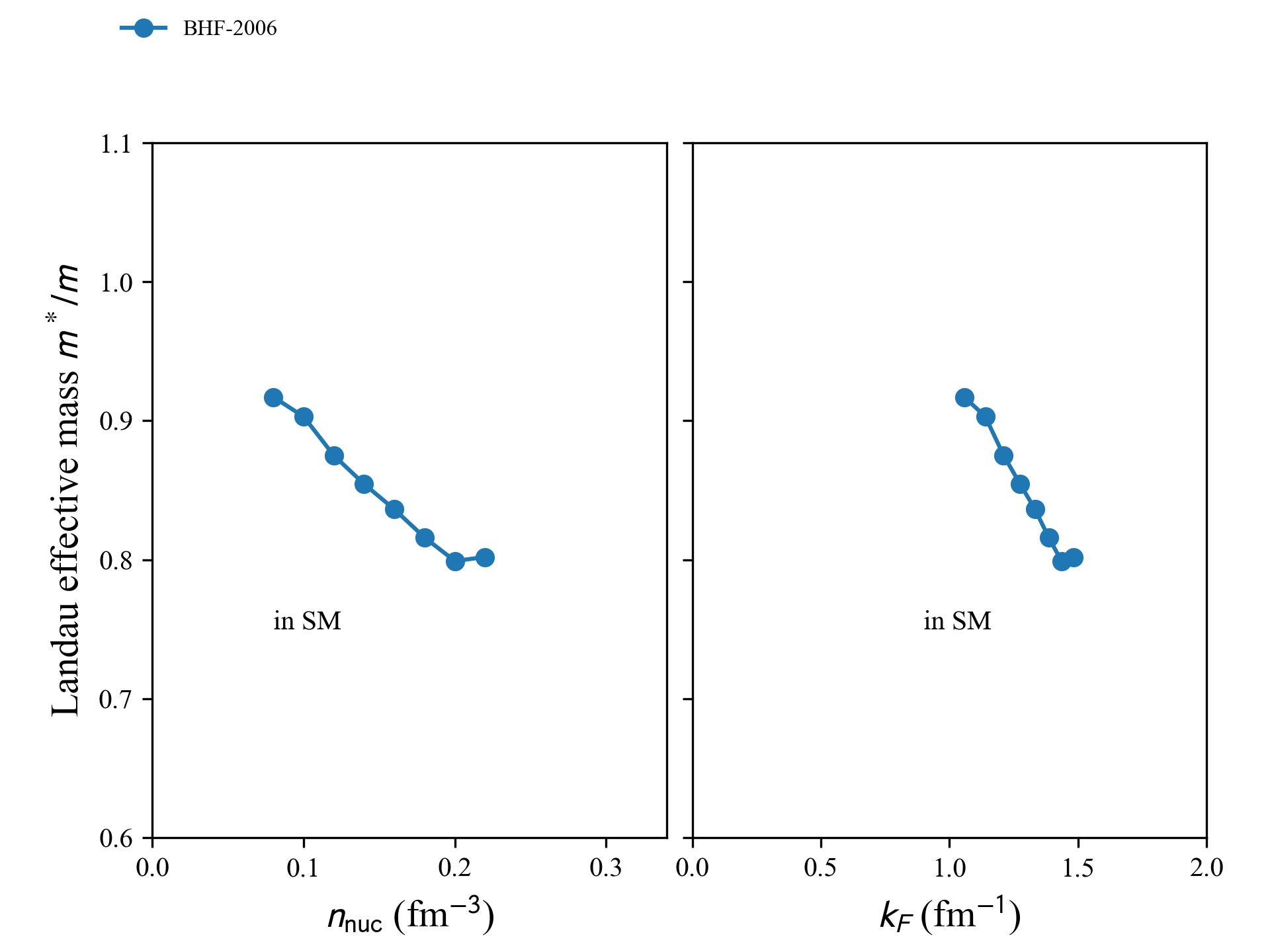}
\caption{Same as Fig.~\ref{fig:effmass:nm} in SM. Figure generated with \texttt{matter\_setupMicro\_effmass\_plot.py}.}
\label{fig:effmass:sm}
\end{figure}

A typical call for given input variables, see above, is:
\begin{lstlisting}[language=Python]
ms = nuda.matter.setupMicroEffmass( model=model, matter=`NM' )
ms.print_outputs()
\end{lstlisting}
where the variable \texttt{matter} can be `NM' (default), `SM' or `AM'. Each calculation can be obtained by fixing the variables \texttt{model} to one of the following values:\\

\noindent
\texttt{model}=`2008-BCS-NM'.\\
Presented in Sec.~\ref{sec:unif:gap}.\\

\noindent
\texttt{model}=`2017-MBPT-NM-GAP-EMG-450-500-N2LO', `2017-MBPT-NM-GAP-EMG-450-500-N3LO', `2017-MBPT-NM-GAP-EMG-450-700-N2LO', `2017-MBPT-NM-GAP-EMG-450-700-N3LO', `2017-MBPT-NM-GAP-EM-500-N2LO', `2017-MBPT-NM-GAP-EM-500-N3LO.\\
Presented in Sec.~\ref{sec:unif:gap}.\\

\noindent
\texttt{model}=`2022-AFDMC-NM'.\\
Presented in Sec.~\ref{sec:unif:gap}.\\

\begin{figure}[t]
\centering
\includegraphics[scale=0.52]{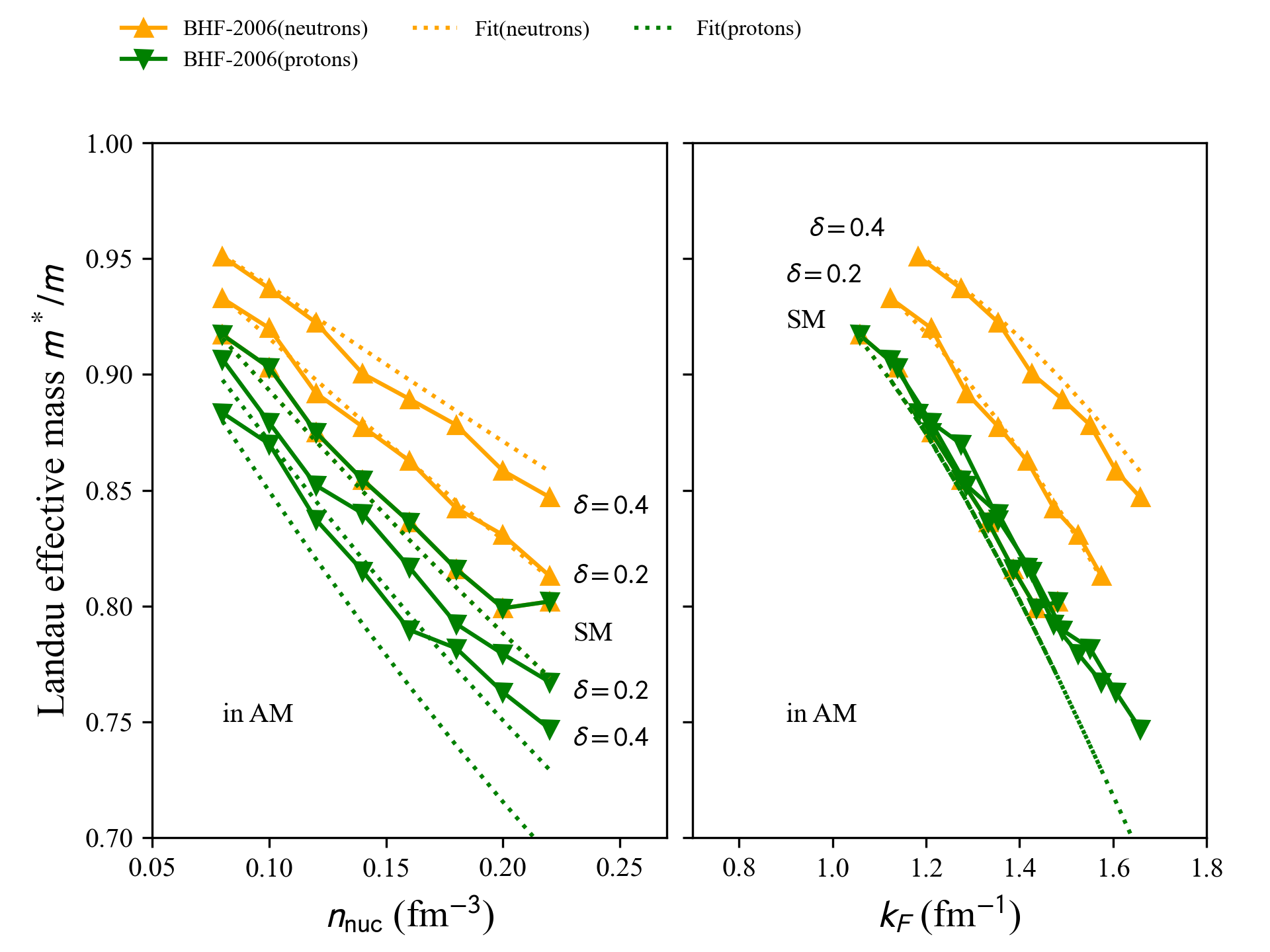}
\caption{Same as Fig.~\ref{fig:effmass:nm} in AM. Figure generated with \texttt{matter\_setupMicro\_effmass\_plot.py}.}
\label{fig:effmass:am}
\end{figure}

The object \texttt{ms} contains the effective masses. The entire list of properties is obtained in the following way: 
\begin{lstlisting}[language=Python]
print(ms.__dict__)
\end{lstlisting}

Output quantities are, for instance: \texttt{ms.nm\_effmass} and \texttt{ms.sm\_effmass} for the nucleon effective masses in NM and SM.
\texttt{ms.am02\_effmass\_n} (\texttt{ms.am02\_effmass\_p}) for the neutron (proton) effective mass with $\delta=0.2$, and
\texttt{ms.am04\_effmass\_n} (\texttt{ms.am04\_effmass\_p}) for $\delta=0.4$.

The Landau effective masses are shown in Fig.~\ref{fig:effmass:nm} (for NM), Fig.~\ref{fig:effmass:sm} (for SM), and Fig.~\ref{fig:effmass:am} (for AM) for a set of microscopic models.
Based on the BHF results in AM, we suggest the following empirical relation:
\begin{equation}
\frac{m^*_q}{m_q}(n_\nuc,\delta) = \left[1+0.2\left(\frac{k_{F\bar{q}}}{k_{F_\sat}}\right)^{3.5}\right]^{-1}\,,
\label{eq:ms:emp}
\end{equation}
where $k_{F\bar{q}}$ is $k_{Fn}$ for $m^*_p/m_p$, and vice-versa. The empirical relation~\eqref{eq:ms:emp} is shown in Fig.~\ref{fig:effmass:am} for $m^*_n/m_n$ and $m^*_p/m_p$ as a function of the density $n_\text{nuc}$ or Fermi momentum $k_F$ and isospin parameter $\delta$ (dotted lines).

The empirical relation~\eqref{eq:ms:emp} can be obtained from \texttt{nuda} toolkit in the following way:
\begin{lstlisting}[language=Python]
den=np.linspace(0.01,0.3,20)
delta=0.0
ms_n, ms_p=nuda.matter.effmass_emp( den, delta, mb=`BHF' )
\end{lstlisting}

\subsection{Landau parameters}

\begin{figure}[t]
\centering
\includegraphics[scale=0.52]{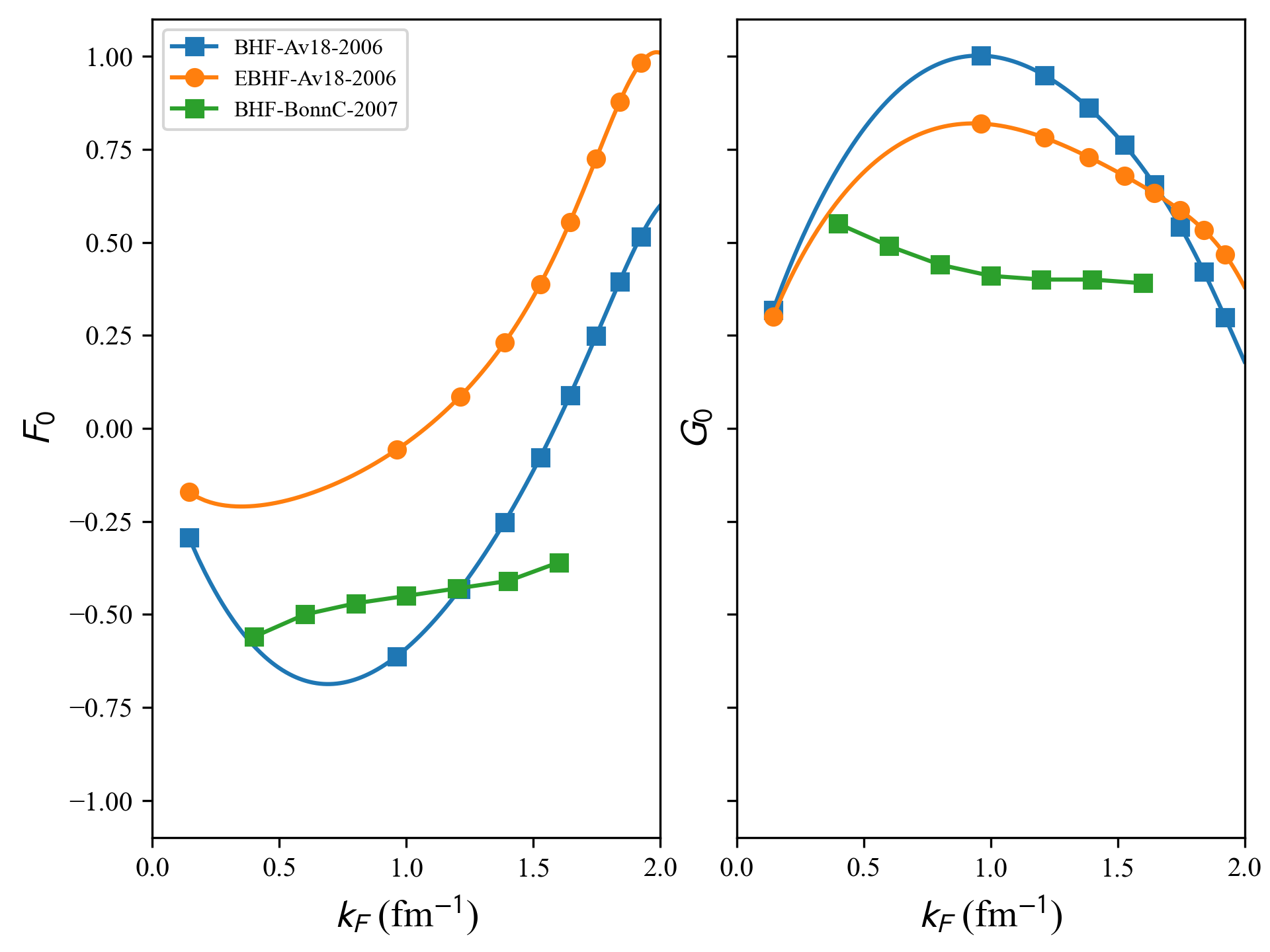}
\caption{$L=0$ Landau parameters in NM predicted by BHF calculations as given by the models available in the \texttt{nuda} toolkit. Figure generated with \texttt{matter\_setupMicro\_LP\_plot.py}.}
\label{fig:matter:NM:LP0}
\end{figure}

\begin{figure*}[t]
\centering
\includegraphics[scale=0.9]{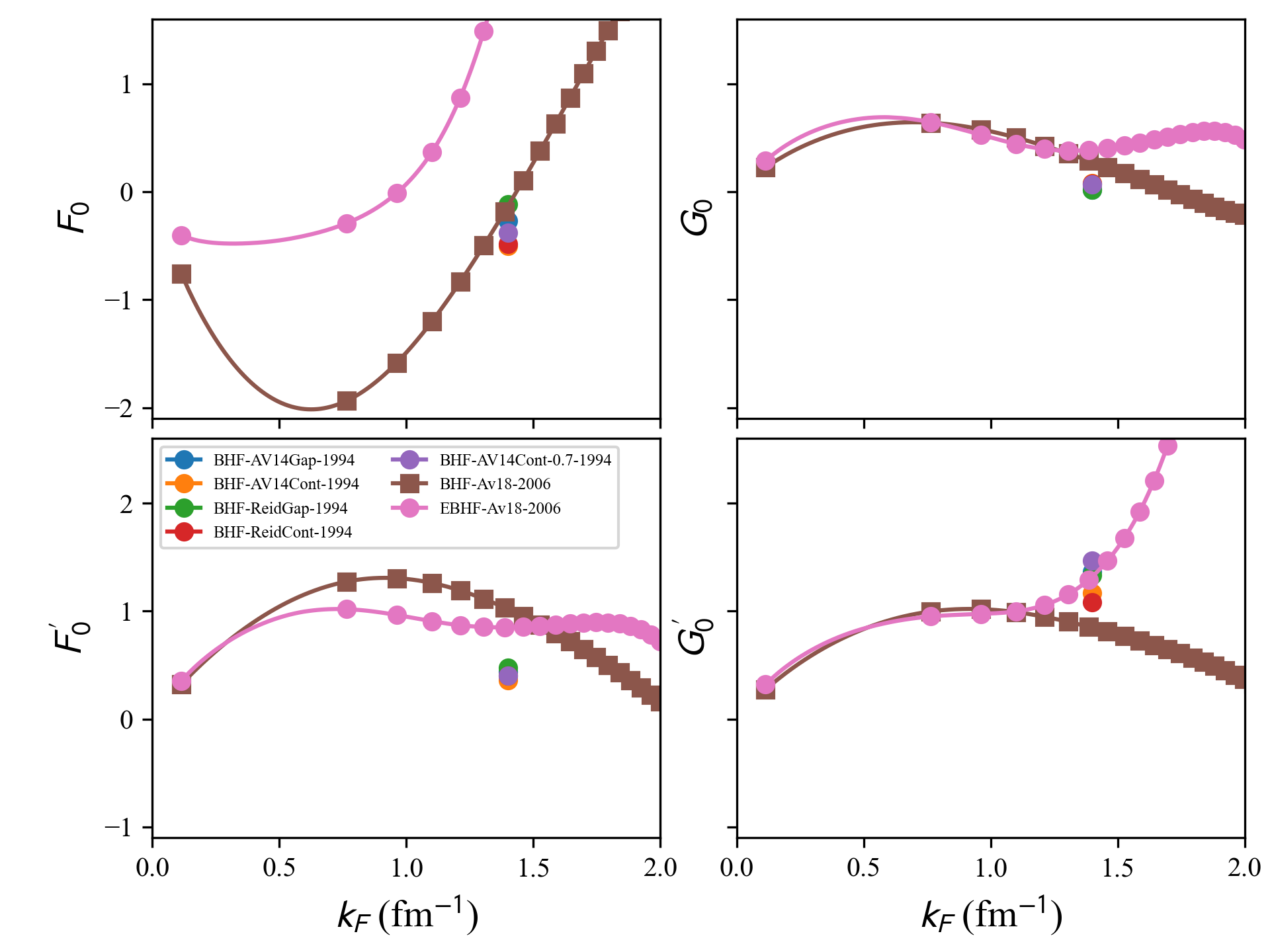}
\caption{$L=0$ Landau parameters in SM predicted by BHF calculations as given by the models available in the \texttt{nuda} toolkit. Figure generated with \texttt{matter\_setupMicro\_LP\_plot.py}.}
\label{fig:matter:SM:LP0}
\end{figure*}

The Landau parameters represent the residual nuclear interaction calculated at the Fermi momentum $k_F$ and decomposed into partial waves. They are given as:
\begin{eqnarray}
V_\text{ph}(k,k^\prime) &=& \frac 1 {N_0} \sum_{L=0}^\infty P_L(\cos \theta) 
\Big[ F_L + F^\prime_L \tau_1\cdot\tau_2 \nonumber \\
&&\hspace{0.5cm}+ G_L \sigma_1\cdot\sigma_2
+ G_L^\prime (\tau_1\cdot\tau_2)(\sigma_1\cdot\sigma_2) \Big]\,,
\end{eqnarray}
where $P_L$ is the Legendre polynomial for the angular momentum $L$. We have $\vert k\vert=\vert k^\prime \vert=k_F$ and $\cos\theta = k\cdot k^\prime$. The density of states $N_0$ reads
$N_0=g m^* k_F /(2\hbar^2\pi^2)$.

The complete list of available microscopic predictions for the Landau parameters is given with the following instruction:
\begin{lstlisting}[language=Python]
print( nuda.matter.models_micro_LP( ) )
\end{lstlisting}
A typical call for given input variables, see above, is:
\begin{lstlisting}[language=Python]
lp=nuda.matter.setupMicroLP( model=`1994-BHF-SM-LP-AV14-GAP' )
lp.print_outputs()
\end{lstlisting}
Results can be obtained by fixing the variables \texttt{model} to one of the following values:\\

\noindent
\texttt{model}=`1994-BHF-SM-LP-AV14-GAP'~\cite{MBaldo:1994}, `1994-BHF-SM-LP-AV14-CONT'~\cite{MBaldo:1994}, `1994-BHF-SM-LP-REID-GAP'~\cite{MBaldo:1994}, `1994-BHF-SM-LP-REID-CONT'~\cite{MBaldo:1994}, `1994-BHF-SM-LP-AV14-CONT-0.7'~\cite{MBaldo:1994}, `2006-BHF-SM-Av18', `2006-EBHF-SM-Av18'.\\
In SM.\\

\noindent
\texttt{model}=`2007-BHF-NM-BONNC'~\cite{ASedrakian:2007}, `2006-BHF-SM-Av18', `2006-EBHF-SM-Av18'.\\
In NM.\\

The object \texttt{lp} contains the Landau parameters, some of the attributes listed by the command \texttt{lp.print\_outputs()}. The entire list of properties is obtained in the following way: 
\begin{lstlisting}[language=Python]
print(lp.__dict__)
\end{lstlisting}

Output quantities are, for instance: \texttt{lp.sm\_LP[cha][ell]} where \texttt{cha} can be `F', `G', `Fp' or `Gp' for the different channels in SM and \texttt{ell} is the angular momentum 0,1, etc. In NM, we have: \texttt{lp.nm\_LP[cha][ell]} with \texttt{cha}=`F' or `G'.

Predictions for the Landau parameters in NM are shown in Fig.~\ref{fig:matter:NM:LP0} and in Fig.~\ref{fig:matter:SM:LP0} for SM. These predictions have been performed for BHF many-body approach, and one prediction is made for an extended BHF approach.

\subsection{Experimental constraints from heavy-ion collisions}
\label{sec:nuc:hic}

Heavy-ion collisions are reactions between heavy nuclei, which are described and analyzed using transport simulations, see, for instance, a recent white paper~\cite{ASorensen:2024}. In the following, we do not describe HIC performed in laboratory experiments and their analysis, but simply provide results in terms of EoS inference performed by various authors.

The complete list of available inferences is given with the following instructions:
\begin{lstlisting}[language=Python]
print(nuda.matter.hic_inferences( ))
\end{lstlisting}

\begin{figure*}[t]
\centering
\includegraphics[scale=0.9]{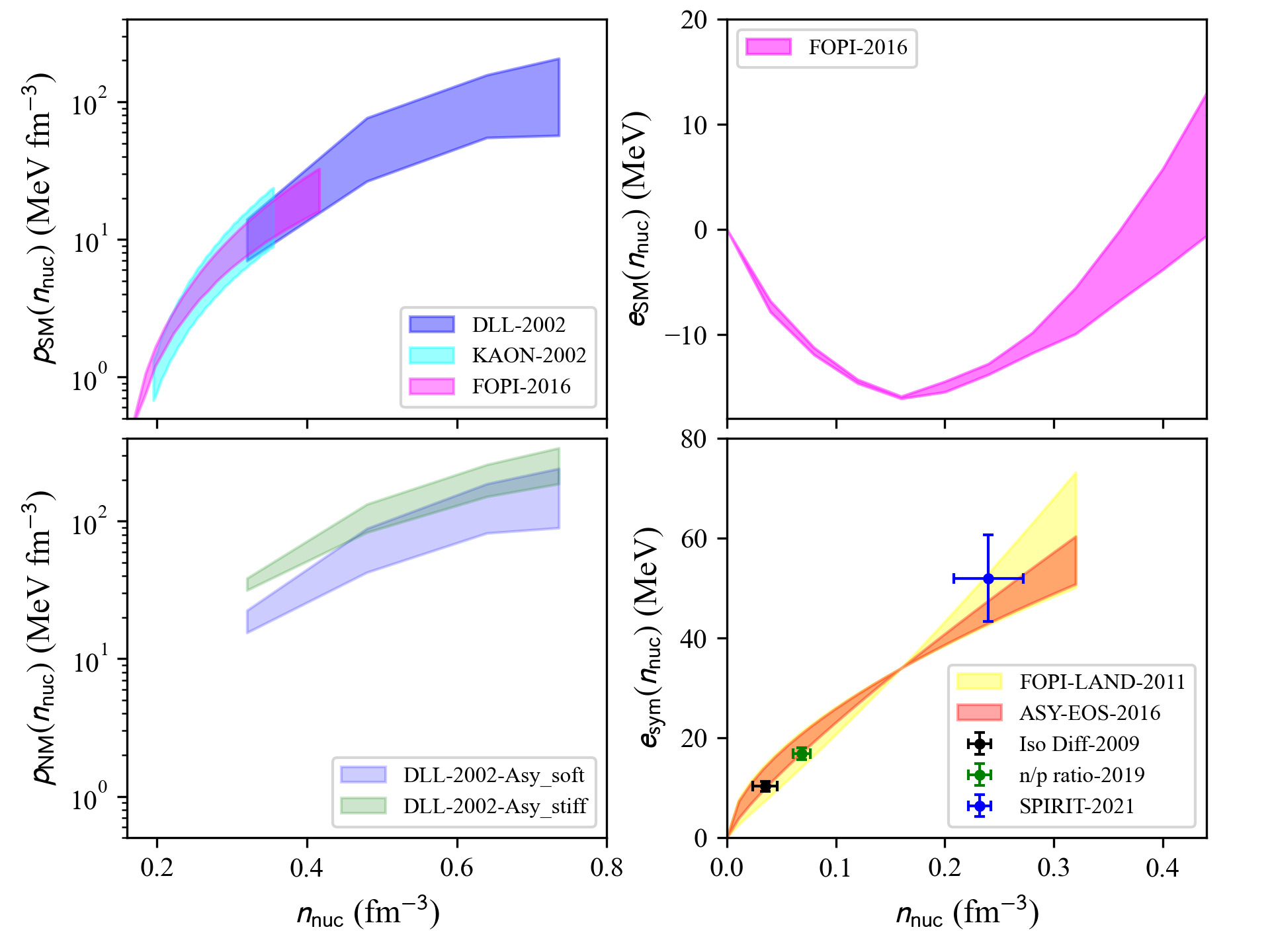}
\caption{HIC experimental inferences for the pressure and energy per particle (left) in SM (left top) and NM (left bottom), and the internal energy per nucleon (right top) and symmetry energy (right bottom) as a function of the nucleon density for different inferences available in the \texttt{nuda} toolkit. Figure generated with \texttt{matter\_setupHIC\_plot.py}.}
\label{fig:hic}
\end{figure*}

Once the variable \texttt{inference} is chosen in the previous list, the call for the EOS inferences can be performed in the following way:
\begin{lstlisting}[language=Python]
hic = nuda.matter.setupHIC( inference=`2002-DLL' )
hic.print_outputs()
\end{lstlisting}
We have eight different inferences available in the \texttt{nuda} toolkit ('2002-DLL', `2002-KAON', `2016-FOPI', `2009-ISO-DIFF', `2011-FOPI-LAND', `2016-ASY-EOS', `2019-NP-RATIO', `2021-SPIRIT'). These inferences are made for various terms of the EoS: the first three inferences are for SM, and the others are for the symmetry energy. We now detail these inferences:\\

\noindent
\texttt{inference}=`2002-DLL'. \\
This is the constraint on the pressure obtained from the collective flow of matter~\cite{PDanielewicz:2002} for densities going from $2.2n_\sat$ up to $4.6n_\sat$. The constraints are obtained from the studies on sideward and elliptical anisotropies of protons in Au + Au collisions at energies ranging from low incident energy, $E_\text{inc} \approx 0.15-2$~$A$GeV, (at SIS) up to high energies, $E_\text{inc} \approx 2-11$~$A$GeV, (at AGS). These constraints rule out strongly repulsive nuclear equations of state ($K_\sat > 300$~MeV) from relativistic mean field theory and weakly repulsive equations of state ($K_\sat < 167$~MeV) with phase transitions at densities less than three times saturation density $n_\sat$. Note, however, that EoS softens by the onset of a phase transition at high density is possible. Results are available for the pressure in SM and NM. Note that in NM, there are two predictions based on two different parametrizations for the symmetry energy (soft and stiff in Fig.~\ref{fig:hic}). Outputs are given in \texttt{hic.sm\_pre\_up} and \texttt{hic.sm\_pre\_lo} for the upper and lower boundaries of the pressure in SM, and in  \texttt{hic.sm\_pre\_so\_up} (\texttt{hic.sm\_pre\_st\_up}) and \texttt{hic.sm\_pre\_so\_lo} (\texttt{hic.sm\_pre\_st\_lo}) for the soft (stiff) boundary prediction in NM. The densities are given in \texttt{hic.den\_pre}. These inferences are shown in Fig.~\ref{fig:hic}.\\

\noindent
\texttt{inference}=`2006-KAON'. \\
This is the inference for the pressure in SM obtained using the excitation function of $K^+$ multiplicities obtained in Au+Au over C+C systems~\cite{CFuchs:2006,WGLynch:2009}. The inferences are obtained by comparing data from KaoS experiments for incident energy ranging from 0.8 to 1.5~$A$GeV using QMD and IQMD transport models. The inference extends from $1.2n_{\sat}$ to $2.2n_{\sat}$ densities~\cite{WGLynch:2009}. Outputs are given in \texttt{hic.sm\_pre\_up} and \texttt{hic.sm\_pre\_lo} for the upper and lower boundaries of the pressure in SM, and the densities are given in \texttt{hic.den\_pre}. \\   

\noindent
\texttt{inference}=`2016-FOPI'. \\
This inference is for the internal energy per particle and for the pressure in SM. The data from FOPI experiments on the elliptic flow of protons and light clusters up to mass 3 for the Au+Au collisions are analyzed for incident energy ranging from 0.4 and 1.5~$A$GeV. Inferences for the EoS of compressed SM up to $3n_\sat$ are extracted using the transport simulation IQMD~\cite{ALefevre:2016}. Outputs are given in \texttt{hic.sm\_e2a\_int\_up} (\texttt{hic.sm\_pre\_up}) and \texttt{hic.sm\_e2a\_int\_lo} (\texttt{hic.sm\_pre\_lo}) for the upper and lower boundaries of the energy per nucleon (pressure) in SM, and the densities are given in \texttt{hic.den\_e2a} (\texttt{hic.den\_pre}). \\ 

\noindent
\texttt{inference}=`2009-ISO-DIFF'. \\
This inference for the symmetry energy for densities $0.22 \pm 0.07 n_{\sat}$~\cite{MBTsang:2009} has been obtained from the
experimental data on isospin diffusion in collisions of Sn+Sn isotopes at 0.05~$A$GeV incident energy. Constraints are extracted using the ImQMD transport model. Outputs are given in \texttt{hic.esym} and \texttt{hic.esym\_err} for the centroid and uncertainty of the symmetry energy at the density \texttt{hic.den\_esym} with the uncertainty \texttt{hic.den\_esym\_err}. \\ 

\noindent
\texttt{inference}=`2011-FOPI-LAND'. \\
The symmetry energy is inferred from the studies of the ratio of the elliptic flow of neutrons and hydrogen isotopes measured for Au+Au collisions at incident energies of 0.4~$A$GeV and using UrQMD transport simulation~\cite{PRussotto:2011}. This inference goes up $2n_{\sat}$ and indicates moderately soft symmetry energy.
Outputs are given in \texttt{hic.esym} and \texttt{hic.esym\_err} for the centroid and uncertainty of the symmetry energy at the density \texttt{hic.den\_esym} with the uncertainty \texttt{hic.den\_esym\_err}. \\ 

\noindent
\texttt{inference}=`2016-ASY-EOS'. \\
The directed and elliptical flows of neutrons and light-charged clusters were measured for Au+Au at 0.4 AGeV by the ASY-EOS experimental campaign~\cite{PRussotto:2016}. The comparison of this data with the UrQMD transport model helped to get an inference for the symmetry energy up to $2n_\sat$. Again, the moderately soft to linear density dependence was seen as consistent with FOPI-LAND results but with reduced uncertainties. Outputs are given in \texttt{hic.esym\_up} and \texttt{hic.esym\_lo} for the upper and lower boundaries of the symmetry energy at the density \texttt{hic.den\_esym}. \\ 

\noindent
\texttt{inference}=`2019-NP-RATIO'. \\
The symmetry energy for densities $0.43 \pm 0.05 n_{\sat}$~\cite{PMorfouace:2019} is inferred using the experimental data on spectral ratios of neutron and proton spectra in collisions of Sn+Sn isotopes at 0.12~$A$GeV incident energy and using the ImQMD transport model. Outputs are given in \texttt{hic.esym} and \texttt{hic.esym\_err} for the centroid and uncertainty of the symmetry energy at the density \texttt{hic.den\_esym} with the uncertainty \texttt{hic.den\_esym\_err}. \\ 

\noindent
\texttt{inference}=`2021-SPIRIT'. \\
The study of the spectral ratios of pions in the Sn+Sn reactions at an incident energy of 270~AMeV~\cite{JEstee:2021} provides an inference for the symmetry energy, 52$\pm$13 MeV, and for the symmetry pressure $p_\sym$, 10.9$\pm$8.7 MeV/fm$^3$, at the density of 1.45$\pm0.2~n_\sat$ using the transport model dcQMD. Outputs are given in \texttt{hic.esym} (\texttt{hic.psym}) and \texttt{hic.esym\_err} (\texttt{hic.psym\_err}) for the centroid and uncertainty of the symmetry energy (symmetry pressure), at the density \texttt{hic.den\_esym} with the uncertainty \texttt{hic.den\_esym\_err}. \\

\begin{figure*}[t]
\centering
\includegraphics[scale=0.9]{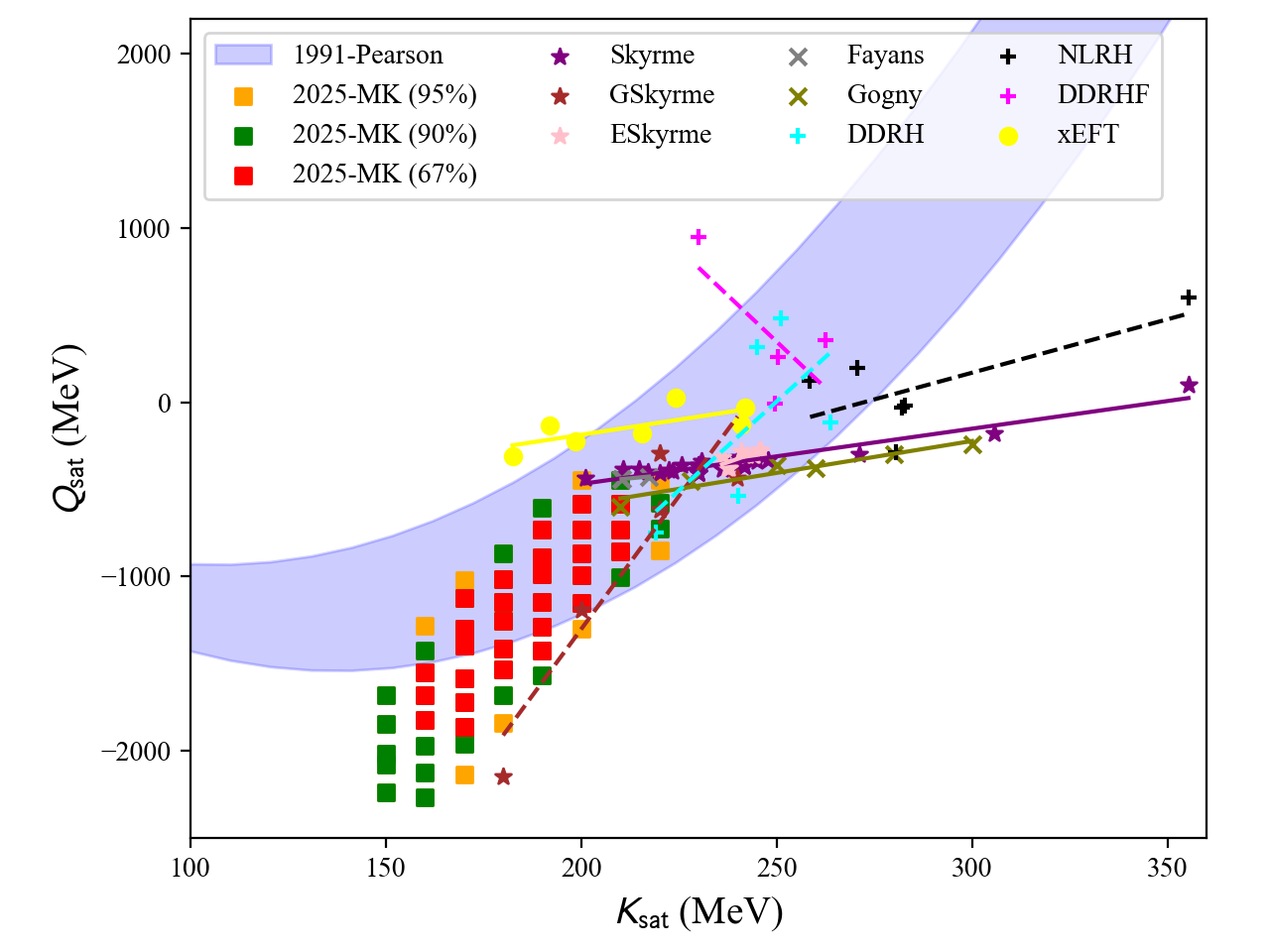}
\caption{$K_{\sat}$-$Q_{\sat}$ correlation from the different constraints provided in the \texttt{nuda} toolkit. Figure generated with \texttt{corr\_setupKsatQsat\_plot.py}.} 
\label{fig:KsatQsat}
\end{figure*}

The HIC constraints for the symmetry energy are shown in Fig.~\ref{fig:hic} (bottom right panel) for the different inferences presented above. It is interesting to see that these different inferences are quite consistent with each other, although they come from different analyses based on different HIC and different transport models.

\section{Correlation diagrams: the \texttt{corr} module.}
\label{sec:corr}

We now introduce the model \texttt{corr} of the \texttt{nuda} toolkit that is aimed at producing correlations among NEPs originating from experimental or theoretical constraints.

\subsection{The correlation between $K_\sat - Q_\sat$}
\label{sec:corr:KsatQsat}

The values for the NEP $K_\sat$ and $Q_\sat$ are very instructive for the inference of the density dependence of the EoS in SM around saturation density. While $K_\sat$ is determined by the analysis of the isoscalar Giant Monopole Resonance (ISGMR) in heavy nuclei, providing $K_\sat=240\pm 20$~MeV~\cite{UGarg:2018}, following the procedure suggested by Blaizot~\cite{JPBlaizot:1980}, the value for $Q_\sat$ remains quite unknown. It was, furthermore, observed that most EDFs induce strong correlations between $K_\sat$ and $Q_\sat$, see Fig.~\ref{fig:KsatQsat} and Refs.~\cite{EKhan:2012,EKhan:2013}. It is also known that the accurate determination of $K_\sat$ is related to a narrow exploration of $Q_\sat$~\cite{JMargueron:2019}. So one may wonder if the small uncertainty in the extraction of $K_\sat$, $\pm20$~MeV, is  underestimated~\cite{JMargueron:2025}. The \texttt{nuda} toolkit provides a method to easily extract and represent the correlation between $K_\sat$ and $Q_\sat$, see Fig.~\ref{fig:KsatQsat} and the following.

The complete list of constraints is given with the following instructions:
\begin{lstlisting}[language=Python]
constraints, constraints_lower = nuda.corr.KsatQsat_constraints( )
print('List of constraints:',constraints)
\end{lstlisting}

A typical call for given input variable \texttt{constraint}, see above, is:
\begin{lstlisting}[language=Python]
KQ = nuda.corr.setupKsatQsat( constraint=`EDF-SKY' )
KQ.print_outputs()
\end{lstlisting}
Results can be obtained by fixing the variable \texttt{constraint} to one of the following values:\\

\noindent
\texttt{constraint}=`1991-Pearson'. \\
Experimental band for $K_\sat$ and $Q_\sat$ given in Ref.~\cite{SRudaz:1992} and obtained from Ref.\cite{JMPearson:1991}. Outputs are given in \texttt{KQ.Ksat\_band} and the lower and upper boundary of the contour are given in \texttt{KQ.Qsat\_band\_lo} and \texttt{KQ.Qsat\_band\_up}.\\

\noindent
\texttt{constraint}=`2025-MK-67', `2025-MK-90', `2025-MK-95'. \\
The 67\%, 90\%, and 95\% confidence levels for a set of extended Skyrme interactions reproducing the binding energy, charge radii, and isoscalar giant monopole resonance in $^{120}$Sn and ${208}$Pb, more details in Ref.~\cite{JMargueron:2025}. Outputs are given in \texttt{KQ.Ksat} and \texttt{KQ.Qsat}.\\

\noindent
\texttt{constraint}=`EDF-SKY', `EDF-GSKY', `EDF-ESKY', `EDF-Gogny', `EDF-xEFT', `EDF-NLRH', `EDF-DDRH', `EDF-DDRHF'. \\
Values for $K_\sat$ and $Q_\sat$ predicted by a set of nuclear interactions~\cite{EKhan:2012,EKhan:2013,JMargueron:2025}. For each one of these constraints, outputs are given in \texttt{KQ.Ksat} and \texttt{KQ.Qsat}, and a linear regression is provided in \texttt{KQ.Ksat\_lin} and \texttt{KQ.Qsat\_lin}.

We show constraints for the $K_\sat$-$Q_\sat$ correlation plot in Fig.~\ref{fig:KsatQsat}.
Some models are very correlated (solid lines) while others are more loosely correlated (dashed lines). However, the region of $Q_\sat$ explored by the different models is narrow. For each model, the value for $Q_\sat$ is quite constrained by the value of $K_\sat$. The experimental domain for Ref.~\cite{MFarine:1997} is shown in blue, and the extended Skyrme models obtained in Ref.~\cite{JMargueron:2025} are also shown, with different colors reflecting the confidence level as shown in the legend. Based on this correlation plot, it is argued in Ref.~\cite{JMargueron:2025} that the experimental uncertainty in $K_\sat$ is underestimated. While the microscopic approach suggested in Ref.~\cite{JPBlaizot:1980} is the most accurate way to extract the value of $K_\sat$ from the analysis of experimental data for the ISGMR, the use of models presenting strong correlations among parameters is to be taken with caution. More flexible microscopic models exploring the parameter space widely for $K_\sat$ and $Q_\sat$ may be more appropriate to better represent the experimental uncertainty.

\subsection{The correlation between $E_{\sym,2} - L_{\sym,2}$}
\label{sec:corr:Esym2Lsym2}

\begin{figure*}[t]
\centering
\includegraphics[scale=0.9]{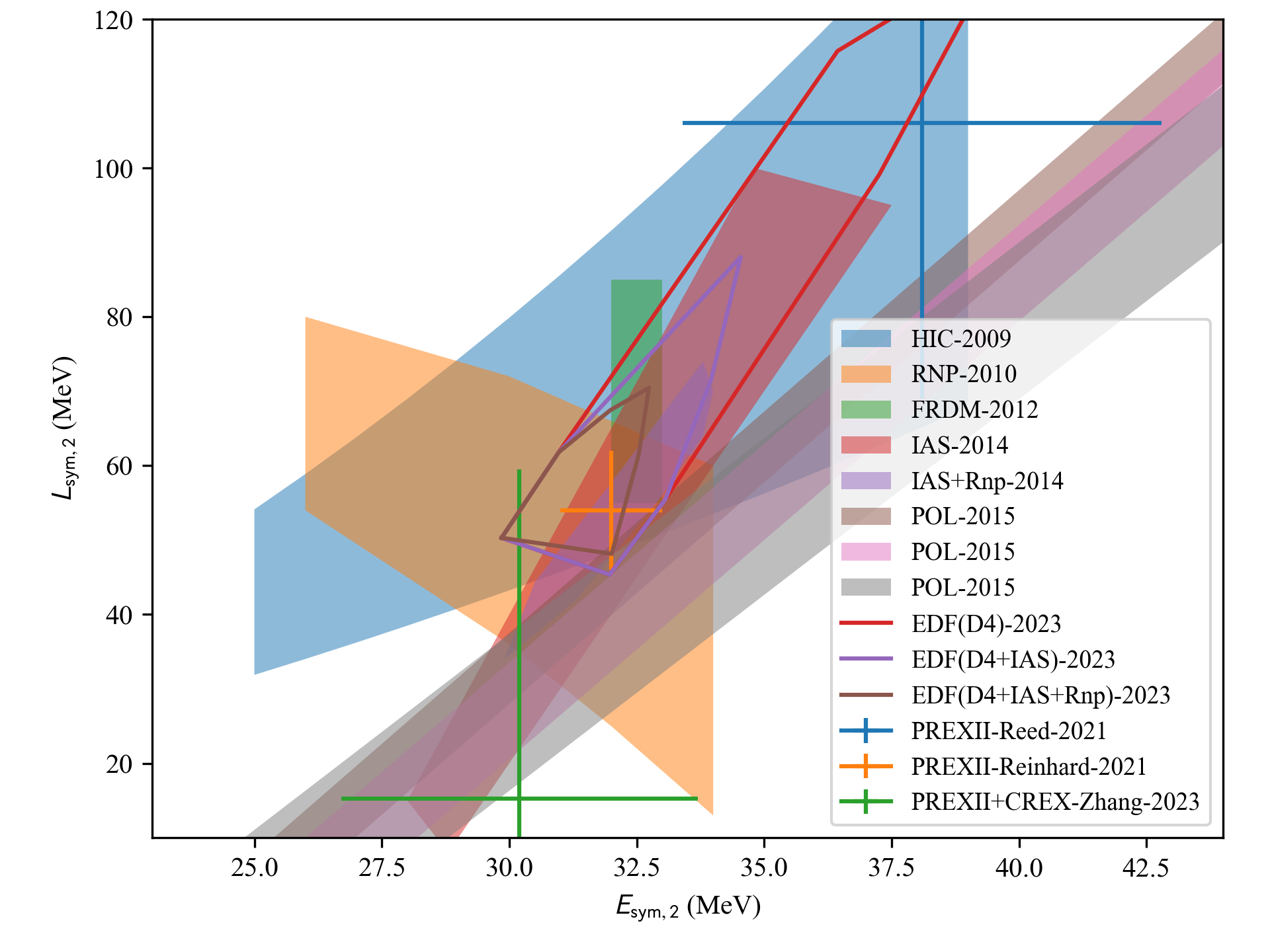}
\caption{$E_{\sym,2}$-$L_{\sym,2}$ correlation from the different constraints provided in the \texttt{nuda} toolkit. Figure generated with \texttt{corr\_setupEsym2Lsym2\_plot.py}.} 
\label{fig:Esym2Lsym2}
\end{figure*}

In the following, we differentiate the correlation originating from measurements in finite nuclei, which probe the symmetry energy around SM, from the one originating from dense matter and neutron star observation, which probe the very asymmetric behavior of the symmetry energy ($\beta$-equilibrium). Since the first one (finite nuclei) constrains the quadratic contribution to the symmetry energy, we will represent the constraint in terms of $E_{\sym,2}$ and $L_{\sym,2}$, while the second one (neutron star) provide constraints for the symmetry energy in very asymmetric matter and NEP $E_{\sym,2}$ and $L_{\sym,2}$.

The correlation between $E_{\sym,2}$ and $L_{\sym,2}$ has been investigated by many different approaches over the last decade, see for instance Refs.~\cite{JMLattimer:2013,JBWei:2020} for more details. Note that since most of these constraints are extracted from analyses of experimental nuclear data, they concern the quadratic contribution of the symmetry energy $e_{\sym,2}$. The \texttt{nuda} toolkit provides a large list of these approaches. The complete list of available constraints is given with the following instruction:
\begin{lstlisting}[language=Python]
print( nuda.corr.EsymLsym_constraints( ) )
\end{lstlisting}

Once the variable \texttt{constraint} is chosen in the previous list, the call for the correlation induced by this constraint in the $E_{\sym,2}$-$L_{\sym,2}$ correlation plot can be performed in the following way:
\begin{lstlisting}[language=Python]
EL = nuda.corr.setupEsymLsym( constraint=`2009-HIC' )
EL.print_outputs()
\end{lstlisting}


Results can be obtained by fixing the variable \texttt{constraint} to one of the following values:\\

\noindent
\texttt{constraint}=`2009-HIC'. \\
Constraints inferred from isospin diffusion in heavy ion collisions (HICs)~\cite{MBTsang:2009}. Outputs are given in \texttt{EL.Esym} and \texttt{EL.Lsym} with \texttt{EL.Lsym\_err}. \\ 


\noindent
\texttt{constraint}=`2010-RNP'.\\ 
Constraints deduced from the analysis of neutron-skin thickness $\Delta r_{np}$(Sn) in Sn isotopes~\cite{LWChen:2010}. Outputs are given in \texttt{EL.Esym} and \texttt{EL.Lsym} with \texttt{EL.Lsym\_err}. \\ 


\noindent
\texttt{constraint}=`2012-FRDM'.\\ 
Constraint from the finite-range droplet mass model (FRDM) calculations~\cite{PMoller:2012}. Outputs are given in \texttt{EL.Esym} and \texttt{EL.Lsym} with \texttt{EL.Lsym\_err}. \\ 


\noindent
\texttt{constraint}=`2014-IAS'.\\ 
Constraint deduced from the analysis of the excitation energy of the isobaric analog state (IAS) based on Skyrme-Hartree-Fock calculations~\cite{PDanielewicz:2014}. Outputs are given in \texttt{EL.Esym} and \texttt{EL.Lsym} with \texttt{EL.Lsym\_err}. \\ 


\noindent
\texttt{constraint}=`2014-IAS+Rnp'.\\ 
Combination of the IAS constraint and neutron skin $\Delta r_{np}$ in $^{208}$Pb~\cite{PDanielewicz:2014}. Outputs are given in \texttt{EL.Esym} and \texttt{EL.Lsym} with \texttt{EL.Lsym\_err}. \\ 


\begin{figure*}[t]
\centering
\includegraphics[scale=0.9]{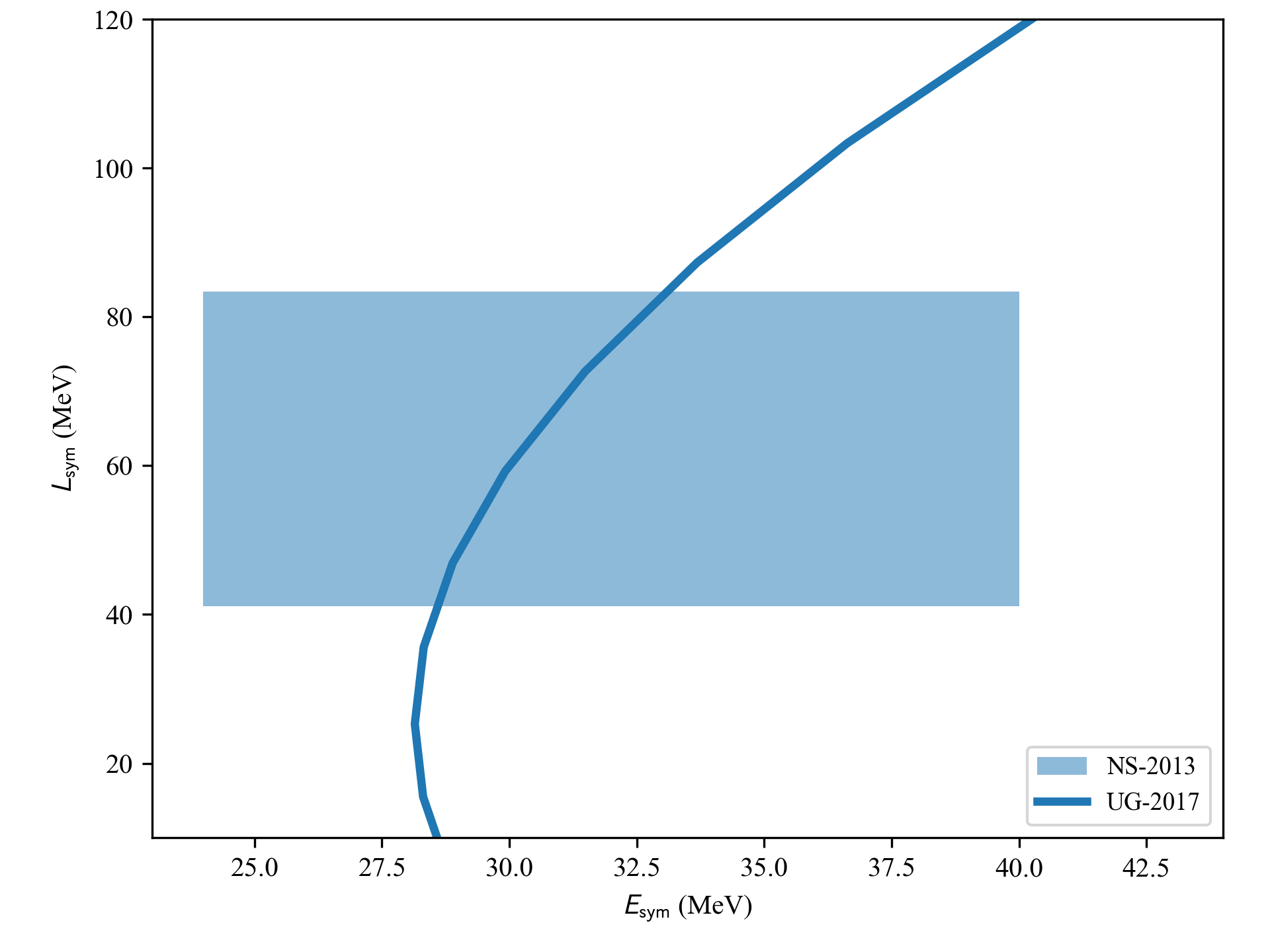}
\caption{$E_\sym$-$L_\sym$ correlation from the different constraints provided in the \texttt{nuda} toolkit. Figure generated with \texttt{corr\_setupEsymLsym\_plot.py}.} 
\label{fig:EsymLsym}
\end{figure*}

\noindent
\texttt{constraint}=`2015-POL-68NI', `2015-POL-120SN', `2015-POL-208PB'.\\
Constraints on the electric dipole polarizability of $^{208}$Pb, $^{120}$Sn and $^{68}$Ni~\cite{XRocaMaza:2015}. Outputs are given in \texttt{EL.Lsym} and \texttt{EL.Esym} with \texttt{EL.Esym\_err}. \\ 


\noindent
\texttt{constraint}=`2021-PREXII-Reed'~\cite{BTReed:2021}, `2021-PREXII-Reinhard'~\cite{PGReinhard:2021}, `2023-PREXII+CREX-Zhang'~\cite{ZZhang:2023}.\\
PREX-II~\cite{DAdhikari:2021} and CREX~\cite{DAdhikari:2022} are parity-violating electron scattering (PVES) experiments that have been performed at JLab. There are big differences between the original analysis by Reed et al.~\cite{BTReed:2021} suggesting
$E_{\sym,2}=38.1\pm 4.7$~MeV, $L_{\sym,2}=106\pm 37$~MeV, and the one by Reinhard et al.~\cite{PGReinhard:2021}, which also includes the constraint from electric dipole polarizability, suggesting $E_{\sym,2}=32\pm 1$~MeV, $L_{\sym,2}=54\pm 8$~MeV. Another analysis by Zhang and Chen~\cite{ZZhang:2023} combining PREX-II and CREX using a Bayesian inference find a very low centroid for $L_{\sym,2}$ ($E_{\sym,2}=30.2^{+3.0}_{-4.1}$~MeV, $L_{\sym,2}=15.3^{+41.5}_{-46.8}$~MeV). It has indeed been pointed out that the results of PREX-II and CREX are in disagreement~\cite{PGReinhard:2022,EYuksel:2022}. Outputs are given in \texttt{EL.Esym} with \texttt{EL.Esym\_err} and \texttt{EL.Lsym} with \texttt{EL.Lsym\_err}. \\ 

\noindent
\texttt{constraint}=`2023-EDF-D4', `2023-EDF-D4-IAS', `2023-EDF-D4-IAS-Rnp'.\\ 
The capacity of a total of 415 Skyrme and relativistic energy density functional (EDF) models is analyzed in Ref.~\cite{BVCarlson:2023}. The constraint `2023-EDF-D4' requires that EDF models describe $N=Z$ and $N\ne Z$ spherical nuclei with the same accuracy (binding energy, charge radii and ISGMR). The constraint `2023-EDF-D4-IAS' implies that the isobaric analog state (IAS) is reproduced as well, i.e., the contour for $e_\sym(n_\nuc)$ suggested in Ref.~\cite{PDanielewicz:2014}. Finally, the constraint `2023-EDF-D4-IAS-Rnp' implies that the neutron radius is reproduced on top of previous constraints, i.e., the contour for $e_\sym(n_\nuc)$ suggested in Ref.~\cite{PDanielewicz:2014}.\\

All these constraints are shown in Fig.~\ref{fig:Esym2Lsym2}. This figure is similar to previous ones, see for instance Refs.~\cite{JMLattimer:2013,JBWei:2020}. The constraints imposed by the different analyses of PREX-II and CREX parity-violating electron scattering (PVES) experiments are shown as crosses in Fig.~\ref{fig:Esym2Lsym2}. There is almost no overlap between these crosses, but it should be remembered that the uncertainties are given for only 67\% confidence interval. It is however interesting to note that there is a small region in the correlation plot $E_{\sym,2}$-$L_{\sym,2}$, coinciding with the one suggested by the constraint `2021-PREXII-Reinhard'~\cite{PGReinhard:2021},  which satisfy all constraints, except for the constraints related to electric dipole polarisability, which shifts $E_{\sym,2}$ towards slightly larger values and/or $L_{\sym,2}$ towards lower values.

\subsection{The correlation between $E_{\sym} - L_{\sym}$}
\label{sec:corr:EsymLsym}

As discussed in the previous section, the constraints from finite nuclei from those from dense matter and neutron stars are differentiated. In this section, we focus on the constraints from neutron stars.

The variable \texttt{constraint} can be `2013-NS' or `2017-UG' as detailed in the following:\\


\noindent
\texttt{constraint}=`2013-NS'.\\ 
The constraint on $L_\sym$ has been obtained from a Bayesian analysis of mass and radius observations of NSs by considering the 95\% confidence level for $L_{\sym}$~\cite{AWSteiner:2013}. Outputs are given in \texttt{EL.Esym} and \texttt{EL.Lsym} with \texttt{EL.Lsym\_err}. \\ 

\noindent
\texttt{constraint}=`2017-UG'.\\ 
The analysis of the unitary gas predictions for the symmetry energy parameters~\cite{ITews:2017} permits the values to the right of the curve. Outputs are given in \texttt{EL.Esym} and \texttt{EL.Lsym}. \\ 

These two constraints are shown in Fig.~\ref{fig:EsymLsym}. There is still a large domain in $E_\sym$ and $L_\sym$ compatible with these constraints. It would therefore be interesting to investigate further the constraints on $E_\sym$ and $L_\sym$ imposed by the observation data from neutron stars.

\subsection{The symmetry energy: $e_{\sym,2}(n_\nuc)$}
\label{sec:corr:Esym_n}

\begin{figure}[t]
\centering
\includegraphics[scale=0.52]{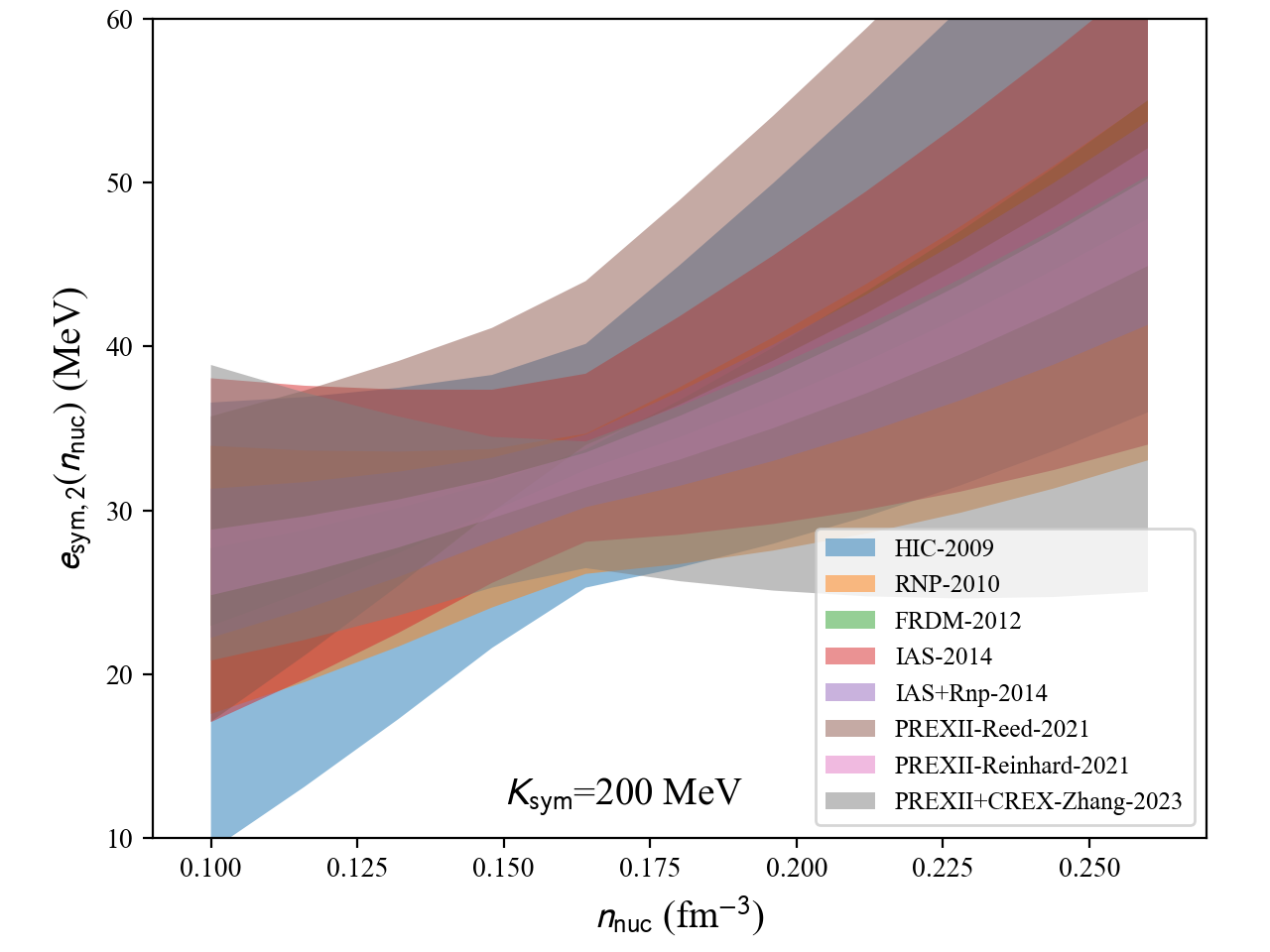}
\includegraphics[scale=0.52]{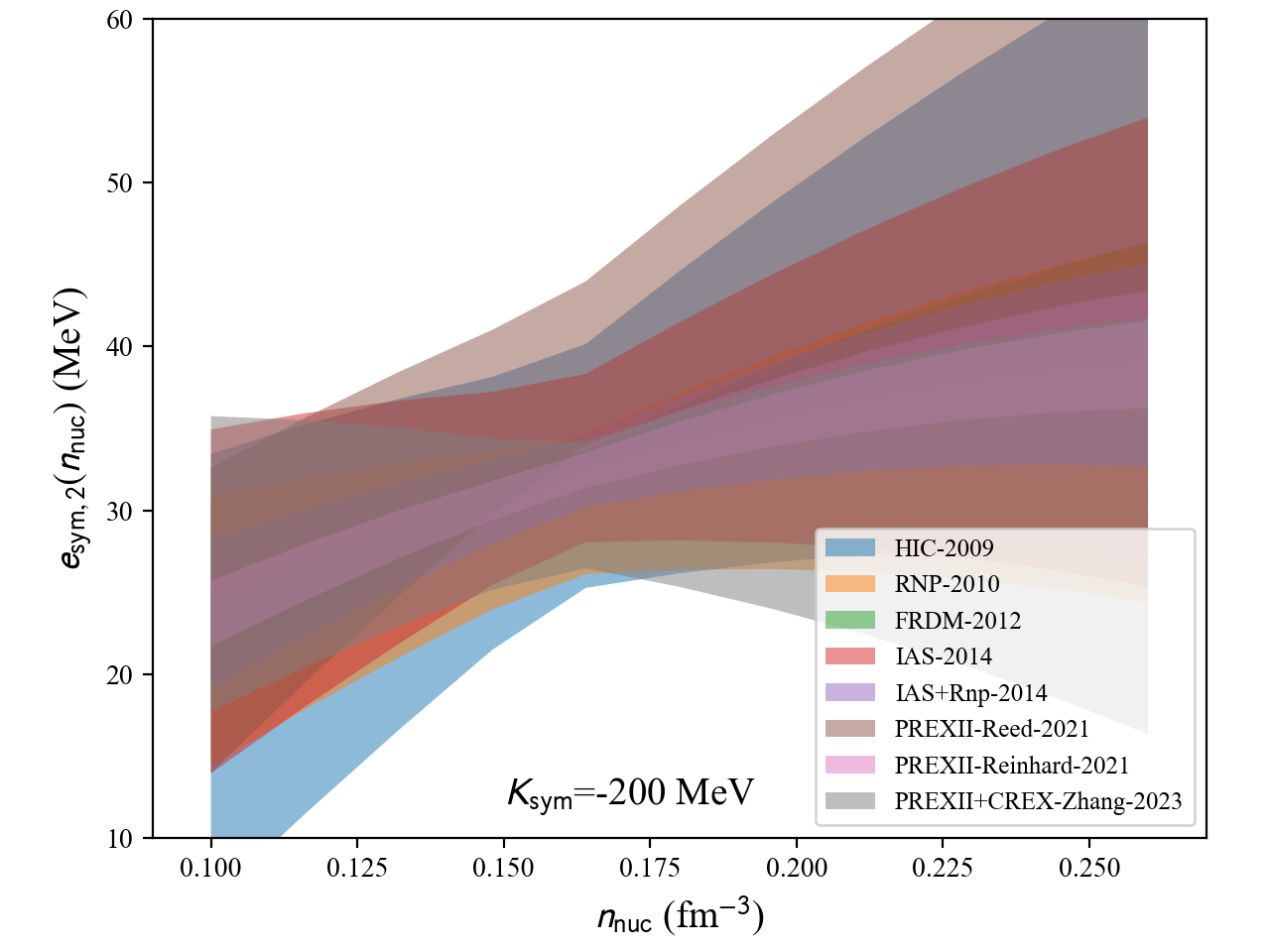}
\caption{The density dependent $e_{\sym}(n_\nuc)$ from the different constraints provided in the \texttt{nuda} toolkit and for three arbitrary choices of $K_\sym$=200 (top) and -200~MeV (bottom). Figure generated with \texttt{corr\_setupEsymDen\_plot.py}.} 
\label{fig:EsymDen}
\end{figure}

From the constraints on $E_{\sym,2}$ and $L_{\sym,2}$ provided by the different constraints previously listed, one can reconstruct the quadratic contribution to the symmetry energy $e_{\sym,2}(n_\nuc)$ around $n_\sat$ as:
\begin{eqnarray}
e_{\sym,2}(n_\nuc)&=&E_{\sym,2} + L_{\sym,2}\frac{n_\nuc-n_\sat}{3n_\sat}  \nonumber \\
&&\hspace{0.1cm}+\frac 1 2 K_{\sym,2}\left(\frac{n_\nuc-n_\sat}{3n_\sat}\right)^2 + \dots
\end{eqnarray}
The call for the density-dependent symmetry energy for a given \texttt{constraint} can be done in the following way:
\begin{lstlisting}[language=Python]
Ksym = 100
Esym = nuda.eos.setupEsymDen( constraint=`2014-IAS' , Ksym=Ksym )
Esym.print_outputs()
\end{lstlisting}
where the variable \texttt{constraint} takes the same input values as the class \texttt{corr.setupEsymLsym}. They are listed in the previous subsection.

The value for the NEP $K_{\sym,2}$ is largely unknown, see the discussion in Sec.~\ref{sec:mat:nep}. From Sec.~\ref{sec:mat:nep}, it can, however, be safely varied from $K_{\sym,2}$=-200, to 200~MeV, as suggested as well in Ref.~\cite{JMargueron:2018a}. For these two values for $K_\sym$, it is interesting to explore the prediction for the density dependence of the symmetry energy $e_{\sym,2}(n_\nuc)$ allowed by the previously presented constraints, see Fig.~\ref{fig:EsymDen}. 

\section{Data from finite nuclei: the \texttt{nuc} module.}
\label{sec:nuc}

In this section, we describe how the \texttt{nuda} toolkit can be employed to obtain some finite nuclei properties, e.g., binding energies, charge radii, ISGMR energies, etc. It is then possible, for instance, to use these experimental data for finite nuclei to calibrate models for the crust of neutron stars.

\subsection{Experimental binding energies over the nuclear chart}
\label{sec:nuc:expchart}

To obtain experimental binding energies for finite nuclei, a mass table should first be selected. The complete list of available experimental mass tables in \texttt{nuda} toolkit can be obtained in the following way:
\begin{lstlisting}[language=Python]
print( nuda.nuc.be_exp_tables( ) )
\end{lstlisting}
In the present version of the \texttt{nuda} toolkit, there is only one mass table available: `AME'. The toolkit is however ready to read more mass tables.

The `AME' mass table has been released several times, with more and more nuclei. The different versions of the table are associated with the year of their release. The complete list of available versions of the experimental mass tables is given with the following instructions:
\begin{lstlisting}[language=Python]
print( nuda.nuc.be_exp_versions( table=`AME' ) )
\end{lstlisting}
For the table `AME', the \texttt{nuda} toolkit provide the following versions: `2012'~\cite{GAudi:2012},  `2016'~\cite{GAudi:2017}, `2020'~\cite{WJHuang:2021}

\begin{figure*}[t]
\centering
\includegraphics[scale=0.9]{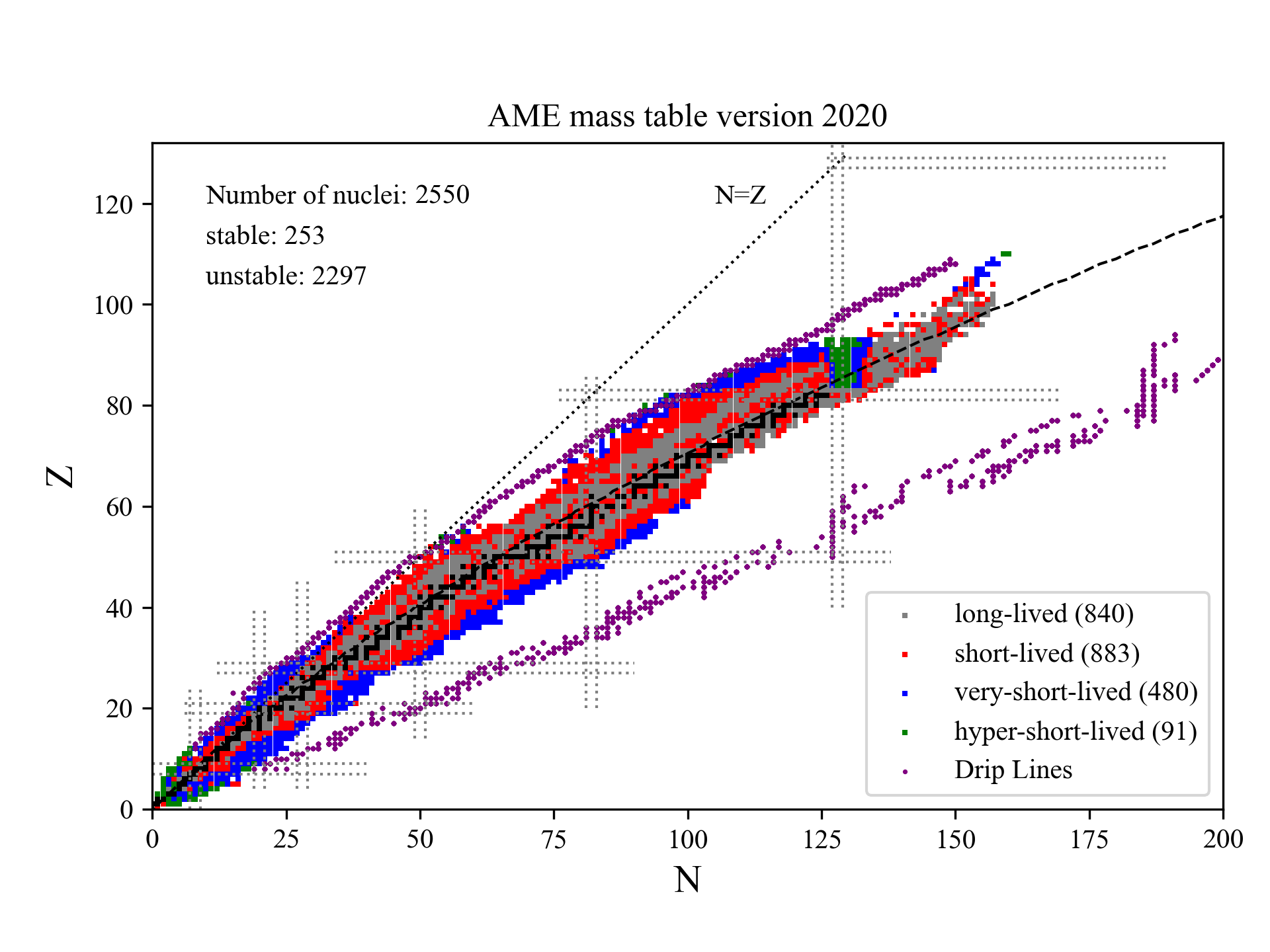}
\caption{Nuclear chart from AME 2020 nuclear chart available in the \texttt{nuda} toolkit. The driplines (in magenta) are calculated for the following set of models: `2013-HFB22', `2013-HFB23', `2013-HFB24', `2013-HFB25', `2013-HFB26', `2021-BSkG1', `2022-BSkG2', `2023-BSkG3'. See text for more details. Figure generated with \texttt{nuc\_setupBEExp\_chart\_lt\_plot.py}.}
\label{fig:exp:nchart}
\end{figure*}

Once the variables \texttt{table} and \texttt{version} are chosen, the call for the experimental nuclear chart can be performed in the following way:
\begin{lstlisting}[language=Python]
chart = nuda.nuc.setupBEExp( table=`AME', version=`2020' )
chart.print_outputs()
\end{lstlisting}

We now provide more details about the experimental nuclear tables available in the \texttt{nuda} toolkit. Here are the present options available:\\

\noindent
\texttt{table}=`AME'.\\
\texttt{version}= `2012'~\cite{GAudi:2012}, `2016'~\cite{GAudi:2017}, `2020'~\cite{WJHuang:2021}.\\

The content of several output attributes is provided with the command \texttt{chart.print\_outputs()}, and their complete list can be obtained as: 
\begin{lstlisting}[language=Python]
chart = nuda.nuc.setupBEExp( table=`AME', version=`2020' )
print(chart.__dict__)
\end{lstlisting}
We now provide more details on these outputs. They are provided as a Numpy array containing nuclear properties provided by the selected mass table. The \texttt{nuda} toolkit gives \texttt{chart.nucA} for the mass number $A=N+Z$, \texttt{chart.nucZ} for the nuclear charge $Z$, \texttt{chart.nucN} for the neutron number $N$,  \texttt{chart.nucI} for the isospin parameter $I$ defined as $I=(n-Z)/A$, and the binding energy (experimental uncertainty) is in \texttt{chart.nucBE} (\texttt{chart.nucBE\_err}). Additionally, the nuclear symbol is given in \texttt{chart.nucSymb}, the interpolation flag is stored in \texttt{chart.flagInterp} (it is set to `n' if the nucleus is not interpolated (true measurement) and `y' if the nucleus is not measured but provided from the interpolation provided by the authors of the mass table), \texttt{chart.flagI} is identical to the variable \texttt{i} in the original mass table: \texttt{i=0} for nuclei measured in their ground-state, the attribute \texttt{chart.nucStbl} reflect if the nucleus is stable ('y') or unstable ('n'), the measured half-time in seconds is given in \texttt{chart.nucHT}, the discovery year is in \texttt{chart.nucYear}, and the maximum charge in the table is in \texttt{chart.Zmax}.

The method \texttt{select} can be applied to the object \texttt{chart} to select a smaller list of nuclei respecting some requirements. The command is the following:
\begin{lstlisting}[language=Python]
chart2 = chart.select(Amin, Zmin, interp=`n', state=`gs', nucleus = `unstable', every=1)
\end{lstlisting}
where all input variables are optional. The input \texttt{Amin} and \texttt{Zmin} allows the user to select nuclei above the limit they impose (default value is 0, \texttt{Amin} and \texttt{Zmin} can be fixed independently from each other), the input variable \texttt{interp} select nuclei that are not interpolated ('n') or only interpolated nuclei ('y'), the input variable \texttt{state} can be `gs' to select nuclei measured in their ground-state (default), the input variable \texttt{nucleus} can be `unstable' (default), `stable', `longlive', `shortlive' or `veryshortlive', as described below, and finally the input variable \texttt{every} select nuclei by steps defined by \texttt{every} (default is 1, no selection).

The attributes of the object \texttt{chart2} are the same as the ones associated with the object \texttt{chart}. They have a name with \texttt{sel\_} distinguishing them. With obvious notations, see above, the attributes are: \texttt{chart2.sel\_nucA}, \texttt{chart2.sel\_nucZ}, \texttt{chart2.sel\_nucN}, \texttt{chart2.sel\_nucI}, \texttt{chart2.sel\_nucBE}, \texttt{chart2.sel\_BE\_err}, \texttt{chart2.sel\_nucSymb}, \texttt{chart2.sel\_flagInterp}, \texttt{chart2.sel\_flagI}, \texttt{chart2.sel\_nucHT}, \texttt{chart2.sel\_nucYear}, \texttt{chart2.sel\_Zmax}.

The nuclear chart for the 2020 AME mass table~\cite{WJHuang:2021} contains 2550 nuclei (with 253 stable nuclei) in their ground state. It is shown in Fig.~\ref{fig:exp:nchart}, where we display the nuclear chart as a function of the number of neutrons ($x$-axis) and protons ($y$-axis). The dotted lines associated with shell closure are obtained from the function \texttt{plot\_shells(axs)} for \texttt{axs} a figure created by \texttt{matplotlib}. Nuclei are colored according to their half-life: stable nuclei (black) have a half-life greater than Earth's age and define the so-called valley of stability (in dashed line, the fit is provided by the function \texttt{stable\_fit\_Z(Zmin,Zmax)} or \texttt{stable\_fit\_N(Nmin,Nmax)} defined as
\begin{eqnarray}
N(Z) &=& Z(1+6\,10^{-3} Z) \, \\
Z(N) &=& N ( 1 - 3.8\,10^{-3} N + 9\,10^{-6} N^2 ) \, ,
\end{eqnarray}
and providing as output \texttt{N} and \texttt{Z}), long-lived nuclei (grey) are nuclei for which the half-life is larger than 1-day, short-lived nuclei (red) are the ones that live more than 1-second, very-short-lived nuclei (green) are the ones that live more than 1-millisecond, and the other are hyper-short-lived nuclei (green). Theoretical drip lines (purple) are also shown for several EDFs, see the caption of Fig.~\ref{fig:exp:nchart} for the list of EDFs, and Sec.~\ref{sec:nuc:theochart} for the use of theoretical models. 

The 1-day frontier between long-lived and short-lived nuclei is a typical time separating nuclear chemistry and nuclear physics, and the 1-second frontier between short-lived and very short-lived nuclei represents the frontier above which nuclei shall be studied online, just after their production. Note that 1 millisecond is the time for a nucleus to travel 3~km at the speed of 1\% of $c$ (no relativistic time dilation). Very few nuclei in the mass table live shorter than 1 millisecond (green color). The 1 millisecond frontier is the new frontier of research for modern facilities exploring rare isotopes, e.g., RIKEN, FRIB, FAIR, SPIRAL2, etc. In the future, more and more hyper-short-lived nuclei will be produced and characterized on the way to the drip lines. The four domains shown in Fig.~\ref{fig:exp:nchart} are therefore reflecting the progress in technology: using chemistry to identify artificial (long-lived) isotopes, using physics detectors to identify rare (short-lived) isotopes, on-line facilities dedicated to (very-short-lived) exotic nuclei, and finally, (hyper-short-lived) nuclei requiring secondary beams to be produced and compact facilities for their detection. They are at the frontier of research.

Note that there is a large empty space between the nuclei experimentally produced on Earth and the predicted drip lines (magenta dots), except for light nuclei and for some neutron-poor nuclei. The dispersion in green points reflects an uncertainty in the model prediction for the location of the drip lines. Note also that Fig.~\ref{fig:exp:nchart} is one of the many representations possible with the \texttt{nuda} toolkit. It is not the purpose of the present work to list them all, and we simply illustrate the capacities of \texttt{nuda}.

\begin{figure}[t]
\centering
\includegraphics[scale=0.52]{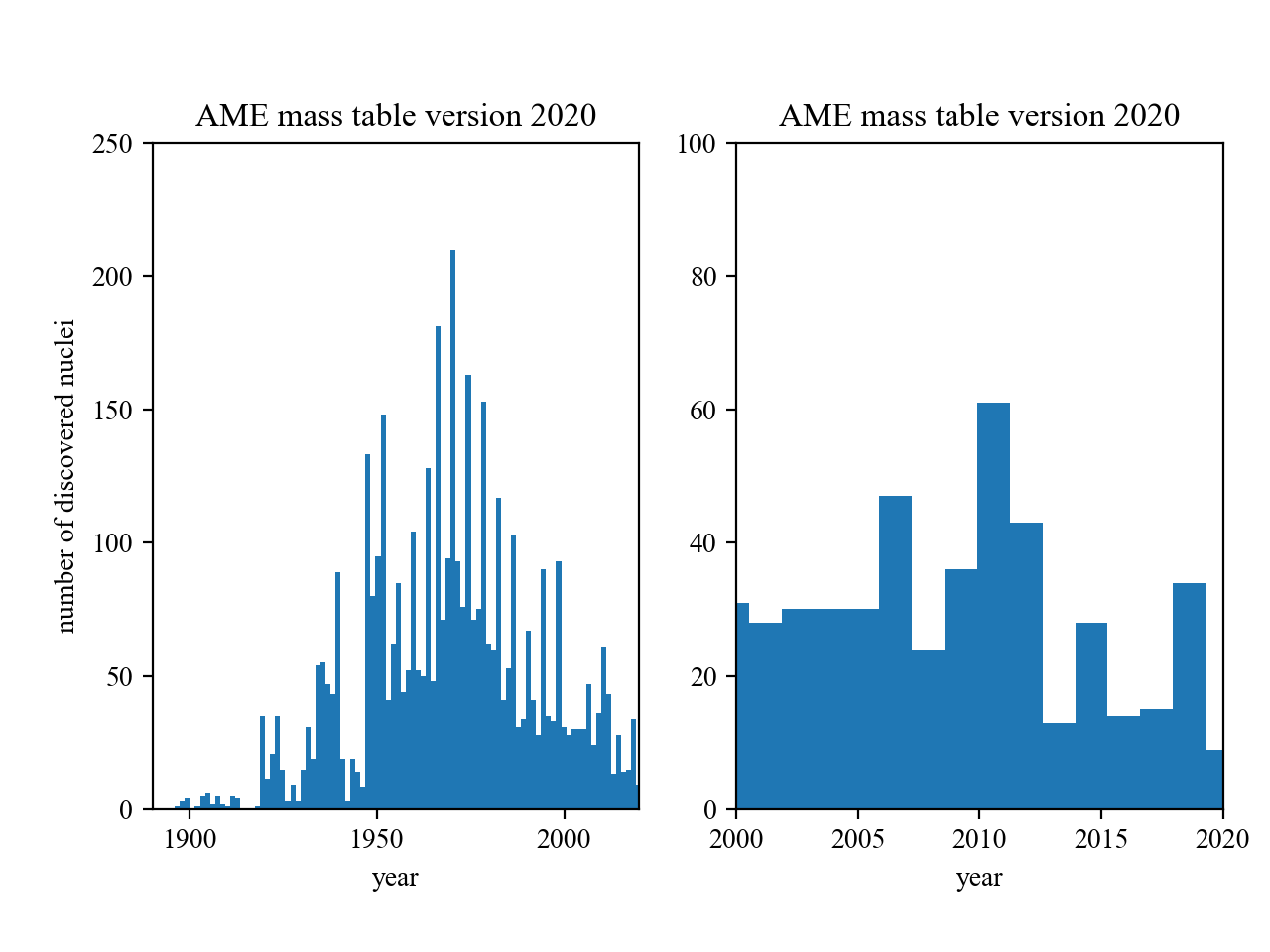}
\caption{Number of nuclei function of their discovery year. Figure generated with \texttt{nuc\_setupBEExp\_year\_plot.py}.}
\label{fig:exp:year}
\end{figure}

Another illustration of the capacities of \texttt{nuda} is shown in Fig.~\ref{fig:exp:year}, where experimental nuclei are distributed according to the year of their discovery. On the left is shown the distribution since the first one in 1897, and on the right is a zoom-in for the recent period between 2000 and 2020. The peak of discoveries is located in the 60ies, when a large number of facilities across the world contributed to the blooming of discoveries. More recently, the cost to produce new nuclei is such (all modern facilities produce secondary beams) that there a fewer exotic nuclei factories.

Fig.~\ref{fig:exp:year} is obtained by using the method \texttt{select\_year()} that is called in the following way:
\begin{lstlisting}[language=Python]
chartY = chart.select_year( year_min, year_max, state=`gs')
\end{lstlisting}
where \texttt{year\_min} and \texttt{year\_max} are the boundaries for the discovery years and \texttt{state} is `gs' for ground-state nuclei (default). The attributes of \texttt{chartY} are the same as the ones of \texttt{chart2}.

Once nuclei are selected in \texttt{chart2}, selecting nuclei in their ground state for instance, there are two additional methods that can be applied to the object \texttt{chart2}, \texttt{isotopes(Zref)} and \texttt{isotones(Nref)}, to fix the boundaries for a isotopic or isotonic chain. These methods are called in the following way:
\begin{lstlisting}[language=Python]
chart3 = chart2.isotopes( Zref=20 )
chart4 = chart2.isotones( Nref=20 )
\end{lstlisting}
The object \texttt{chart3} (\texttt{chart4}) contains the boundary of the isotopic (isotonic) chains for \texttt{Zref} (\texttt{Nref}). For the present example, \texttt{chart3} contains the boundaries \texttt{chart3.itp\_Nmin} and \texttt{chart3.itp\_Nmax} for calcium isotopes existing in the selected nuclear chart while \texttt{chart4} contains boundaries \texttt{chart4.itn\_Zmin} and \texttt{chart4.itn\_Zmax} for $N=20$ isotones.

\subsection{Nuclear chart from theoretical models}
\label{sec:nuc:theochart}

Predictions for nuclear binding energies from a set of theoretical models are also available in \texttt{nuda} toolkit. The complete list of available theoretical tables is given with the following instructions:
\begin{lstlisting}[language=Python]
print( nuda.nuc.be_theo_tables( ) )
\end{lstlisting}
among which a value for the variable \texttt{table} can be chosen. The call for the theoretical nuclear energies can be performed in the following way:
\begin{lstlisting}[language=Python]
mas = nuda.nuc.setupBETheo( table=`2007_HFB14' )
mas.print_outputs()
\end{lstlisting}
where we have chosen the theoretical mass model `2007\_HFB14'. The input variable \texttt{table} can be selected among the following values.\\

\noindent
\texttt{table}=`1988-GK'.\\
This empirical calculation is based on known masses of neighboring nuclei from transverse Garvey-Kelson (GK) mass relations for approximately 5800 atomic masses are reported in Ref.~\cite{JJaenecke:1988}. While this table is given in the \texttt{nuda} toolkit, we will not show results since they are quite poor.\\

\noindent
\texttt{table}=`1988-MJ'.\\
This empirical calculation is based on an inhomogeneous partial difference equation, incorporating shell-dependent symmetry energy expressions, Coulomb energy corrections, and higher-order isospin contributions, for approximately 4385 nuclear masses are reported in Ref.~\cite{PJMasson:1988}. \\

\noindent
\texttt{table}=`1995-DZ'.\\
This phenomenological calculation is based on the shell model, incorporating renormalized monopole and multipole interactions with empirical fits to known masses, for approximately 1751 nuclear masses with $N, Z  \geq 8$ are reported in Ref.~\cite{JDuflo:1995}.\\

\noindent
\texttt{table}=`1995-ETFSI'.\\
This semi-classical calculation is based on the extended Thomas-Fermi plus Strutinsky integral method, including shell and BCS pairing corrections with a Skyrme force. Approximately 1680 nuclear masses with $36 \leq A \leq 300$ are reported in Ref.~\cite{YAboussir:1995}.\\

\noindent
\texttt{table}=`1995-FRDM'.\\
This macroscopic-microscopic calculation is based on the finite-range droplet macroscopic model (FRDM) and the folded-Yukawa single-particle microscopic model. The results of~\cite{PMoller:1995} present 8979 nuclei ranging from $Z, N \geq 8$ to $A = 339$.\\

\noindent
\texttt{table}=`2005-KTUY'.\\
This Macroscopic-microscopic mass model is based on a gross macroscopic term and a microscopic shell correction derived from spherical single-particle potentials. The model includes an improved even-odd term to account for pairing effects. The results of~\cite{HKoura:2005} present 9437 nuclei ranging from $Z, N \geq 2$ to $Z \leq 130$, $N \leq 200$. \\

\noindent
\texttt{table}=`2007-HFB14'.\\
This microscopic mass model is based on a nuclear EDF of Skyrme type. It solves HFB equations on a harmonic oscillator basis and imposes axial symmetry. HFB14~\cite{SGoriely:2007} is fitted to experimental masses and fission data by adjusting a vibrational term in the phenomenological collective correction.\\

\noindent
\texttt{table}=`2010-HFB21'.\\
This microscopic mass model is based on a nuclear EDF of extended-Skyrme type. It solves HFB equations on a harmonic oscillator basis and imposes axial symmetry. HFB21~\cite{SGoriely:2010} is fitted to experimental masses and realistic calculations for neutron matter energy in homogeneous matter. \\

\noindent
\texttt{table}=`2010-WS*'.\\
This macroscopic-microscopic mass model is based on a liquid-drop macroscopic term and a Strutinsky shell correction. The model WS$^*$~\cite{NWang:2010} incorporates a mirror nuclei constraint to improve isospin symmetry and refines the Coulomb energy term. \\

\noindent
\texttt{table}=`2011-WS3'.\\
The macroscopic-microscopic mass model WS3~\cite{MLiu:2011} based on a liquid-drop macroscopic term and a Strutinsky shell correction was further refined by incorporating mirror nuclei constraints, Wigner-like effects and additional residual corrections. \\

\noindent
\texttt{table}=`2013-HFB22', `2013-HFB23', `2013-HFB24', `2013-HFB25', `2013-HFB26'.\\
Extended-Skyrme Hartree-Fock-Bogoliubov (HFB) mass models are reported in Ref.~\cite{SGoriely:2013}. The models are fitted to experimental masses AME12, realistic calculations for neutron matter energy in homogeneous matter while varying the symmetry coefficient $E_\sym$ from 29 to 32 MeV. \\

\noindent
\texttt{table}=`2021-BSkG1'.\\
This microscopic mass model is based on a nuclear EDF of Skyrme type. The BSkGs models solve HFB calculations on a three-dimensional coordinate-space representation. BSkG1~\cite{GScamps:2021} incorporates for the first time the possibility of triaxial ground-state deformation on global nuclear structure models.\\

\noindent
\texttt{table}=`2022-BSkG2'.\\
This microscopic mass model is based on an EDF of Skyrme type. In addition to the features of BSkG1, BSkG2~\cite{WRyssens:2022} allows for the effects of time-reversal symmetry breaking. This enables the model to access the spin and current densities in the ground states of odd-mass and odd-odd nuclei. 
BSkG2 also includes information on fission barrier heights of actinide nuclei in the parameter adjustment. \\

\noindent
\texttt{table}=`2023-BSkG3'.\\
This microscopic mass model is based on an EDF of extended-Skyrme type. The improvements of BSkG3~\cite{GGrams:2023} for BSkG1 and BSkG2 are: i) break reflection symmetry, allowing for both triaxial and octupole-deformed ground states at the same time; ii) produce stiff EoS, ensuring the model can accommodate the existence of heavy pulsars; iii) replace the phenomenological pairing interaction of previous models by a more microscopically grounded interaction designed to match the $^1 S_0$ pairing gaps in NM and SM deduced from ab-initio calculations.\\

\noindent
\texttt{table}=`2025-BSkG4'.\\
This microscopic mass model is based on an EDF of extended-Skyrme type. In addition to the features of BSkG3, BSkG4~\cite{GGrams:2025} offers an improved treatment of $^1 S_0$ nucleon pairing gaps in asymmetric nuclear matter, inspired by advanced many-body calculations.\\

\begin{figure}[t]
\centering
\includegraphics[scale=0.52]{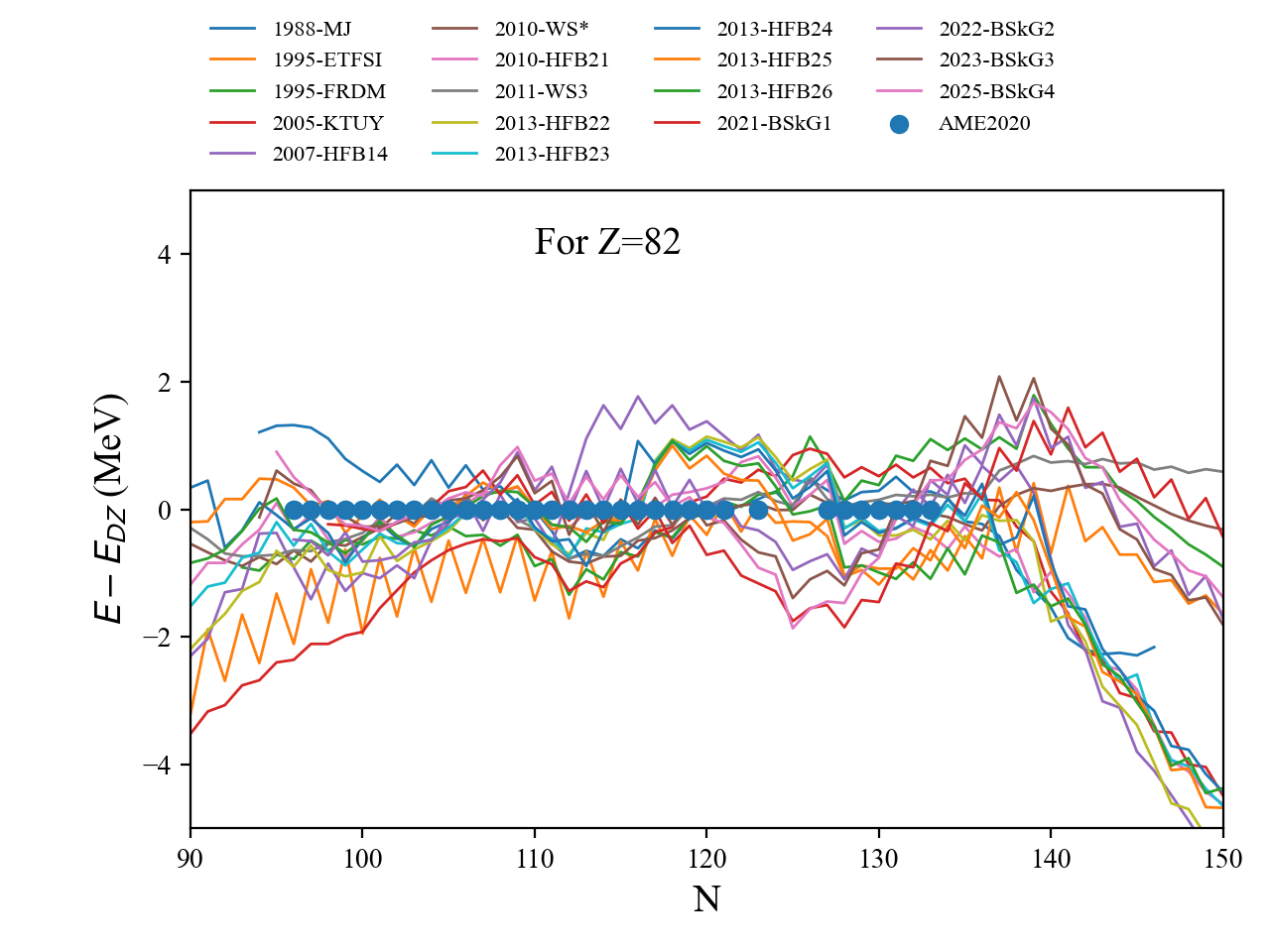}
\includegraphics[scale=0.52]{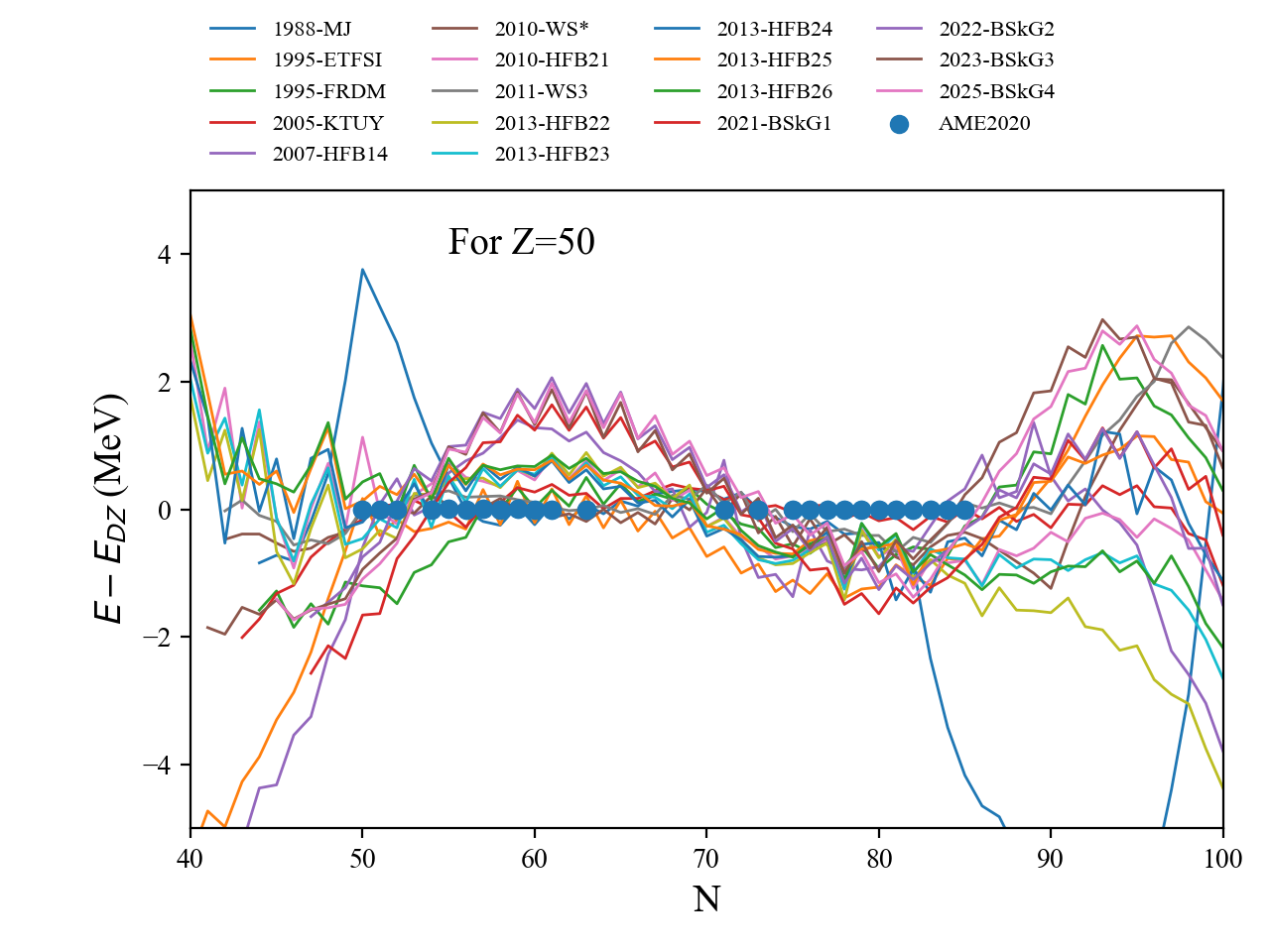}
\caption{Differences between binding energies predicted by different models with respect to the one predicted by Duflo-Zuker ('1995-DZ') for $Z=82$ (top) and $Z=50$ (bottom). The experimental measurements reported in AME 2020 mass table are also shown. Figure generated with \texttt{nuc\_setupBETheo\_diff\_plot.py}.}
\label{fig:mass:theo:Z82Z50}
\end{figure}

The outputs provided by these theoretical tables are the following:
\texttt{mas.nucA}, \texttt{mas.nucZ}, \texttt{mas.nucN}, \texttt{mas.nucI}, \texttt{mas.nucBE}, and \texttt{mas.Zmax} with obvious notations (same as for the experimental mass table detailed in Sec.~\ref{sec:nuc:expchart}). The attribute \texttt{mas.nucBE2A} contains the binding energy per nucleon.

An illustration of the use of the theoretical mass models is shown in Fig.~\ref{fig:mass:theo:Z82Z50} for Sn and Pb isotopic chains. The difference between the theoretical mass model and the Duflo-Zuker~\cite{JDuflo:1995} one is shown in Fig.~\ref{fig:mass:theo:Z82Z50}. The difference between the experimental data from AME 2020~\cite{WJHuang:2021} and the Duflo-Zuker mass model is also shown in blue dots. These dots are very flat and close to zero, although they are not identical to zero. However, this remark legitimates taking the Duflo-Zuker mass model as a reference one, even where there are no experimental data. Note that the most significant differences occur for neutron number for which there are no experimental data (low and high neutron numbers). All mass models have parameters adjusted to experimental nuclear masses, therefore, the extreme isospin asymmetries are unconstrained by nuclear experimental data.

The comparison between different models is obtained from \texttt{mas.diff} function in the following way:
\begin{lstlisting}[language=Python]
ref=nuda.nuc.setupBETheo( table=`1995_DZ' )
N_diff, A_diff, BE_diff, BE2A_diff=ref.diff( table=`2007_HFB21', Zref=50 )
\end{lstlisting}
where the object \texttt{ref} contains the reference mass table, the variable \texttt{table} contains the name of the mass table to compare with, and finally, the variable \texttt{Zref} contains the reference charge of the isotopic chain.

\subsection{Two-neutron and two-proton separation energies}
\label{sec:nuc:S2}

The experimental mass tables as well as the theoretical ones can be employed to extract the two-neutron separation energies, $S_{2n}$, and the two-proton one, $S_{2p}$, defined, respectively, as:
\begin{eqnarray}
S_{2n} &=& E(Z,N)-E(Z,N-2) \, , \\
S_{2p} &=& E(Z,N)-E(Z-2,N) \, .
\end{eqnarray}
From the theoretical mass models, $S_{2n}$ and $S_{2p}$ can be obtained in the following way:
\begin{lstlisting}[language=Python]
mas = nuda.nuc.setupBETheo( table=`1995_DZ' )
Nref=50; Zref=50;
mas2=mas.isotopes( Zref=Zref )
s2n=mas2.S2n( Zref=Zref )
mas2=mas.isotones( Nref=Nref )
s2p = mas2.S2p( Nref=Nref )
\end{lstlisting}
From the experimental mass tables, they can be obtained in the following way:
\begin{lstlisting}[language=Python]
mas = nuda.nuc.setupBEExp( table=`AME', version=`2020' )
mas2 = mas.select( state=`gs', interp=`n' )
Nref=50; Zref=50;
mas3=mas2.isotopes( Zref=Zref )
s2n=mas3.S2n( Zref=Zref )
mas3=mas2.isotones( Nref=Nref )
s2p=mas3.S2p( Nref=Nref )
\end{lstlisting}
There is an additional selection of the nuclei in their ground state and an exclusion of the interpolation that shall be applied to the experimental mass table.

The attributes of the objects  are Numpy arrays: \texttt{s2n.s2n\_N} and \texttt{s2n.s2n\_E} for $S_{2n}$ and \texttt{s2p.s2p\_Z} and \texttt{s2p.s2p\_E} for $S_{2p}$.

\begin{figure}[t]
\centering
\includegraphics[scale=0.52]{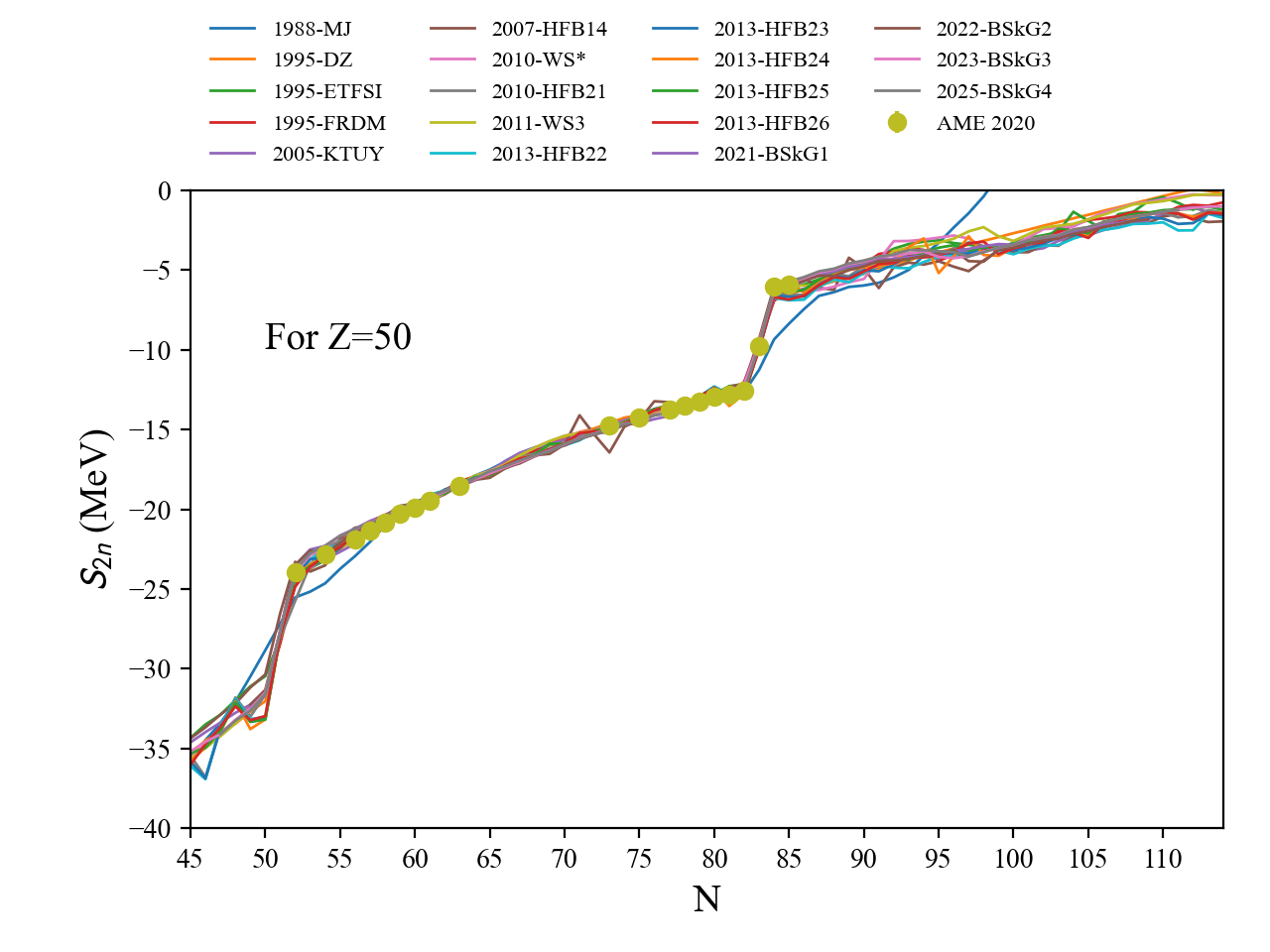}
\includegraphics[scale=0.52]{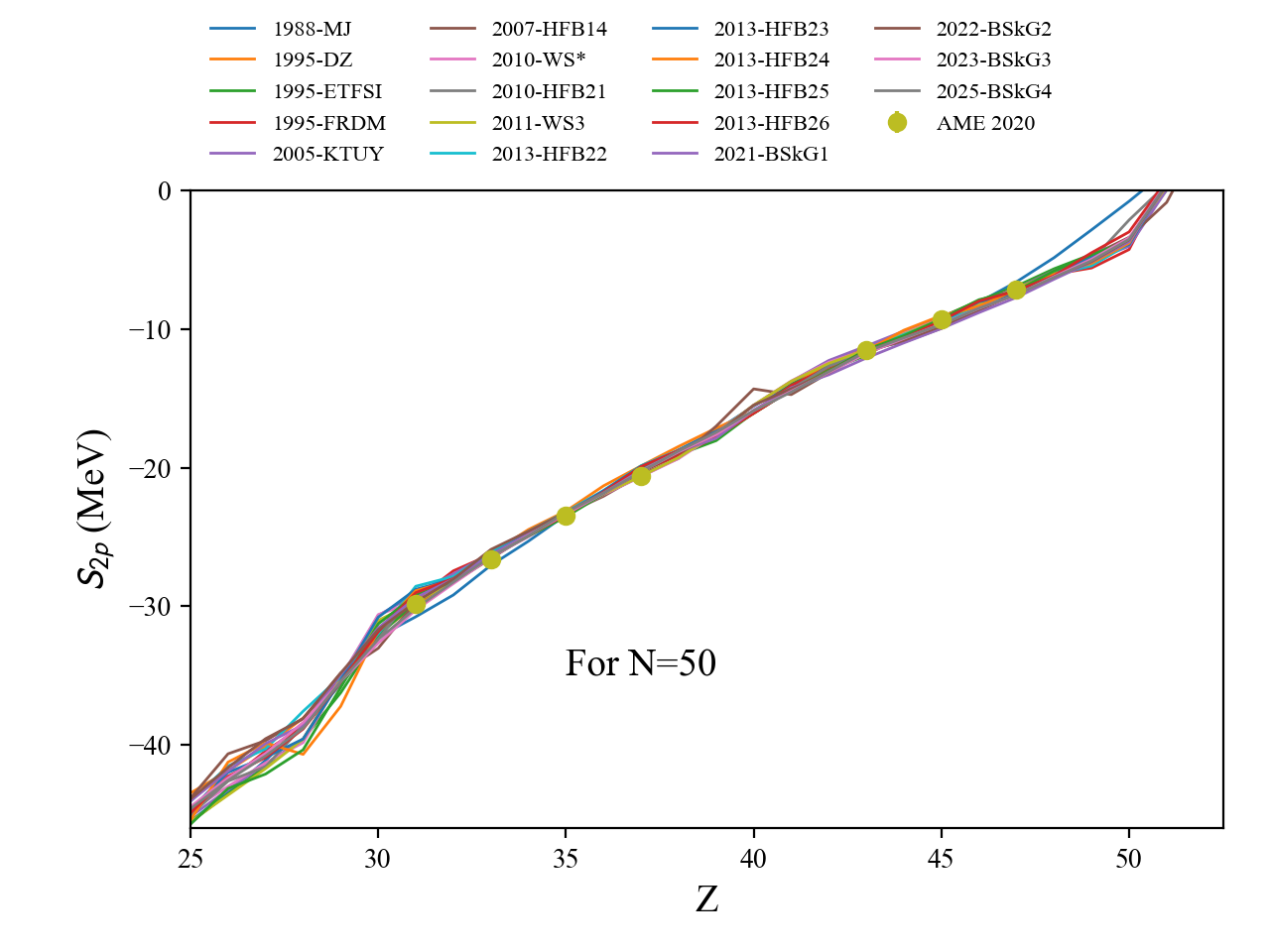}
\caption{Two-neutron $S_{2n}$ and two-proton $S_{2p}$ separation energies for $Z=50$ (top) and $N=50$ (bottom).
The experimental measurements reported in AME 2020 are also shown. Figure generated with \texttt{nuc\_setupBETheo\_S2n\_plot.py} and \texttt{nuc\_setupBETheo\_S2p\_plot.py}.}
\label{fig:mass:theo:S2}
\end{figure}

A comparison of model predictions and experimental data for the  $S_{2n}$ ($S_{2p}$) is shown Fig.~\ref{fig:mass:theo:S2} on the top (bottom) for $Z=50$ ($N=50$). The $S_{2n}$ ($S_{2p}$) are negative for bound systems and change their sign at the drip lines. The plots for $S_{2n}$ and $S_{2p}$ exhibit a smooth trend, except close to shell closure. There are changes of slopes for typical magic numbers: 28, 50, 82. Note however the difference between $S_{2n}$ and $S_{2p}$: shell effects are more pronounced for $S_{2n}$ compared to $S_{2p}$. This difference arises because neutron shell effects more directly influence the total energy, while the influence of proton shells is smoothed by Coulomb repulsion and proton-neutron interactions.

\subsection{Odd-even mass staggering}
\label{sec:nuc:oems}

There is another interesting quantity which can be extracted from the nuclear mass: The odd-even mass staggering. It can be obtained from the 3-point formula for isotopes,
\begin{equation}
\Delta_{3n} = \frac{(-1)^N}{2}\Big( E(Z,N-1)-2E(Z,N)+E(Z,N+1) \Big) \, ,
\end{equation}
and, for isotones,
\begin{equation}
\Delta_{3p} = \frac{(-1)^Z}{2}\Big( E(Z-1,N)-2E(Z,N)+E(Z+1,N) \Big) \, .
\end{equation}

From the theoretical mass models, $\Delta_{3n}$ and $\Delta_{3p}$ can be obtained in the following way:
\begin{lstlisting}[language=Python]
mas = nuda.nuc.setupBETheo( table=`1995_DZ' )
Nref=50; Zref=50;
mas2=mas.isotopes( Zref=Zref )
D3n=mas2.D3n( Zref=Zref )
mas2=mas.isotones( Nref=Nref )
D3p=mas2.D3p( Nref=Nref )
\end{lstlisting}
From the experimental mass tables, they can be obtained in the following way:
\begin{lstlisting}[language=Python]
mas = nuda.nuc.setupBEExp( table=`AME', version=`2020' )
mas2=mas.select( state=`gs', interp=`n' )
Nref=50; Zref=50;
mas3=mas2.isotopes( Zref=Zref )
D3n=mas3.D3n( Zref=Zref )
mas3=mas2.isotones( Nref=Nref )
D3p=mas3.D3p( Nref=Nref )
\end{lstlisting}
Similarly to the calculation of the separation energies, there is an additional selection of the nuclei in their ground state and an exclusion of the interpolation that shall be applied to the experimental mass table.

The attributes of the objects are Numpy arrays: \texttt{D3n.D3n\_N\_even} and \texttt{D3n.D3n\_E\_even} for $\Delta_{3n}$ for even $N$ and \texttt{D3p.D3p\_Z\_even} and \texttt{D3p.D3p\_E\_even} for $\Delta_{3p}$ for even $Z$. For odd $N$ and $Z$, simply replace \texttt{even} by \texttt{odd} in the name of the attributes.

The smooth $Z$ and $N$ behavior of the odd-even mass staggering is employed to define empirical expressions. These empirical relations are encoded in the \texttt{nuda} toolkit in the following way:
\begin{lstlisting}[language=Python]
print( nuda.nuc.delta_emp( A, Z, formula ) )
\end{lstlisting}
where the variable \texttt{formula} can be one of the following choice:\\

\noindent
\texttt{formula}=`classic'.\\
Classical formula from Ref.~\cite{BohrMottelson:1998}:
\begin{equation}
\Delta(A) = 12 A^{-1/2}\hbox{ MeV} \, .
\end{equation}\\

\noindent
\texttt{formula}=`Vogel'.\\
By studying the OEMS for $50<Z<82$ and $82<N<126$, the following empirical formula  best reproduces nuclear data for isotopic and isotonic chains, see Ref.~\cite{PVogel:1984},
\begin{equation}
\Delta(A,Z) = \left( 7.2 -44 [(N-Z)/A]^2 \right) A^{-1/3}\hbox{ MeV} \, .
\end{equation}

\begin{figure}[t]
\centering
\includegraphics[scale=0.52]{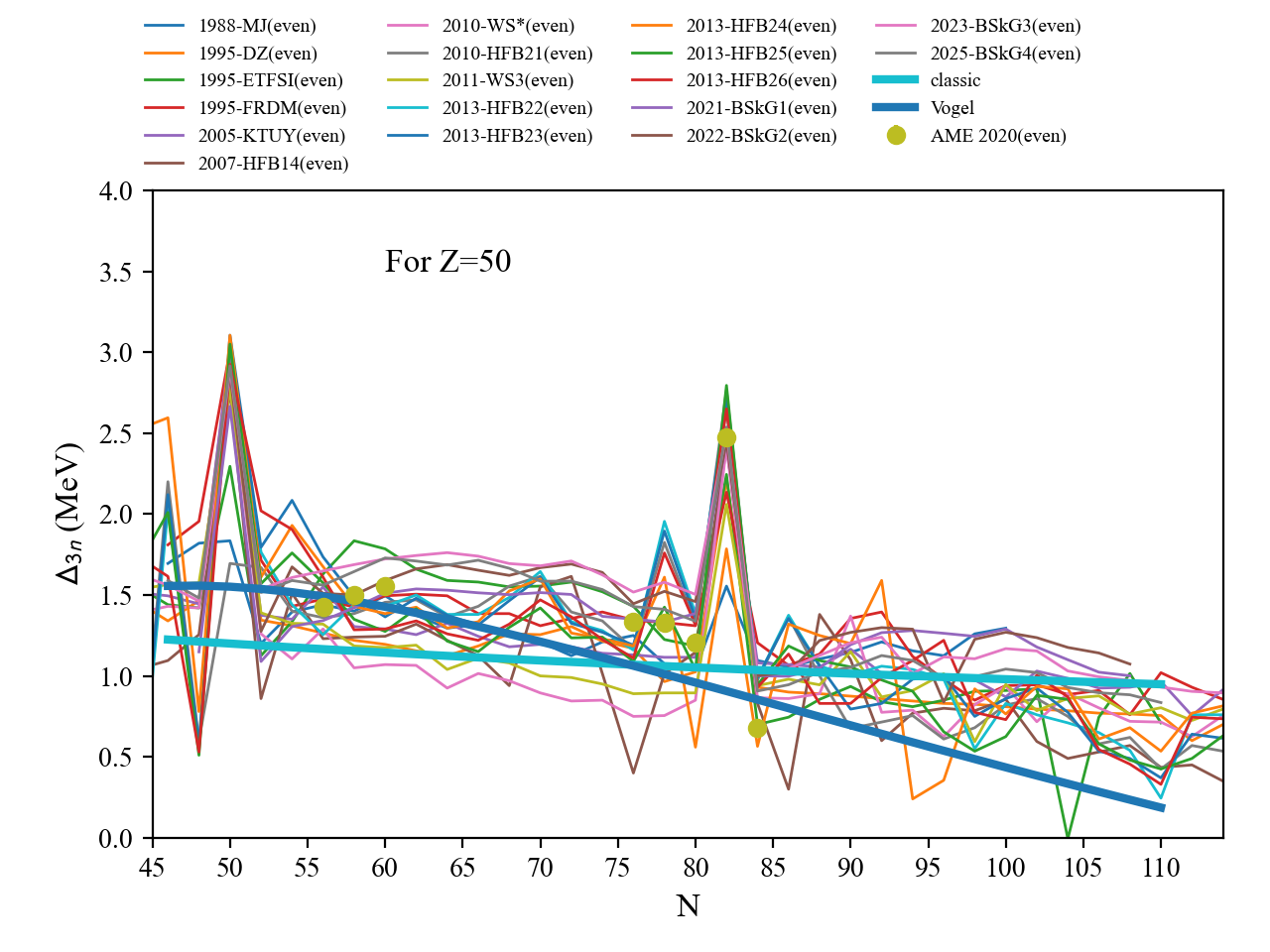}
\includegraphics[scale=0.5]{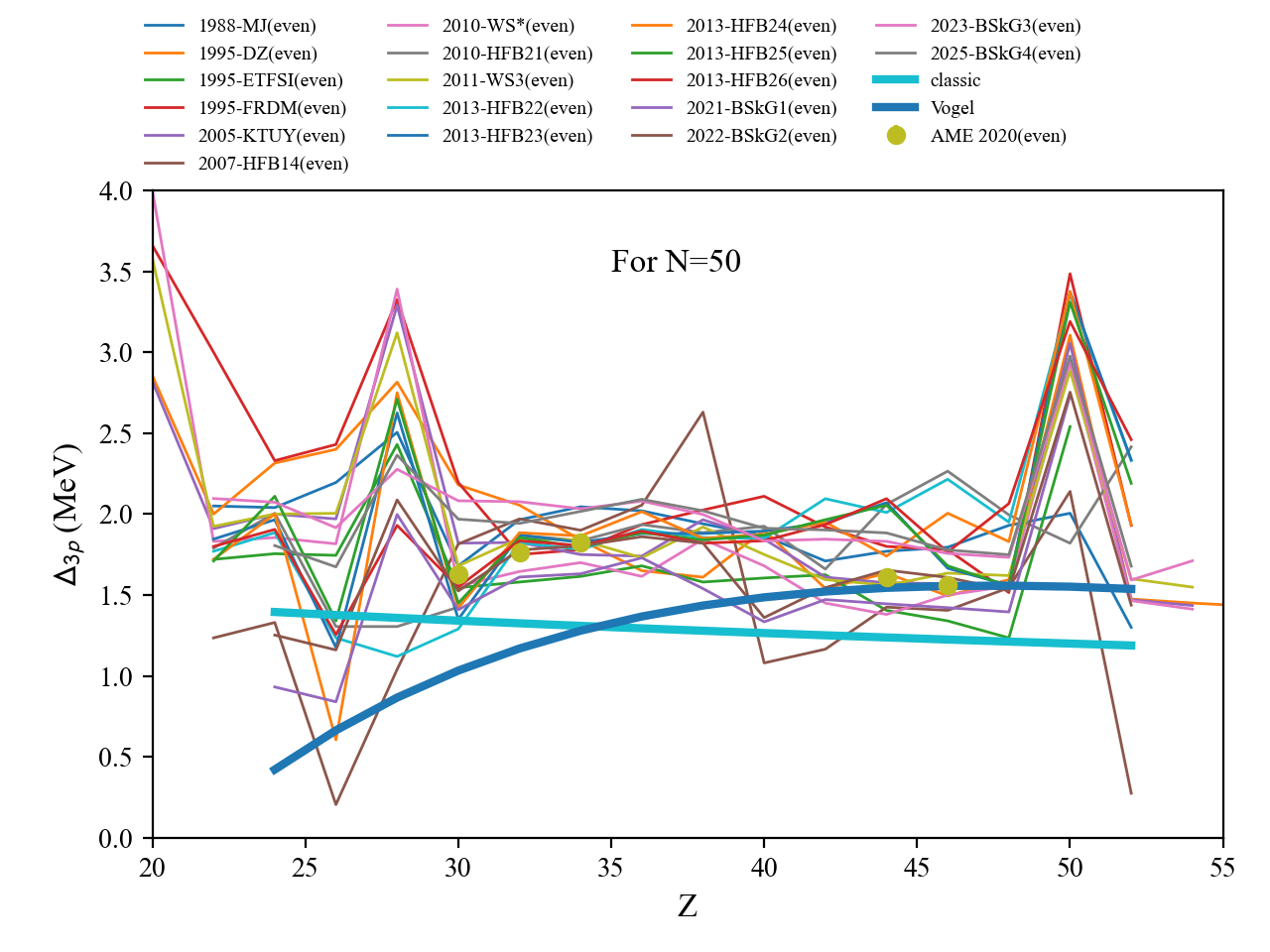}
\caption{Odd-even mass staggering $\Delta_{3n}$ for $Z=50$ (top) and $\Delta_{3p}$ for $N=50$ (bottom).
The experimental measurements reported in AME 2020 mass table are also shown. Figure generated with \texttt{nuc\_setupBETheo\_D3p\_plot.py}.}
\label{fig:mass:theo:oems}
\end{figure}

\begin{figure*}[t]
\centering
\includegraphics[scale=0.9]{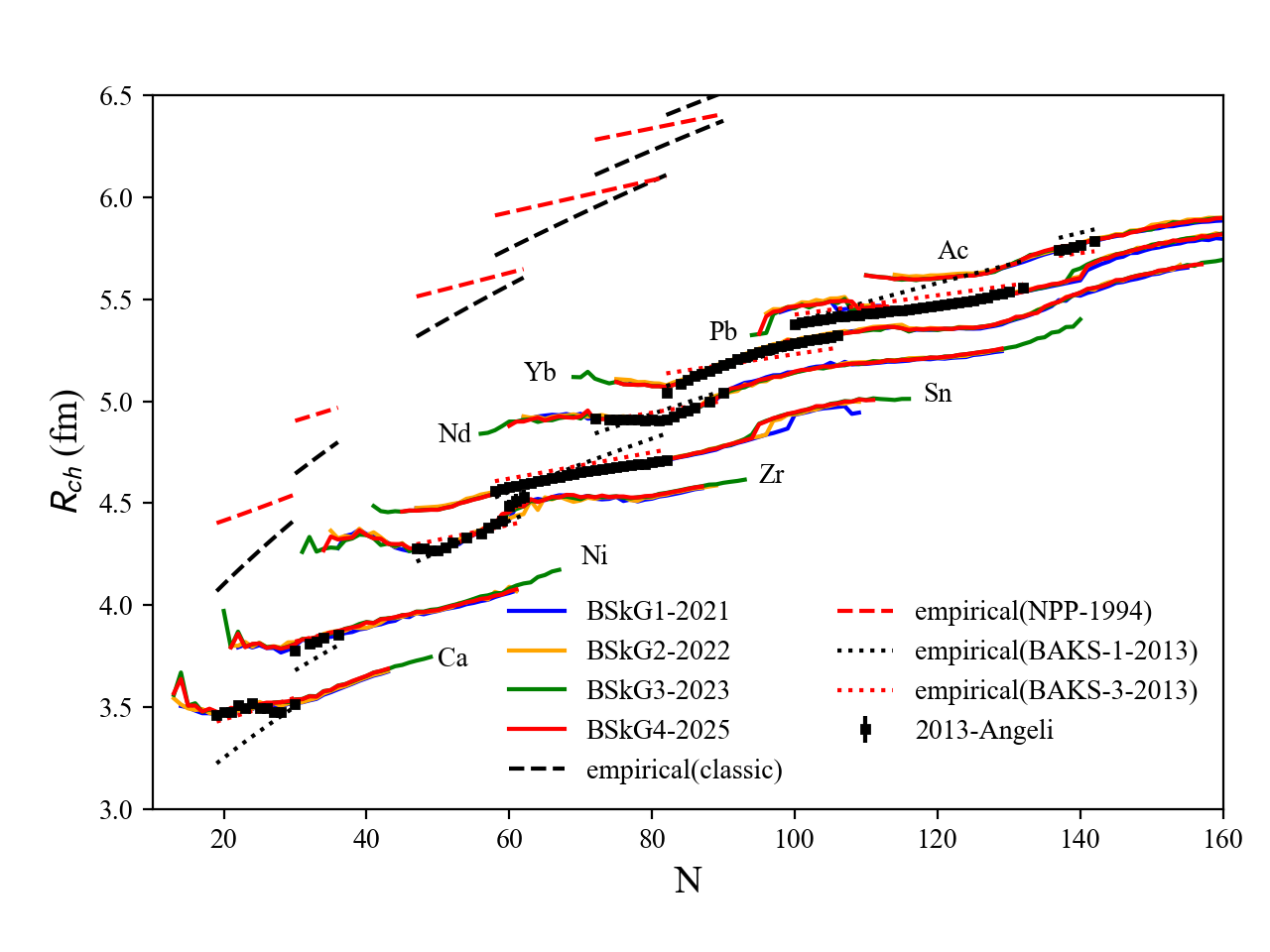}
\caption{Charge radii for Ca, Ni, Zr, Sn, Nd, Yb, Pb and Ac isotopes and for the models available in the \texttt{nuda} toolkit. Figure generated with \texttt{nuc\_setupRchTheo\_plot.py}.}
\label{fig:Rch}
\end{figure*}

The odd-even mass staggering are shown in Fig.~\ref{fig:mass:theo:oems} for a set of theoretical models and for the experimental data (dots) for even $N$ or $Z$. The Sn isotopic chain (top) and $N=50$ isotonic chain (bottom) are shown. The empirical formulae are shown with thick solid lines. The peaks at shell closure are visible for magic numbers. The dispersion among theoretical models is larger than the experimental uncertainties. The empirical formulae are adjusted on the isotopic chains and reproduce rather well the smooth behavior of the data shown on the top panel of Fig.~\ref{fig:mass:theo:oems}. For the isotonic chain shown on the bottom panel of Fig.~\ref{fig:mass:theo:oems}, the empirical formulae are slightly below the experimental data.

\subsection{Nuclear charge radii}
\label{sec:nuc:rch}

Nuclear charge radii play a key role in determining nuclear shell structure and provide direct information on the Coulomb energy in nuclei. It impacts nuclear astrophysics and imposes strong constraints on the saturation properties of nuclear interactions~\cite{FBuchinger:1994}. The latest compilation of nuclear charge radii for stable and unstable nuclei is the one by Angeli and Marinova in 2013~\cite{IAngeli:2013}. Since then, laser spectroscopy has been employed to determine the charge radii of exotic nuclei, but no new table yet exists.

The complete list of available experimental tables is given with the following instructions:
\begin{lstlisting}[language=Python]
print( nuda.nuc.rch_exp_tables( ) )
\end{lstlisting}
This instruction provides only one output since there is only one mass table available for the moment: `2013-Angeli'~\cite{IAngeli:2013}.
Once the variable \texttt{table} is chosen, the call for the experimental table can be performed in the following way:
\begin{lstlisting}[language=Python]
rch = nuda.nuc.setupRchExp( table=`2013-Angeli' )
rch.print_outputs()
\end{lstlisting}

We now provide more details about the experimental nuclear tables available in the \texttt{nuda} toolkit.\\

\noindent
\texttt{table}=`2013-ANGELI'~\cite{IAngeli:2013}.\\

The object \texttt{rch} contains the whole experimental table. In the following, we are interested in a subset of data, for the isotopic chain, for instance. The charge radii for an isotopic chain identified by the selection of a nuclear charge in the variable \texttt{Zref} are obtained in the following way:
\begin{lstlisting}[language=Python]
rch_itp = nuda.nuc.setupRchExpIsotopes( rch, Zref=Zref )
\end{lstlisting}
which attributes are the following Numpy arrays: \texttt{rch\_itp.A}, \texttt{rch\_itp.Z}, \texttt{rch\_itp.N}, \texttt{rch\_itp.Rch}, and \texttt{rch\_itp.Rch\_err}.

Some of the mass models given in the \texttt{nuda} toolkit also provide the charge radii that they predict. The list of theoretical mass models providing charge radii is provided by the following command:
\begin{lstlisting}[language=Python]
print( nuda.nuc.rch_theo_tables( ) )
\end{lstlisting}
One can choose the variable \texttt{table} among the following options, `2021-BSkG1', `2022-BSkG2', `2023-BSkG3', `2025-BSkG4', which are described in Sec.~\ref{sec:nuc:theochart}.
The theoretical table is loaded in the following way:
\begin{lstlisting}[language=Python]
rch = nuda.nuc.setupRchTheo( table=`2025-BSkG4' )
rch.print_outputs()
\end{lstlisting}
for the choice \texttt{table = `2025-BSkG4'}, and the results for a given isotopic chain is obtained as,
\begin{lstlisting}[language=Python]
rch_itp=nuda.nuc.setupRchTheoIsotopes( rch, Zref=Zref )
\end{lstlisting}
The attributes are the same as the ones which have been given for the experimental table: \texttt{rch\_itp.A}, \texttt{rch\_itp.Z}, \texttt{rch\_itp.N}, \texttt{rch\_itp.Rch}, and \texttt{rch\_itp.Rch\_err}.

\begin{table}[t]
\begin{center}
\caption{Parameters of Eq.~\eqref{eq:rch:empiric} for different choices for the variable \texttt{formula}.}
\label{table:rch:empirical}
\tabcolsep=0.24cm
\def\arraystretch{1.5}
\begin{tabular}{lllll}
\hline\noalign{\smallskip}
\texttt{formula} & $r_0$ & $b$ & $c$ & Ref.\\
 & fm &  &  \\
\hline\noalign{\smallskip}
'classic' & 1.2 & 0 & 0 & \cite{BohrMottelson:1998} \\
'1994-NPP' & 1.24 & 0.191 & 1.646 & \cite{BNerloPomorska:1994} \\
'2013-BAKS-1' & 0.951 & 0 & 0 & \cite{TBayram:2013}\\ 
'2013-BAKS-2' & 0.996 & 0.278 & 0 &\cite{TBayram:2013}\\
'2013-BAKS-3' & 0.966 & 0.182 & 1.652 & \cite{TBayram:2013}\\
\noalign{\smallskip}\hline
\end{tabular}
\end{center}
\end{table}

In Fig.~\ref{fig:Rch} we show a comparison between the experimental table and the theoretical predictions for various isotopic chains: Ca, Ni, Zr, Sn, Nd, Yb, Pb, and Ac. We also show some empirical formulae for nuclear charge radii that have been proposed in the literature.
They can be obtained from the following function:
\begin{lstlisting}[language=Python]
nuda.nuc.rch_emp( A, Z, formula )
\end{lstlisting}
where the variable \texttt{formula} can be one of the following choice:\\

\noindent
\texttt{formula}=`classic'.\\
Classical formula from Ref.~\cite{BohrMottelson:1998}:
\begin{equation}
r_{ch}(A) = 1.2 A^{1/3}\hbox{ fm} \, .
\label{eq:rch:classic}
\end{equation}

\noindent
\texttt{formula}=`1994-NPP'~\cite{BNerloPomorska:1994}, `2013-BAKS-1'~\cite{TBayram:2013}, `2013-BAKS-2'~\cite{TBayram:2013}, or `2013-BAKS-3'~\cite{TBayram:2013}.\\
Other empirical parameterizations have considered a generalization of Eq.~\eqref{eq:rch:classic} in the following form:
\begin{equation}
r_{ch}(A,Z) = r_0 * \left( 1-b\frac{N-Z}{A} +c\frac{1}{A} \right) A^{1/3}\hbox{ fm} \, .
\label{eq:rch:empiric}
\end{equation}
Note that Eq.~\eqref{eq:rch:classic} is a peculiar case of Eq.~\eqref{eq:rch:empiric} for which $r_0=1.2$ MeV and $b=c=0$.

The value of the parameters $r_0$, $b$, and $c$ for the different empirical adjustments are given in Table~\ref{table:rch:empirical}.

\subsection{Neutron-skin thickness}
\label{sec:nuc:nskin}

For the neutron-skin thickness, the \texttt{nuda} toolkit provides a set of experimental measurements and theoretical predictions.
The complete nuclei list in the \texttt{nuda} toolkit is given with the following instructions:
\begin{lstlisting}[language=Python]
print( nuda.rnp_exp( ) )
\end{lstlisting}

The available experimental/analysis data for the neutron-skin thickness ($R_{\rm np}=R_n-R_p$) closely follow the organization given in Ref.~\cite{FSchupp:2024}, in which the authors summarize a comprehensive set of results in Tables III and IV of the Appendix.

\begin{figure}[t]
\centering
\includegraphics[scale=0.32]{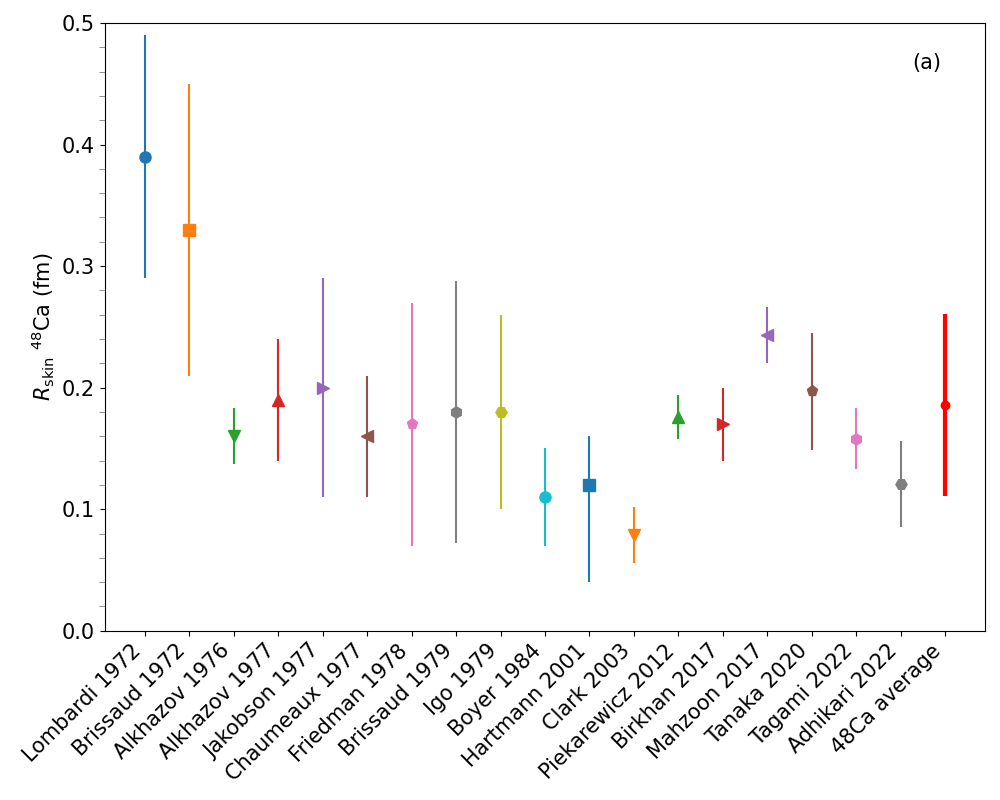}\\
\includegraphics[scale=0.32]{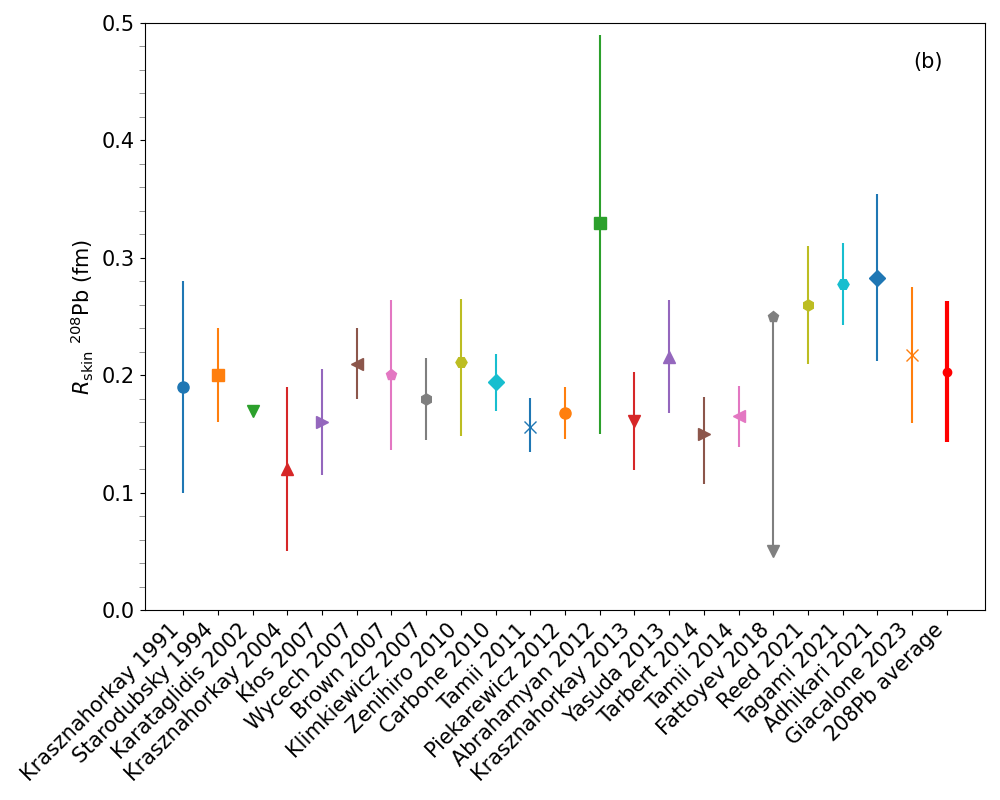}
\caption{Data deduced by different experiments and analyses for neutron skin thickness of (a) \(^{48}\mathrm{Ca}\) and (b) \(^{208}\mathrm{Pb}\), based on the available datasets in the \texttt{nuda} toolkit. Figures generated with \texttt{nuc\_setupRnpExp\_plot.py}.}
\label{fig:RnpExp}
\end{figure}

We currently have two data sources available in the toolkit, namely, `48Ca' and `208Pb', corresponding to the respective nuclei. Once the variable \texttt{source} is chosen, the information regarding the experimental/analysis is performed in the following way:
\begin{lstlisting}[language=Python, breaklines=true]
Rnp = nuda.setupRnpExp( source=`48Ca', cal=`1' )
Rnp.print_outputs()
\end{lstlisting}

In the source `48Ca', we have a total of 18 data points available extracted from a variety of experimental and analysis methods. Each of the 18 data points is assigned to the variable \texttt{cal = 1, \dots, 18}. The references corresponding to these points are \cite{JCLombardi:1972,IBrissaud:1972,GDAlkhazov:1976,GDAlkhazov:1977,MJJakobson:1977,AChaumeaux:1977,EFriedman:1978,IBrissaud:1979,GIgo:1979,KGBoyer:1984,FJHartmann:2001,BCClark:2003,JPiekarewicz:2012,JBirkhan:2017,MHMahzoon:2017,MTanaka:2020,STagami:2022,DAdhikari:2022}, and data are shown in Fig.~\ref{fig:RnpExp}(a). For the source `208Pb', we have a total of 22 data points available with similar structure (\texttt{cal = 1, \dots, 22}). The respective references are \cite{AKrasznahorkay:1991,VEStarodubsky:1994,SKarataglidis:2002,AKrasznahorkay:2004,BKlos:2007,SWycech:2007,BABrown:2007,AKlimkiewicz:2007,JZenihiro:2010,ACarbone:2010,ATamii:2011,JPiekarewicz:2012,SAbrahamyan:2012,AKrasznahorkay:2013,JYasuda:2013,CMTarbert:2014,ATamii:2014,FJFattoyev:2018,BTReed:2021,STagami:2021,DAdhikari:2021,GGiacalone:2023}, and in Fig.~\ref{fig:RnpExp}(b) we display the data.

The theoretical predictions for $R_{\rm np}$ are calculated by using some parametrizations implemented in the \texttt{nuda} toolkit ($R_p$ and $R_n$ are obtained from the method used in Ref.~\cite{BVCarlson:2023}). The complete list is accessed from the command:
\begin{lstlisting}[language=Python]
print( nuda.nuc.rnp_theo_params( model ) )
\end{lstlisting}
We have three models in this calculation: `Skyrme', `NLRH', and `DDRH'. Once the model is chosen, $R_{\rm np}$ for $^{48}$Ca and $^{208}$Pb are listed in the \texttt{nuda} toolkit from the command:
\begin{lstlisting}[language=Python, breaklines=true]
Rnp_theo = nuda.nuc.setupRnpTheo( model=`Skyrme', param=`BSK14', nucleus=`48Ca' )
Rnp_theo.print_outputs()
\end{lstlisting}

\begin{figure}[t]
\centering
\includegraphics[scale=0.32]{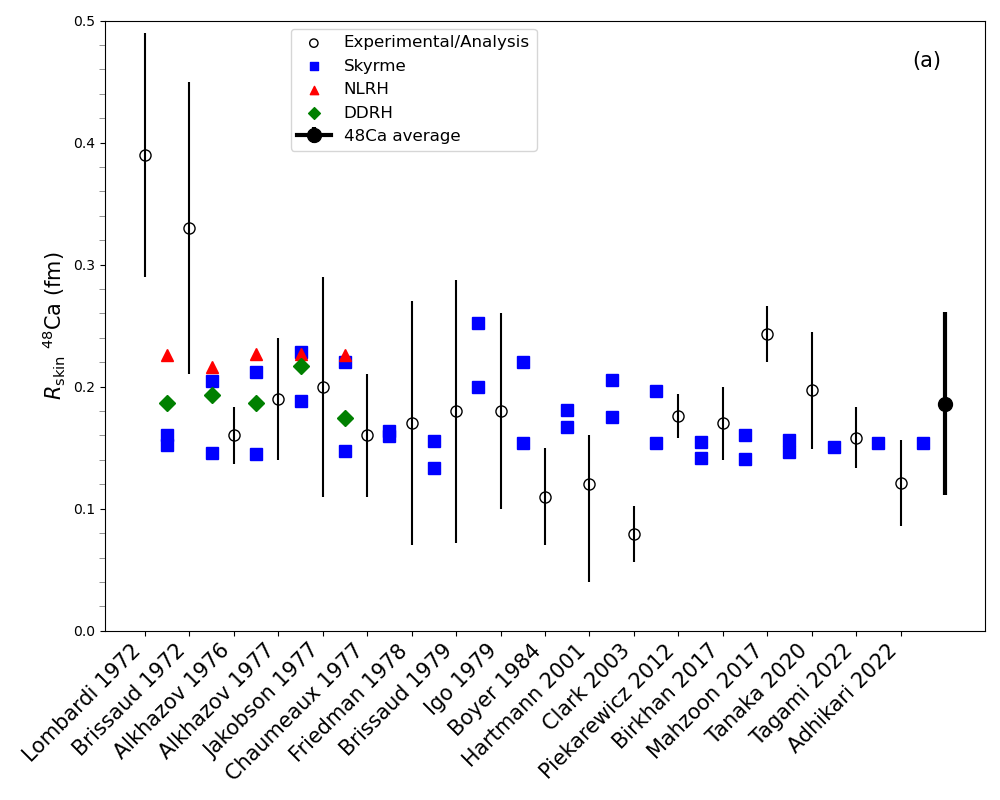}
\includegraphics[scale=0.32]{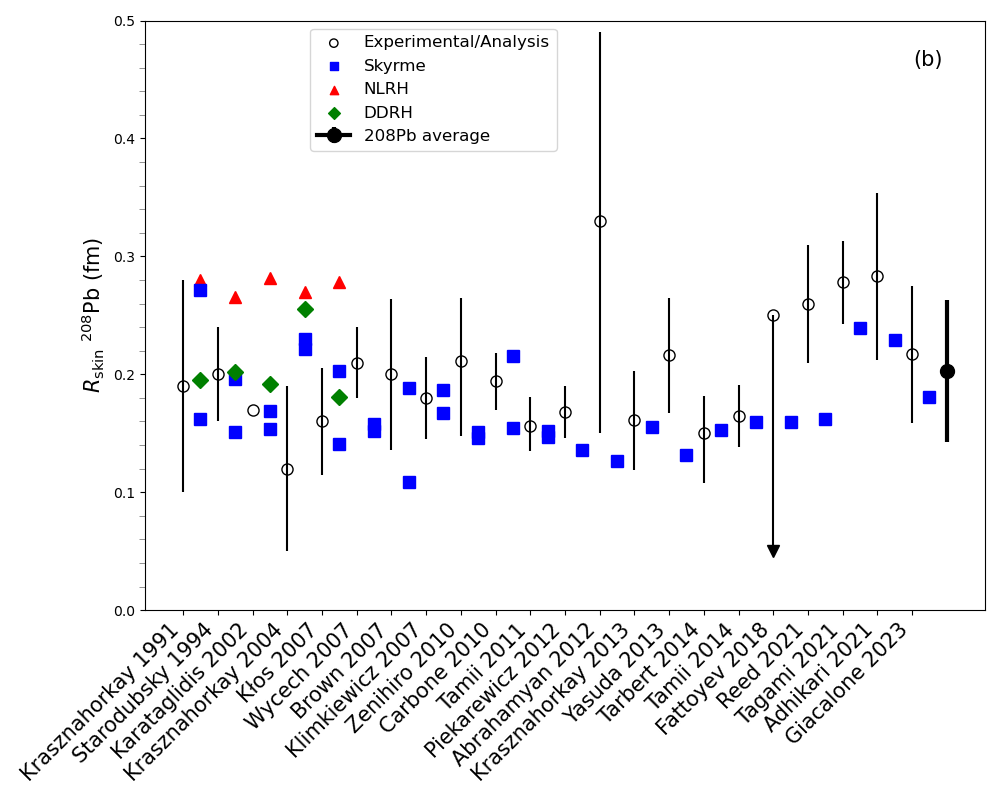}
\caption{Theoretical neutron skin thickness for (a) \(^{48}\mathrm{Ca}\) and (b) \(^{208}\mathrm{Pb}\), as predicted by models implemented in the \texttt{nuda} toolkit. The experimental/analysis data from Fig.~\ref{fig:RnpExp} are also shown. Figures generated with \texttt{nuc\_setupRnpTheo\_plot.py}.}
\label{fig:RnpTheo}
\end{figure}

The theoretical predictions of $R_{\rm np}$ for \(^{48}\mathrm{Ca}\) and \(^{208}\mathrm{Pb}\) are shown in Fig.~\ref{fig:RnpTheo}. They can be directly compared to the experimental ones shown in Fig.~\ref{fig:RnpExp}.

\subsection{Experimental centroid energy for the ISGMR}
\label{sec:nuc:ka}

The ISGMR, or breathing mode, is often employed to evaluate the goodness of the nuclear modeling. The complete list of available experimental tables is given with the following instructions:
\begin{lstlisting}[language=Python]
print( nuda.nuc.isgmr_exp_tables( ) )
\end{lstlisting}
We have two tables available for the moment: `2018-ISGMR-LI'~\cite{TLi:2010} and `2018-ISGMR-GARG'~\cite{UGarg:2018}. Once the variable \texttt{table} is chosen, the call for the experimental ISGMR can be performed in the following way:
\begin{lstlisting}[language=Python]
gmr = nuda.nuc.setupISGMRExp( table=`2018-ISGMR-GARG' )
gmr.print_outputs()
\end{lstlisting}

We now provide more details about the experimental ISGMR tables available in the \texttt{nuda} toolkit.\\

\noindent
\texttt{table}= `2018-ISGMR-LI'.\\
Table from Ref.~\cite{TLi:2010}.\\

\noindent
\texttt{table}= `2018-ISGMR-GARG'.\\
Table from Ref.~\cite{UGarg:2018}.\\

\noindent
\texttt{table}= `2018-ISGMR-GARG-LATEX'.\\
Original table from Ref.~\cite{UGarg:2018}.\\

\noindent
\texttt{table}= `2022-average'.\\
Average table employed in Ref.~\cite{GGrams:2022}.\\

Note that the \texttt{nuda} toolkit provides only experimental data for the ISGMR energy. To select some isotopes from the table, there is a \texttt{select()} function that can be called in the following way:
\begin{lstlisting}[language=Python]
gmrs = gmr.select( Zref=50, obs=`M12Mm1' )
\end{lstlisting}
where \texttt{Zref} selects the isotope charge and \texttt{obs} the observable, that can be taken from the following options: `M12M0' ($m_1/m_0$), `M12Mm1' ($\sqrt{m_1/m_{-1}}$) or `M32M1' ($\sqrt{m_3/m_1}$).
The output arrays are: \texttt{gmrs.nucA} for the mass $A$ of the isotopes, \texttt{gmrs.cent} for the centroid of the GMR energy (as defined from \texttt{obs}), \texttt{gmrs.errp}, \texttt{gmrs.errm}, \texttt{gmrs.erra} for the positive, negative, and average uncertainties.

\begin{figure}[t]
\centering
\includegraphics[scale=0.52]{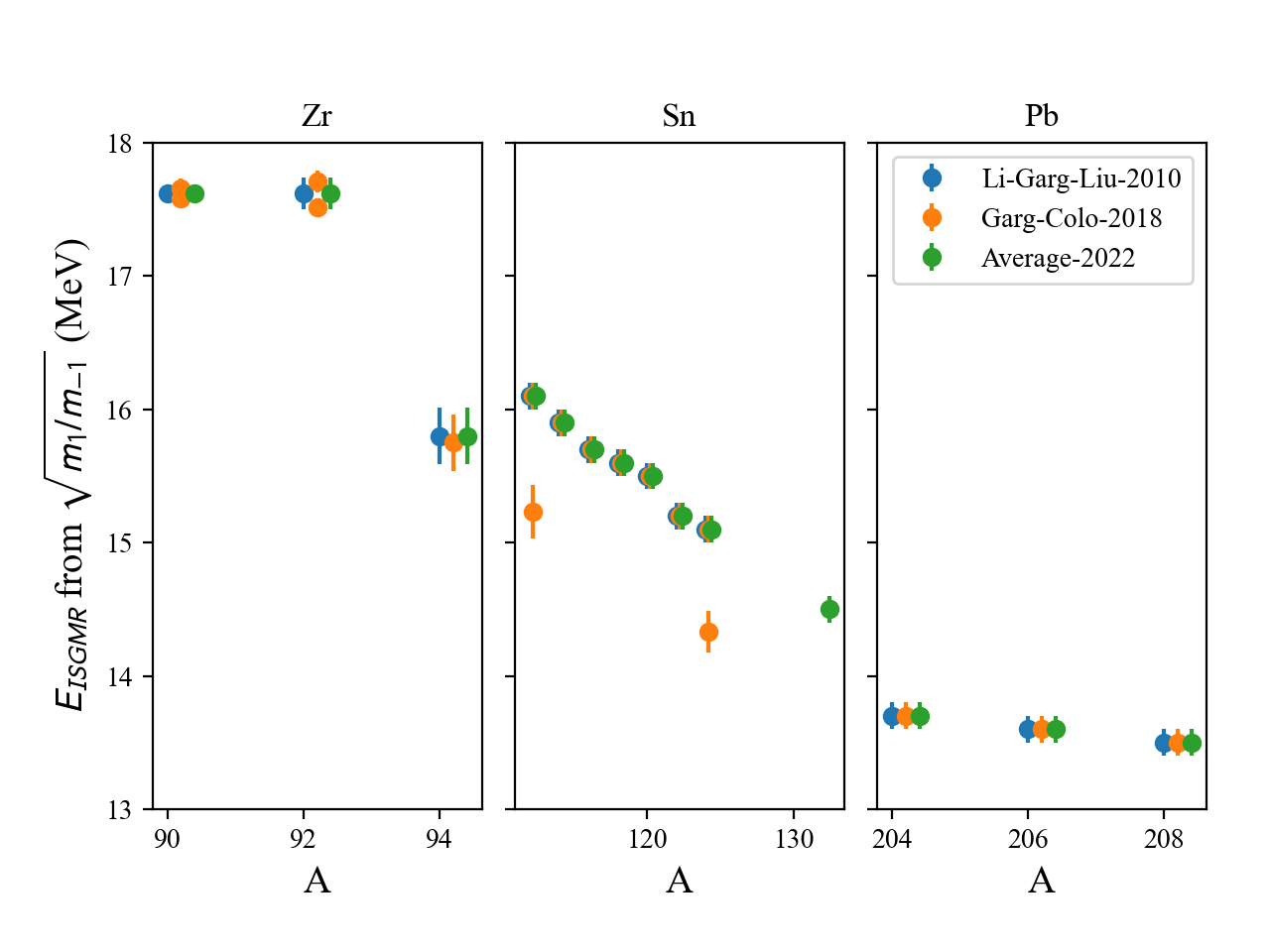}
\caption{ISGMR energies available in the \texttt{nuda} toolkit extracted from the sum rule $\sqrt{m_1/m_{-1}}$. Figure generated with \texttt{nuc\_setupISGMRExp\_plot.py}.}
\label{fig:exp:isgmr}
\end{figure}

We show in Fig.~\ref{fig:exp:isgmr} a comparison between the three tables available in the \texttt{nuda} toolkit. Except for the nuclei in the Sn isotopic chain, the two tables provide comparable results for the isotopes shown in Fig.~\ref{fig:exp:isgmr}.

\section{Data for hypernuclei: the \texttt{hnuc} module.}
\label{sec:hypernuclei}

Hypernuclei are bound systems composed of nucleons and one or more hyperons (baryons with strangeness content) such as $\Lambda$, $\Sigma$, $\Xi$, or $\Omega$. Since their discovery in 1951, nowadays more than 40 single-$\Lambda$ hypernuclei, and few double-$\Lambda$ and single-$\Xi^-$ ones have been identified. In contrast, it has not been possible to prove without any ambiguity the existence of $\Sigma$-hypernuclei.

\begin{figure}[t]
\centering
\includegraphics[scale=0.52]{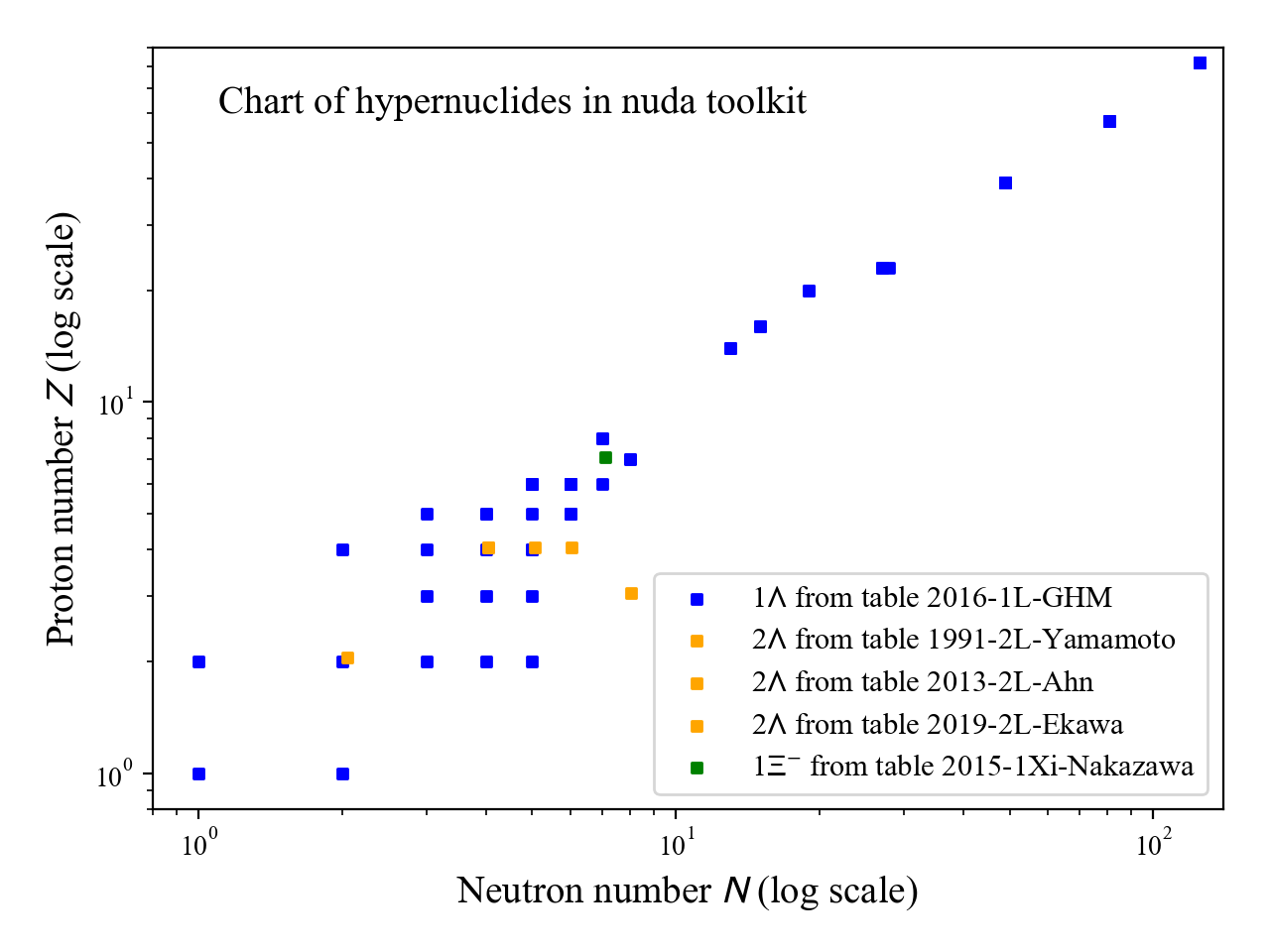}
\caption{Hypernuclear chart representing the hypernuclides available in the \texttt{nuda} toolkit as a function of $N$ and $Z$ (log scale). The hypernuclei from tables '1991-2L-Yamamoto' and '2019-2L-Ekawa' are ambigous and should be undertaken with due caution (see text for more details). Figure generated with \texttt{hnuc\_setupChart\_plot.py}.}
\label{fig:hnuc:masses}
\end{figure}

We adopt the usual nomenclature for single $\Lambda$, $\Sigma$ or $\Xi$ hypernuclei, $^{A}_Y Z$, with $A=N+Z+1$, $Y=\Lambda$, $\Sigma$ or $\Xi$, and $Z$ the atomic number of the pure nuclear core that forms the hypernuclei. For double $\Lambda$ hypernuclei, the nomenclature is $^{A}_{YY} Z$ with $A=N+Z+2$. The total energy for the hypernucleus
$^{A}_{N_Y} Z$ is defined as:
\begin{equation}
E_\tot(Z,N,N_Y) = Z m_p + N m_n + N_Y m_Y
+ BE(^{A}_{N_Y}Z)\, ,
\end{equation}
where $BE(^{A}_{N_Y}Z)$ is the binding energy and is defined as a negative number for bound nuclei.

The hypernuclear chart, see Fig.~\ref{fig:hnuc:masses}, shows, as a function of $N$ and $Z$, the hypernuclides available in the \texttt{nuda} toolkit tables that we describe in the next subsections. These nuclei have a single $\Lambda$, a double $\Lambda$, or a single $\Xi^{-}$; see legend for more details. 
Over the past six decades, only a small number of double-$\Lambda$ hypernuclei have been experimentally observed, mainly through emulsion and hybrid-emulsion techniques at CERN, KEK, and J-PARC. The most unambiguous and universally accepted identification is the NAGARA event~\cite{HTakahashi:2001}, corresponding to $^6_{\Lambda\Lambda}$He, which provided a precise $\Lambda\Lambda$ binding energy of $\Delta B_{\Lambda\Lambda}= 0.67 \pm 0.17$ MeV~\cite{JKAhn:2013}. Several additional candidate events, such as 
$^{10}_{\Lambda\Lambda}$Be and $^{13}_{\Lambda\Lambda}$B reported in Ref.~\cite{YYamamoto:1991},
and the more recent MINO event from the J-PARC E07 experiment~\cite{HEkawa:2019}, have been interpreted as other possible double-$\Lambda$ systems.
However, in these cases, the parent hypernucleus assignments remain ambiguous, owing to incomplete decay-chain reconstruction, uncertainties in the kinematics of captured $\Xi^-$ hyperons, or the possible involvement of excited nuclear states. As a consequence, although several $\Lambda\Lambda$-hypernuclear candidate events have now been
only the NAGARA event can be regarded as uniquely identified. Therefore, any quantitative or phenomenological use of the data provided in the toolkit for the additional and ambiguously identified double-$\Lambda$ hypernuclei should be undertaken with due caution.

\subsection{Single-$\Lambda$ hypernuclei}

Single-$\Lambda$ hypernuclei can be produced by several reaction  mechanisms such as 
$K^-+^AZ\rightarrow ^A_\Lambda Z + \pi^-$ 
strangeness exchange reactions, where a neutron hit by a $K^-$ is changed into a $\Lambda$ emitting a $\pi^-$; 
$\pi^+ + ^AZ \rightarrow ^A_\Lambda Z + K^+$ associated production reactions, where an $s\bar{s}$ pair is created from the vacuum and $K^+$ and a $\Lambda$ are produced in the final state; electro-production 
$e^-+^AZ\rightarrow e^-+K^+ + ^A_\Lambda (Z-1) $; or by using stable and unstable heavy ion beams. The binding energies of the produced single-$\Lambda$ hypernuclei can be accurately determined by measuring the momenta of the incoming and outgoing kaons and pions with the help of magnetic spectrometers. Here we provide an analysis published in Ref.~\cite{AGal:2016} considering the results of several probes, see the list hereafter. 

\begin{figure}[t]
\centering
\includegraphics[scale=0.52]{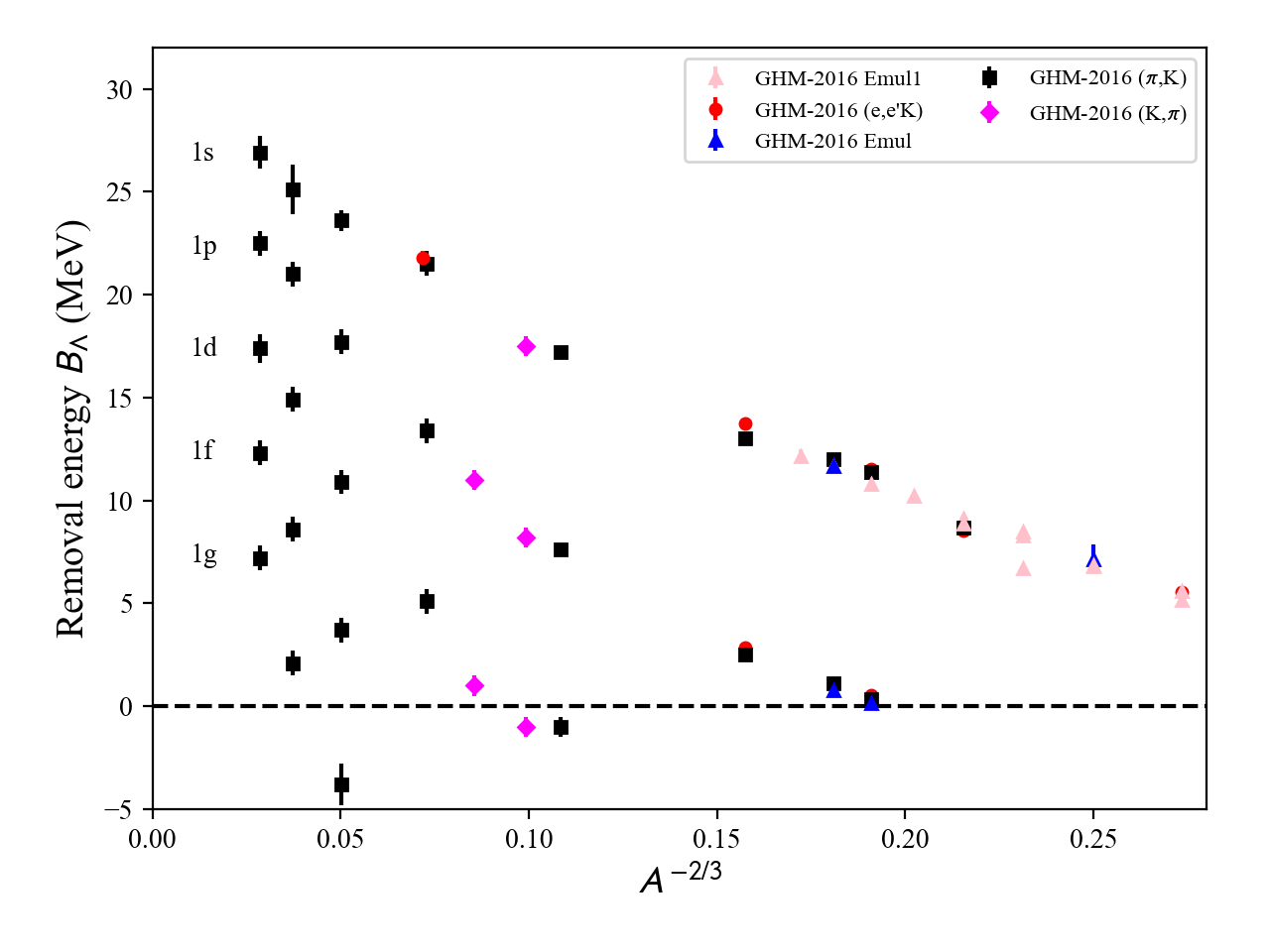}
\caption{Experimental removal energies for single-$\Lambda$ hypernuclei available in the \texttt{nucleardatapy} toolkit. 
Figure generated with \texttt{hnuc\_setupRE1LExp\_plot.py}.}
\label{fig:hnuc:re1L:exp}
\end{figure}

The complete list of available tables is given with the following instructions:
\begin{lstlisting}[language=Python]
print( nuda.hnuc.re1L_exp_tables( ) )
\end{lstlisting}
In the present release of \texttt{nuda} toolkit, there is only one table which has been implemented: \texttt{table}=`2016-1L-GHM'. The data are obtained from tables I and IV of Ref.~\cite{AGal:2016}. The experimental table is loaded in the following way:
\begin{lstlisting}[language=Python]
hyp = nuda.hnuc.setupRE1LExp( table=`2016-1L-GHM' )
hyp.print_outputs()
\end{lstlisting}
This table provides the $\Lambda$ removal energy $B_\Lambda$, defined as
\begin{eqnarray}
B_{\Lambda} &=& E_\tot(Z,N-1) + m_\Lambda - E_\tot(Z,N-1,N_{\Lambda}=1) \, , \nonumber \\
&=& BE(^{A-1} Z)- BE(^{A}_\Lambda Z ) \, ,
\end{eqnarray}
for the hypernucleus $^{A}_\Lambda Z$. This energy is associated with a single-particle state and measure its energy (with opposite sign), see Fig.~\ref{fig:hnuc:re1L:exp}. The other properties of this single particle state, e.g., its angular momentum, are obtained by the experimental analysis.

The attributes are the following Numpy arrays: \texttt{hyp.A} the baryon number ($A=N+Z-S$), \texttt{hyp.N} the neutron number $N$, \texttt{hyp.Z} the proton number $Z$, \texttt{hyp.Q} the nuclear charge ($Q=Z-q_SS$), where $q_S$ is the charge of the hyperon particule ($q_\Lambda=0$ for instance), \texttt{hyp.S} the strangeness number ($S=-1$ for 1 $\Lambda$ for instance), and \texttt{hyp.symb} the symbol representing the hypernucleus. 

Since measurements have been performed with various excitation energies, states with different angular momenta have been identified, corresponding to different single-particle energies in the ground state of single-$\Lambda$ hypernuclei. These states are shown in Fig.~\ref{fig:hnuc:re1L:exp} and are provided as attributes to the object \texttt{hyp} in the following way: \texttt{hyp.sps} the single-particle state, \texttt{hyp.ell} its associated angular momentum, \texttt{hyp.lre} and \texttt{hyp.lre\_err} the removal energy and its experimental uncertainty, and finally \texttt{hyp.probe} the probe employed to extract the experimental information: `piK' for $(\pi^+,K^+)$, `eeK' for $(e,e^\prime K^+)$, `emul1' for emulsion obtained from table~I, `emul' for emulsion obtained from table~IV, and `Kpi' for $(K^-,\pi^-)$.

\subsection{Double-$\Lambda$ hypernucleus}

Double-$\Lambda$ hypernuclei are nowadays the best systems to investigate the properties of the strangeness S=-2 baryon-baryon interaction. Contrary to single-$\Lambda$ hypernuclei, double-$\Lambda$ hypernuclei cannot be produced in a single reaction. To produce them, first a $\Xi^-$ has to the created, through reactions like $K^-+p\rightarrow \Xi^-+K^+$ or $p+\bar{p}\rightarrow \Xi^+ + \bar{\Xi}^+$. Then, provided the $\Xi^-$ is captured in an atomic orbit, it can interact in a second step with the nuclear core, producing two $\Lambda$ hyperons via processes such as, e.g., $\Xi^-+p\rightarrow \Lambda + \Lambda + 28$~MeV, where the 28~MeV energy is equally shared by the two $\Lambda$'s. 
Earlier emulsion experiments reported the formation of a few double-$\Lambda$ hypernuclei: $^{\,\,\,6}_{\Lambda\Lambda}$He, $^{\,\,10}_{\Lambda\Lambda}$Be and $^{\,\,13}_{\Lambda\Lambda}$B. However, the identification of some of these double-$\Lambda$ hypernuclei was ambiguous. In 2001, a new $^{\,\,\,6}_{\Lambda\Lambda}$He candidate was unambiguously observed at KEK in Japan. Further experiments are planned in the future at BNL, KEK, and J-PARC with $K^-$ beams, and at FAIR/GSI with proton and antiproton beams.

The list of available tables is given with the following instructions:
\begin{lstlisting}[language=Python]
print( nuda.hnuc.re2L_exp_tables( ) )
\end{lstlisting}
There are several tables for double-$\Lambda$ hypernuclei in the toolkit: \texttt{table = `1991-2L-Yamamoto'}~\cite{YYamamoto:1991}, \texttt{table = `2013-2L-Ahn'}~\cite{JKAhn:2013}, and \texttt{table = `2019-2L-Ekawa'}~\cite{HEkawa:2019}. These tables contain the binding energy $B_{\Lambda\Lambda}$ and the bond energy $\Delta B_{\Lambda\Lambda}$. The $\Lambda\Lambda$ excess binding energy, also called the bond energy, $\Delta B_{\Lambda\Lambda}$ for the nucleus $(Z,N,2\Lambda)$ is defined as~\cite{AGal:2016}:
\begin{eqnarray}
\Delta B_{\Lambda\Lambda} &=& B_{\Lambda\Lambda}(^A_{\Lambda\Lambda}Z) - 2B_{\Lambda}(^{A-1}_{\Lambda}Z) \, , \\
&=&BE(^{A}_{\Lambda\Lambda}Z)+BE(^{A-2}Z) -2 BE(^{A-1}_{\Lambda}Z) \, ,
\end{eqnarray}
where the $\Lambda\Lambda$ binding energy is defined as,
\begin{equation}
B_{\Lambda\Lambda}(^A_{\Lambda\Lambda}Z)=BE(^{A-2}Z)- BE(^A_{\Lambda\Lambda}Z) \, .
\end{equation}
Note that the bond energy can be defined equivalently as~\cite{JCugnon:2000,IVidana:2001}:
\begin{eqnarray}
\Delta B_{\Lambda\Lambda} &=& E_\tot(^{A}_{\Lambda\Lambda}Z)+E_\tot(^{A-2}Z) -2 E_\tot(^{A-1}_{\Lambda}Z) \, .
\end{eqnarray}

The instantiation of the object \texttt{hyp} with the mass table is performed in the following way:
\begin{lstlisting}[language=Python]
hyp = nuda.hnuc.setupRE2LExp( table=`2013-2L-Ahn' )
hyp.print_outputs()
\end{lstlisting}

\begin{table}[t]
\begin{center}
\caption{Data from tables available in the \texttt{nuda} toolkit, see Refs.~\cite{AGal:2016,YYamamoto:1991,JKAhn:2013,KNakazawa:2015,HEkawa:2019}: $B_{Y}$ and $B_{YY}$ are the removal energies, and $\Delta B_{YY}$ is the bond energy, see text for more details. As discussed in the text, we remind that among the double-$\Lambda$ hypernuclei, only the NAGARA event is unambiguous, the identication of the others remain still uncertain. The use of these data should therefore be taken with care. This table has been generated with \texttt{hnuc\_setupRE1LExp\_script.py}, \texttt{hnuc\_setupRE2LExp\_script.py} and \texttt{hnuc\_setupRE1XiExp\_script.py}.}
\label{table:hnuc}
\tabcolsep=0.04cm
\def\arraystretch{1.5}
\begin{tabular}{llllllll}
\hline\noalign{\smallskip}
Z & N & S & Q & Name & & & Ref. \\
 & & & & & $B_Y$ (MeV) & & \\
\multicolumn{8}{l}{single-$\Lambda$ hypernucleus} \\
2 & 2 & -1 & 2 & He & $3.12\pm 0.020$ & & \cite{AGal:2016} \\
\multicolumn{8}{l}{single-$\Xi^{-}$ hypernucleus} \\
7 & 7 & -2 & 6 & N & $4.378\pm 0.250$ & & \cite{KNakazawa:2015} \\[0.3cm]
 & & & & & $B_{YY}$ (MeV) & $\Delta B_{YY}$ (MeV) & \\
\multicolumn{8}{l}{double-$\Lambda$ hypernucleus} \\
4 & 4 & -2 & 4 & Be & $8.50\pm0.70$ & $-4.90\pm0.70$ & \cite{YYamamoto:1991} \\
3 & 8 & -2 & 3 & B & $27.60\pm0.70$ & $4.80\pm0.70$ & \cite{YYamamoto:1991} \\
2 & 2 & -2 & 2 & He & $6.910\pm 0.160$ & $0.670\pm 0.170$ & NAGARA~\cite{JKAhn:2013}  \\
4 & 4 & -2 & 4 & Be & $15.05\pm0.11$ & $1.63\pm0.11$ & MINO~\cite{HEkawa:2019} \\
4 & 5 & -2 & 4 & Be & $19.07\pm0.11$ & $1.87\pm0.37$ & MINO~\cite{HEkawa:2019} \\
4 & 6 & -2 & 4 & Be & $13.68\pm0.11$ & $-2.70\pm1.00$ & MINO~\cite{HEkawa:2019} \\
\noalign{\smallskip}\hline
\end{tabular}
\end{center}
\end{table}

The content of \texttt{nuda} toolkit for double-$\Lambda$ is shown in table~\ref{table:hnuc}, together with the $B_Y$ for 1-$\Lambda$ and 1-$\Xi$ hypernuclei. We remind that the strangeness number of $\Lambda$ and $\Xi$ are $S_\Lambda=-1$ and $S_\Xi=-2$.

The attributes of the object \texttt{hyp} are: \texttt{hyp.A} the baryon number ($A=N+Z-S$), \texttt{hyp.N} the neutron number $N$, \texttt{hyp.Z} the proton number $Z$, \texttt{hyp.Q} the nuclear charge $Q$, \texttt{hyp.S} the strangeness number $S$, and \texttt{hyp.symb} the symbol representing the hypernucleus. Are also provided the $\Lambda\Lambda$ removal energy $B_{\Lambda\Lambda}$ as \texttt{hyp.llre} and its experimental error \texttt{hyp.llre\_err} (in MeV), and the bond energy $\Delta B_{\Lambda\Lambda}$ as \texttt{hyp.lldre}, with experimental uncertainty \texttt{hyp.lldre\_err} (in MeV). Finally, the attribute \texttt{hyp.probe} contains the probe employed to extract the experimental information.

\subsection{Single-$\Xi^{-}$ hypernucleus}

Single-$\Xi^{-}$ hypernuclei can be produced via the reactions $K^-+p\rightarrow \Xi^-+K^+$ or $p+\bar{p}\rightarrow \Xi^+ + \bar{\Xi}^+$. Nowadays, very few 
single-$\Xi$ hypernuclei have been identified, although future production experiments of $\Xi^-$ hypernuclei are being planned at the J-PARC facility in Japan.\\

The list of available tables is given with the following instructions:
\begin{lstlisting}[language=Python]
print( nuda.hnuc.re1Xi_exp_tables( ) )
\end{lstlisting}
The toolkit provides a single table: \texttt{table = `2015-1Xi-Nakazawa'}. This table is constructed from Ref.~\cite{KNakazawa:2015} where the so-called `Kiso' event related to the creation of $^{15}_{\Xi^-}N$ is reported. The instantiation of the object \texttt{hyp} with the table is performed in the following way:
\begin{lstlisting}[language=Python]
hyp = nuda.hnuc.setupRE1XiExp( table=`2015-1Xi-Nakazawa' )
hyp.print_outputs()
\end{lstlisting}

The $\Xi^{-}$ removal energy is defined as
\begin{equation}
B_{\Xi^{-}}=BE(^{A-1} Z)-BE(^A_{\Xi^{-}} Z) \, ,
\end{equation}
and the content of \texttt{nuda} toolkit for the single-$\Xi^{-}$ event is shown in Table~\ref{table:hnuc}. 

The attributes of the object \texttt{hyp} are: \texttt{hyp.A} the baryon number ($A=N+Z-S$), \texttt{hyp.N} the neutron number $N$, \texttt{hyp.Z} the proton number $Z$, \texttt{hyp.Q} the nuclear charge $Q$, \texttt{hyp.S} the strangeness number $S$, and \texttt{hyp.symb} the symbol representing the hypernucleus. 
The $\Xi$ removal energy $B_{\Xi}$ and its experimental error are also provided in \texttt{hyp.xire} and \texttt{hyp.xire\_err} (in MeV).
Finally, the attribute \texttt{hyp.probe} contains the probe employed to extract the experimental information.

\section{Neutron star crust: the \texttt{crust} module.}
\label{sec:crust}

The crust of NSs is a non-uniform system composed of nuclear clusters embedded in an electron gas (outer crust) and with additional contributions from a neutron fluid (inner crust). There are several calculations for the crust of NSs, and some of them are provided by the \texttt{nuda} toolkit.

The complete list of available crust model predictions is given with the following instructions:
\begin{lstlisting}[language=Python]
print( nuda.crust.crust_models( ) )
\end{lstlisting}
The object \texttt{crust} is instantiated in the following way:
\begin{lstlisting}[language=Python]
crust = nuda.crust.setupCrust( model = `2022-crustGMSR_H4' )
crust.print_outputs()
\end{lstlisting}
where the variable \texttt{model} is fixed to one of the following crust models.\\

\noindent
\texttt{model}=`1973-Negele-Vautherin'.\\
This provides results of the Hartree-Fock calculation for the inner crust of NS from Ref.~\cite{JWNegele:1973}.\\

\noindent
\texttt{model}=`2018-PCPFDDG-BSK22', `2018-PCPFDDG-BSK24', `2018-PCPFDDG-BSK25', `2018-PCPFDDG-BSK26'.\\
The crust is obtained from BSK22, BSK24, BSK25, and BSK26 extended Skyrme EDF and based on the 4th-order Extended Thomas-Fermi (ETF) method with proton shell correction via the Strutinsky integral (SI)~\cite{MPearson:2018}. The nucleon distributions are parametrized using damped Fermi profiles.\\

\noindent
\texttt{model}=`2020-MVCD-D1S', `2020-MVCD-D1M', `2020-MVCD-D1M$^*$'.\\
The crust is obtained from D1S, D1M, and  D1M$^*$ Gogny interactions using a semiclassical variational Wigner-Kirkwood method along with shell and pairing corrections calculated with the Strutinsky integral method and the BCS approximation~\cite{CMondal:2020}.\\

\noindent
\texttt{model}=`2022-crustGMSR\_BSK14', `2022-crustGMSR\_BSK16', `2022-crustGMSR\_DHSL59',
'2022-crustGMSR\_DHSL69', `2022-crustGMSR\_F0',
'2022-crustGMSR\_H1', `2022-crustGMSR\_H2', 
'2022-crustGMSR\_H3', `2022-crustGMSR\_H4', 
'2022-crustGMSR\_H5', `2022-crustGMSR\_H7', 
'2022-crustGMSR\_LNS5', `2022-crustGMSR\_RATP',
'2022-crustGMSR\_SGII', `2022-crustGMSR\_SLY5'.\\
The GMSR crust~\cite{GGrams:2022b} is computed with the meta-model~\cite{JMargueron:2018b,RSomasundaram:2021} calibrated to different parametrizations of the Skyme force (BSK14~\cite{SGoriely:2007}, BSK16~\cite{NChamel:2008}, F0~\cite{TLesinski:2006}, LNS5~\cite{DGambacurta:2011}, RATP~\cite{MRayet:1982}, SGII~\cite{NVanGiai:1981}, SLy5~\cite{EChabanat:1998}) and to Chiral EFT Hamiltonians(H1-H7~\cite{CDrischler:2016} (except H6), and DHS$_{L59}$-DHS$_{L69}$~\cite{CDrischler:2021})) using a compressible liquid drop model (CLDM) approach as detailed in Ref.~\cite{GGrams:2022a}.\\

\begin{figure}[t]
\centering
\includegraphics[scale=0.52]{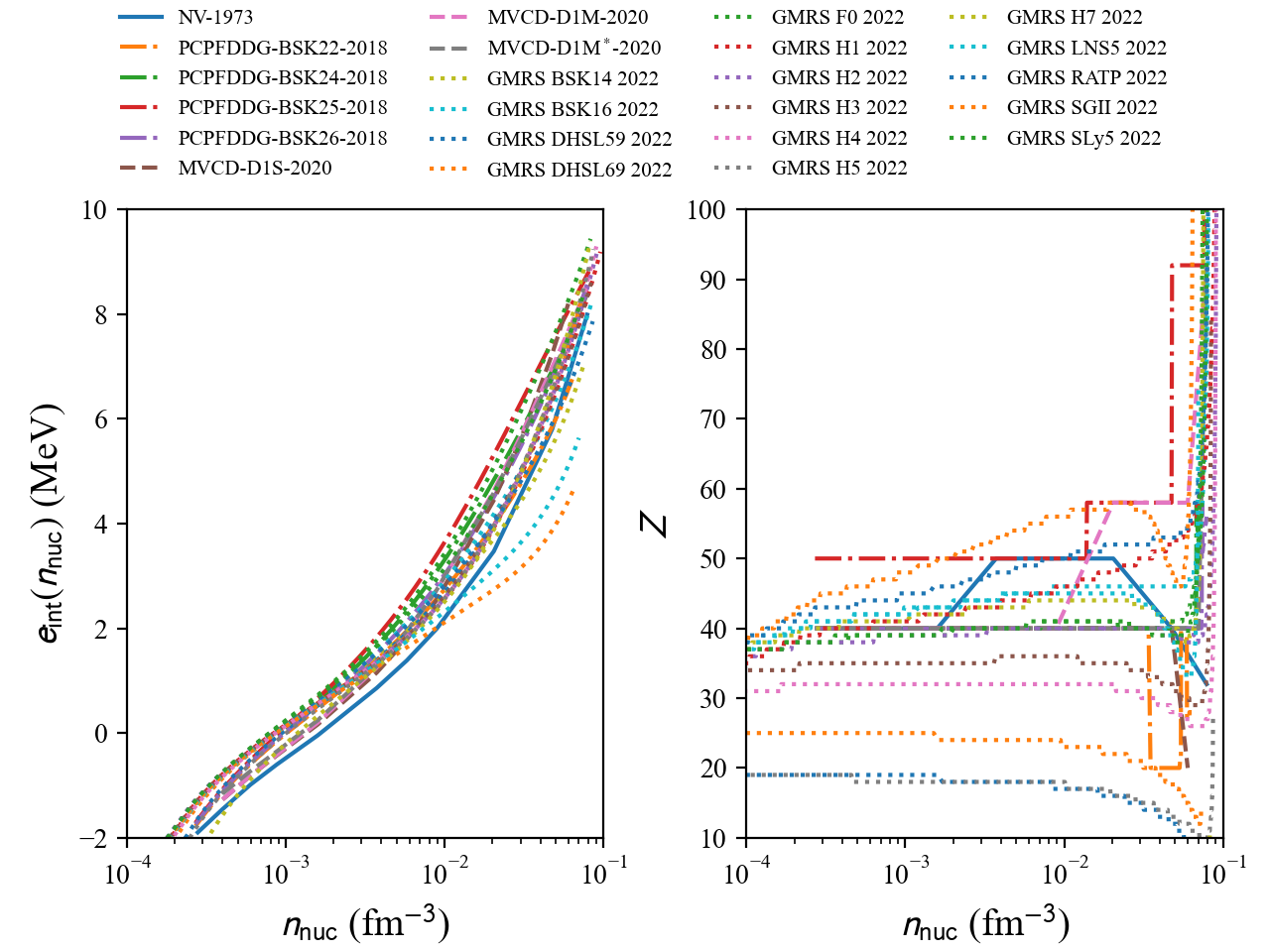}
\caption{Properties of the crust, internal energy per nucleon $e_\intn$ and proton number $Z$ a a function of the nucleon density $n_\nuc$ provided by the models available in the \texttt{nuda} toolkit. Figure generated with \texttt{crust\_setupCrust\_plot.py}.}
\label{fig:crust}
\end{figure}

The attributes of the object \texttt{crust} are Numpy arrays: \texttt{crust.den} the nucleon density $n_\text{nuc}$ in fm$^{-3}$ (\texttt{crust.den\_cgs} in g~cm$^{-3}$), \texttt{crust.A} and \texttt{crust.A\_bound} the total number of nucleons $A$ in a Wigner-Seitz cell with radius $R_{WS}$ and volume $V_{WS}$ (in \texttt{crust.RWS} and \texttt{crust.VWS}) and the number of bound nucleons $A_\text{bound}$ in a cluster with radius $R_{cl}$ and volume $V_{cl}$ (in \texttt{crust.Rcl} and \texttt{crust.Vcl}) , \texttt{crust.Z} and \texttt{crust.Z\_bound} the total number of protons $Z$ and the number of bound protons $Z_\text{bound}$, \texttt{crust.N} and \texttt{crust.N\_bound} the total number of neutrons $N$ and the number of bound neutrons $N_\text{bound}$, \texttt{crust.N\_f} the number of free neutrons ($N_f=N-N_\text{bound}$). Note that \texttt{crust.N\_bound} and \texttt{crust.N\_f} are not provided for the crust models with prefix `2020-MVCD'. \texttt{crust.den\_f} is the free neutron density $n_f$, \texttt{crust.u} is the volume fraction $u=V_{cl}/V_{WS}$, and finally \texttt{crust.e2a\_tot}, \texttt{crust.eps\_tot} and \texttt{crust.e2a\_int} are the total energy per nucleon, the total energy density and the internal energy per nucleon.
More attributes can be obtained in the following way:
\begin{lstlisting}[language=Python]
print(crust.__dict__)
\end{lstlisting}

A few properties in the NS crust, internal energy per nucleon and proton number, are represented in Fig.~\ref{fig:crust} for the models available in the \texttt{nuda} toolkit. The densities are selected to represent the properties of the inner crust. There is a large dispersion in the internal energies and in the proton number, which is related to the fact that most of the EDF do not reproduce neutron fluid properties accurately. This point is well discussed in Refs.~\cite{GGrams:2022a,GGrams:2022b}.

\begin{figure*}[t]
\centering
\includegraphics[scale=0.9]{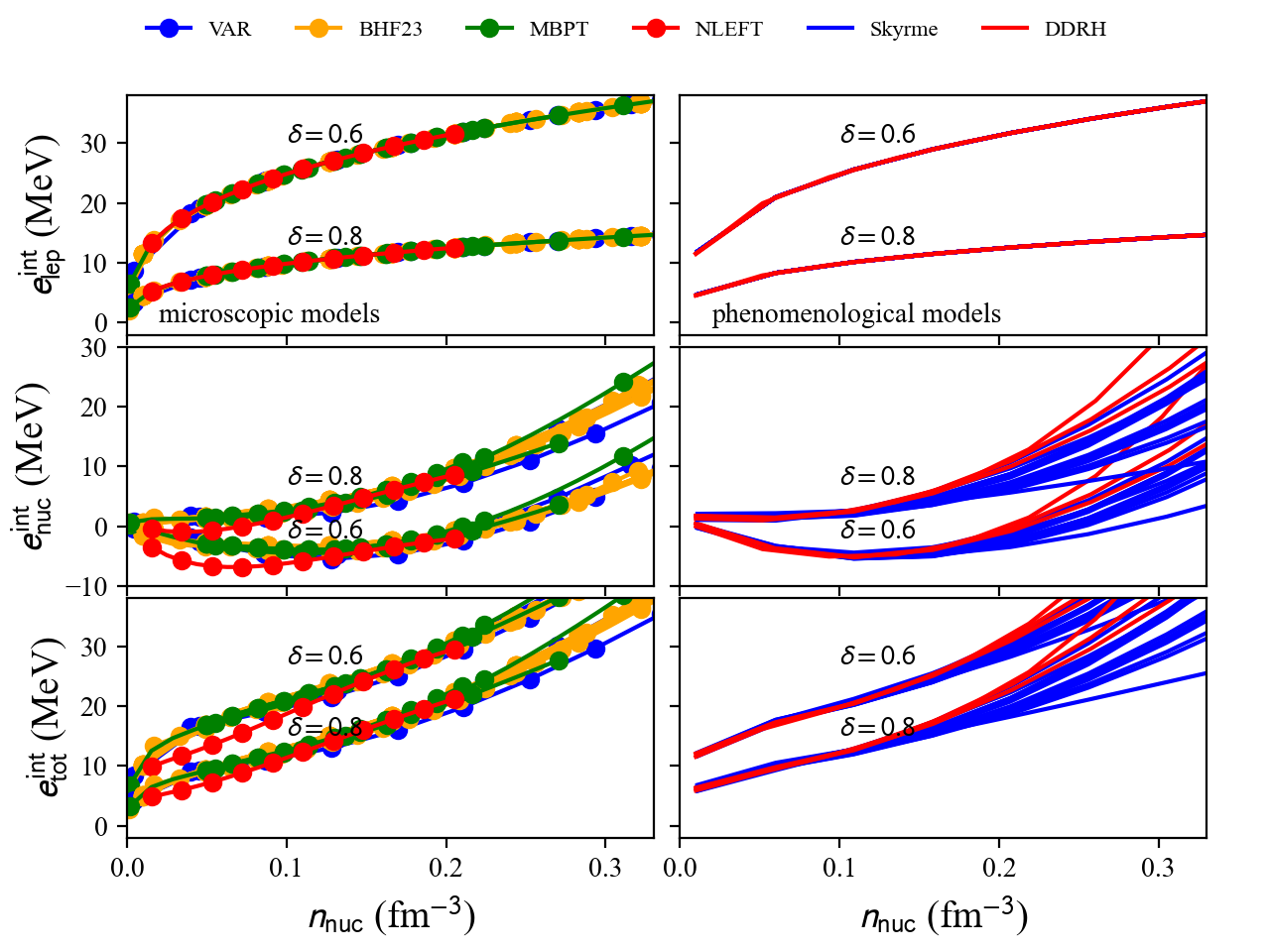}
\caption{Energy per particle in neutron-rich matter (with $\delta=0.6$ and $0.8$) for the list of microscopic (left) and phenomenological (right) models available in the \texttt{nuda} toolkit. Figure generated with \texttt{eos\_setupAM\_e2a\_plot.py}.}
\label{fig:eos:e2a}
\end{figure*}

\section{Equation of state: the \texttt{eos} module.}
\label{sec:eos}

We now describe the \texttt{eos} module, where nucleon and lepton contributions are considered together.

The classes employed in this section are located in the \texttt{eos} module. The new feature in the \texttt{eos} class is that lepton contribution is added to the nucleon thermodynamic quantities, adding two variables: the electron density $n_e$ and the muon density $n_\mu$. At $T=0$, four components contribute to dense matter: $n$, $p$, $e$, and $\mu$. We also consider the electro-neutrality condition: $n_p=n_e+n_\mu$. There are, therefore, only three independent components.

In the following, we consider asymmetric matter (AM), controlled by the nucleon density, the proton, and the muon fractions; the lepton-equilibrated matter (also imposing $\mu_e=\mu_\mu$) controlled by the nucleon density and the proton fraction; and the beta-equilibrated matter (adding the $\beta$-equilibrium condition $\mu_n-\mu_p=\mu_e$) controlled only by the nucleon density.

\subsection{Ground state of asymmetric matter}
\label{sec:eos:am}

For the models predicting SM and NM, the energy in AM is defined in the following way:
\begin{equation}
e_\nuc(n,\delta) \approx e_{\SM}(n)+e_{\sym,2}(n) \delta^2 \, ,
\label{eq:eos:e}
\end{equation}
where $e_{\sym,2}\approx e_{\sym}$, defined by Eq.~\eqref{eq:esym}.

Similarly, the pressure reads
\begin{equation}
p_\nuc(n,\delta) \approx p_{\SM}(n)+p_{\sym}(n) \delta^2 \, ,
\label{eq:eos:p}
\end{equation}
where $p_{\SM}=n^2(\partial e_\SM/\partial n)$ and $p_{\sym}=n^2(\partial e_\sym/\partial n)$.

The ground state of uniform matter at fixed electron and muon fractions is governed by the following equations:
\begin{itemize}
\item nucleon number: $n_\nuc=n_n+n_p$;
\item charge neutrality: $n_p=n_e$ if only electrons, while $n_p=n_e+n_\mu$ with electrons and muons;
\item no neutrinos: $\mu_\nu=0$.
\end{itemize}

In the following, we investigate only models providing results in SM and NM, since the symmetry energy $e_\sym$ requires SM and NM to be defined. We therefore disregard all models providing results in only NM. The list of models suitable for calculating AM can be obtained in the following way:
\begin{lstlisting}[language=Python]
models, models_lower = \
nuda.eos.micro_esym_models()
print(models)
\end{lstlisting}

The call of the EoS in AM for a given \texttt{model}, \texttt{param}, and a given \texttt{kind} (which can be either `micro' or `pheno') can be done in the following way:
\begin{lstlisting}[language=Python]
eos = nuda.eos.setupAM( model, kind, asy, xmu )
eos.print_outputs()
\end{lstlisting}
where the scalar variable \texttt{asy} is the isospin asymmetric parameter (\texttt{asy}=0 by default, for SM), and the scalar variable \texttt{xmu} is the muon fraction (\texttt{xmu}=0 by default, only electrons). Note that the density mesh is the one fixed by the class
\texttt{nuda.matter.setupMicroEsym} or
\texttt{nuda.matter.setupPhenoEsym}. It corresponds to the densities where the symmetry energy is calculated. If \texttt{kind=`micro'}, then the variable \texttt{param} is not used and can be set to \texttt{param=None}.

We show in Fig.~\ref{fig:eos:e2a} the internal energy per nucleon for the lepton contribution $E_\text{lep}^\text{int}/A$, for the nuclear contribution $E_\text{nuc}^\text{int}/A$, and for the total energy per nucleon $E_\text{tot}^\text{int}/A$, running over all microscopic (left) and phenomenological (right) models available in the toolkit. The asymmetry parameter \texttt{asy} is fixed to two values: \texttt{asy}=0.6, 0.8. 
All models shown are the ones passing inside the reference band in NM, that we previously described in Sec.~\ref{sec:unif:band}. Note the opposite behavior of the energy per nucleon for the lepton contribution and for the nuclear contribution as a function of the asymmetry parameter $\delta$: as $\delta$ increases, the number of leptons decreases and the energy per nucleon decreases as well. Inversely, the nucleon energy per particle increases as a function of $\delta$. Since the reduction of the lepton energy per nucleon is larger than the increase of the nuclear component, the total energy per particle decreases as a function of $\delta$. This phenomenon is observed for microscopic and phenomenological models.

Attributes are the following (Numpy arrays): the nucleon density $n_\nuc$ is given by \texttt{eos.den}, the neutron and proton densities $n_n$ and $n_p$ ($n_\nuc=n_n+n_p$) are given in \texttt{eos.n\_n} and \texttt{eos.n\_p}. The electron, muon and lepton densities are in \texttt{eos.n\_el}, \texttt{eos.n\_mu}, and \texttt{eos.n\_lep}. The thermodynamic properties are: the total energy per nucleon \texttt{eos.e2a\_tot}, its nucleon contribution \texttt{eos.e2a\_nuc} and lepton contribution \texttt{eos.e2a\_lep}, the symmetry energy \texttt{eos.esym}, the total energy density \texttt{eos.eps\_tot}, the total pressure \texttt{eos.pre\_tot} and the sound speed \texttt{eos.cs2\_tot}. More attributes can be obtained in the following way:
\begin{lstlisting}[language=Python]
print(eos.__dict__)
\end{lstlisting}

\subsection{Ground state of lepton-equilibrated matter}
\label{sec:eos:Leq}

The ground state of lepton-equilibrated matter at fixed proton fraction is governed by the following equations:
\begin{itemize}
\item baryon number: $n_\nuc=n_n+n_p$;
\item charge neutrality: $n_p=n_e$ if only electrons, $n_p=n_e+n_\mu$ with electrons and muons;
\item chemical potentials: $\mu_\mu=\mu_e$ when muons are present, $\mu_\mu=0$ otherwise.
\item no neutrinos: $\mu_\nu=0$;
\end{itemize}

At $T=0$, if $\mu_e \leq m_\mu c^2$, only electrons are present, $n_\mu=0$, and $x_p=x_e$, while if $\mu_e>m_\mu c^2$ there are electrons and muons in asymmetric matter, $x_p=x_e+x_\mu$ and $n_l=k_{Fl}^3/(3\pi^2)$ with $l=e$, $\mu$, and the following relation between lepton Fermi momenta imposes the equilibrium:
\begin{equation}
\left( m_\mu c^2\right)^2 + \left( \hbar c k_{F\mu} \right)^2 = \left( \hbar c k_{Fe} \right)^2 \, .
\end{equation}

\begin{figure}[t]
\centering
\includegraphics[scale=0.52]{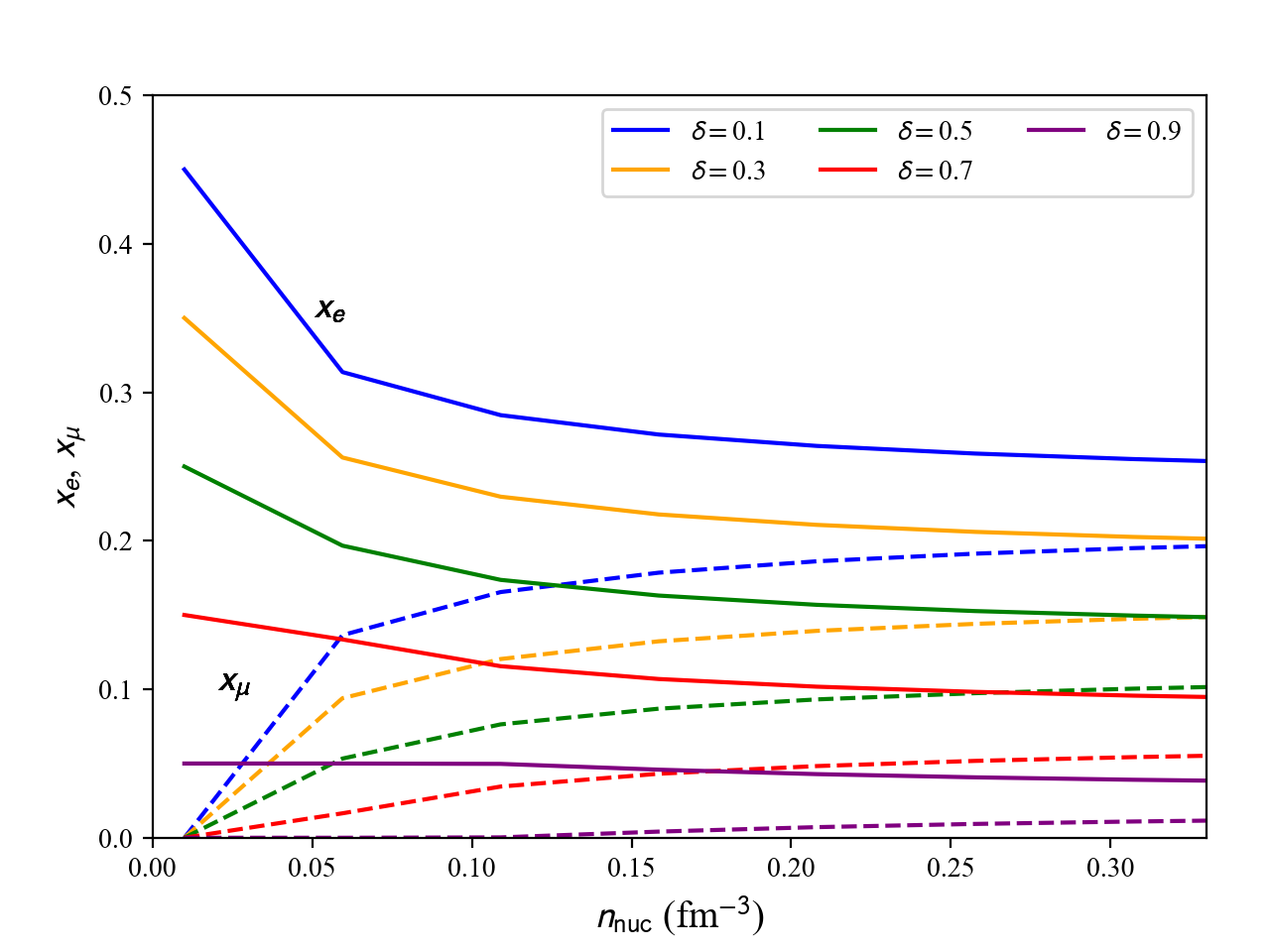}
\caption{Lepton fractions $x_e$ (solid lines) and $x_\mu$ (dashed lines) for lepton-equilibrated AM with various values for $\delta$ from 0.1 up to 0.9 (see legend). Figure generated with \texttt{eos\_setupAMLeq\_plot.py}.}
\label{fig:eos:Leq:xmu}
\end{figure}

Lepton-equilibrated EoS in AM, for a given nuclear \texttt{model} and \texttt{kind} (which can be either `micro' or `pheno'), is contained in the object \texttt{Leq} instantiated in the following way:
\begin{lstlisting}[language=Python]
Leq = nuda.eos.setupAMLeq( model, param, kind, asy )
Leq.print_outputs()
\end{lstlisting}
where the (scalar) variable \texttt{asy} is the isospin asymmetric parameter, with $-1<$\texttt{asy}$<1$.

The lepton fractions $x_e$ and $x_\mu$ for lepton-equilibrated AM as functions of the nucleon density $n_\text{nuc}$ are shown in Fig.~\ref{fig:eos:Leq:xmu}. The asymmetry parameter $\delta$ is varied from 0.1 up to 0.9, as shown in the legend. Note that these fractions are independent of the nuclear EoS, so the quantities shown in Fig.~\ref{fig:eos:Leq:xmu} are identical for all nuclear models.

Attributes are the same as in AM (Numpy arrays): the nucleon density $n_\nuc$ is given by \texttt{Leq.den}, the neutron and proton densities $n_n$ and $n_p$ ($n_\nuc=n_n+n_p$) are given in \texttt{Leq.n\_n} and \texttt{Leq.n\_p}. The electron, muon and lepton densities are in \texttt{Leq.n\_el}, \texttt{Leq.n\_mu}, and \texttt{Leq.n\_lep}. The thermodynamic properties are: the total energy per nucleon \texttt{Leq.e2a\_tot}, its nucleon contribution \texttt{Leq.e2a\_nuc} and lepton contribution \texttt{Leq.e2a\_lep}, the symmetry energy \texttt{Leq.esym}, the total energy density \texttt{Leq.eps\_tot}, the total pressure \texttt{Leq.pre\_tot} and the sound speed \texttt{Leq.cs2\_tot}. More attributes can be obtained in the following way:
\begin{lstlisting}[language=Python]
print(Leq.__dict__)
\end{lstlisting}

\subsection{Ground state of $\beta$-equilibrated matter}
\label{sec:eos:beta}

The ground state of matter at beta-equilibrium is governed by the following equations:
\begin{itemize}
\item baryon number: $n_\nuc=n_n+n_p$;
\item charge neutrality: $n_p=n_e$ if only electrons, $n_p=n_e+n_\mu$ with electrons and muons;
\item chemical potentials: $\mu_e+\mu_p=\mu_n$ if only electrons; we have to add the following relation: $\mu_e=\mu_\mu$ when muons are present;
\item no neutrinos: $\mu_\nu=0$.
\end{itemize}
There is only one additional equation in $\beta$-equilibrated matter compared to the lepton-equilibrated case, $\mu_e+\mu_p=\mu_n$, fixing the asymmetry parameter $\delta$. As a consequence, there is only one independent variable that we choose to be the nucleon density $n_\nuc$.

Considering the quadratic approximation~\eqref{eq:eos:e}, we have $\mu_n-\mu_p\approx 4\, \delta\, e_{\sym,2}(n_\nuc)$ and since $\mu_e=\hbar c k_{Fe}=\hbar c( 3\pi^2 n_e)^{1/3}$ for ultra-relativistic electrons, we obtain the following equation to be solved at $\beta$-equilibrium:
\begin{equation}
4\, e_{\sym,2}(n_\nuc) \left( 1-2x_p \right) \approx \hbar c (3\pi^2 x_e n_\nuc)^{1/3}
\label{eq:beta}
\end{equation}
where $x_p=n_p/n_\nuc$, $x_e=n_e/n_\nuc$, and $n_\nuc$ is fixed.
Eq.~\eqref{eq:beta} can be solved with the initial solution:
\begin{equation}
x_e\approx (4e_{\sym,2}/\hbar c)^3/[3\pi^2 n_\nuc+6(4e_{\sym,2}/\hbar c)^3] \, ,
\end{equation}
assuming $x_e\ll 1$ and $x_p=x_e$, for low density matter.

The EoS at $\beta$-equilibrium for a given \texttt{model}, \texttt{param}, and \texttt{kind} (which can be either `micro' or `pheno') is contained in the object \texttt{beta} that is instantiated in the following way:
\begin{lstlisting}[language=Python]
beta = nuda.eos.setupBeta( model, param, kind )
beta.print_outputs()
\end{lstlisting}

\begin{figure}[t]
\centering
\includegraphics[scale=0.52]{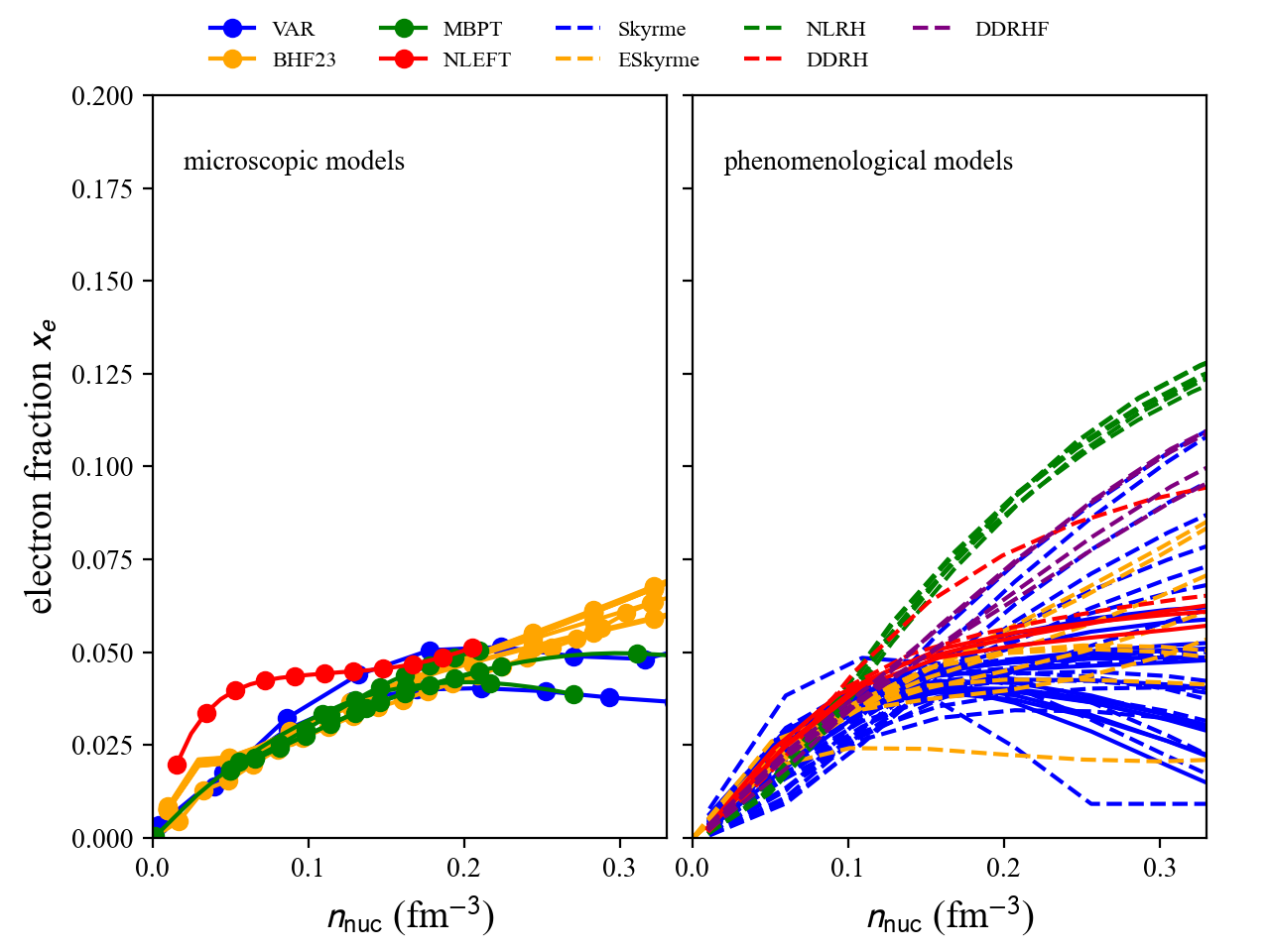}
\caption{Electron fraction for microscopic models (left) and phenomenological models (right) in $\beta$-equilibrium matter. Figure generated with \texttt{eos\_setupAMBeq\_plot.py}.}
\label{fig:beta:xe}
\end{figure}

\begin{figure}[t]
\centering
\includegraphics[scale=0.52]{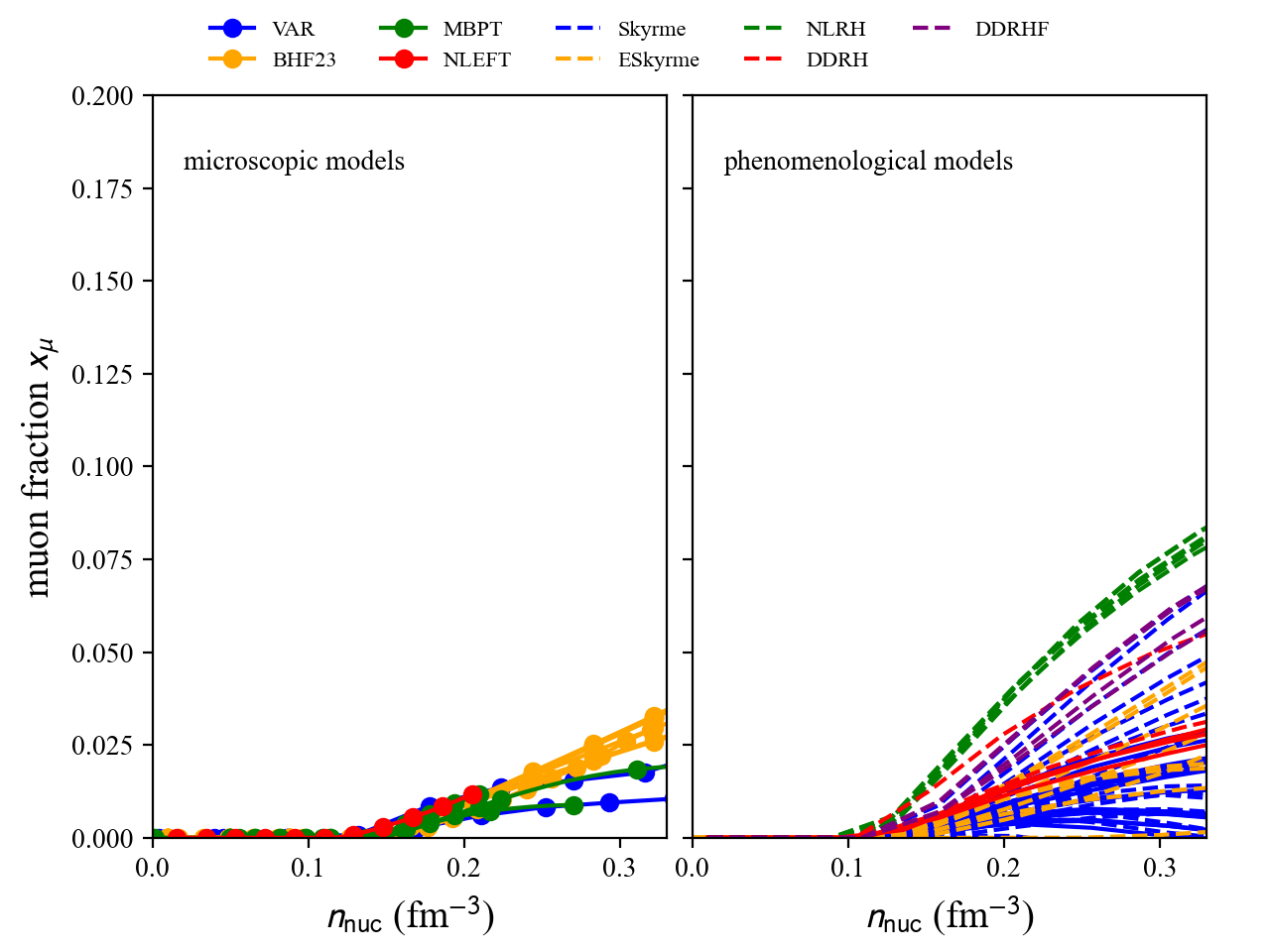}
\caption{Same as Fig.~\ref{fig:beta:xe} for the muon fraction.}
\label{fig:beta:xmu}
\end{figure}

Running over all microscopic and phenomenological models available in the toolkit, we show the electron fraction in Fig.~\ref{fig:beta:xe}, and the muon fraction in Fig.~\ref{fig:beta:xmu}. For all the microscopic and phenomenological models considered in our meta-analysis, muons appear for densities above 0.1~fm$^{-3}$ (for phenomenological models) and 0.15~fm$^{-3}$ (for microscopic models).

\begin{figure}[t]
\centering
\includegraphics[scale=0.52]{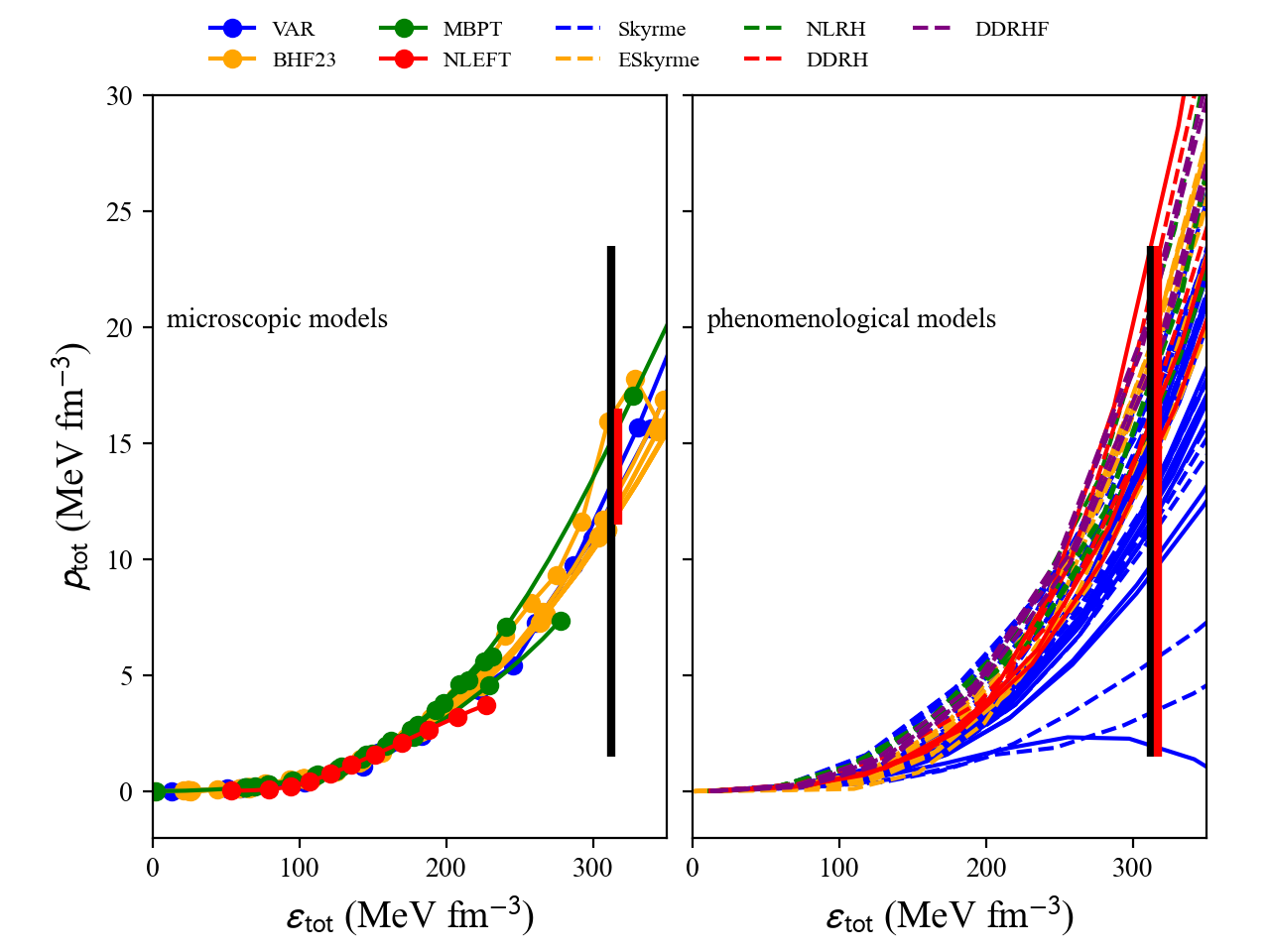}
\caption{Pressure in matter at beta-equilibrium for the list of microscopic (left) and phenomenological (right) models available in the \texttt{nuda} toolkit. Figure generated with \texttt{eos\_setupAMBeq\_plot.py}.}
\label{fig:pbeta}
\end{figure}

The pressure in nuclear matter at beta-equilibrium as a function of the energy density $\epsilon_\text{tot}$ is shown in Fig.~\ref{fig:pbeta}. The vertical lines represent the dispersion in the microscopic models, phenomenological models, and total at twice the saturation energy density. The dispersion for the phenomenological models is large and is the widest one. The uncertainties are estimated for the models that match the reference band in NM, and they are shown in solid lines. It can be remarked that the models that are not consistent with the reference band (in dashed lines) would not contribute to increasing the uncertainties for the total pressure at twice the saturation energy density. The large dispersion in the prediction observed for the phenomenological models cannot be reduced by imposing the models to be consistent with the reference band. In particular, the very soft phenomenological model largely contributes to the uncertainty in the pressure. This model predicts the collapse of nuclear matter, but it could not be disregarded if there is a low-density phase transition (just above the saturation density) bringing the necessary repulsion to avoid the collapse.

Attributes are the same as in AM (Numpy arrays): the nucleon density $n_\nuc$ is given by \texttt{beta.den}, the neutron and proton densities $n_n$ and $n_p$ ($n_\nuc=n_n+n_p$) are given in \texttt{beta.n\_n} and \texttt{beta.n\_p}. The electron, muon, and lepton densities are in \texttt{beta.n\_el}, \texttt{beta.n\_mu}, and \texttt{beta.n\_lep}. The thermodynamic properties are: the total energy per nucleon \texttt{beta.e2a\_tot}, its nucleon contribution \texttt{beta.e2a\_nuc} and lepton contribution \texttt{beta.e2a\_lep}, the symmetry energy \texttt{beta.esym}, the total energy density \texttt{beta.eps\_tot}, the total pressure \texttt{beta.pre\_tot} and the sound speed \texttt{beta.cs2\_tot}. More attributes can be obtained in the following way:
\begin{lstlisting}[language=Python]
print(beta.__dict__)
\end{lstlisting}

\begin{table}[t]
\begin{center}
\caption{Pressure and uncertainties in MeV~fm$^{-3}$ for NM, SM, and for matter at beta-equilibrium and at $2\epsilon_{\sat}$ (with $\epsilon_\sat=$156~MeV~fm$^{-3}$ and $\rho_{\sat}=2.8\times 10^{14}$~g~cm$^{-3}$). These quantities are also represented in Figs.~\ref{fig:pnm}, \ref{fig:psm}, and \ref{fig:pbeta}. They are compared to the inference from gravitational wave analysis presented in Ref.~\cite{BPAbbott:2018}, see text for more discussion.}
\label{table:pre}
\tabcolsep=0.6cm
\def\arraystretch{1.5}
\begin{tabular}{lll}
\hline
micro & pheno & total \\
\multicolumn{3}{l}{In NM:} \\
$[8.5:25.5]$ & $[8.5:37.5]$ & $[8.5:37.5]$\\
\multicolumn{3}{l}{In SM:} \\
$[4.5:17.5]$ & $[11.0:33.0]$ & $[4.5:33.0]$\\
\multicolumn{3}{l}{In matter at beta-equilibrium:} \\
$[11.5:16.5]$ & $[1.5:23.5]$ & $[1.5:23.5]$ \\
\multicolumn{3}{l}{From gravitational wave detection (90\% C.I.):\cite{BPAbbott:2018}} \\
 &  & $21.9^{+16.9}_{-10.6}$ \\
\hline
\end{tabular}
\end{center}
\end{table}

Finally, in Table~\ref{table:pre} are given the uncertainties for the pressure in NM, SM, and for AM at $\beta$-equilibrium shown in Figs.~\ref{fig:pnm}, \ref{fig:psm}, and \ref{fig:pbeta}. The dispersion in the pressure is obtained from meta-analyses running over all microscopic models (micro), phenomenological models (pheno), or assembling all predictions together (total). We remind that the dispersion is estimated only for the models in agreement with the reference band in NM. It is a pure nuclear prediction since results for these nuclear models for neutron stars are not considered here. These nuclear predictions are also compared with the one inferred from the gravitational-wave detection GW170817~\cite{BPAbbott:2018}. There is a pretty good overlap between the nuclear-physics prediction and the one inferred from gravitational-wave measurements. It can, however, be noted that the predictions from nuclear physics favor the lower values of pressure as predicted by gravitational-wave analyses. In conclusion of the present meta-analysis, there is a good agreement between nuclear physics and gravitational-wave inferences for the pressure at twice the saturation energy density, with a preference, for nuclear models, for the half lower uncertainty of the gravitational-wave estimation of the pressure. 

Improvements from gravitational-wave analyses require new observations of an event similar to GW170817, while the improvement from nuclear physics requires tighter constraints, such as, for instance, those associated with heavy-ion collisions or microscopic models.

\subsection{Connecting crust and core EoS}

Since the \texttt{nuda} toolkit contains a set of crust and core equations of state at $\beta$-equilibrium, it is possible to connect the crust and the core EoS and provide a complete EoS describing neutron stars in their ground state.

The construction of an EoS for NSs requires: i) a crust EoS, ii) a core EoS at $\beta$-equilibrium, and iii) a method to connect the crust and the core. This is usually the case since a given equation of state does not work for all densities. The following class provides such a construction:
\begin{lstlisting}[language=Python]
eos = nuda.eos.setupCC( crust_model, core_kind, core_model, core_param, connect, boundaries, emp )
eos.print_outputs()
\end{lstlisting}
where the variable \texttt{crust\_model} fixes the crust EoS by choosing among the models described in Sec.~\ref{sec:crust}, for memory, the list of crust models is available as
\begin{lstlisting}[language=Python]
print( nuda.crust.crust_models( ) )
\end{lstlisting}

The variable \texttt{core\_kind} fixes the kind of model for the core: `micro' or `pheno'. The variable \texttt{core\_model} fixes the model for the core and the variable \texttt{core\_param} fixes the parameter set for the core in case \texttt{core\_kind=`pheno'}. If \texttt{core\_kind=`micro'}, the user can fix \texttt{core\_param=None}.
The list of microscopic models for the core model can be accessed by
\begin{lstlisting}[language=Python]
print( nuda.crust.micro_esym_models( ) )
\end{lstlisting}
For phenomenological models, the list of core models is available as
\begin{lstlisting}[language=Python]
print( nuda.crust.pheno_esym_models( ) )
\end{lstlisting}
For a chosen \texttt{model}, the list of parameter sets can be printed as
\begin{lstlisting}[language=Python]
print( nuda.crust.pheno_esym_params( model=model ) )
\end{lstlisting}

To connect the core and the crust equations of state, we use the variable \texttt{connect}. One can choose to connect them in nucleonic density (\texttt{connect=`density'}), in energy-density (\texttt{connect=`epsilon'}) or in pressure (\texttt{connect=`pressure'}). In all cases, the upper boundary of the crust has to be fixed, which also coincides with the lower limit for the connection (\texttt{den\_lo}, \texttt{rho\_lo}, \texttt{pre\_lo}), as well as the lower boundary of the core, coinciding with the upper limit of the connection (\texttt{den\_up}, \texttt{rho\_up}, \texttt{pre\_up}). The variable \texttt{boundaries} contains the doublet value (\texttt{den\_lo}, \texttt{den\_up}) if \texttt{connect=`density'}. If \texttt{connect=`epsilon'}, then \texttt{boundaries=(\texttt{eps\_lo}, \texttt{eps\_up})} and if \texttt{connect=`pressure'}, then \texttt{boundaries=(\texttt{pre\_lo}, \texttt{pre\_up})}.

The new EoS connecting the crust and the core is discretized on log-log scale, and a linear interpolation is considered between the connection boundaries.

\begin{figure}[t]
\centering
\includegraphics[scale=0.52]{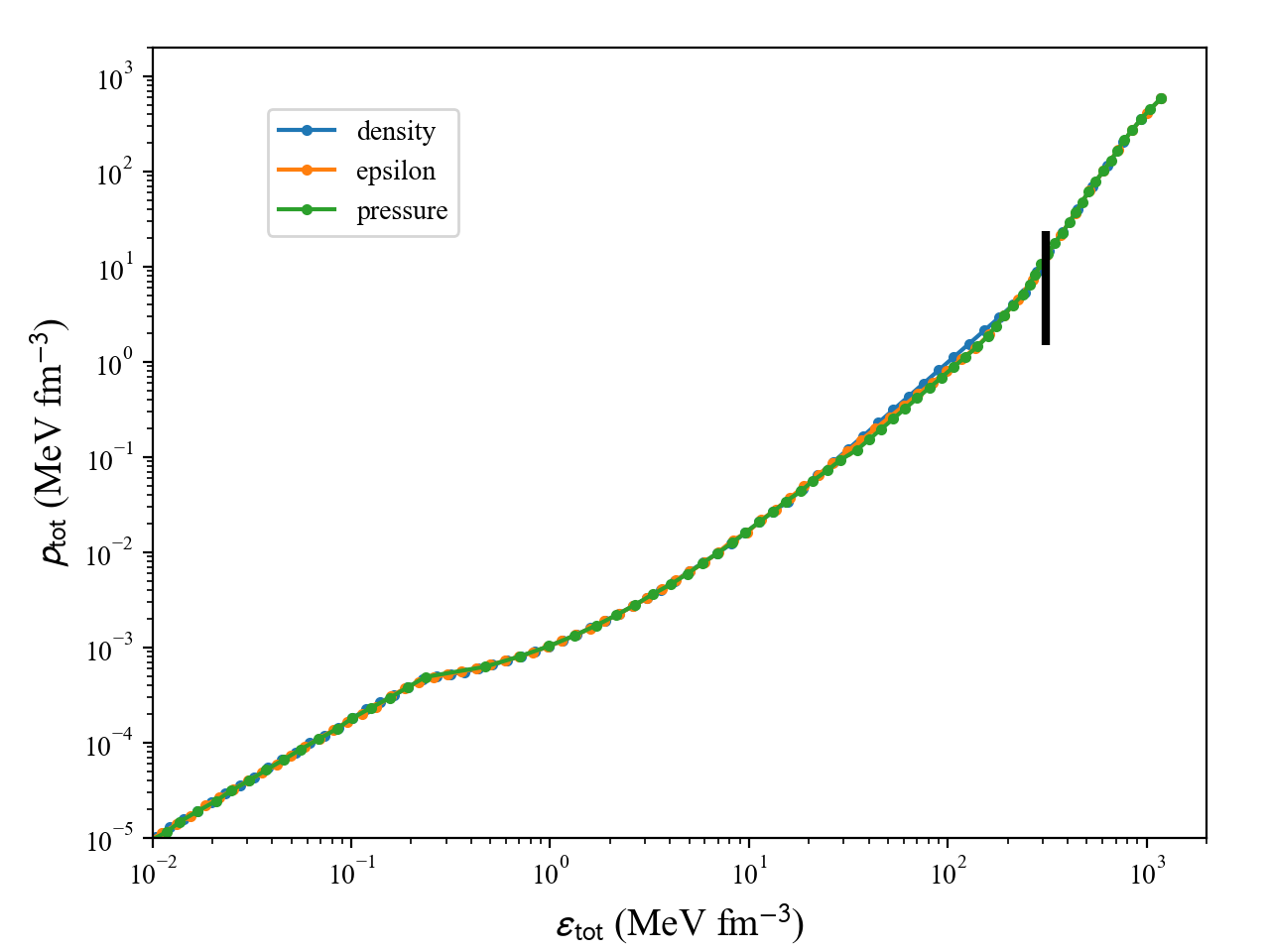}
\caption{Crust+Core EoS: pressure $p$ as a function of the energy density $\epsilon$ for three connections: `density', `epsilon' and `pressure', see text. For the crust, we took `2022-GMRS-H4' and for the core `1998-VAR-AM-APR'. The vertical bar represents the pressure uncertainty at $2\epsilon_\sat$ from nuclear physics, see table~\ref{table:pre}. This figure is generated with \texttt{eos\_setupCC\_eos\_plot.py}.}
\label{fig:eos:cc}
\end{figure}

The attributes of the object \texttt{eos} are: \texttt{eos.den}, \texttt{eos.pre}, \texttt{eos.eps} and \texttt{eos.cs2} for the particle density, the pressure, the energy density, and the square of the sound speed.

The transition between the core and the crust can be fixed by the user as explained previously, by fixing the lower and upper limits of the variable \texttt{boundary}. It can also be fixed by using empirical relations that are encoded in the \texttt{nuda} toolkit in the following way:
\begin{lstlisting}[language=Python]
den_cc, pre_cc = nuda.eos.denCC_emp( nsat, Esym, Lsym, Ksym, Qsym, emp )
print( den_cc )
\end{lstlisting}
providing \texttt{den\_cc} in fm$^{-3}$, and for the input variable \texttt{emp} which can be set to:\\

\noindent
\texttt{emp}=`Simple'\\
It simply assumes the crust-core transition density to be half that of the saturation density:
\begin{equation}
n_{cc} = \frac{n_\sat}{2} \,
\end{equation}
or \\
\noindent
\texttt{emp}=`Ducoin'\\
This empirical relation is suggested in Ref.~\cite{CDucoin:2011}:
\begin{eqnarray}
n_{cc} &=& 0.0802 + 3.23\times10^{-4}\big( L_{\sym,0.1} \nonumber \\
&&\hspace{3cm}+ 0.426 K_{\sym,0.1}\big) \,  ,\\
&\approx& 0.0802 + 3.23\times10^{-4}\big( L_\sym + (x_{0.1}+0.426) K_\sym \nonumber \\
&&\hspace{3cm} + 0.426 Q_\sym x_{0.1}\big) \, \\
p_{cc} &=& -0.328 + 9.59\times10^{-3} \big( L_{\sym,0.1}\,, \nonumber\\
&&\hspace{3cm}- 0.343 K_{\sym,0.1} \big)  \, , \\
&\approx&  -0.328 + 9.59\times10^{-3} \big( L_\sym +(x_{0.1}-0.343) K_\sym \nonumber \\
&&\hspace{3cm}- 0.343 Q_\sym x_{0.1} \big)  \, ,
\end{eqnarray}
where $x_{0.1}=x(n_\nuc=0.1$~fm$^{-3})$ and $x=(n_\nuc-n_\sat)/(3n_\sat)$. The NEP
$L_{\sym,0.1}$ and $K_{\sym,0.1}$ are empirical parameters defined at 0.1~fm$^{-3}$ reference density~\cite{CDucoin:2011}.
\\

\noindent
\texttt{emp}=`Newton'\\
This empirical relation is suggested in Ref.~\cite{WGNewton:2013}:
\begin{eqnarray}
n_{cc} &=& E_{\sym,30}\big( 0.135 - 0.098L_{\sym,70} \nonumber \\
&& \hspace{3cm}+ 0.026L_{\sym,70}^2  \big) \, , \\
p_{cc} &=& -0.724 + 0.0157 \left( L_{\sym,0.1} - 0.343 K_{\sym,0.1}^2 \right)  \, , \\
&\approx&  -0.724 + 0.0157 \big( L_\sym - 0.343 K_\sym^2 \nonumber \\
&& \hspace{1.5cm} +(1-2\times 0.343 Q_\sym)x_{0.1} K_\sym \big) \, ,
\end{eqnarray}
where $E_{\sym,30}=E_\sym/30$ and $L_{\sym,70}=L_\sym/70$, $E_\sym$ and $L_\sym$ being the NEP (in MeV) of the crust EDF.\\

\noindent
\texttt{emp}=`Steiner'\\
This empirical relation has been employed in Refs.~\cite{AWSteiner:2015, SLalit:2019}:
\begin{equation}
n_{cc} = E_{\sym,30}\left( 0.1327 - 0.0898L_{\sym,70} + 0.0228L_{\sym,70}^2  \right). 
\end{equation}
\\

To take the values given in the input array \texttt{boundaries}, instead of an empirical relation, one should fix \texttt{emp=None}. Otherwise, the choice for the variable \texttt{emp} is considered above all. We also mention that in the \texttt{nuda} toolkit, boundaries are still considered if an empirical relation is chosen. In the `density' connection, the lower boundary is taken as $0.8 n_{cc}$ and the upper boundary is $1.2 n_{cc}$.

We show a result of the connection between an EoS for the crust and another for the core in Fig.~\ref{fig:eos:cc}. Three cases are investigated: a connection in `density' between $0.016$ and $0.16$~fm$^{-3}$, in energy density `epsilon' between $15$ and $150$~MeV~fm$^{-3}$, and finally a connection in `pressure' between $0.1$ and $1$~MeV~fm$^{-3}$ (note that the boundaries are well separated from each other). A linear interpolation on a log-log scale is considered in the gap region where the two EoS are connected. The connection method has a weak influence on the EoS, but it would still be interesting to study more systematically the impact of varying the crust and the core EoS. It would also be interesting to analyse the impact of the EoS and of the connection on the NS, as suggested in Ref.~\cite{MFortin:2016}. This requires solving the general relativistic hydrostatic equations (TOV), which is currently beyond the scope of the \texttt{nuda} toolkit.

\section{Astrophysical observations: the \texttt{astro} module.}
\label{sec:astro}

Astrophysical observations have become more and more accurate over the last decades. They can now be employed to select among various EoS. The \texttt{nuda} toolkit, therefore, provides a set of astrophysical observations that are interesting to evaluate models. We first present individual measurements for the masses of neutron stars (NSs) from radio-astronomy and then the total mass of binary NSs (BNSs) from gravitational-wave (GW) observations. We then use these observations to construct the probability profile for the maximum mass of isolated, non-rotating, and non-magnetic NS (M$_\tov$).

\subsection{Masses of neutron stars}
\label{sec:astro:masses}

The observation of massive neutron stars from radio astronomy provides a challenge for many EoS models. We provide a list of several observations of massive neutron stars in table~\ref{tab:astro:mmax1} together with several measurements that are provided in the variable \texttt{obs}.

\begin{table}[t]
\centering
\renewcommand{\arraystretch}{1.5}
\setlength{\tabcolsep}{4pt}
\caption{Observed masses ($M_{\obs,i}$) and associated uncertainties ($\sigma_{M,i}$) at 68\% CL for five of the most massive pulsars. For two of these objects, there are several measurements of the masses, which do not often coincide.}
\begin{tabular}{cccccc}
\hline
source & obs & $M_{\obs}$ & $\sigma_{M}^{+}$ & $\sigma_{M}^{-}$ & Ref. \\
 & & [$M_{\odot}$] & [$M_{\odot}$] & [$M_{\odot}$] & \\
\hline
PSR J1614–2230 & 1 & 1.970 & 0.040 & 0.040 & \cite{PDemorest:2010} \\
               & 2 & 1.928 & 0.017 & 0.017 & \cite{EFonseca:2016} \\
               & 3 & 1.908 & 0.016 & 0.016 & \cite{ZArzoumanian:2018} \\
               & 4 & 1.922 & 0.015 & 0.015 & \cite{MdFAlam:2021} \\
               & 5 & 1.937 & 0.014 & 0.014 & \cite{GAgazie:2023} \\
               & av & 1.933 & 0.031 & 0.031 & \\
PSR J0348+0432 & 1 & 2.01 & 0.04 & 0.04 & \cite{JAntoniadis:2013} \\
PSR J2215+5135 & 1 & 2.27 & 0.15 & 0.15 & \cite{MLinares:2018}\\
PSR J1600+3053 & 1 & 2.5 & 0.9 & 0.7 & \cite{ZArzoumanian:2018} \\
               & av & 2.579 & 0.792 & 0.792 & \\
MSP J0740+6620 & 1 & 2.14 & 0.10 & 0.09 & \cite{HTCromartie:2019}\\
               & 2 & 2.08 & 0.07 & 0.07 & \cite{EFonseca:2021} \\
               & 3 & 1.99 & 0.07 & 0.07 & \cite{GAgazie:2023} \\
               & av & 2.071 & 0.101 & 0.101 & \\
\hline
\end{tabular}
\label{tab:astro:mmax1}
\end{table}

The complete list of available sources is given with the following instructions:
\begin{lstlisting}[language=Python]
print( nuda.astro.masses_sources( ) )
\end{lstlisting}

The list of observations for a given source is given with the following instructions:
\begin{lstlisting}[language=Python]
print( nuda.astro.mass_obss( source = `J1614-2230' ) )
\end{lstlisting}
This list of observations is further detailed in table~\ref{tab:astro:mmax1}.

Once the variables \texttt{source} and \texttt{obs} are chosen in the previous list, the mass measurement is instantiated in the object \texttt{m} with the following instruction:
\begin{lstlisting}[language=Python]
m = nuda.astro.setupMasses( source=
`J1614-2230', obs=4 )
m.print_outputs()
\end{lstlisting}

We now provide more details about neutron star masses available in \texttt{nuda} toolkit. There are several sources, some of which have multiple measurements providing different values for neutron star masses.\\

\noindent
\texttt{source}=`J1614–2230'. \\
\texttt{obs}= 1~\cite{PDemorest:2010}, 2~\cite{EFonseca:2016}, 3~\cite{ZArzoumanian:2018}, 4~\cite{MdFAlam:2021}, 5~\cite{GAgazie:2023}. \\
Measured mass for \texttt{source} PSR J1614–2230 obtained from different observations (\texttt{obs}).\\

\noindent
\texttt{source}=`J0348+0432'. \\
\texttt{obs}= 1~\cite{JAntoniadis:2013}. Measured mass for \texttt{source} PSR J0348+0432 obtained from a single observation (\texttt{obs}).\\

\noindent
\texttt{source}=`J2215+5135'. \\
\texttt{obs}= 1~\cite{MLinares:2018}. Measured mass for \texttt{source} PSR J2215+5135 obtained from a single observation (\texttt{obs}).\\

\noindent
\texttt{source}=`J1600+3053'. \\
\texttt{obs}= 1~\cite{ZArzoumanian:2018}. Measured mass for \texttt{source} PSR J1600+3053 obtained from a single observation (\texttt{obs}).\\

\noindent
\texttt{source}=`J0740+6620'. \\
\texttt{obs}= 1~\cite{HTCromartie:2019}, 2~\cite{EFonseca:2021}, 3~\cite{GAgazie:2023}. \\
Measured mass for \texttt{source} MSP J0740+6620 obtained from different observations (\texttt{obs}).\\

For a given measurement, the attributes are: \texttt{m.mass} is the centroid of the measured mass, \texttt{m.sig\_up} and \texttt{m.sig\_lo} are the reported upper and lower uncertainty. For a single source, the different measurements can be averaged together in the following way:
\begin{lstlisting}[language=Python]
mav = nuda.astro.setupMassesAverage( source=`J1614-2230' )
mav.print_outputs()
\end{lstlisting}
The attributes are the centroid \texttt{mav.mass\_cen} and the standard deviation \texttt{mav.mass\_std} of the reconstructed measurement (assuming a Gaussian distribution for each measurement).

\begin{figure}[t]
\centering
\includegraphics[scale=0.52]{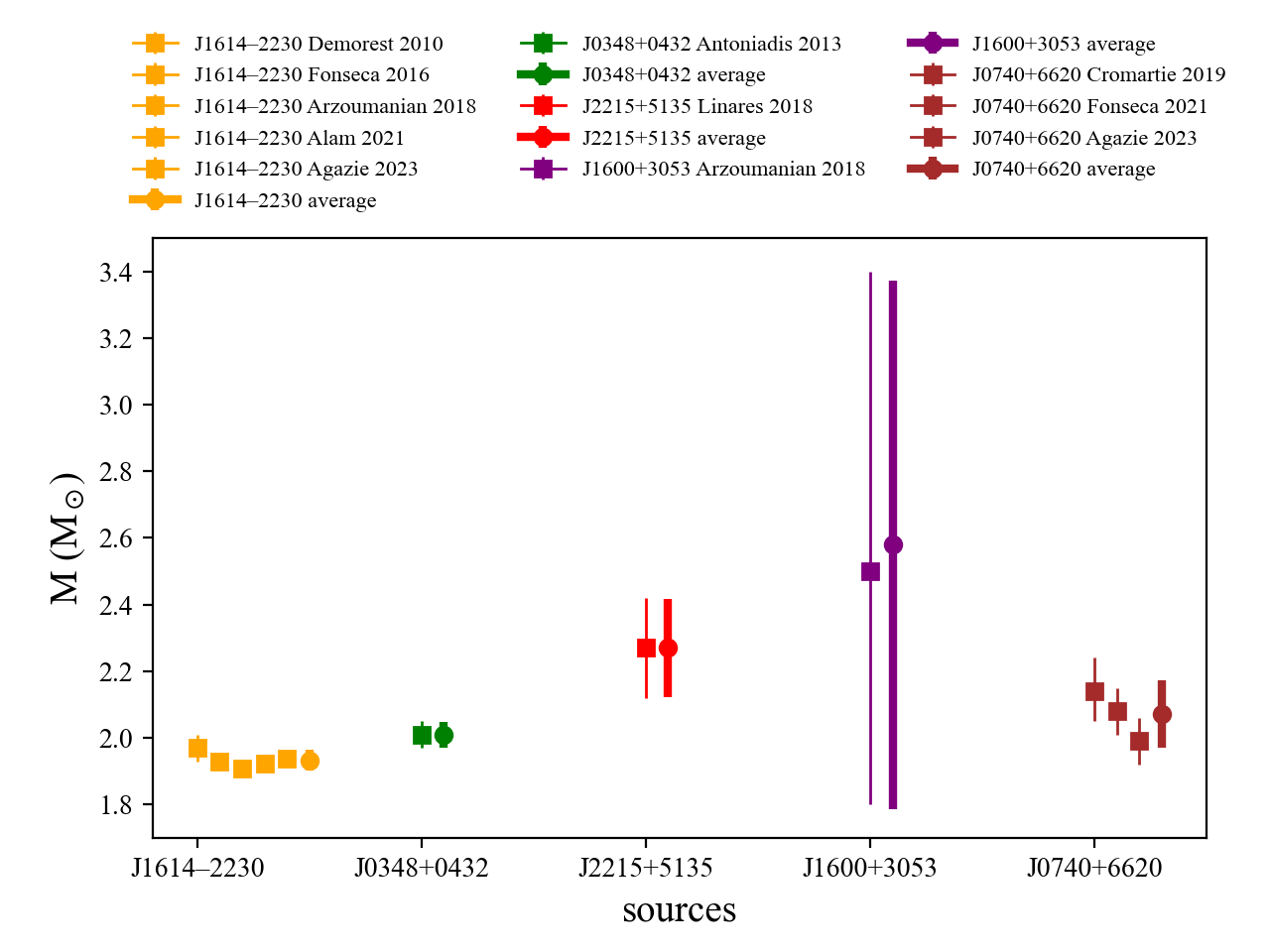}
\caption{Distribution of measured masses (thin lines) for massive neutron stars and average values (thick lines) defined as the centroid and standard deviation of the cumulative observations per source. Figure generated with \texttt{astro\_setupMasses\_plot.py}.} 
\label{fig:astro:masses}
\end{figure}

All the measured masses (thin lines) as well as average values of the measurements per source (thick lines) are shown in Fig.~\ref{fig:astro:masses}. Note that for PSR J1600+3053, the average is not exactly identical to the original measure. It is due to the fact that the original measurement is asymmetric, while the average symmetrizes it. Exact numbers are given in Table~\ref{tab:astro:mmax1}.

The average quantities provided by the toolkit shall be considered for convenience and are not meant to replace the exact measurements. Note, however, that it may be difficult to choose one measurement instead of another, and the interesting feature of the average masses, illustrated for instance in Fig.~\ref{fig:astro:masses} is that differences in the measured masses for different sources is larger than the measured uncertainties, except for PSR J1600+3053. The average as we compute it represents well the measurements and suggest an interesting reference quantity.

\subsection{Upper masses from gravitational-wave observations}
\label{sec:astro:upmasses}

GW emitted from BNSs provides an estimate of the total masses (M$_\tot$) of the two merging NSs with a good accuracy. If the final state is a Black Hole (BH), then the total mass can be taken as an estimate of the upper limit of the TOV mass (defined as the maximum mass of non-rotating and non-magnetized NS).

\begin{table}[t]
\centering
\renewcommand{\arraystretch}{1.5}
\setlength{\tabcolsep}{2pt}
\caption{Masses for GW observation of BNS and for different hypotheses on the spin of the BNSs and the waveform model.}
\begin{tabular}{ccccccc}
\hline
source & hyp & hypothesis & $M$ & $\sigma_{M}^{+}$ & $\sigma_{M}^{-}$ & Ref. \\
 &  & & [$M_{\odot}$] & [$M_{\odot}$] & [$M_{\odot}$] & \\
\hline
GW170817 & 1 & low-spin & 2.74 & 0.04 & 0.01 & \cite{BPAbbott:2017} \\
 & & + TaylorF2 & & & & \\
GW170817 & 2 & high-spin & 2.82 & 0.47 & 0.09 & \cite{BPAbbott:2017} \\
 & & + TaylorF2 & & & & \\
GW170817 & 3 & low-spin & 2.73 & 0.04 & 0.01 & \cite{BPAbbott:2019} \\
 & & + PhenomPNRT & & & & \\
GW170817 & 4 & high-spin & 2.77 & 0.22 & 0.05 & \cite{BPAbbott:2019} \\
 & & + PhenomPNRT & & & & \\
GW170817 & av & & 2.825 & 0.189 & 0.189 & \\
GW190814 & 1 & & 2.59 & 0.08 & 0.09 & \cite{BPAbbott:2020}\\
GW190814 & av & & 2.585 & 0.085 & 0.085 & \\
\hline
\end{tabular}
\label{tab:astro:mmax2}
\end{table}

The complete list of available GW sources is given with the following instructions:
\begin{lstlisting}[language=Python]
print( nuda.astro.mup_sources( ) )
\end{lstlisting}

The list of hypotheses for a given source is given with the following instructions:
\begin{lstlisting}[language=Python]
print( nuda.astro.mup_hyps( source=`GW170817' ) )
\end{lstlisting}
These hypotheses are given in Tab.~\ref{tab:astro:mmax2}, where the list of masses for GW observations are reported. For the so-called golden event, GW170817 reported in the toolkit, the total mass $M_\tot$ is given based on different hypotheses, e.g., low-spin versus high-spin or gravitational waveforms employed. For the other event GW190814, only one analysis has been carried out.

The object \texttt{mup} contains a measured mass for a given choice for the variables \texttt{source} and \texttt{hyp}. It is instantiated in the  following way:
\begin{lstlisting}[language=Python]
mup = nuda.astro.setupMup( source=`GW170817', hyp=4 )
mup.print_outputs()
\end{lstlisting}
There are several sources, and for some sources, there are several hypotheses in the analysis providing different values for the measured total masses.\\

\noindent
\texttt{source}=`GW170817'. \\
\texttt{hyp}= 1~\cite{BPAbbott:2017}, 2~\cite{BPAbbott:2017}, 3~\cite{BPAbbott:2019}, 4~\cite{BPAbbott:2019}. \\
Observed mass for \texttt{source} GW170817 obtained from different hypotheses (\texttt{hyp}), see tab.~\ref{tab:astro:mmax2} for more details.\\

\noindent
\texttt{source}=`GW190814'. \\
\texttt{hyp}= 1~\cite{BPAbbott:2020}. \\
Observed mass for \texttt{source} GW190814 obtained from different hypotheses (\texttt{hyp}), see tab.~\ref{tab:astro:mmax2} for more details.\\

\begin{figure}[t]
\centering
\includegraphics[scale=0.52]{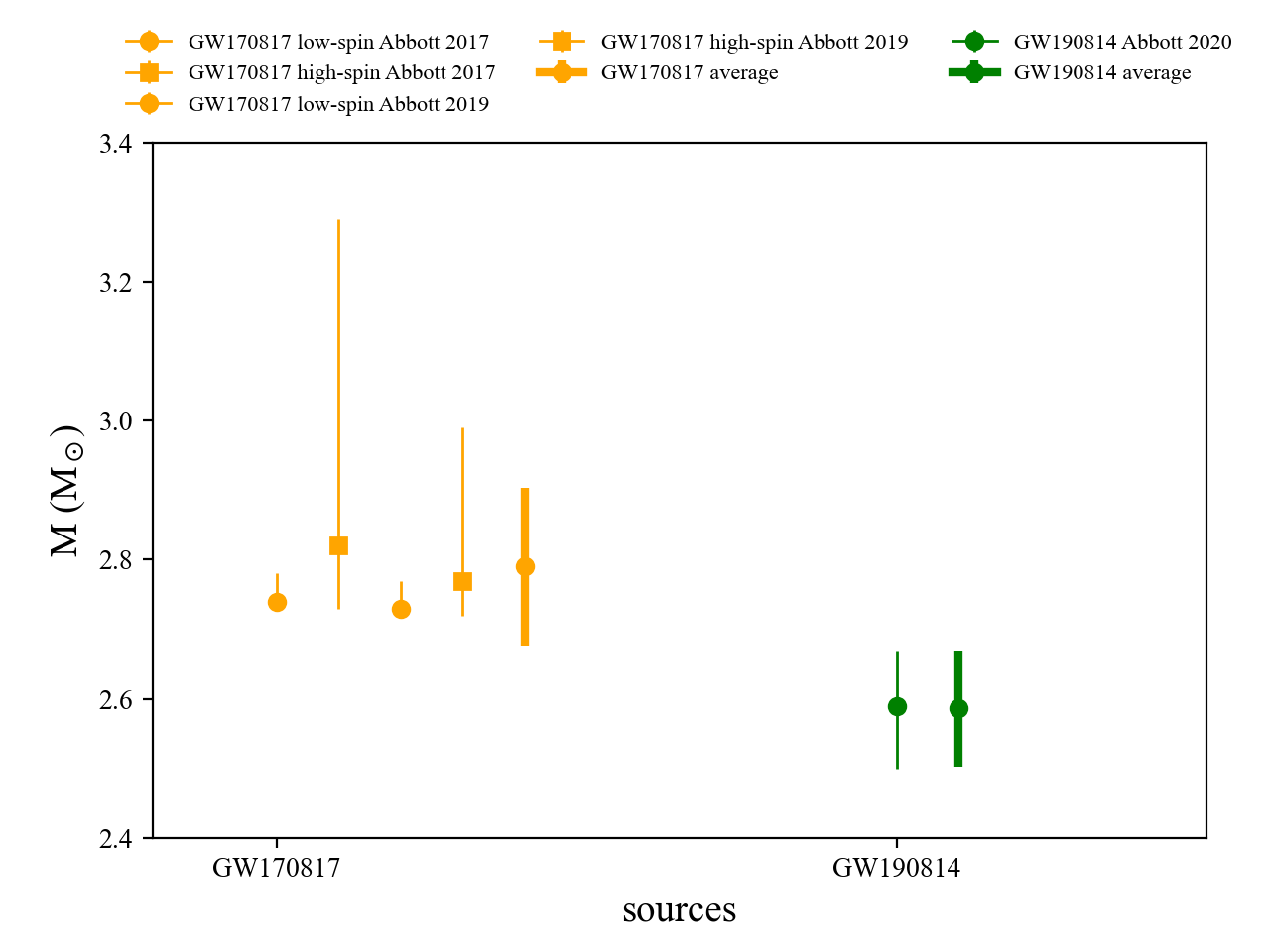}
\caption{Distribution of upper masses from different GW sources: GW170817 and GW190814. The average for GW170817 is obtained considering only the two most recent analyses, fixing \texttt{hyps = [3, 4]} in the instantiation, as in the example provided in the text. Figure generated with \texttt{astro\_setupMup\_plot.py}.} 
\label{fig:astro:mup}
\end{figure}

For a given analysis, the attributes are: \texttt{mup.mup} is the centroid of the measured mass, \texttt{mup.sig\_up} and \texttt{mup.sig\_lo} are the reported upper and lower uncertainty respectively. For a given source, the different measurements can be averaged together in the following way:
\begin{lstlisting}[language=Python]
mupav = nuda.astro.setupMupAverage( source=`GW170817', hyps=[3, 4] )
mupav.print_outputs()
\end{lstlisting}
The attributes are the centroid \texttt{mupav.mass\_cen} and the standard deviation \texttt{mupav.mass\_std} of the reconstructed measurement (assuming a Gaussian distribution for each measurement).

All the observed total masses are shown in Fig.~\ref{fig:astro:mup}, as well as average values of the observations, defined as the centroid and standard deviation of the reconstructed observation (assuming a Gaussian distribution for each hypothesis).

\subsection{Probability distribution function for $M_\tov$}
\label{sec:astro:mtov}

The set of observed pulsar masses provides lower limits for $M_\tov$, while the quantity $M_\tot$ determined from GW emission from BNS provides an upper limit for $M_\tov$, provided the BNS collapses to a BH after the merger. It is, therefore, possible to estimate the boundaries for $M_\tov$ from astrophysical observations. Moreover, considering the uncertainties in the measurements, it is preferable to determine a distribution of probabilities as a function of $M_\tov$. This is what we are now constructing.

We consider a function of the TOV mass, $z_\obs(M_\tov)$, defined in the following way:
\begin{eqnarray}
z_\obs(M_\tov) &=& \frac{M_\tov-M_{\obs}}{\sqrt{2} \, \sigma_{M}^{-}}\Theta(M_{\obs}-M_\tov) \nonumber \\
&&+ \frac{M_\tov-M_{\obs}}{\sqrt{2} \, \sigma_{M}^{+}}\Theta(M_\tov-M_{\obs})
\, ,
\end{eqnarray}
where $\Theta$ is the Heaviside step function and
$M_\obs$ and $\sigma_M^{\pm}$ are the observed mass and the associated uncertainties given in tables~\ref{tab:astro:mmax1} (for the lower limit of $M_\tov$) and \ref{tab:astro:mmax2} (for the upper limit).

\begin{figure}[t]
\centering
\includegraphics[scale=0.52]{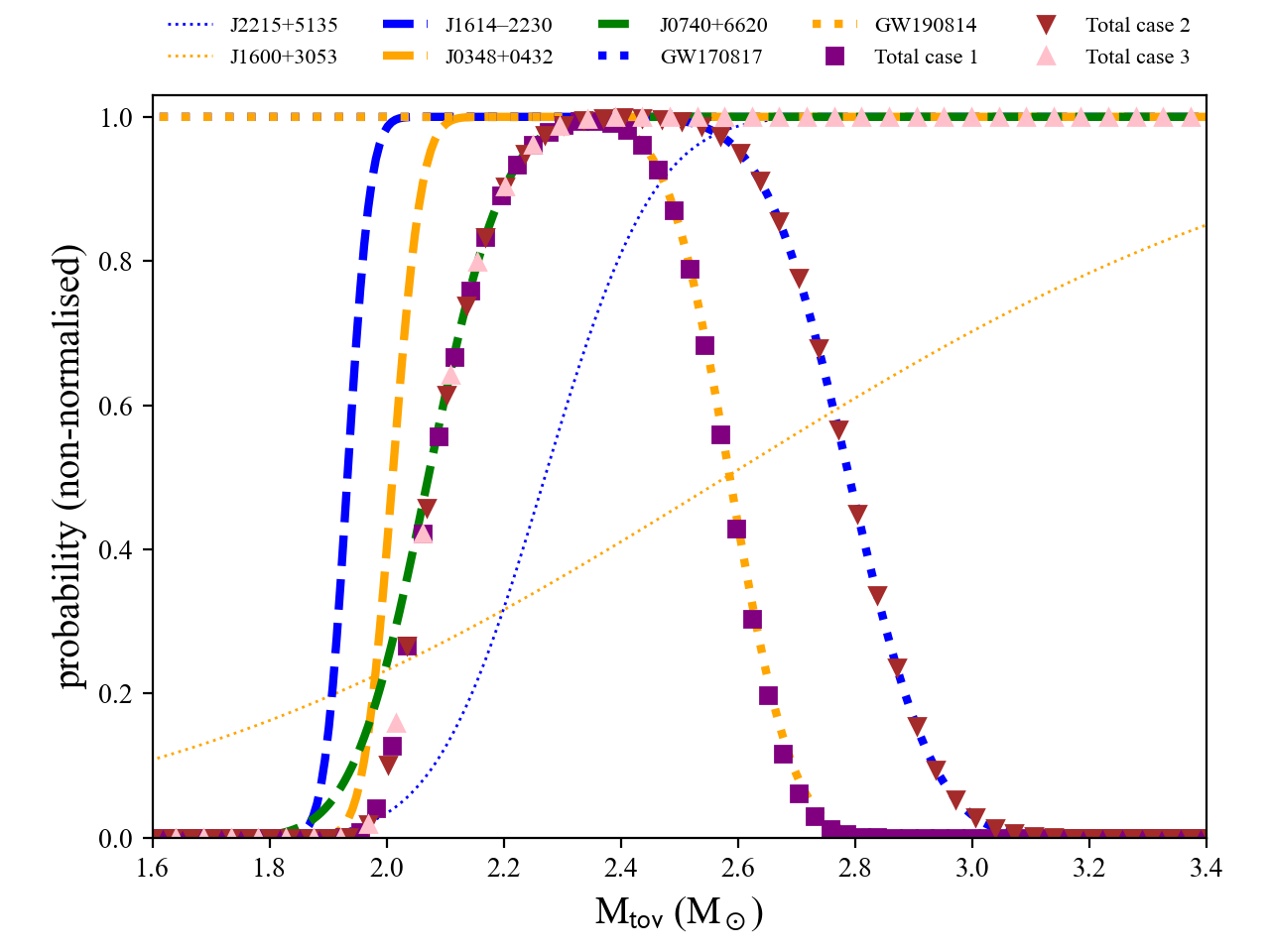}
\caption{Probability distribution function for the TOV mass (Total, thick solid line or dashed line or dotted line) and the contributions from the different sources (see legend in the figure). Figure generated with \texttt{astro\_setupMtov\_plot.py}.} 
\label{fig:astro:mtov}
\end{figure}

The probability associated with $M_\tov$ considering a given source is calculated as follows 
\begin{equation}
P_\obs(M_\tov) = \frac{\text{erf}[z_\obs(M_\tov)]+1}{2}, 
\label{eq:astro:plower}
\end{equation}
where $\mathrm{erf}(z)$ is the error function defined as  $\mathrm{erf}(z)=(2/\sqrt{\pi})\int_0^z \exp(-t^2) dt$. The probabilities associated with each observation are shown in Fig.~\ref{fig:astro:mtov}, see the five dashed curves for the sources listed in Tab.~\ref{tab:astro:mmax1}.

The lower bound probability distribution $P_{\text{lower}}(M_\tov)$ is defined as,
\begin{equation}
P_\mathrm{lower}(M_\tov) = \prod_{\obs} P_\obs(M_\tov) \, .
\end{equation}
$P_\mathrm{lower}$ is shown in Fig.~\ref{fig:astro:mtov} as a set of circle symbols. We see that the probability distribution is largely impacted by the highest mass distribution with small uncertainties (here PSR J2215-5135). The role of PSR J1600-3053 is not negligible but remains weak because of the large uncertainties.

The upper limit for non-rotating and non-magnetized neutron stars is more difficult to obtain. GW170817, however, provides an estimation of the upper limit of neutron stars, since it is possible that the two neutron stars have collapsed into a black hole after their merger. This limit depends, however, on the prior for the spin of the neutron stars, see Tab.~\ref{tab:astro:mmax2}.

The probability associated with $M_\tov$ considering a given source is calculated as follows,
\begin{equation}
P_\mathrm{Mtot}(M_\tov) = \frac{1-\text{erf}[z_\obs(M_\tov)]}{2}\, ,
\label{eq:astro:pmtot}
\end{equation}
where we consider the average value including the uncertainties in the hypothesis for the spin priors and the waveforms. The upper bound probability distribution $P_{\text{upper}}(M_\tov)$ is defined as,
\begin{equation}
P_\mathrm{upper}(M_\tov) = \prod_\mathrm{Mtot} P_\mathrm{Mtot}(M_\tov) \, .
\end{equation}
The toolkit is ready for several sources, e.g., GW170817 and GW190814, and new sources can be added easily. The dashed line (for GW190814) and the upper probability distribution (square symbols) coincide in Fig.~\ref{fig:astro:mtov}. The main contributions come from PSR J2215+5135 and J1600+3053 for the lower boundary and GW190814 for the upper boundary.

Finally, the total probability distribution for the TOV mass is defined as the product of the distribution for the lower masses and the upper masses, as
\begin{equation}
P_{\text{mass}}(M_\max) = P_\mathrm{lower}(M_\max) \times P_\mathrm{upper}(M_\max) \, ,
\label{eq:astro:ptot}
\end{equation}
and is shown in thick solid line in Fig.~\ref{fig:astro:mtov}.

The variables \texttt{source\_lo} and \texttt{source\_up} have to be fixed. The existing sources can be obtained in the following way:
\begin{lstlisting}[language=Python]
sources_lo=nuda.astro.masses_sources( )
print('sources lo:',sources_lo)
sources_up=nuda.astro.mtot_sources( )
print('sources up:',sources_up)
\end{lstlisting}

Once the variables \texttt{source\_lo} and \texttt{source\_up} are chosen in the previous list, the call for the mass observation is performed in the following way:
\begin{lstlisting}[language=Python]
mtov = nuda.astro.setupMtov( sources_lo=sources_lo, sources_up=sources_up) )
mtov.print_outputs()
\end{lstlisting}

The mass distributions are represented in Fig.~\ref{fig:astro:mtov}. It illustrates the impact of the different measurements for the lower and upper boundaries of the TOV mass.

The attributes are given in the form of Numpy arrays: \texttt{mtov.mass} is a vector with a set of masses for which values the distribution is provided, \texttt{mtov.proba\_lo[i]} provide the lower mass distribution of probabilities corresponding to each of the source given in the array \texttt{sources\_lo} with index \texttt{i}, see Eq.~\eqref{eq:astro:plower}, \texttt{mtov.proba\_up[j]} provide the upper mass distribution of probabilities corresponding to each of the source given in the array \texttt{sources\_up} with index \texttt{j}, see Eq.~\eqref{eq:astro:pmtot}, and finally,  \texttt{mtov.proba\_tot} contains the product of lower and upper probabilities, see Eq.~\eqref{eq:astro:ptot}.

\subsection{Mass-radius measurement by x-ray emission}
\label{sec:astro:nicer}

The NICER observatory~\cite{KCGendreau:2016} is providing a large amount of data which are analyzed to determine accurately the mass and the radius of pulsars. A summary of these masses and radii measurements is shown in table~\ref{tab:astro:nicer}.

\begin{table*}[t]
\centering
\renewcommand{\arraystretch}{1.8}
\setlength{\tabcolsep}{12pt}
\caption{Observational radii and masses for different pulsars. $C$ stands for compactness and is written when it is provided by the authors. We also complete the table with our calculation of the compactness obtained from the reported distribution of masses and radii (in bold).}
\begin{tabular}{ccccccc}
\hline
source & obs & telescope & $R$ & $M$ & $C$ & Ref. \\
 &  & & [km] & [$M_{\odot}$] \\
\hline
PSR J0030+0451 & 1 & NICER & $13.02^{+1.24}_{-1.06}$ & $1.440^{+0.150}_{-0.140}$ & $0.163^{+0.008}_{-0.009}$ & \cite{MCMiller:2019} \\
& 2 & NICER & $12.71^{+1.14}_{-1.19}$ & $1.340^{+0.150}_{-0.160}$ & $0.156^{+0.008}_{-0.010}$ & \cite{TERiley:2019}\\
& 3 & NICER & $14.44^{+0.88}_{-1.05}$ & $1.700^{+0.180}_{-0.190}$ & $0.179^{+0.011}_{-0.022}$ & \cite{SVinciguerra:2024}\\
& 4 & NICER & $11.71^{+0.88}_{-0.83}$ & $1.400^{+0.130}_{-0.120}$ & $0.1773^{+0.0056}_{-0.0074}$ & \cite{SVinciguerra:2024}\\
& av & & $12.89\pm1.17$ & $1.390\pm0.158$ & $0.159\pm0.009$ \\
MSP J0740+6620 & 1 & NICER & $13.70^{+2.60}_{-1.50}$ & $2.080^{+0.070}_{-0.070}$ & & \cite{MCMiller:2021} \\
& 2 & NICER & $12.39^{+1.30}_{-0.98}$ & $2.072^{+0.067}_{-0.066}$ &  & \cite{TERiley:2021}\\
& 3 & NICER & ${12.49}^{+1.28}_{-0.88}$ & $2.073^{+0.069}_{-0.069}$ &  & \cite{TSalmi:2024} \\
& av & & $13.10\pm1.66$ & $2.075\pm0.068$ & \\
PSR J0437-4715 & 1 & NICER & $11.360^{+0.945}_{-0.629}$ & $1.418^{+0.037}_{-0.037}$ & $0.1847^{+0.0097}_{-0.0143}$ & \cite{DChoudhury:2024} \\
\hline
\end{tabular}
\label{tab:astro:nicer}
\end{table*}

The complete list of available sources is given with the following instructions:
\begin{lstlisting}[language=Python]
print( nuda.astro.mr_sources( ) )
\end{lstlisting}

The list of observations, see table~\ref{tab:astro:nicer}, for a given source is given with the following instruction:
\begin{lstlisting}[language=Python]
print( nuda.astro.mr_obss( source=`J0030+0451' ) )
\end{lstlisting}

The object \texttt{mr} is instantiated in the following way:
\begin{lstlisting}[language=Python]
mr=nuda.astro.setupMR( source=`J0030+0451', obs=1 )
mr.print_outputs()
\end{lstlisting}
where the variables \texttt{source} and \texttt{obs} are chosen in the previous list, see also table~\ref{tab:astro:nicer}. We now provide more details for each of the sources encoded in the \texttt{nuda} toolkit.\\

\noindent
\texttt{source}=`J0030+0451'. \\
Observed mass and radii for \texttt{source} J0030+0451 obtained from different analyses of the observational data (\texttt{obs}=1, 2, 3, or 4), see tab.~\ref{tab:astro:nicer} for more details.\\
\texttt{obs}= 1 is the result of the fit to NICER with three oval hot-spots from Ref.~\cite{MCMiller:2019}, \\
\texttt{obs}= 2 is the result of the fit to NICER for ST+PST hot-spot model from Ref.~\cite{TERiley:2019}. \\
\texttt{obs}= 3 is the result of the fit to joint NICER and XMM-Newton data for PDT-U hot-spot model from Ref.~\cite{SVinciguerra:2024}. \\
\texttt{obs}= 4 is the result of the fit to joint NICER and XMM-Newton data for ST+PDT hot-spot model from Ref.~\cite{SVinciguerra:2024}. \\

\noindent
\texttt{source}=`J0740+6620'. \\
Observed mass and radii for 
\texttt{source} J0740+6620 obtained from different analyses of the observational data (\texttt{obs}=1, 2 or 3), see tab.~\ref{tab:astro:nicer} for more details.\\
\texttt{obs}= 1 is the result of the fits to NICER and XMM-Newton Data with a Parameterized Normalization for the XMM-Newton Calibration obtained in Ref.~\cite{MCMiller:2021}, \\
\texttt{obs}= 2 is the result of the fit to NICER considering ST-U model with a conditional constraint on NICER XTI pulse-profile modeling, joint NANOGrav and CHIME/Pulsar wideband radio timing, and XMM EPIC spectroscopy obtained in Ref.~\cite{TERiley:2021}. \\

\noindent
\texttt{source}=`J0437-4715'. \\
Observed mass and radii for 
\texttt{source} J0437-4715 obtained from different analyses of the observational data (\texttt{obs}=1), see tab.~\ref{tab:astro:nicer} for more details.\\
\texttt{obs}= 1 is the result of the fits to NICER considering the high MULTINEST resolution CST+PDT runs and in the presence of lower and upper background constraints involving the instrument background and 3C50 AGN spectrum obtained in Ref.~\cite{DChoudhury:2024}.\\

The attributes of the object \texttt{mr} are: \texttt{mr.rad}, \texttt{mr.rad\_sig\_lo}, and \texttt{mr.rad\_sig\_up} for the radius and its lower and upper uncertainties, as well as \texttt{mr.mass}, \texttt{mr.mass\_sig\_lo}, and \texttt{mr.mass\_sig\_up} for the mass measurement.

\begin{figure}[t]
\centering
\includegraphics[scale=0.52]{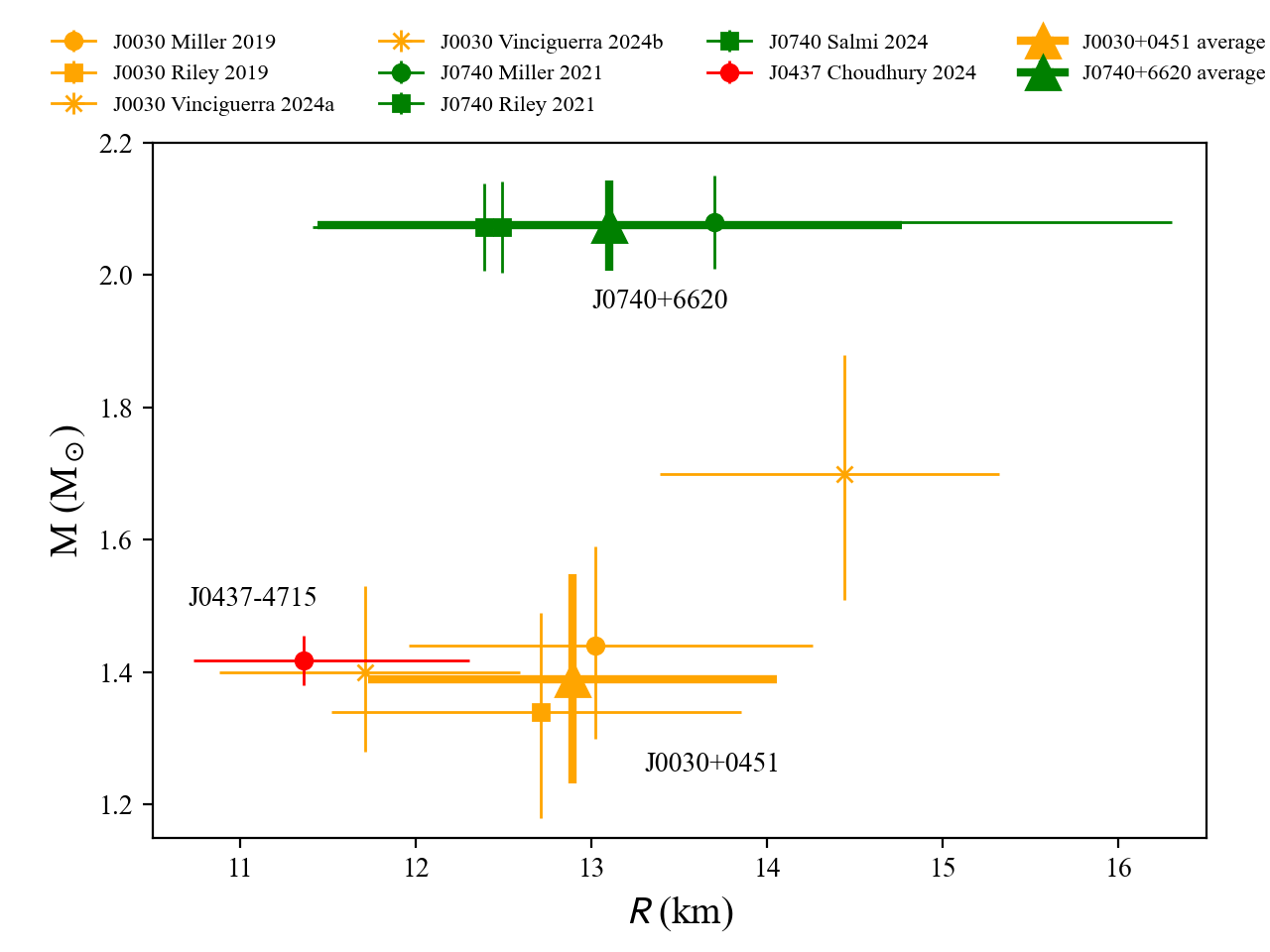}
\caption{Observational measurement of masses and radii by the NICER observatory~\cite{KCGendreau:2016}. Figure generated with \texttt{astro\_setupMR\_plot.py}.} 
\label{fig:astro:nicer}
\end{figure}

The different measurements are represented in Fig.~\ref{fig:astro:nicer}, where each color is specific to a given pulsar. For a given source, the different measurements can be averaged together in the following way:
\begin{lstlisting}[language=Python]
mrav = nuda.astro.setupMRAverage( source=`J1614-2230', obss=[ 1, 2 ] )
mrav.print_outputs()
\end{lstlisting}
The attributes are the centroid \texttt{mrav.mass\_cen} and the standard deviation \texttt{mrav.mass\_std} for the masses and \texttt{mrav.rad\_cen} and the standard deviation \texttt{mrav.rad\_std} for the radii.

\subsection{Tidal deformabilities from gravitational waves}
\label{sec:astro:tidal}

At leading order, the tidal effects are imprinted in the gravitational waveform through the effective tidal deformability $\tilde{\Lambda}$~\cite{KThorne:1967, THinderer:2008,TBinnington:2009,TDamour:2009,THinderer:2010, SPostnikov:2010}, which is defined as 
\begin{equation}
\tilde{\Lambda} = \frac{16}{13}
\frac{(12q+1)\Lambda_1+(12+q)q^4\Lambda_2}{(1+q)^5}\,,
\end{equation}
where $\Lambda_1$ and $\Lambda_2$ are the dimensionless tidal deformability for the two merging stars, related to the tidal Love number $k_i$ as,
\begin{equation}
\Lambda_i=\frac 2 3 k_i \left( \frac{R_i c^2}{G M_i}\right)^5 \,.
\end{equation}
$R_i$ and $M_i$ being the radii and the mass if the star $i$, $i=1$, $2$. The binary mass ratio is $q=M_2/M_1\leq 1$.
The chirp mass is
\begin{equation}
M_\chirp=\frac{(M_1 M_2)^{3/5}}{M_\tot^{-1/5}}
\, ,
\end{equation}
with $M_\tot=M_1+M_2$.

\begin{table*}[t]
\centering
\renewcommand{\arraystretch}{1.8}
\setlength{\tabcolsep}{10pt}
\caption{The lower and upper boundary at 90\% CI of the effective tidal deformability obtained from various analyses of GW sources.}
\begin{tabular}{ccccccc}
\hline
source & hyp & hypothesis & $M_\chirp$ [M$_{\odot}$] & $q$ & $\tilde{\Lambda}_{90\%}$ & Ref. \\
\hline
GW170817 & 1 & low-spin & 1.188$^{+0.004}_{-0.004}$ & [0.7:1] & 400$^{+400}_{-400}$ & \cite{BPAbbott:2017} \\
 & & + TaylorF2 & & & & \\
GW170817 & 2 & high-spin & 1.188$^{+0.004}_{-0.004}$ & [0.4:1] & 350$^{+350}_{-350}$ & \cite{BPAbbott:2017} \\
 & & + TaylorF2 & & & & \\
GW170817 & 3 & low-spin & 1.186735$^{+0.000119}_{-0.000095}$ & [0.639:0.982] & 222$^{+420}_{-138}$ & \cite{SDe:2018} \\
 & & + TaylorF2 & & & & \\
GW170817 & 4 & low-spin & 1.186$^{+0.001}_{-0.001}$ & [0.73:1] & 300$^{+420}_{-230}$ & \cite{BPAbbott:2019} \\
 & & + PhenomPNRT & & & & \\
GW170817 & 5 & high-spin & 1.186$^{+0.001}_{-0.001}$ & [0.53:1] & 315$^{+315}_{-315}$ & \cite{BPAbbott:2019} \\
 & & + PhenomPNRT & & & & \\
  & av & & & & 355$^{+338}_{-338}$ & \\
 GW190425 & 1 & low-spin & 1.44$^{+0.02}_{-0.02}$ & [0.8:1.0] & 300$^{+300}_{-300}$ & \cite{BPAbbott:2020}\\
 GW190425 & 2 & high-spin & 1.44$^{+0.02}_{-0.02}$ & [0.8:1.0] & 550$^{+550}_{-550}$ & \cite{BPAbbott:2020}\\
  & av & & & & 425$^{+456}_{-425}$ & \\
\hline
\end{tabular}
\label{tab:astro:tidal}
\end{table*}

Several analyses have extracted the tidal deformabilities from the GW signal of GW170817 and GW190425. They are reported in Tab.~\ref{tab:astro:tidal}.

The complete list of available sources is given with the following instructions:
\begin{lstlisting}[language=Python]
print( nuda.astro.gw_sources( ) )
\end{lstlisting}

The list of hypotheses for a given source is given with the following instruction:
\begin{lstlisting}[language=Python]
print( nuda.astro.gw_hyps( source=`GW170817' ) )
\end{lstlisting}

The object \texttt{gw} is instantiated in the following way:
\begin{lstlisting}[language=Python]
gw = nuda.astro.setupGW( source=`GW170817', hyp=1 )
gw.print_outputs()
\end{lstlisting}
where the variables \texttt{source} and \texttt{hyp} are chosen in the previous list.

We now provide more details about the observed masses available in \texttt{nuda} toolkit. There are several sources, and for some sources, there are several observations providing different values for the observed masses.\\

\noindent
\texttt{source}=`GW170817'. \\
Observed mass for \texttt{source} GW170817 obtained from different hypotheses (\texttt{hyp}=1, 2, 3, 4 or 5).\\
\texttt{hyp}=1 low spin prior from Ref.~\cite{BPAbbott:2017}, 2 high spin prior from Ref.~\cite{BPAbbott:2017}, \\
\texttt{hyp}=3 from Ref.~\cite{SDe:2018}, \\
\texttt{hyp}=4 low prior prior from Ref.~\cite{BPAbbott:2019}, or 5 high spin prior from Ref.~\cite{BPAbbott:2019}. \\

\noindent
\texttt{source}=`GW190425'. \\
Observed mass for \texttt{source} GW170817 obtained from a single hypothesis (\texttt{hyp}=1 or 2).\\
\texttt{hyp}= 1 low spin prior from Ref.~\cite{BPAbbott:2020}, or 2 high spin prior from Ref.~\cite{BPAbbott:2020}. 

The attributes of the object \texttt{gw} are: \texttt{gw.mchirp}, \texttt{gw.mchirp\_sig\_lo}, \texttt{gw.mchirp\_sig\_up} for the chirp mass $M_\chirp$, \texttt{gw.q\_lo} and \texttt{gw.q\_up} for the mass ratio $q$, and \texttt{gw.lam}, \texttt{gw.lam\_sig\_lo}, and \texttt{gw.lam\_sig\_up} for the effective tidal deformability $\tilde{\Lambda}$.

\begin{figure}[t]
\centering
\includegraphics[scale=0.52]{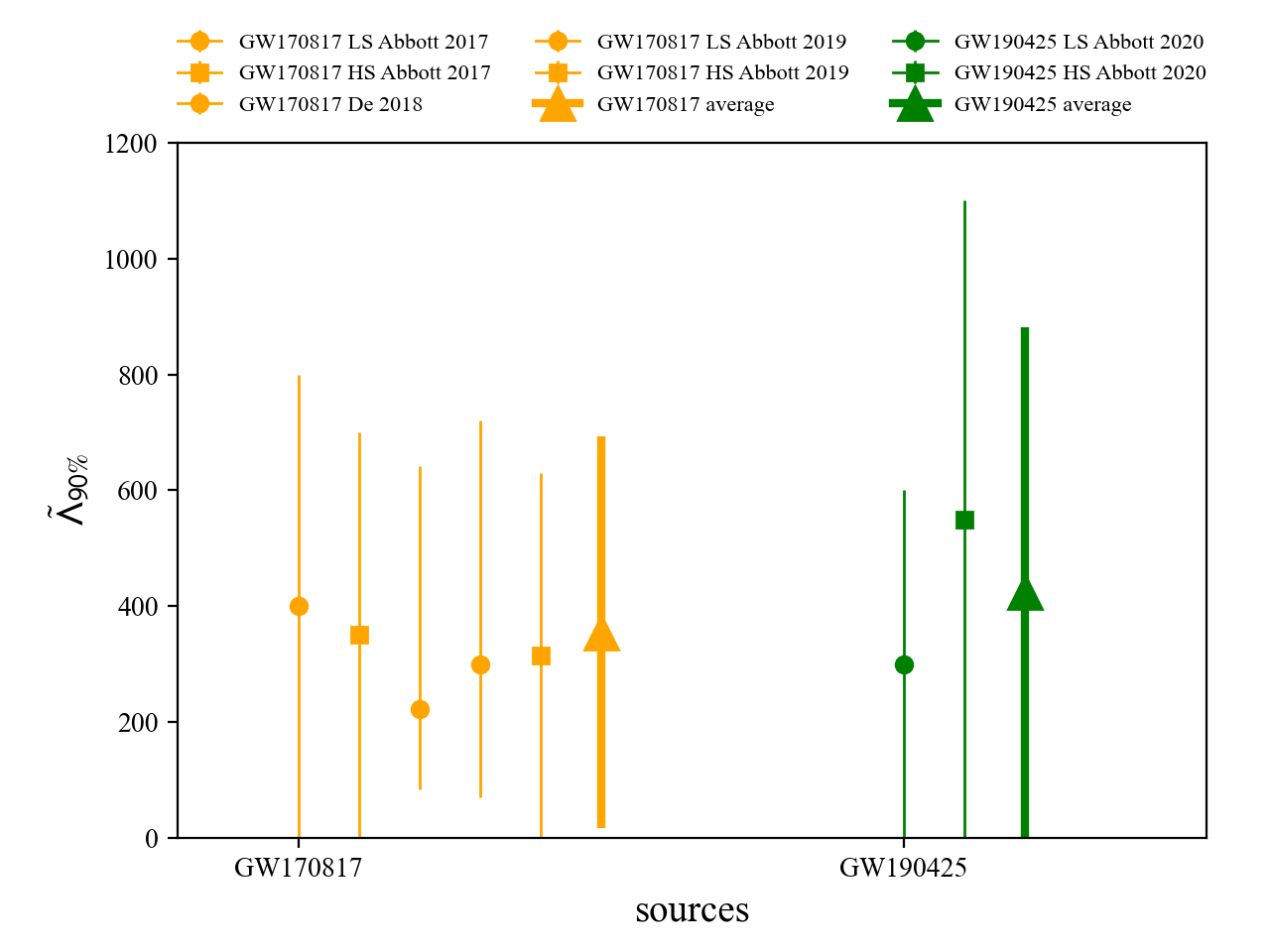}
\caption{Distribution of effective tidal deformabilities at 90\% confidence level for the two gravitational wave source presently known GW170817 and GW190425. Figure generated with \texttt{astro\_setupGW\_plot.py}.} 
\label{fig:astro:gw}
\end{figure}

The GW data in the \texttt{nuda} toolkit are shown in Fig.~\ref{fig:astro:gw} for the two sources GW170817 and GW190425. For a given source, the different measurements can be averaged together in the following way:
\begin{lstlisting}[language=Python]
gwav=nuda.astro.setupGWAverage( source=`GW170817' )
gwav.print_outputs()
\end{lstlisting}
The attributes are the centroid \texttt{gwav.lam\_cen} and the standard deviation \texttt{gwav.lam\_std} for the effective tidal deformability $\tilde{\Lambda}$.

\section{Conclusions}
\label{sec:conclusions}

In this paper, we have described in detail the use of the \texttt{nucleardatapy} toolkit for uniform matter, nuclear experimental data, correlations, and astrophysical data. The goal of this numerical tool is to provide an easy way to access most of the available theoretical,  experimental, and astrophysical data. The amount of data included in the present version of the toolkit represents only a fraction of the data existing in the literature, and we hope that our colleagues will find it an interesting tool to contribute to the database by sharing with us their results and making them easily accessible to the community. Another advantage of the toolkit is that it offers modular functions that average over several predictions, for which the user is free to decide which one to consider. Through the present paper, we have shown at different places in tables and figures how the original data compares to the averaged data, or how to build a reference band from the user's choice.

This paper illustrates the use of the toolkit for research and provides many examples. For more details, please consult the documentation~\cite{NudaDocumentation2025} and the tutorials~\cite{NudaTutorials2025}. The scripts generating the figures shown in this paper are all available in the toolkit, and the GitHub repository~\cite{nuda:rep} is public. All the data provided in this toolkit are accompanied by the proper citations to the original paper, and this toolkit should not supersede them. 

In the future, we plan to update the toolkit with newly available data, as well as extend it to new results.
We are therefore willing to continue to enrich the toolkit in the future. Any suggestions from the community or requests for extensions are welcome. Any of the co-authors can be contacted for this purpose.

\begin{acknowledgements}
We deeply thank our colleagues who agreed to share their results in \texttt{nucleardatapy} toolkit. In particular, Ligang Cao, Gianluca Col\`o, Umberto Lombardo, Armen Sedrakian.
This document is released under LA-UR-25-20848.
JM is supported by CNRS-IN2P3 MAC masterproject, and this work benefited from the support of the project RELANSE ANR-23-CE31-0027-01 of the French National Research Agency (ANR) and the LABEX Lyon Institute of Origins (ANR-10-LABX-0066). MD and OL are members of the project INCT-FNA Proc. No. 464898/2014-5, and acknowledge the support provided by Conselho Nacional de Desenvolvimento Cient\'ifico e Tecnol\'ogico (CNPq) under Grants No. 308528/2021-2, No. 307255/2023-9, and No. 401565/2023-8 (Universal). 
C.~D. acknowledges support from the National Science Foundation under award PHY~2339043.
This material is based upon work supported by the U.S. Department of Energy, Office of Science, Office of Nuclear Physics, under the FRIB Theory Alliance award DE-SC0013617. 
A.G. acknowledges support by the Natural Sciences and Engineering Research Council
(NSERC) of Canada, the Canada Foundation for Innovation (CFI), and Compute Ontario through the Digital Research Alliance of Canada.
S.~L. acknowledges support from the U.S. Department of Energy,
Office of Science, Nuclear Physics, under Award DE-SC0023688. 
S.G, R.S., and I.T. are supported by the Laboratory-Directed Research and Development program of LANL under project number 20230315ER and by the U.S. Department of Energy, Office of Science, Office of Advanced Scientific Computing Research, Scientific Discovery through Advanced Computing (SciDAC) NUCLEI program. 
S.G., and I.T. are also supported by the U.S. Department of Energy, Office of Nuclear Physics, under contract No. DE-AC52-06NA25396.
\end{acknowledgements}

\appendix

\section{List of abbreviations}
\label{app:acronyms}

\begin{tabular}{lp{5.2cm}}
Abbreviation & Meaning \\
\hline
AFDMC & Auxiliary Field Diffusion Monte Carlo \\
AM    & Asymmetric (infinite) Matter \\
BCS  & Bardeen-Cooper-Schrieffer \\
BHF  & Brueckner-Hartree-Fock \\
DDRH  & Density-Dependent Relativistic Hartree \\
DDRHF & Density-Dependent Relativistic Hartree-Fock \\
EDF   & Energy Density Functional \\
EFT   & Effective Field Theory \\ 
EoS   & Equation of State \\
FFG   & Free Fermi Gas \\
GW    & Gravitational wave \\
HFB   & Hartree-Fock-Bogoliubov \\
HIC   & Heavy-Ion Collision  \\
MBPT  & Many-Body Perturbation Theory \\ 
NEP   & Nuclear Empirical Parameters \\
NLEFT & Nuclear Lattice Effective Field
Theory  \\ 
NLRH   & Non-Linear Relativistic Hartree \\
NM     & Neutron (infinite) Matter \\
NR    & Non-Relativistic \\
NRFFG & Non-Relativistic Free Fermi Gas \\
NS    & Neutron Star  \\
N2(3)LO  & next-to-next-to-(next-to)-leading order \\
PVES  & Parity-Violating Electron Scattering  \\
QMC   & Quantum Monte Carlo \\
RFFG & Relativistic Free Fermi Gas \\
RH & Relativistic Hartree or RMF\\
{\it r.m.}  & rest mass \\
RMF   & Relativistic Mean Field \\
SM     & Symmetric (infinite) Matter  \\
\hline
\end{tabular}

\section{Install \texttt{nuda} toolkit on an iPad}
\label{app:ipad}

\subsection{Using \texttt{a-Shell}}

The free app \texttt{a-Shell} emulates a terminal for the iOS of your iPad. and can do locally most of the things that a terminal can do. It can be downloaded directly from the App Store. Once installed, you can launch it and write in the prompt: 
\begin{lstlisting}[language=bash]
$ help
\end{lstlisting}
A list of commands and some explanation are provided on the terminal screen.

To install \texttt{nucleardatapy}, write in \texttt{a-Shell}:
\begin{lstlisting}[language=bash]
$ pip install nucleardatapy
\end{lstlisting}
Maybe you will have to install some additional libraries:
\begin{lstlisting}[language=bash]
python -m pip install numpy
python -m pip install scipy
\end{lstlisting}

You can start by importing the toolkit as:
\begin{lstlisting}[language=Python]
import nucleardatapy as nuda
\end{lstlisting}

You can launch any samples.

Alternatively, you can clone the git repository with the following command:
\begin{lstlisting}[language=bash]
lg2 clone https://github.com/jeromemargueron/nucleardatapy.git
\end{lstlisting}

Note: a-Shell has a Vim editor. Press i to write into the file, and if you do not have an Esc key on your keyboard, then use cmd+., and write \texttt{:wq} to write and quit. 

\subsection{Using \texttt{Carnets-Plus}}

The iPad app \texttt{Carnets-Plus} is a free app providing a Jupyter Notebook running locally on an iPad. 

To start, \texttt{create a new File} will create a Jupyter notebook. Then install the toolkit as before:
\begin{lstlisting}[language=bash]
$ pip install nucleardatapy
\end{lstlisting}
You can start by importing the toolkit as:
\begin{lstlisting}[language=Python]
import nucleardatapy as nuda
\end{lstlisting}

You can now start running the toolkit, and follow, for instance, the instructions given in this paper.

\bibliographystyle{spphys}       
\bibliography{biblio-paper,biblio-nuda-micro,biblio-nuda-pheno,biblio-nuda-exp,biblio-nuda-corr,biblio-nuda-crust,biblio-nuda-nuc,biblio-nuda-astro,biblio-nuda-hypernuclei,biblio-nuda-hic,biblio-nuda-eos}

@article{PDemorest:2010,
author = "Demorest, Paul and Pennucci, Tim and Ransom, Scott and Roberts, Mallory and Hessels, Jason",
title = "{Shapiro Delay Measurement of A Two Solar Mass Neutron Star}",
eprint = "1010.5788",
archivePrefix = "arXiv",
primaryClass = "astro-ph.HE",
doi = "10.1038/nature09466",
journal = "Nature",
volume = "467",
pages = "1081--1083",
year = "2010"
}

@article{JAntoniadis:2013,
Archiveprefix = {arXiv},
Author = {Antoniadis, John and Freire, Paulo C.C. and Wex, Norbert and Tauris, Thomas M. and others},
Doi = {10.1126/science.1233232},
Journal = {Science},
Pages = {6131},
Title = {{A Massive Pulsar in a Compact Relativistic Binary}},
Volume = {340},
Year = {2013}
}

@article{EFonseca:2016,
author = "Fonseca, Emmanuel and others",
title = "{The NANOGrav Nine-year Data Set: Mass and Geometric Measurements of Binary Millisecond Pulsars}",
eprint = "1603.00545",
archivePrefix = "arXiv",
primaryClass = "astro-ph.HE",
doi = "10.3847/0004-637X/832/2/167",
journal = "Astrophys. J.",
volume = "832",
number = "2",
pages = "167",
year = "2016"
}

@article{ZArzoumanian:2018,
author = "Arzoumanian, Zaven and others",
collaboration = "NANOGrav",
title = "{The NANOGrav 11-year Data Set: High-precision timing of 45 Millisecond Pulsars}",
eprint = "1801.01837",
archivePrefix = "arXiv",
primaryClass = "astro-ph.HE",
doi = "10.3847/1538-4365/aab5b0",
journal = "Astrophys. J. Suppl.",
volume = "235",
number = "2",
pages = "37",
year = "2018"
}

@article{MLinares:2018,
author = "Linares, Manuel and Shahbaz, Tariq and Casares, Jorge",
title = "{Peering into the dark side: Magnesium lines establish a massive neutron star in PSR J2215+5135}",
eprint = "1805.08799",
archivePrefix = "arXiv",
primaryClass = "astro-ph.HE",
doi = "10.3847/1538-4357/aabde6",
journal = "Astrophys. J.",
volume = "859",
number = "1",
pages = "54",
year = "2018"
}

@article{HTCromartie:2019,
author = "Cromartie, H. Thankful and Fonseca, Emmanuel and Ransom, Scott M. and Demorest, Paul B. and others",
title = "{Relativistic Shapiro delay measurements of an extremely massive millisecond pulsar}",
eprint = "1904.06759",
archivePrefix = "arXiv",
primaryClass = "astro-ph.HE",
doi = "10.1038/s41550-019-0880-2",
journal = "Nature Astron.",
volume = "4",
number = "1",
pages = "72--76",
year = "2019"
}

@ARTICLE{EFonseca:2021,
author = {{Fonseca}, E. and {Cromartie}, H.~T. and {Pennucci}, T.~T. and {Ray}, P.~S. and others},
title = "{Refined Mass and Geometric Measurements of the High-mass PSR J0740+6620}",
journal = {\apjl},
year = 2021,
month = jul,
volume = {915},
number = {1},
eid = {L12},
pages = {L12},
doi = {10.3847/2041-8213/ac03b8},
archivePrefix = {arXiv},
eprint = {2104.00880},
primaryClass = {astro-ph.HE},
adsurl = {https://ui.adsabs.harvard.edu/abs/2021ApJ...915L..12F}
}

@article{MdFAlam:2021,
title = "The NANOGrav 12.5 yr data set: Observations and narrowband timing of 47 millisecond pulsars",
author = "Alam, {Md F.} and Zaven Arzoumanian and Baker, {Paul T.} and Harsha Blumer and others",
year = "2021",
month = jan,
doi = "10.3847/1538-4365/abc6a0",
language = "English",
volume = "252",
journal = "Astrophysical Journal Supplement Series",
issn = "0067-0049",
publisher = "American Astronomical Society",
number = "1",
}

@article{GAgazie:2023,
doi = {10.3847/2041-8213/acda9a},
url = {https://dx.doi.org/10.3847/2041-8213/acda9a},
year = {2023},
month = {jun},
publisher = {The American Astronomical Society},
volume = {951},
number = {1},
pages = {L9},
author = {Gabriella Agazie and Md Faisal Alam and Akash Anumarlapudi and Anne M. Archibald and others},
title = {The NANOGrav 15 yr Data Set: Observations and Timing of 68 Millisecond Pulsars},
journal = {The Astrophysical Journal Letters}
}

@article{THinderer:2008,
doi = {10.1086/533487},
url = {https://dx.doi.org/10.1086/533487},
year = {2008},
month = {apr},
publisher = {},
volume = {677},
number = {2},
pages = {1216},
author = {Hinderer, Tanja},
title = {Tidal Love Numbers of Neutron Stars},
journal = {The Astrophysical Journal}
}

@article{TBinnington:2009,
title = {Relativistic theory of tidal Love numbers},
author = {Binnington, Taylor and Poisson, Eric},
journal = {Phys. Rev. D},
volume = {80},
issue = {8},
pages = {084018},
numpages = {30},
year = {2009},
month = {Oct},
publisher = {American Physical Society},
doi = {10.1103/PhysRevD.80.084018},
url = {https://link.aps.org/doi/10.1103/PhysRevD.80.084018}
}

@article{TDamour:2009,
title = {Relativistic tidal properties of neutron stars},
author = {Damour, Thibault and Nagar, Alessandro},
journal = {Phys. Rev. D},
volume = {80},
issue = {8},
pages = {084035},
numpages = {21},
year = {2009},
month = {Oct},
publisher = {American Physical Society},
doi = {10.1103/PhysRevD.80.084035},
url = {https://link.aps.org/doi/10.1103/PhysRevD.80.084035}
}

@article{THinderer:2010,
title = {Tidal deformability of neutron stars with realistic equations of state and their gravitational wave signatures in binary inspiral},
author = {Hinderer, Tanja and Lackey, Benjamin D. and Lang, Ryan N. and Read, Jocelyn S.},
journal = {Phys. Rev. D},
volume = {81},
issue = {12},
pages = {123016},
numpages = {12},
year = {2010},
month = {Jun},
publisher = {American Physical Society},
doi = {10.1103/PhysRevD.81.123016},
url = {https://link.aps.org/doi/10.1103/PhysRevD.81.123016}
}

@article{KThorne:1967,
       author = {{Thorne}, Kip S. and {Campolattaro}, Alfonso},
        title = "{Non-Radial Pulsation of General-Relativistic Stellar Models. I. Analytic Analysis for L $>=$ 2}",
         year = 1967,
        month = sep,
        pages = {591},
        journal = {Astrophys. J.},
        vol = {140},
          doi = {10.1086/149288},
    publisher = {IOP},
       url = {https://ui.adsabs.harvard.edu/abs/1967ApJ...149..591T},
}

@article{SPostnikov:2010,
  title = {Tidal Love numbers of neutron and self-bound quark stars},
  author = {Postnikov, Sergey and Prakash, Madappa and Lattimer, James M.},
  journal = {Phys. Rev. D},
  volume = {82},
  issue = {2},
  pages = {024016},
  numpages = {12},
  year = {2010},
  month = {Jul},
  publisher = {American Physical Society},
  doi = {10.1103/PhysRevD.82.024016},
  url = {https://link.aps.org/doi/10.1103/PhysRevD.82.024016}
}

@ARTICLE{BPAbbott:2017,
author = {Abbott, B. P. and Abbott, R. and Abbott, T. D. and others}, 
collaboration = {LIGO Scientific Collaboration and Virgo Collaboration},
title = "{GW170817: Observation of Gravitational Waves from a Binary Neutron Star Inspiral}",
journal = {\prl},
year = 2017,
month = oct,
volume = {119},
number = {16},
eid = {161101},
pages = {161101},
doi = {10.1103/PhysRevLett.119.161101},
archivePrefix = {arXiv},
eprint = {1710.05832},
primaryClass = {gr-qc},
adsurl = {https://ui.adsabs.harvard.edu/abs/2017PhRvL.119p1101A}
}

@article{SDe:2018,
title = {Tidal Deformabilities and Radii of Neutron Stars from the Observation of GW170817},
author = {De, Soumi and Finstad, Daniel and Lattimer, James M. and Brown, Duncan A. and Berger, Edo and Biwer, Christopher M.},
journal = {Phys. Rev. Lett.},
volume = {121},
issue = {9},
pages = {091102},
numpages = {6},
year = {2018},
month = {Aug},
publisher = {American Physical Society},
doi = {10.1103/PhysRevLett.121.091102},
url = {https://link.aps.org/doi/10.1103/PhysRevLett.121.091102}
}

@article{BPAbbott:2018,
title = {GW170817: Measurements of Neutron Star Radii and Equation of State},
author = {Abbott, B. P. and Abbott, R. and Abbott, T. D. and Acernese, F. and Ackley, K. and others},
collaboration = {The LIGO Scientific Collaboration and the Virgo Collaboration},
journal = {Phys. Rev. Lett.},
volume = {121},
issue = {16},
pages = {161101},
numpages = {16},
year = {2018},
month = {Oct},
publisher = {American Physical Society},
doi = {10.1103/PhysRevLett.121.161101},
url = {https://link.aps.org/doi/10.1103/PhysRevLett.121.161101}
}

@article{BPAbbott:2019,
title = {Properties of the Binary Neutron Star Merger GW170817},
author = {Abbott, B. P. and Abbott, R. and Abbott, T. D. and Acernese, F. and others},
collaboration = {LIGO Scientific Collaboration and Virgo Collaboration},
journal = {Phys. Rev. X},
volume = {9},
issue = {1},
pages = {011001},
numpages = {32},
year = {2019},
month = {Jan},
publisher = {American Physical Society},
doi = {10.1103/PhysRevX.9.011001},
url = {https://link.aps.org/doi/10.1103/PhysRevX.9.011001}
}

@article{BPAbbott:2020,
doi = {10.3847/2041-8213/ab75f5},
url = {https://dx.doi.org/10.3847/2041-8213/ab75f5},
year = {2020},
month = {mar},
publisher = {The American Astronomical Society},
volume = {892},
number = {1},
pages = {L3},
author = {B. P. Abbott and R. Abbott and T. D. Abbott and S. Abraham and others},
title = {GW190425: Observation of a Compact Binary Coalescence with Total Mass∼3.4 $M_\odot$},
journal = {The Astrophysical Journal Letters}
}

@INPROCEEDINGS{KCGendreau:2016,
author = {{Gendreau}, Keith C. and {Arzoumanian}, Zaven and {Adkins}, Phillip W. and {Albert}, Cheryl L. and others},
title = "{The Neutron star Interior Composition Explorer (NICER): design and development}",
booktitle = {Space Telescopes and Instrumentation 2016: Ultraviolet to Gamma Ray},
year = 2016,
editor = {{den Herder}, Jan-Willem A. and {Takahashi}, Tadayuki and {Bautz}, Marshall},
series = {Society of Photo-Optical Instrumentation Engineers (SPIE) Conference Series},
volume = {9905},
month = jul,
eid = {99051H},
pages = {99051H},
doi = {10.1117/12.2231304},
adsurl = {https://ui.adsabs.harvard.edu/abs/2016SPIE.9905E..1HG}
}

@article{MCMiller:2019,
doi = {10.3847/2041-8213/ab50c5},
url = {https://dx.doi.org/10.3847/2041-8213/ab50c5},
year = {2019},
month = {dec},
publisher = {The American Astronomical Society},
volume = {887},
number = {1},
pages = {L24},
author = {M. C. Miller and F. K. Lamb and A. J. Dittmann and S. Bogdanov and others},
title = {PSR J0030+0451 Mass and Radius from NICER Data and Implications for the Properties of Neutron Star Matter},
journal = {The Astrophysical Journal Letters}
}

@article{TERiley:2019,
doi = {10.3847/2041-8213/ab481c},
url = {https://dx.doi.org/10.3847/2041-8213/ab481c},
year = {2019},
month = {dec},
publisher = {The American Astronomical Society},
volume = {887},
number = {1},
pages = {L21},
author = {T. E. Riley and A. L. Watts and S. Bogdanov and P. S. Ray and others},
title = {A NICER View of PSR J0030+0451: Millisecond Pulsar Parameter Estimation},
journal = {The Astrophysical Journal Letters}
}

@article{MCMiller:2021,
doi = {10.3847/2041-8213/ac089b},
url = {https://dx.doi.org/10.3847/2041-8213/ac089b},
year = {2021},
month = {sep},
publisher = {The American Astronomical Society},
volume = {918},
number = {2},
pages = {L28},
author = {M. C. Miller and F. K. Lamb and A. J. Dittmann and S. Bogdanov and others},
title = {The Radius of PSR J0740+6620 from NICER and XMM-Newton Data},
journal = {The Astrophysical Journal Letters}
}

@article{TERiley:2021,
doi = {10.3847/2041-8213/ac0a81},
url = {https://dx.doi.org/10.3847/2041-8213/ac0a81},
year = {2021},
month = {sep},
publisher = {The American Astronomical Society},
volume = {918},
number = {2},
pages = {L27},
author = {Thomas E. Riley and Anna L. Watts and Paul S. Ray and Slavko Bogdanov and others},
title = {A NICER View of the Massive Pulsar PSR J0740+6620 Informed by Radio Timing and XMM-Newton Spectroscopy},
journal = {The Astrophysical Journal Letters}
}

@ARTICLE{SVinciguerra:2024,
author = {{Vinciguerra}, Serena and {Salmi}, Tuomo and {Watts}, Anna L. and {Choudhury}, Devarshi and others},
title = "{An Updated Mass-Radius Analysis of the 2017-2018 NICER Data Set of PSR J0030+0451}",
journal = {The Astrophysical Journal},
year = 2024,
month = jan,
volume = {961},
number = {1},
eid = {62},
pages = {62},
doi = {10.3847/1538-4357/acfb83},
archivePrefix = {arXiv},
eprint = {2308.09469},
primaryClass = {astro-ph.HE},
adsurl = {https://ui.adsabs.harvard.edu/abs/2024ApJ...961...62V}
}

@ARTICLE{TSalmi:2024,
author = {{Salmi}, Tuomo and {Choudhury}, Devarshi and {Kini}, Yves and {Riley}, Thomas E. and others},
title = "{The Radius of the High-mass Pulsar PSR J0740+6620 with 3.6 yr of NICER Data}",
journal = {\apj},
year = 2024,
month = oct,
volume = {974},
number = {2},
eid = {294},
pages = {294},
doi = {10.3847/1538-4357/ad5f1f},
archivePrefix = {arXiv},
eprint = {2406.14466},
primaryClass = {astro-ph.HE},
adsurl = {https://ui.adsabs.harvard.edu/abs/2024ApJ...974..294S}
}

@article{DChoudhury:2024,
doi = {10.3847/2041-8213/ad5a6f},
url = {https://dx.doi.org/10.3847/2041-8213/ad5a6f},
year = {2024},
month = {aug},
publisher = {The American Astronomical Society},
volume = {971},
number = {1},
pages = {L20},
author = {Devarshi Choudhury and Tuomo Salmi and Serena Vinciguerra and Thomas E. Riley and others},
title = {A NICER View of the Nearest and Brightest Millisecond Pulsar: PSR J0437–4715},
journal = {The Astrophysical Journal Letters}
}

@article{EKhan:2012,
title = {Constraining the Nuclear Equation of State at Subsaturation Densities},
author = {Khan, E. and Margueron, J. and Vida\~na, I.},
journal = {Phys. Rev. Lett.},
volume = {109},
issue = {9},
pages = {092501},
numpages = {4},
year = {2012},
month = {Aug},
publisher = {American Physical Society},
doi = {10.1103/PhysRevLett.109.092501},
url = {https://link.aps.org/doi/10.1103/PhysRevLett.109.092501}
}

@article{EKhan:2013,
title = {Determination of the density dependence of the nuclear incompressibility},
author = {Khan, E. and Margueron, J.},
journal = {Phys. Rev. C},
volume = {88},
issue = {3},
pages = {034319},
numpages = {9},
year = {2013},
month = {Sep},
publisher = {American Physical Society},
doi = {10.1103/PhysRevC.88.034319},
url = {https://link.aps.org/doi/10.1103/PhysRevC.88.034319}
}

@article{JMargueron:2019,
title = {Effect of high-order empirical parameters on the nuclear equation of state},
author = {Margueron, J\'er\^ome and Gulminelli, Francesca},
journal = {Phys. Rev. C},
volume = {99},
issue = {2},
pages = {025806},
numpages = {10},
year = {2019},
month = {Feb},
publisher = {American Physical Society},
doi = {10.1103/PhysRevC.99.025806},
url = {https://link.aps.org/doi/10.1103/PhysRevC.99.025806}
}

@article{JMargueron:2025,
author = {Margueron, J\'er\^ome and Khan, Elias},
journal = {In preparation},
year = {2025},
}

@article{JMPearson:1991,
title = {The incompressibility of nuclear matter and the breathing mode},
journal = {Physics Letters B},
volume = {271},
number = {1},
pages = {12-16},
year = {1991},
issn = {0370-2693},
doi = {https://doi.org/10.1016/0370-2693(91)91269-2},
url = {https://www.sciencedirect.com/science/article/pii/0370269391912692},
author = {J.M. Pearson}
}

@article{SRudaz:1992,
title = {Anharmonicity of the nuclear matter ground state},
journal = {Physics Letters B},
volume = {285},
number = {3},
pages = {183-186},
year = {1992},
issn = {0370-2693},
doi = {https://doi.org/10.1016/0370-2693(92)91449-J},
url = {https://www.sciencedirect.com/science/article/pii/037026939291449J},
author = {Serge Rudaz and Paul J. Ellis and Erik K. Heide and M. Prakash},
}

@article{MFarine:1997,
title = {Nuclear-matter incompressibility from fits of generalized Skyrme force to breathing-mode energies},
journal = {Nuclear Physics A},
volume = {615},
number = {2},
pages = {135-161},
year = {1997},
issn = {0375-9474},
doi = {https://doi.org/10.1016/S0375-9474(96)00453-8},
url = {https://www.sciencedirect.com/science/article/pii/S0375947496004538},
author = {Michel Farine and J.M. Pearson and F. Tondeur}
}

@article{MBTsang:2009,
title = {Constraints on the Density Dependence of the Symmetry Energy},
author = {Tsang, M. B. and Zhang, Yingxun and Danielewicz, P. and Famiano, M. and Li, Zhuxia and Lynch, W. G. and Steiner, A. W.},
journal = {Phys. Rev. Lett.},
volume = {102},
issue = {12},
pages = {122701},
numpages = {4},
year = {2009},
month = {Mar},
publisher = {American Physical Society},
doi = {10.1103/PhysRevLett.102.122701},
url = {https://link.aps.org/doi/10.1103/PhysRevLett.102.122701}
}

@article{LWChen:2010,
title = {Density slope of the nuclear symmetry energy from the neutron skin thickness of heavy nuclei},
author = {Chen, Lie-Wen and Ko, Che Ming and Li, Bao-An and Xu, Jun},
journal = {Phys. Rev. C},
volume = {82},
issue = {2},
pages = {024321},
numpages = {7},
year = {2010},
month = {Aug},
publisher = {American Physical Society},
doi = {10.1103/PhysRevC.82.024321},
url = {https://link.aps.org/doi/10.1103/PhysRevC.82.024321}
}

@article{PMoller:2012,
title = {New Finite-Range Droplet Mass Model and Equation-of-State Parameters},
author = {M\"oller, Peter and Myers, William D. and Sagawa, Hiroyuki and Yoshida, Satoshi},
journal = {Phys. Rev. Lett.},
volume = {108},
issue = {5},
pages = {052501},
numpages = {4},
year = {2012},
month = {Jan},
publisher = {American Physical Society},
doi = {10.1103/PhysRevLett.108.052501},
url = {https://link.aps.org/doi/10.1103/PhysRevLett.108.052501}
}

@article{AWSteiner:2013,
doi = {10.1088/2041-8205/765/1/L5},
url = {https://dx.doi.org/10.1088/2041-8205/765/1/L5},
year = {2013},
month = {feb},
publisher = {The American Astronomical Society},
volume = {765},
number = {1},
pages = {L5},
author = {Andrew W. Steiner and James M. Lattimer and Edward F. Brown},
title = {THE NEUTRON STAR MASS–RADIUS RELATION AND THE EQUATION OF STATE OF DENSE MATTER},
journal = {The Astrophysical Journal Letters}
}

@article{JMLattimer:2013,
doi = {10.1088/0004-637X/771/1/51},
url = {https://dx.doi.org/10.1088/0004-637X/771/1/51},
year = {2013},
month = {jun},
publisher = {The American Astronomical Society},
volume = {771},
number = {1},
pages = {51},
author = {James M. Lattimer and Yeunhwan Lim},
title = {CONSTRAINING THE SYMMETRY PARAMETERS OF THE NUCLEAR INTERACTION},
journal = {The Astrophysical Journal}
}

@article{PDanielewicz:2014,
title = {Symmetry energy II: Isobaric analog states},
journal = {Nuclear Physics A},
volume = {922},
pages = {1-70},
year = {2014},
issn = {0375-9474},
doi = {https://doi.org/10.1016/j.nuclphysa.2013.11.005},
url = {https://www.sciencedirect.com/science/article/pii/S0375947413007872},
author = {Pawel Danielewicz and Jenny Lee}
}

@article{XRocaMaza:2015,
title = {Neutron skin thickness from the measured electric dipole polarizability in $^{68}\text{Ni}$, $^{120}\text{Sn}$, and $^{208}\text{Pb}$},
author = {Roca-Maza, X. and Vi\~nas, X. and Centelles, M. and Agrawal, B. K. and Col\`o, G. and Paar, N. and Piekarewicz, J. and Vretenar, D.},
journal = {Phys. Rev. C},
volume = {92},
issue = {6},
pages = {064304},
numpages = {11},
year = {2015},
month = {Dec},
publisher = {American Physical Society},
doi = {10.1103/PhysRevC.92.064304},
url = {https://link.aps.org/doi/10.1103/PhysRevC.92.064304}
}

@article{ITews:2017,
doi = {10.3847/1538-4357/aa8db9},
url = {https://dx.doi.org/10.3847/1538-4357/aa8db9},
year = {2017},
month = {oct},
publisher = {The American Astronomical Society},
volume = {848},
number = {2},
pages = {105},
author = {Ingo Tews and James M. Lattimer and Akira Ohnishi and Evgeni E. Kolomeitsev},
title = {Symmetry Parameter Constraints from a Lower Bound on Neutron-matter Energy},
journal = {The Astrophysical Journal}
}

@article{JBWei:2020,
author = {{Wei, Jin-Biao} and {Lu, Jia-Jing} and {Burgio, G. F.} and {Li, Zeng-Hua} and {Schulze, H.-J.}},
title = {Are nuclear matter properties correlated to neutron star observables?},
DOI= "10.1140/epja/s10050-020-00058-3",
url= "https://doi.org/10.1140/epja/s10050-020-00058-3",
journal = {Eur. Phys. J. A},
year = 2020,
volume = 56,
number = 2,
pages = "63",
}

@article{BTReed:2021,
title = {Implications of PREX-2 on the Equation of State of Neutron-Rich Matter},
author = {Reed, Brendan T. and Fattoyev, F. J. and Horowitz, C. J. and Piekarewicz, J.},
journal = {Phys. Rev. Lett.},
volume = {126},
issue = {17},
pages = {172503},
numpages = {5},
year = {2021},
month = {Apr},
publisher = {American Physical Society},
doi = {10.1103/PhysRevLett.126.172503},
url = {https://link.aps.org/doi/10.1103/PhysRevLett.126.172503}
}

@article{PGReinhard:2021,
title = {Information Content of the Parity-Violating Asymmetry in $^{208}\mathrm{Pb}$},
author = {Reinhard, Paul-Gerhard and Roca-Maza, Xavier and Nazarewicz, Witold},
journal = {Phys. Rev. Lett.},
volume = {127},
issue = {23},
pages = {232501},
numpages = {7},
year = {2021},
month = {Nov},
publisher = {American Physical Society},
doi = {10.1103/PhysRevLett.127.232501},
url = {https://link.aps.org/doi/10.1103/PhysRevLett.127.232501}
}

@article{DAdhikari:2021,
title = {Accurate Determination of the Neutron Skin Thickness of $^{208}\mathrm{Pb}$ through Parity-Violation in Electron Scattering},
author = {Adhikari, D. and Albataineh, H. and Androic, D. and Aniol, K. and Armstrong and others},
collaboration = {PREX Collaboration},
journal = {Phys. Rev. Lett.},
volume = {126},
issue = {17},
pages = {172502},
numpages = {7},
year = {2021},
month = {Apr},
publisher = {American Physical Society},
doi = {10.1103/PhysRevLett.126.172502},
url = {https://link.aps.org/doi/10.1103/PhysRevLett.126.172502}
}

@article{PGReinhard:2022,
title = {Combined Theoretical Analysis of the Parity-Violating Asymmetry for $^{48}\mathrm{Ca}$ and $^{208}\mathrm{Pb}$},
author = {Reinhard, Paul-Gerhard and Roca-Maza, Xavier and Nazarewicz, Witold},
journal = {Phys. Rev. Lett.},
volume = {129},
issue = {23},
pages = {232501},
numpages = {7},
year = {2022},
month = {Dec},
publisher = {American Physical Society},
doi = {10.1103/PhysRevLett.129.232501},
url = {https://link.aps.org/doi/10.1103/PhysRevLett.129.232501}
}

@article{EYuksel:2022,
title = {Implications of parity-violating electron scattering experiments on 48Ca (CREX) and 208Pb (PREX-II) for nuclear energy density functionals},
journal = {Physics Letters B},
volume = {836},
pages = {137622},
year = {2023},
issn = {0370-2693},
doi = {https://doi.org/10.1016/j.physletb.2022.137622},
url = {https://www.sciencedirect.com/science/article/pii/S0370269322007560},
author = {Esra Yüksel and Nils Paar}
}

@article{DAdhikari:2022,
title = {Precision Determination of the Neutral Weak Form Factor of $^{48}\mathrm{Ca}$},
author = {Adhikari, D. and Albataineh, H. and Androic, D. and Aniol, K. A. and Armstrong and others},
collaboration = {CREX Collaboration},
journal = {Phys. Rev. Lett.},
volume = {129},
issue = {4},
pages = {042501},
numpages = {8},
year = {2022},
month = {Jul},
publisher = {American Physical Society},
doi = {10.1103/PhysRevLett.129.042501},
url = {https://link.aps.org/doi/10.1103/PhysRevLett.129.042501}
}

@article{ZZhang:2023,
  title = {Bayesian inference of the symmetry energy and the neutron skin in $^{48}\mathrm{Ca}$ and $^{208}\mathrm{Pb}$ from CREX and PREX-2},
  author = {Zhang, Zhen and Chen, Lie-Wen},
  journal = {Phys. Rev. C},
  volume = {108},
  issue = {2},
  pages = {024317},
  numpages = {8},
  year = {2023},
  month = {Aug},
  publisher = {American Physical Society},
  doi = {10.1103/PhysRevC.108.024317},
  url = {https://link.aps.org/doi/10.1103/PhysRevC.108.024317}
}

@article{BVCarlson:2023,
title = {Low-energy nuclear physics and global neutron star properties},
author = {Carlson, Brett V. and Dutra, Mariana and Louren\ifmmode \mbox{\c{c}}\else \c{c}\fi{}o, Odilon and Margueron, J\'er\^ome},
journal = {Phys. Rev. C},
volume = {107},
issue = {3},
pages = {035805},
numpages = {24},
year = {2023},
month = {Mar},
publisher = {American Physical Society},
doi = {10.1103/PhysRevC.107.035805},
url = {https://link.aps.org/doi/10.1103/PhysRevC.107.035805}
}

@article{JWNegele:1973,
title = {Neutron star matter at sub-nuclear densities},
journal = {Nuclear Physics A},
volume = {207},
number = {2},
pages = {298-320},
year = {1973},
issn = {0375-9474},
doi = {https://doi.org/10.1016/0375-9474(73)90349-7},
url = {https://www.sciencedirect.com/science/article/pii/0375947473903497},
author = {J.W. Negele and D. Vautherin}
}

@article{CMondal:2020,
title = {Structure and composition of the inner crust of neutron stars from Gogny interactions},
author = {Mondal, C. and Vi\~nas, X. and Centelles, M. and De, J. N.},
journal = {Phys. Rev. C},
volume = {102},
issue = {1},
pages = {015802},
numpages = {18},
year = {2020},
month = {Jul},
publisher = {American Physical Society},
doi = {10.1103/PhysRevC.102.015802},
url = {https://link.aps.org/doi/10.1103/PhysRevC.102.015802}
}

@article{MPearson:2018,
    author = {Pearson, J M and Chamel, N and Potekhin, A Y and Fantina, A F and Ducoin, C and Dutta, A K and Goriely, S},
    title = {Unified equations of state for cold non-accreting neutron stars with Brussels–Montreal functionals – I. Role of symmetry energy},
    journal = {Monthly Notices of the Royal Astronomical Society},
    volume = {481},
    number = {3},
    pages = {2994-3026},
    year = {2018},
    month = {09},
    issn = {0035-8711},
    doi = {10.1093/mnras/sty2413},
    url = {https://doi.org/10.1093/mnras/sty2413},
    eprint = {https://academic.oup.com/mnras/article-pdf/481/3/2994/25817956/sty2413.pdf},
}

@article{GGrams:2022a,
author = {Grams, G. and Somasundaram, R. and Margueron, J. and Reddy, S.},
journal = {Phys. Rev. C},
volume = {105},
issue = {3},
pages = {035806},
numpages = {21},
year = {2022},
month = {Mar},
publisher = {American Physical Society},
doi = {10.1103/PhysRevC.105.035806},
url = {https://link.aps.org/doi/10.1103/PhysRevC.105.035806}
}

@article{GGrams:2022b,
author = {Grams, G. and Margueron, J. and Somasundaram, R. and Reddy, S.},
journal = {Eur. Phys. J. A},
volume = {58},
pages = {56},
year = {2022},
doi = {10.1140/epja/s10050-022-00706-w},
url = { https://doi.org/10.1140/epja/s10050-022-00706-w}
}

@ARTICLE{CDucoin:2011,
       author = {{Ducoin}, Camille and {Margueron}, J{\'e}r{\^o}me and {Provid{\^e}ncia}, Constan{\c{c}}a and {Vida{\~n}a}, Isaac},
        title = "{Core-crust transition in neutron stars: Predictivity of density developments}",
      journal = {\prc},
         year = 2011,
        month = apr,
       volume = {83},
       number = {4},
          eid = {045810},
        pages = {045810},
          doi = {10.1103/PhysRevC.83.045810},
archivePrefix = {arXiv},
       eprint = {1102.1283},
 primaryClass = {nucl-th},
       adsurl = {https://ui.adsabs.harvard.edu/abs/2011PhRvC..83d5810D}
}

@article{WGNewton:2013,
Author = {W. G. Newton and M. Gearheart and Bao-An Li},
Journal = {\apjs},
Number = {1},
Pages = {9},
Title = {A Survey of the Parameter Space of the Compressible Liquid Drop Model as Applied to the Neutron Star Inner Crust},
Volume = {204},
Year = {2013}
}

@article{AWSteiner:2015,
Author = {Steiner, A. W. and Gandolfi, S. and Fattoyev, F. J. and Newton, W. G.},
Doi = {10.1103/PhysRevC.91.015804},
Issue = {1},
Journal = {\prc},
Month = {Jan},
Numpages = {7},
Pages = {015804},
Publisher = {American Physical Society},
Title = {Using neutron star observations to determine crust thicknesses, moments of inertia, and tidal deformabilities},
Volume = {91},
Year = {2015}
}

@article{MFortin:2016,
title = {Neutron star radii and crusts: Uncertainties and unified equations of state},
author = {Fortin, M. and Provid\^encia, C. and Raduta, Ad. R. and Gulminelli, F. and Zdunik, J. L. and Haensel, P. and Bejger, M.},
journal = {Phys. Rev. C},
volume = {94},
issue = {3},
pages = {035804},
numpages = {21},
year = {2016},
month = {Sep},
publisher = {American Physical Society},
doi = {10.1103/PhysRevC.94.035804},
url = {https://link.aps.org/doi/10.1103/PhysRevC.94.035804}
}

@article{SLalit:2019,
doi = {10.3847/1538-4357/ab338c},
url = {https://dx.doi.org/10.3847/1538-4357/ab338c},
year = {2019},
month = {sep},
publisher = {The American Astronomical Society},
volume = {882},
number = {2},
pages = {91},
author = {Lalit, Sudhanva and Meisel, Zach and Brown, Edward F.},
title = {Crust-cooling Models Are Insensitive to the Crust–Core Transition Pressure for Realistic Equations of State},
journal = {The Astrophysical Journal}
}

@article{GAudi:2012,
doi = {10.1088/1674-1137/36/12/002},
url = {https://dx.doi.org/10.1088/1674-1137/36/12/002},
year = {2012},
month = {dec},
publisher = {},
volume = {36},
number = {12},
pages = {1287},
author = {and  and  and  and  and  and},
title = {The Ame2012 atomic mass evaluation},
journal = {Chinese Physics C}
}

@article{GAudi:2017,
doi = {10.1088/1674-1137/41/3/030001},
url = {https://dx.doi.org/10.1088/1674-1137/41/3/030001},
year = {2017},
month = {mar},
publisher = {},
volume = {41},
number = {3},
pages = {030001},
author = {and F. G. Kondev and  and  and S. Naimi},
title = {The NUBASE2016 evaluation of nuclear properties},
journal = {Chinese Physics C}
}

@article{WJHuang:2021,
doi = {10.1088/1674-1137/abddb0},
url = {https://dx.doi.org/10.1088/1674-1137/abddb0},
year = {2021},
month = {mar},
publisher = {Chinese Physical Society and the Institute of High Energy Physics of the Chinese Academy of Sciences and the Institute of Modern Physics of the Chinese Academy of Sciences and IOP Publishing Ltd},
volume = {45},
number = {3},
pages = {030002},
author = {W.J. Huang and Meng Wang and F.G. Kondev and G. Audi and S. Naimi},
title = {The AME 2020 atomic mass evaluation (I). Evaluation of input data, and adjustment procedures*},
journal = {Chinese Physics C}
}

@book{BohrMottelson:1998,
title={Nuclear Structure},
author={Bohr, A. and Mottelson, B.R.},
number={vol.~1},
isbn={9789810239794},
lccn={97051929},
series={Nuclear Structure},
url={https://books.google.com/books?id=bDXgCO3Z4bIC},
year={1998},
publisher={World Scientific}
}

@article{PVogel:1984,
title = {Is there neutron-proton pairing in medium heavy nuclei?},
journal = {Physics Letters B},
volume = {139},
number = {4},
pages = {227-230},
year = {1984},
issn = {0370-2693},
doi = {https://doi.org/10.1016/0370-2693(84)91068-2},
url = {https://www.sciencedirect.com/science/article/pii/0370269384910682},
author = {P. Vogel and B. Jonson and P.G. Hansen}
}

@article{IAngeli:2013,
title = {Table of experimental nuclear ground state charge radii: An update},
journal = {Atomic Data and Nuclear Data Tables},
volume = {99},
number = {1},
pages = {69-95},
year = {2013},
issn = {0092-640X},
doi = {https://doi.org/10.1016/j.adt.2011.12.006},
url = {https://www.sciencedirect.com/science/article/pii/S0092640X12000265},
author = {I. Angeli and K.P. Marinova}
}

@article{TLi:2010,
title = {Isoscalar giant resonances in the Sn nuclei and implications for the asymmetry term in the nuclear-matter incompressibility},
author = {Li, T. and Garg, U. and Liu, Y. and Marks, R. and others},
journal = {Phys. Rev. C},
volume = {81},
issue = {3},
pages = {034309},
numpages = {11},
year = {2010},
month = {Mar},
publisher = {American Physical Society},
doi = {10.1103/PhysRevC.81.034309},
url = {https://link.aps.org/doi/10.1103/PhysRevC.81.034309}
}

@article{UGarg:2018,
title = {The compression-mode giant resonances and nuclear incompressibility},
journal = {Progress in Particle and Nuclear Physics},
volume = {101},
pages = {55-95},
year = {2018},
issn = {0146-6410},
doi = {https://doi.org/10.1016/j.ppnp.2018.03.001},
url = {https://www.sciencedirect.com/science/article/pii/S0146641018300322},
author = {Umesh Garg and Gianluca Colò}
}

@article{GGrams:2022,
title = {Nuclear incompressibility and speed of sound in uniform matter and finite nuclei},
author = {Grams, G. and Somasundaram, R. and Margueron, J. and Khan, E.},
journal = {Phys. Rev. C},
volume = {106},
issue = {4},
pages = {044305},
numpages = {20},
year = {2022},
month = {Oct},
publisher = {American Physical Society},
doi = {10.1103/PhysRevC.106.044305},
url = {https://link.aps.org/doi/10.1103/PhysRevC.106.044305}
}

@article{PDanielewicz:2002,
author = {Danielewicz, Pawe{\l} and Lacey, Roy and Lynch, William G},
journal = {Science},
volume = "298",
pages = {1592--1596},
publisher = {American Association for the Advancement of Science},
title = {Determination of the equation of state of dense matter},
year = {2002},
doi ={10.1126/science.1078070}
}

@article{ALefevre:2016,
 author = {Le Fevre, Arnaud and Leifels, Yvonne and Reisdorf, Willibrord and Aichelin, Joerg and Hartnack, Ch},
 journal = {Nuclear Physics A},
 pages = {112--133},
 publisher = {Elsevier},
 title = {Constraining the nuclear matter equation of state around twice saturation density},
 volume = {945},
 year = {2016},
doi ={https://doi.org/10.1016/j.nuclphysa.2015.09.015}
}

@article{CFuchs:2006,
author = {Christian Fuchs},
title = {Kaon production in heavy ion reactions at intermediate energies},
journal = {Progress in Particle and Nuclear Physics},
volume = {56},
pages = {1-103},
year = {2006},
doi = {https://doi.org/10.1016/j.ppnp.2005.07.004}
}

@article{WGLynch:2009,
author = {W.G. Lynch and M.B. Tsang and Y. Zhang and P. Danielewicz and M. Famiano and Z. Li and A.W. Steiner},
title = {Probing the symmetry energy with heavy ions},
journal = {Progress in Particle and Nuclear Physics},
volume = {62},
pages = {427-432},
year = {2009},
doi = {https://doi.org/10.1016/j.ppnp.2009.01.001}
}

@article{PRussotto:2011,
author = {P. Russotto and P.Z. Wu and M. Zoric and M. Chartier and Y. Leifels and R.C. Lemmon and Q. Li and J. Łukasik and A. Pagano and P. Pawłowski and W. Trautmann},
title = {Symmetry energy from elliptic flow in $^{197}$Au+$^{197}$Au},
journal = {Physics Letters B},
volume = {697},
pages = {471-476},
year = {2011},
issn = {0370-2693},
doi = {https://doi.org/10.1016/j.physletb.2011.02.033}
}

@article{PRussotto:2016,
title = {Results of the ASY-EOS experiment at GSI: The symmetry energy at suprasaturation density},
author = {Russotto, P. and Gannon, S. and Kupny, S. and Lasko, P. and Acosta, L. and Adamczyk, M. and Al-Ajlan, A. and Al-Garawi, M. and others},
journal = {Phys. Rev. C},
volume = {94},
pages = {034608},
year = {2016},
doi = {10.1103/PhysRevC.94.034608}
}

@article{PMorfouace:2019,
title = {Constraining the symmetry energy with heavy-ion collisions and Bayesian analyses},
journal = {Physics Letters B},
volume = {799},
pages = {135045},
year = {2019},
doi = {https://doi.org/10.1016/j.physletb.2019.135045},
author = {P. Morfouace and C.Y. Tsang and Y. Zhang and W.G. Lynch and M.B. Tsang and D.D.S. Coupland and M. Youngs and Z. Chajecki and M.A. Famiano and T.K. Ghosh and G. Jhang and Jenny Lee and H. Liu and A. Sanetullaev and R. Showalter and J. Winkelbauer}
}

@article{JEstee:2021,
title = {Probing the Symmetry Energy with the Spectral Pion Ratio},
author = {Estee, J. and Lynch, W. G. and Tsang, C. Y. and Barney, J. and Jhang, G. and Tsang, M. B. and Wang, R. and Kaneko, M. and Lee, J. W. and Isobe, T. and Kurata-Nishimura, M. and Murakami, T. and Ahn, D. S. and Atar, L. and Aumann, T. and others},
journal = {Phys. Rev. Lett.},
volume = {126},
issue = {16},
pages = {162701},
numpages = {8},
year = {2021},
doi = {10.1103/PhysRevLett.126.162701}
}

@article{ASorensen:2024,
title = {Dense nuclear matter equation of state from heavy-ion collisions},
journal = {Progress in Particle and Nuclear Physics},
volume = {134},
pages = {104080},
year = {2024},
issn = {0146-6410},
doi = {https://doi.org/10.1016/j.ppnp.2023.104080},
url = {https://www.sciencedirect.com/science/article/pii/S0146641023000613},
author = {Agnieszka Sorensen and Kshitij Agarwal and Kyle W. Brown and Zbigniew Chajęcki and Paweł Danielewicz and Christian Drischler and Stefano Gandolfi and Jeremy W. Holt and Matthias Kaminski and Che-Ming Ko and Rohit Kumar and Bao-An Li and William G. Lynch and Alan B. McIntosh and William G. Newton and Scott Pratt and Oleh Savchuk and Maria Stefaniak and Ingo Tews and ManYee Betty Tsang and Ramona Vogt and Hermann Wolter and Hanna Zbroszczyk and Navid Abbasi and Jörg Aichelin and Anton Andronic and Steffen A. Bass and Francesco Becattini and David Blaschke and Marcus Bleicher and Christoph Blume and Elena Bratkovskaya and B. Alex Brown and David A. Brown and Alberto Camaiani and Giovanni Casini and Katerina Chatziioannou and Abdelouahad Chbihi and Maria Colonna and Mircea Dan Cozma and Veronica Dexheimer and Xin Dong and Travis Dore and Lipei Du and José A. Dueñas and Hannah Elfner and Wojciech Florkowski and Yuki Fujimoto and Richard J. Furnstahl and Alexandra Gade and Tetyana Galatyuk and Charles Gale and Frank Geurts and Fabiana Gramegna and Sašo Grozdanov and Kris Hagel and Steven P. Harris and Wick Haxton and Ulrich Heinz and Michal P. Heller and Or Hen and Heiko Hergert and Norbert Herrmann and Huan Zhong Huang and Xu-Guang Huang and Natsumi Ikeno and Gabriele Inghirami and Jakub Jankowski and Jiangyong Jia and José C. Jiménez and Joseph Kapusta and Behruz Kardan and Iurii Karpenko and Declan Keane and Dmitri Kharzeev and Andrej Kugler and Arnaud {Le Fèvre} and Dean Lee and Hong Liu and Michael A. Lisa and William J. Llope and Ivano Lombardo and Manuel Lorenz and Tommaso Marchi and Larry McLerran and Ulrich Mosel and Anton Motornenko and Berndt Müller and Paolo Napolitani and Joseph B. Natowitz and Witold Nazarewicz and Jorge Noronha and Jacquelyn Noronha-Hostler and Grażyna Odyniec and Panagiota Papakonstantinou and Zuzana Paulínyová and Jorge Piekarewicz and Robert D. Pisarski and Christopher Plumberg and Madappa Prakash and Jørgen Randrup and Claudia Ratti and Peter Rau and Sanjay Reddy and Hans-Rudolf Schmidt and Paolo Russotto and Radoslaw Ryblewski and Andreas Schäfer and Björn Schenke and Srimoyee Sen and Peter Senger and Richard Seto and Chun Shen and Bradley Sherrill and Mayank Singh and Vladimir Skokov and Michał Spaliński and Jan Steinheimer and Mikhail Stephanov and Joachim Stroth and Christian Sturm and Kai-Jia Sun and Aihong Tang and Giorgio Torrieri and Wolfgang Trautmann and Giuseppe Verde and Volodymyr Vovchenko and Ryoichi Wada and Fuqiang Wang and Gang Wang and Klaus Werner and Nu Xu and Zhangbu Xu and Ho-Ung Yee and Sherry Yennello and Yi Yin}
}

@article{YYamamoto:1991,
author = {Yamamoto, Yasuo and Takaki, Hideo and Ikeda, Kiyomi},
title = {Newly Observed Double-Λ Hypernucleus and ΛΛ Interaction},
journal = {Progress of Theoretical Physics},
volume = {86},
number = {4},
pages = {867-875},
year = {1991},
month = {10},
issn = {0033-068X},
doi = {10.1143/ptp/86.4.867},
url = {https://doi.org/10.1143/ptp/86.4.867},
eprint = {https://academic.oup.com/ptp/article-pdf/86/4/867/5329253/86-4-867.pdf},
}

@article{HTakahashi:2001,
title = {Observation of a $_{\ensuremath{\Lambda}\ensuremath{\Lambda}}^{6}He$ Double Hypernucleus},
author = {Takahashi, H. and Ahn, J. K. and Akikawa, H. and Aoki, S. and others},
journal = {Phys. Rev. Lett.},
volume = {87},
issue = {21},
pages = {212502},
numpages = {5},
year = {2001},
month = {Nov},
publisher = {American Physical Society},
doi = {10.1103/PhysRevLett.87.212502},
url = {https://link.aps.org/doi/10.1103/PhysRevLett.87.212502}
}

@article{JKAhn:2013,
title = {Double-$\ensuremath{\Lambda}$ hypernuclei observed in a hybrid emulsion experiment},
author = {Ahn, J. K. and Akikawa, H. and Aoki, S. and Arai, K. and others},
collaboration = {E373 (KEK-PS) Collaboration},
journal = {Phys. Rev. C},
volume = {88},
issue = {1},
pages = {014003},
numpages = {10},
year = {2013},
month = {Jul},
publisher = {American Physical Society},
doi = {10.1103/PhysRevC.88.014003},
url = {https://link.aps.org/doi/10.1103/PhysRevC.88.014003}
}

@article{KNakazawa:2015,
author = {Nakazawa, K. and Endo, Y. and Fukunaga, S. and Hoshino, K. and others},
title = {The first evidence of a deeply bound state of Xi−–14N system},
journal = {Progress of Theoretical and Experimental Physics},
volume = {2015},
number = {3},
pages = {033D02},
year = {2015},
month = {03},
issn = {2050-3911},
doi = {10.1093/ptep/ptv008},
url = {https://doi.org/10.1093/ptep/ptv008},
eprint = {https://academic.oup.com/ptep/article-pdf/2015/3/033D02/9720010/ptv008.pdf},
}

@article{AGal:2016,
  title = {Strangeness in nuclear physics},
  author = {Gal, A. and Hungerford, E. V. and Millener, D. J.},
  journal = {Rev. Mod. Phys.},
  volume = {88},
  issue = {3},
  pages = {035004},
  numpages = {58},
  year = {2016},
  month = {Aug},
  publisher = {American Physical Society},
  doi = {10.1103/RevModPhys.88.035004},
  url = {https://link.aps.org/doi/10.1103/RevModPhys.88.035004}
}

@article{HEkawa:2019,
author = {Ekawa, H and Agari, K and Ahn, J K and Akaishi, T and others},
title = {Observation of a Be double-Lambda hypernucleus in the J-PARC E07 experiment},
journal = {Progress of Theoretical and Experimental Physics},
volume = {2019},
number = {2},
pages = {021D02},
year = {2019},
month = {02},
issn = {2050-3911},
doi = {10.1093/ptep/pty149},
url = {https://doi.org/10.1093/ptep/pty149},
eprint = {https://academic.oup.com/ptep/article-pdf/2019/2/021D02/27970468/pty149.pdf},
}

@article{JCugnon:2000,
title = {Hypernuclei in the Skyrme-Hartree-Fock formalism with a microscopic hyperon-nucleon force},
author = {Cugnon, J. and Lejeune, A. and Schulze, H.-J.},
journal = {Phys. Rev. C},
volume = {62},
issue = {6},
pages = {064308},
numpages = {11},
year = {2000},
month = {Nov},
publisher = {American Physical Society},
doi = {10.1103/PhysRevC.62.064308},
url = {https://link.aps.org/doi/10.1103/PhysRevC.62.064308}
}

@article{IVidana:2001,
title = {Hypernuclear structure with the new Nijmegen potentials},
author = {Vida\~na, I. and Polls, A. and Ramos, A. and Schulze, H.-J.},
journal = {Phys. Rev. C},
volume = {64},
issue = {4},
pages = {044301},
numpages = {7},
year = {2001},
month = {Sep},
publisher = {American Physical Society},
doi = {10.1103/PhysRevC.64.044301},
url = {https://link.aps.org/doi/10.1103/PhysRevC.64.044301}
}

@manual{BUQEYEsoftware,
author="{{BUQEYE collaboration}}",
year="2022",
title="Software",
note={\url{https://buqeye.github.io/software/}},
url={\mbox{https://buqeye.github.io/software/}}
}

@article{CDrischler:2024ebw,
author = "Drischler, C. and Giuliani, P. G. and Bezoui, S. and Piekarewicz, J. and Viens, F.",
title = "{Bayesian mixture model approach to quantifying the empirical nuclear saturation point}",
eprint = "2405.02748",
archivePrefix = "arXiv",
primaryClass- = "nucl-th",
doi = "https://doi.org/10.1103/PhysRevC.110.044320",
journal = "Phys. Rev. C",
volume = "110",
number = "4",
pages = "044320",
year = "2024",
URL ="https://journals.aps.org/prc/abstract/10.1103/PhysRevC.110.044320"
}

@article{BFriedman:1981,
title = {Hot and cold, nuclear and neutron matter},
journal = {Nuclear Physics A},
volume = {361},
number = {2},
pages = {502-520},
year = {1981},
issn = {0375-9474},
doi = {https://doi.org/10.1016/0375-9474(81)90649-7},
url = {https://www.sciencedirect.com/science/article/pii/0375947481906497},
author = {B. Friedman and V.R. Pandharipande}
}

@article{MBaldo:1994,
title = {Dependence of the Landau parameters on the single particle potential},
author = {Baldo, M. and Ferreira, L. S.},
journal = {Phys. Rev. C},
volume = {50},
issue = {4},
pages = {1887--1892},
numpages = {0},
year = {1994},
month = {Oct},
publisher = {American Physical Society},
doi = {10.1103/PhysRevC.50.1887},
url = {https://link.aps.org/doi/10.1103/PhysRevC.50.1887}
}

@article{RBWiringa:1995,
  title = {Accurate nucleon-nucleon potential with charge-independence breaking},
  author = {Wiringa, R. B. and Stoks, V. G. J. and Schiavilla, R.},
  journal = {Phys. Rev. C},
  volume = {51},
  issue = {1},
  pages = {38--51},
  numpages = {0},
  year = {1995},
  month = {Jan},
  publisher = {American Physical Society},
  doi = {10.1103/PhysRevC.51.38},
  url = {https://link.aps.org/doi/10.1103/PhysRevC.51.38}
}

@article{APR:1998,
  title = {Equation of state of nucleon matter and neutron star structure},
  author = {Akmal, A. and Pandharipande, V. R. and Ravenhall, D. G.},
  journal = {Phys. Rev. C},
  volume = {58},
  issue = {3},
  pages = {1804--1828},
  numpages = {0},
  year = {1998},
  month = {Sep},
  publisher = {American Physical Society},
  doi = {10.1103/PhysRevC.58.1804},
  url = {https://link.aps.org/doi/10.1103/PhysRevC.58.1804}
}

@article{RBWiringa:2002,
  title = {Evolution of Nuclear Spectra with Nuclear Forces},
  author = {Wiringa, R. B. and Pieper, Steven C.},
  journal = {Phys. Rev. Lett.},
  volume = {89},
  issue = {18},
  pages = {182501},
  numpages = {4},
  year = {2002},
  month = {Oct},
  publisher = {American Physical Society},
  doi = {10.1103/PhysRevLett.89.182501},
  url = {https://link.aps.org/doi/10.1103/PhysRevLett.89.182501}
}

@article{LGCao:2006b,
  title = {Screening effects in superfluid nuclear and neutron matter within Brueckner theory},
  author = {Cao, L. G. and Lombardo, U. and Schuck, P.},
  journal = {Phys. Rev. C},
  volume = {74},
  issue = {6},
  pages = {064301},
  numpages = {7},
  year = {2006},
  month = {Dec},
  publisher = {American Physical Society},
  doi = {10.1103/PhysRevC.74.064301},
  url = {https://link.aps.org/doi/10.1103/PhysRevC.74.064301}
}

@article{ASedrakian:2007,
  title = {Vertex renormalization of weak interactions and Cooper-pair breaking in cooling compact stars},
  author = {Sedrakian, Armen and M\"uther, Herbert and Schuck, Peter},
  journal = {Phys. Rev. C},
  volume = {76},
  issue = {5},
  pages = {055805},
  numpages = {11},
  year = {2007},
  month = {Nov},
  publisher = {American Physical Society},
  doi = {10.1103/PhysRevC.76.055805},
  url = {https://link.aps.org/doi/10.1103/PhysRevC.76.055805}
}

@article{AFabrocini:2008,
title = {S-pairing in neutron matter: I. Correlated basis function theory},
journal = {Nuclear Physics A},
volume = {803},
number = {3},
pages = {137-158},
year = {2008},
issn = {0375-9474},
doi = {https://doi.org/10.1016/j.nuclphysa.2008.01.024},
url = {https://www.sciencedirect.com/science/article/pii/S0375947408000493},
author = {Adelchi Fabrocini and Stefano Fantoni and Alexey Yu. Illarionov and Kevin E. Schmidt}
}

@article{AGezerlis:2008,
title = {Strongly paired fermions: Cold atoms and neutron matter},
author = {Gezerlis, Alexandros and Carlson, J.},
journal = {Phys. Rev. C},
volume = {77},
issue = {3},
pages = {032801},
numpages = {4},
year = {2008},
month = {Mar},
publisher = {American Physical Society},
doi = {10.1103/PhysRevC.77.032801},
url = {https://link.aps.org/doi/10.1103/PhysRevC.77.032801}
}

@article{TAbe:2009,
  title = {Lattice calculation of thermal properties of low-density neutron matter with pionless $\mathit{NN}$ effective field theory},
  author = {Abe, T. and Seki, R.},
  journal = {Phys. Rev. C},
  volume = {79},
  issue = {5},
  pages = {054002},
  numpages = {19},
  year = {2009},
  month = {May},
  publisher = {American Physical Society},
  doi = {10.1103/PhysRevC.79.054002},
  url = {https://link.aps.org/doi/10.1103/PhysRevC.79.054002}
}

@article{AGezerlis:2010,
title = {Low-density neutron matter},
author = {Gezerlis, Alexandros and Carlson, J.},
journal = {Phys. Rev. C},
volume = {81},
issue = {2},
pages = {025803},
numpages = {9},
year = {2010},
month = {Feb},
publisher = {American Physical Society},
doi = {10.1103/PhysRevC.81.025803},
url = {https://link.aps.org/doi/10.1103/PhysRevC.81.025803}
}

@article{Mbaldo:2012,
title = {Comparative study of neutron and nuclear matter with simplified Argonne nucleon-nucleon potentials},
author = {Baldo, M. and Polls, A. and Rios, A. and Schulze, H.-J. and Vida\~na, I.},
journal = {Phys. Rev. C},
volume = {86},
issue = {6},
pages = {064001},
numpages = {13},
year = {2012},
month = {Dec},
publisher = {American Physical Society},
doi = {10.1103/PhysRevC.86.064001},
url = {https://link.aps.org/doi/10.1103/PhysRevC.86.064001}
}

@article{SGandolfi:2012,
title = {Maximum mass and radius of neutron stars, and the nuclear symmetry energy},
author = {Gandolfi, S. and Carlson, J. and Reddy, Sanjay},
journal = {Phys. Rev. C},
volume = {85},
issue = {3},
pages = {032801},
numpages = {5},
year = {2012},
month = {Mar},
publisher = {American Physical Society},
doi = {10.1103/PhysRevC.85.032801},
url = {https://link.aps.org/doi/10.1103/PhysRevC.85.032801}
}

@article{ITews:2013,
title = {Neutron Matter at Next-to-Next-to-Next-to-Leading Order in Chiral Effective Field Theory},
author = {Tews, I. and Kr\"uger, T. and Hebeler, K. and Schwenk, A.},
journal = {Phys. Rev. Lett.},
volume = {110},
issue = {3},
pages = {032504},
numpages = {5},
year = {2013},
month = {Jan},
publisher = {American Physical Society},
doi = {10.1103/PhysRevLett.110.032504},
url = {https://link.aps.org/doi/10.1103/PhysRevLett.110.032504}
}

@article{GWlazlowski:2014,
title = {Auxiliary-Field Quantum Monte Carlo Simulations of Neutron Matter in Chiral Effective Field Theory},
author = {Wlaz\l{}owski, G. and Holt, J. W. and Moroz, S. and Bulgac, A. and Roche, K. J.},
journal = {Phys. Rev. Lett.},
volume = {113},
issue = {18},
pages = {182503},
numpages = {5},
year = {2014},
month = {Oct},
publisher = {American Physical Society},
doi = {10.1103/PhysRevLett.113.182503},
url = {https://link.aps.org/doi/10.1103/PhysRevLett.113.182503}
}

@article{ITews:2016,
title = {Quantum Monte Carlo calculations of neutron matter with chiral three-body forces},
author = {Tews, I. and Gandolfi, S. and Gezerlis, A. and Schwenk, A.},
journal = {Phys. Rev. C},
volume = {93},
issue = {2},
pages = {024305},
numpages = {10},
year = {2016},
month = {Feb},
publisher = {American Physical Society},
doi = {10.1103/PhysRevC.93.024305},
url = {https://link.aps.org/doi/10.1103/PhysRevC.93.024305}
}

@article{CDrischler:2016,
title = {Asymmetric nuclear matter based on chiral two- and three-nucleon interactions},
author = {Drischler, C. and Hebeler, K. and Schwenk, A.},
journal = {Phys. Rev. C},
volume = {93},
issue = {5},
pages = {054314},
numpages = {12},
year = {2016},
month = {May},
publisher = {American Physical Society},
doi = {10.1103/PhysRevC.93.054314},
url = {https://link.aps.org/doi/10.1103/PhysRevC.93.054314}
}

@article{CDrischler:2017,
title = {Pairing in neutron matter: New uncertainty estimates and three-body forces},
author = {Drischler, C. and Kr\"uger, T. and Hebeler, K. and Schwenk, A.},
journal = {Phys. Rev. C},
volume = {95},
issue = {2},
pages = {024302},
numpages = {14},
year = {2017},
month = {Feb},
publisher = {American Physical Society},
doi = {10.1103/PhysRevC.95.024302},
url = {https://link.aps.org/doi/10.1103/PhysRevC.95.024302}
}

@article{ITews:2018,
doi = {10.3847/1538-4357/aac267},
url = {https://dx.doi.org/10.3847/1538-4357/aac267},
year = {2018},
month = {jun},
publisher = {The American Astronomical Society},
volume = {860},
number = {2},
pages = {149},
author = {I. Tews and J. Carlson and S. Gandolfi and S. Reddy},
title = {Constraining the Speed of Sound inside Neutron Stars with Chiral Effective Field Theory Interactions and Observations},
journal = {The Astrophysical Journal}
}

@article{CDrischler:2019,
  title = {Chiral Interactions up to Next-to-Next-to-Next-to-Leading Order and Nuclear Saturation},
  author = {Drischler, C. and Hebeler, K. and Schwenk, A.},
  journal = {Phys. Rev. Lett.},
  volume = {122},
  issue = {4},
  pages = {042501},
  numpages = {6},
  year = {2019},
  month = {Jan},
  publisher = {American Physical Society},
  doi = {10.1103/PhysRevLett.122.042501},
  url = {https://link.aps.org/doi/10.1103/PhysRevLett.122.042501}
}

@article{CDrischler:2020a,
author = "Drischler, C. and Furnstahl, R. J. and Melendez, J. A. and Phillips, D. R.",
title = "{How Well Do We Know the Neutron-Matter Equation of State at the Densities Inside Neutron Stars? A Bayesian Approach with Correlated Uncertainties}",
eprint = "2004.07232",
archivePrefix = "arXiv",
primaryClass = "nucl-th",
doi = "10.1103/PhysRevLett.125.202702",
journal = "Phys. Rev. Lett.",
volume = "125",
number = "20",
pages = "202702",
year = "2020"
}

@article{CDrischler:2020b,
title = {Quantifying uncertainties and correlations in the nuclear-matter equation of state},
author = {Drischler, C. and Melendez, J. A. and Furnstahl, R. J. and Phillips, D. R.},
journal = {Phys. Rev. C},
volume = {102},
issue = {5},
pages = {054315},
numpages = {21},
year = {2020},
month = {Nov},
publisher = {American Physical Society},
doi = {10.1103/PhysRevC.102.054315},
url = {https://link.aps.org/doi/10.1103/PhysRevC.102.054315}
}

@article{MPiarulli:2020,
title = {Benchmark calculations of pure neutron matter with realistic nucleon-nucleon interactions},
author = {Piarulli, M. and Bombaci, I. and Logoteta, D. and Lovato, A. and Wiringa, R. B.},
journal = {Phys. Rev. C},
volume = {101},
issue = {4},
pages = {045801},
numpages = {17},
year = {2020},
month = {Apr},
publisher = {American Physical Society},
doi = {10.1103/PhysRevC.101.045801},
url = {https://link.aps.org/doi/10.1103/PhysRevC.101.045801}
}

@ARTICLE{Rios:2020,
AUTHOR={Rios, Arnau },
TITLE={Green's Function Techniques for Infinite Nuclear Systems},
JOURNAL={Frontiers in Physics},
VOLUME={8},
YEAR={2020},
URL={https://www.frontiersin.org/journals/physics/articles/10.3389/fphy.2020.00387},
DOI={10.3389/fphy.2020.00387},
ISSN={2296-424X}
}

@article{CDrischler:2021,
author = "Drischler, C. and Holt, J.W. and Wellenhofer, C.",
title = "Chiral Effective Field Theory and the High-Density Nuclear Equation of State", 
journal= "Annual Review of Nuclear and Particle Science",
year = "2021",
volume = "71",
number = "Volume 71, 2021",
pages = "403-432",
doi = "https://doi.org/10.1146/annurev-nucl-102419-041903",
url = "https://www.annualreviews.org/content/journals/10.1146/annurev-nucl-102419-041903",
publisher = "Annual Reviews",
issn = "1545-4134",
type = "Journal Article",
}

@article{ALovato:2022,
title = {Benchmark calculations of infinite neutron matter with realistic two- and three-nucleon potentials},
author = {Lovato, A. and Bombaci, I. and Logoteta, D. and Piarulli, M. and Wiringa, R. B.},
journal = {Phys. Rev. C},
volume = {105},
issue = {5},
pages = {055808},
numpages = {14},
year = {2022},
month = {May},
publisher = {American Physical Society},
doi = {10.1103/PhysRevC.105.055808},
url = {https://link.aps.org/doi/10.1103/PhysRevC.105.055808}
}

@Article{SGandolfi:2022,
AUTHOR = {Gandolfi, Stefano and Palkanoglou, Georgios and Carlson, Joseph and Gezerlis, Alexandros and Schmidt, Kevin E.},
TITLE = {The 1S0 Pairing Gap in Neutron Matter},
JOURNAL = {Condensed Matter},
VOLUME = {7},
YEAR = {2022},
NUMBER = {1},
ARTICLE-NUMBER = {19},
URL = {https://www.mdpi.com/2410-3896/7/1/19},
ISSN = {2410-3896},
DOI = {10.3390/condmat7010019}
}

@Article{IVidana:2024,
AUTHOR = {Vidaña, Isaac and Margueron, Jérôme and Schulze, Hans-Josef},
TITLE = {Nuclear Matter Equation of State in the Brueckner–Hartree–Fock Approach and Standard Skyrme Energy Density Functionals},
JOURNAL = {Universe},
VOLUME = {10},
YEAR = {2024},
NUMBER = {5},
ARTICLE-NUMBER = {226},
URL = {https://www.mdpi.com/2218-1997/10/5/226},
ISSN = {2218-1997},
DOI = {10.3390/universe10050226}
}

@Article{SElhatisari:2024,
author = {{Elhatisari}, Serdar and {Bovermann}, Lukas and {Ma}, Yuan-Zhuo and {Epelbaum}, Evgeny and {Frame}, Dillon and {Hildenbrand}, Fabian and {Kim}, Myungkuk and {Kim}, Youngman and {Krebs}, Hermann and {L{\"a}hde}, Timo A. and {Lee}, Dean and {Li}, Ning and {Lu}, Bing-Nan and {Mei{\ss}ner}, Ulf-G. and {Rupak}, Gautam and {Shen}, Shihang and {Song}, Young-Ho and {Stellin}, Gianluca},
title = "{Wavefunction matching for solving quantum many-body problems}",
journal = {Nature},
year = 2024,
month = jun,
volume = {630},
number = {8015},
pages = {59-63},
doi = {10.1038/s41586-024-07422-z},
larchivePrefix = {arXiv},
leprint = {2210.17488},
lprimaryClass = {nucl-th},
url = {https://doi.org/10.1038/s41586-024-07422-z},
ladsnote = {Provided by the SAO/NASA Astrophysics Data System}
}

@misc{ITews:2024,
title={Neutron matter from local chiral EFT interactions at large cutoffs}, 
author={I. Tews and R. Somasundaram and D. Lonardoni and H. Göttling and R. Seutin and J. Carlson and S. Gandolfi and K. Hebeler and A. Schwenk},
year={2024},
eprint={2407.08979},
archivePrefix={arXiv},
primaryClass={nucl-th},
url={https://arxiv.org/abs/2407.08979}, 
}

@article{Marino:2024,
title = {Diagrammatic ab initio methods for infinite nuclear matter with modern chiral interactions},
author = {Marino, F. and Jiang, W. G. and Novario, S. J.},
journal = {Phys. Rev. C},
volume = {110},
issue = {5},
pages = {054322},
numpages = {13},
year = {2024},
month = {Nov},
publisher = {American Physical Society},
doi = {10.1103/PhysRevC.110.054322},
url = {https://link.aps.org/doi/10.1103/PhysRevC.110.054322}
}

@article{Marino:2025,
  author = "Marino, Francesco  and  Barbieri, Carlo  and  Colo, Gianluca  and  Jiang, Weiguang  and  Novario, Samuel J.",
  title = "{Ab initio Green's functions approach for homogeneous nuclear matter}",
  doi = "10.22323/1.465.0180",
  journal = "PoS",
  year = 2025,
  volume = "QNP2024",
  pages = "180"
}

@dataset{Marino:2025:Zenodo,
author       = {Marino, Francesco and
                Jiang, Weiguang and
                Barbieri, Carlo and
                Colò, Gianluca and
                Novario, Samuel},
title        = {Data: Equations of state from chiral interactions with ADC Green's function and coupled-cluster},
month        = sep,
year         = 2025,
publisher    = {Zenodo},
doi          = {10.5281/zenodo.17053839},
url          = {https://doi.org/10.5281/zenodo.17053839},
}

@article{PJMasson:1988,
title = {Masses from an inhomogeneous partial difference equation with higher-order isospin contributions},
journal = {Atomic Data and Nuclear Data Tables},
volume = {39},
number = {2},
pages = {273-280},
year = {1988},
issn = {0092-640X},
doi = {https://doi.org/10.1016/0092-640X(88)90029-0},
url = {https://www.sciencedirect.com/science/article/pii/0092640X88900290},
author = {P.J. Masson and J. Jänecke}
}

@article{JJaenecke:1988,
title = {Mass predictions from the Garvey-Kelson mass relations},
journal = {Atomic Data and Nuclear Data Tables},
volume = {39},
number = {2},
pages = {265-271},
year = {1988},
issn = {0092-640X},
doi = {https://doi.org/10.1016/0092-640X(88)90028-9},
url = {https://www.sciencedirect.com/science/article/pii/0092640X88900289},
author = {J. Jänecke and P.J. Masson}
}

@article{JDuflo:1995,
title = {Microscopic mass formulas},
author = {Duflo, J. and Zuker, A.P.},
journal = {Phys. Rev. C},
volume = {52},
issue = {1},
pages = {R23--R27},
numpages = {0},
year = {1995},
month = {Jul},
publisher = {American Physical Society},
doi = {10.1103/PhysRevC.52.R23},
url = {https://link.aps.org/doi/10.1103/PhysRevC.52.R23}
}

@article{PMoller:1995,
title = {Nuclear Ground-State Masses and Deformations},
journal = {Atomic Data and Nuclear Data Tables},
volume = {59},
number = {2},
pages = {185-381},
year = {1995},
issn = {0092-640X},
doi = {https://doi.org/10.1006/adnd.1995.1002},
url = {https://www.sciencedirect.com/science/article/pii/S0092640X85710029},
author = {P. Moller and J.R. Nix and W.D. Myers and W.J. Swiatecki}
}

@article{YAboussir:1995,
title = {Nuclear mass formula via an approximation to the Hartree—Fock method},
journal = {Atomic Data and Nuclear Data Tables},
volume = {61},
number = {1},
pages = {127-176},
year = {1995},
issn = {0092-640X},
doi = {https://doi.org/10.1016/S0092-640X(95)90014-4},
url = {https://www.sciencedirect.com/science/article/pii/S0092640X95900144},
author = {Y. Aboussir and J.M. Pearson and A.K. Dutta and F. Tondeur}
}

@article{HKoura:2005,
author = {Koura, Hiroyuki and Tachibana, Takahiro and Uno, Masahiro and Yamada, Masami},
title = "{Nuclidic Mass Formula on a Spherical Basis with an Improved Even-Odd Term}",
journal = {Progress of Theoretical Physics},
volume = {113},
number = {2},
pages = {305-325},
year = {2005},
month = {02},
issn = {0033-068X},
doi = {10.1143/PTP.113.305},
url = {https://doi.org/10.1143/PTP.113.305},
eprint = {https://academic.oup.com/ptp/article-pdf/113/2/305/5192381/113-2-305.pdf},
}

@article{NWang:2010,
title = {Mirror nuclei constraint in nuclear mass formula},
author = {Wang, Ning and Liang, Zuoying and Liu, Min and Wu, Xizhen},
journal = {Phys. Rev. C},
volume = {82},
issue = {4},
pages = {044304},
numpages = {6},
year = {2010},
month = {Oct},
publisher = {American Physical Society},
doi = {10.1103/PhysRevC.82.044304},
url = {https://link.aps.org/doi/10.1103/PhysRevC.82.044304}
}

@article{SGoriely:2010,
title = {Further explorations of Skyrme-Hartree-Fock-Bogoliubov mass formulas. XII. Stiffness and stability of neutron-star matter},
author = {Goriely, S. and Chamel, N. and Pearson, J. M.},
journal = {Phys. Rev. C},
volume = {82},
issue = {3},
pages = {035804},
numpages = {18},
year = {2010},
month = {Sep},
publisher = {American Physical Society},
doi = {10.1103/PhysRevC.82.035804},
url = {https://link.aps.org/doi/10.1103/PhysRevC.82.035804}
}

@article{MLiu:2011,
title = {Further improvements on a global nuclear mass model},
author = {Liu, Min and Wang, Ning and Deng, Yangge and Wu, Xizhen},
journal = {Phys. Rev. C},
volume = {84},
issue = {1},
pages = {014333},
numpages = {8},
year = {2011},
month = {Jul},
publisher = {American Physical Society},
doi = {10.1103/PhysRevC.84.014333},
url = {https://link.aps.org/doi/10.1103/PhysRevC.84.014333}
}

@article{SGoriely:2013,
title = {Further explorations of Skyrme-Hartree-Fock-Bogoliubov mass formulas. XIII. The 2012 atomic mass evaluation and the symmetry coefficient},
author = {Goriely, S. and Chamel, N. and Pearson, J. M.},
journal = {Phys. Rev. C},
volume = {88},
issue = {2},
pages = {024308},
numpages = {14},
year = {2013},
month = {Aug},
publisher = {American Physical Society},
doi = {10.1103/PhysRevC.88.024308},
url = {https://link.aps.org/doi/10.1103/PhysRevC.88.024308}
}

@article{GScamps:2021,
author = "Scamps, Guillaume and Goriely, Stephane and Olsen, Erik and Bender, Michael and Ryssens, Wouter",
title = "{Skyrme-Hartree-Fock-Bogoliubov mass models on a 3D mesh: effect of triaxial shape}",
eprint = "2011.07904",
archivePrefix = "arXiv",
primaryClass = "nucl-th",
doi = "10.1140/epja/s10050-021-00642-1",
journal = "Eur. Phys. J. A",
volume = "57",
number = "12",
pages = "333",
year = "2021"
}

@article{WRyssens:2022,
author = "Ryssens, Wouter and Scamps, Guillaume and Goriely, Stephane and Bender, Michael",
title = "{Skyrme\textendash{}Hartree\textendash{}Fock\textendash{}Bogoliubov mass models on a 3D mesh: II. Time-reversal symmetry breaking}",
eprint = "2208.06455",
archivePrefix = "arXiv",
primaryClass = "nucl-th",
reportNumber = "NT@UW-22-08",
doi = "10.1140/epja/s10050-022-00894-5",
journal = "Eur. Phys. J. A",
volume = "58",
number = "12",
pages = "246",
year = "2022"
}

@article{GGrams:2023,
author = "Grams, Guilherme and Ryssens, Wouter and Scamps, Guillaume and Goriely, Stephane and Chamel, Nicolas",
title = "{Skyrme-Hartree-Fock-Bogoliubov mass models on a 3D mesh: III. From atomic nuclei to neutron stars}",
eprint = "2307.14276",
archivePrefix = "arXiv",
primaryClass = "nucl-th",
reportNumber = "NT@UW-23-11",
doi = "10.1140/epja/s10050-023-01158-6",
journal = "Eur. Phys. J. A",
volume = "59",
number = "11",
pages = "270",
year = "2023",
url = {https://doi.org/10.1140/epja/s10050-023-01158-6}
}

@article{GGrams:2025,
author = "Grams, Guilherme and Shchechilin, Nikolai N. and Sanchez-Fernandez, Adrian and Ryssens, Wouter and Chamel, Nicolas and Goriely, Stephane",
title = "{Skyrme–Hartree–Fock–Bogoliubov mass models on a 3D mesh: IV. Improved description of the isospin dependence of pairing}",
eprint = "2411.08007",
archivePrefix = "arXiv",
primaryClass = "nucl-th",
doi = "10.1140/epja/s10050-025-01503-x",
journal = "Eur. Phys. J. A",
volume = "61",
pages = "35",
year = "2025",
url = {https://doi.org/10.1140/epja/s10050-025-01503-x}
}

@article{BNerloPomorska:1994,
author = "Nerlo-Pomorska, B and Pomorski, K.",
title = "{Simple formula for nuclear charge radius,}",
doi = "doi.org/10.1007/BF01291913.",
journal = "Z. Physik A - Hadrons and Nuclei",
volume = "348",
pages = "169",
year = "1994",
url = {https://link.springer.com/article/10.1007/BF01291913}
}

@article{TBayram:2013,
author = "Bayram, Tuncay and Akkoyun, Serkan and Kara, S. Okan and Sinan, Alper",
title = "{New Parameters for Nuclear Charge Radius Formulas}",
doi = "10.5506/APhysPolB.44.1791",
journal = "Acta Phys. Polon. B",
volume = "44",
number = "8",
pages = "1791--1799",
year = "2013"
}

@article{FBuchinger:1994,
title = {Nuclear charge radii in modern mass formulas},
author = {Buchinger, F. and Crawford, J. E. and Dutta, A. K. and Pearson, J. M. and Tondeur, F.},
journal = {Phys. Rev. C},
volume = {49},
issue = {3},
pages = {1402--1411},
numpages = {0},
year = {1994},
month = {Mar},
publisher = {American Physical Society},
doi = {10.1103/PhysRevC.49.1402},
url = {https://link.aps.org/doi/10.1103/PhysRevC.49.1402}
}

@article{JPBlaizot:1980,
author = {J. P. Blaizot },
journal = {Phys. Rep.},
volume = {64},
pages = {171},
year = {1980}
}

@misc{FSchupp:2024,
title={Probing small neutron skin variations in isotope pairs by hyperon-antihyperon production in antiproton--nucleus interactions}, 
author={Falk Schupp and Josef Pochodzalla and Michael Bölting and Martin Christiansen and Theodoros Gaitanos and Horst Lenske and Marcell Steinen},
year={2024},
eprint={2411.13622},
archivePrefix={arXiv},
primaryClass={nucl-th},
url={https://arxiv.org/abs/2411.13622}, 
}

@article{JCLombardi:1972,
title = {Nuclear sizes in 40, 44, 48Ca},
journal = {Nuclear Physics A},
volume = {188},
number = {1},
pages = {103-114},
year = {1972},
issn = {0375-9474},
doi = {https://doi.org/10.1016/0375-9474(72)90186-8},
url = {https://www.sciencedirect.com/science/article/pii/0375947472901868},
author = {J.C. Lombardi and R.N. Boyd and R. Arking and A.B. Robbins}
}

@article{IBrissaud:1972,
title = {Détermination du rayon de la distribution de neutrons de certains noyaux par l'étude de la diffusion élastique de particules alpha de 166 mev},
journal = {Nuclear Physics A},
volume = {191},
number = {1},
pages = {145-165},
year = {1972},
issn = {0375-9474},
doi = {https://doi.org/10.1016/0375-9474(72)90599-4},
url = {https://www.sciencedirect.com/science/article/pii/0375947472905994},
author = {I. Brissaud and Y. {Le Bornec} and B. Tatischeff and L. Bimbot and M.K. Brussel and G. Duhamel}
}

@article{GDAlkhazov:1976,
title = {Elastic and inelastic scattering of 1.044 GeV protons BY 40Ca, 42Ca, 44Ca, 48Ca and 48Ti},
journal = {Nuclear Physics A},
volume = {274},
number = {3},
pages = {443-462},
year = {1976},
issn = {0375-9474},
doi = {https://doi.org/10.1016/0375-9474(76)90212-8},
url = {https://www.sciencedirect.com/science/article/pii/0375947476902128},
author = {G.D. Alkhazov and T. Bauer and R. Beurtey and A. Boudard and G. Bruge and A. Chaumeaux and P. Couvert and G. Cvijanovich and H.H. Duhm and J.M. Fontaine and D. Garreta and A.V. Kulikov and D. Legrand and J.C. Lugol and J. Saudinos and J. Thirion and A.A. Vorobyov}
}

@article{GDAlkhazov:1977,
title = {Elastic and inelastic scattering of 1.37 GeV α-particles from 40, 42, 44, 48Ca},
journal = {Nuclear Physics A},
volume = {280},
number = {2},
pages = {365-376},
year = {1977},
issn = {0375-9474},
doi = {https://doi.org/10.1016/0375-9474(77)90611-X},
url = {https://www.sciencedirect.com/science/article/pii/037594747790611X},
author = {G.D. Alkhazov and T. Bauer and R. Bertini and L. Bimbot and O. Bing and A. Boudard and G. Bruge and H. Catz and A. Chaumeaux and P. Couvert and J.M. Fontaine and F. Hibou and G.J. Igo and J.C. Lugol and M. Matoba}
}

@article{MJJakobson:1977,
  title = {Neutron Radii of Calcium Isotopes from Pion Total Cross Section Measurements},
  author = {Jakobson, M. J. and Burleson, G. R. and Calarco, J. R. and Cooper, M. D. and Hagerman, D. C. and Halpern, I. and Jeppeson, R. H. and Johnson, K. F. and Knutson, L. D. and Marrs, R. E. and Meyer, H. O. and Redwine, R. P.},
  journal = {Phys. Rev. Lett.},
  volume = {38},
  issue = {21},
  pages = {1201--1204},
  numpages = {0},
  year = {1977},
  month = {May},
  publisher = {American Physical Society},
  doi = {10.1103/PhysRevLett.38.1201},
  url = {https://link.aps.org/doi/10.1103/PhysRevLett.38.1201}
}

@article{AChaumeaux:1977,
title = {Neutron densities from 1 GeV proton scattering},
journal = {Physics Letters B},
volume = {72},
number = {1},
pages = {33-36},
year = {1977},
issn = {0370-2693},
doi = {https://doi.org/10.1016/0370-2693(77)90056-9},
url = {https://www.sciencedirect.com/science/article/pii/0370269377900569},
author = {A. Chaumeaux and V. Layly and R. Schaeffer}
}

@article{EFriedman:1978,
title = {$^{48}\mathrm{Ca}$-$^{40}\mathrm{Ca}$ Radius Difference from Elastic Scattering of 104-MeV $\ensuremath{\alpha}$ Particles},
author = {Friedman, E. and Gils, H. J. and Rebel, H. and Majka, Z.},
journal = {Phys. Rev. Lett.},
volume = {41},
issue = {18},
pages = {1220--1224},
numpages = {0},
year = {1978},
month = {Oct},
publisher = {American Physical Society},
doi = {10.1103/PhysRevLett.41.1220},
url = {https://link.aps.org/doi/10.1103/PhysRevLett.41.1220}
}

@article{IBrissaud:1979,
title = {Determination of matter densities of Ca isotopes by 600 MeV and 1 GeV proton elastic scattering},
journal = {Physics Letters B},
volume = {86},
number = {2},
pages = {141-145},
year = {1979},
issn = {0370-2693},
doi = {https://doi.org/10.1016/0370-2693(79)90803-7},
url = {https://www.sciencedirect.com/science/article/pii/0370269379908037},
author = {I. Brissaud and X. Campi}
}

@article{GIgo:1979,
title = {Elastic differential cross sections and analyzing powers for p+40,42,44,48Ca at 0.8 GeV},
journal = {Physics Letters B},
volume = {81},
number = {2},
pages = {151-155},
year = {1979},
issn = {0370-2693},
doi = {https://doi.org/10.1016/0370-2693(79)90510-0},
url = {https://www.sciencedirect.com/science/article/pii/0370269379905100},
author = {G. Igo and G.S. Adams and T.S. Bauer and G. Pauletta and C.A. Whitten and A. Wreikat and G.W. Hoffmann and G.S. Blanpied and W.R. Coker and C. Harvey and R.P. Liljestrand and L. Ray and J.E. Spencer and H.A. Thiessen and C. Glashausser and N.M. Hintz and M.A. Oothoudt and H. Nann and K.K. Seth and B.E. Wood and D.K. McDaniels and M. Gazzaly}
}

@article{KGBoyer:1984,
title = {Pion elastic and inelastic scattering from $^{40,42,44,48}\mathrm{Ca}$ and $^{54}\mathrm{Fe}$},
author = {Boyer, K. G. and Braithwaite, W. J. and Cottingame, W. B. and Greene, S. J. and Smith, L. E. and Moore, C. Fred and Morris, C. L. and Thiessen, H. A. and Blanpied, G. S. and Burleson, G. R. and Davis, J. F. and McCarthy, J. S. and Minehart, R. C. and Goulding, C. A.},
journal = {Phys. Rev. C},
volume = {29},
issue = {1},
pages = {182--194},
numpages = {0},
year = {1984},
month = {Jan},
publisher = {American Physical Society},
doi = {10.1103/PhysRevC.29.182},
url = {https://link.aps.org/doi/10.1103/PhysRevC.29.182}
}

@article{FJHartmann:2001,
title = {Nucleon density in the nuclear periphery determined with antiprotonic x rays: Calcium isotopes},
author = {Hartmann, F. J. and Schmidt, R. and Ketzer, B. and von Egidy, T. and others},
journal = {Phys. Rev. C},
volume = {65},
issue = {1},
pages = {014306},
numpages = {7},
year = {2001},
month = {Dec},
publisher = {American Physical Society},
doi = {10.1103/PhysRevC.65.014306},
url = {https://link.aps.org/doi/10.1103/PhysRevC.65.014306}
}

@article{BCClark:2003,
  title = {Neutron densities from a global analysis of medium-energy proton-nucleus elastic scattering},
  author = {Clark, B. C. and Kerr, L. J. and Hama, S.},
  journal = {Phys. Rev. C},
  volume = {67},
  issue = {5},
  pages = {054605},
  numpages = {13},
  year = {2003},
  month = {May},
  publisher = {American Physical Society},
  doi = {10.1103/PhysRevC.67.054605},
  url = {https://link.aps.org/doi/10.1103/PhysRevC.67.054605}
}

@article{JBirkhan:2017,
  title = {Electric Dipole Polarizability of $^{48}\mathrm{Ca}$ and Implications for the Neutron Skin},
  author = {Birkhan, J. and Miorelli, M. and Bacca, S. and Bassauer, S. and Bertulani, C. A. and Hagen, G. and Matsubara, H. and von Neumann-Cosel, P. and Papenbrock, T. and Pietralla, N. and Ponomarev, V. Yu. and Richter, A. and Schwenk, A. and Tamii, A.},
  journal = {Phys. Rev. Lett.},
  volume = {118},
  issue = {25},
  pages = {252501},
  numpages = {6},
  year = {2017},
  month = {Jun},
  publisher = {American Physical Society},
  doi = {10.1103/PhysRevLett.118.252501},
  url = {https://link.aps.org/doi/10.1103/PhysRevLett.118.252501}
}

@article{MHMahzoon:2017,
  title = {Neutron Skin Thickness of $^{48}\mathrm{Ca}$ from a Nonlocal Dispersive Optical-Model Analysis},
  author = {Mahzoon, M. H. and Atkinson, M. C. and Charity, R. J. and Dickhoff, W. H.},
  journal = {Phys. Rev. Lett.},
  volume = {119},
  issue = {22},
  pages = {222503},
  numpages = {5},
  year = {2017},
  month = {Nov},
  publisher = {American Physical Society},
  doi = {10.1103/PhysRevLett.119.222503},
  url = {https://link.aps.org/doi/10.1103/PhysRevLett.119.222503}
}

@article{MTanaka:2020,
title = {Swelling of Doubly Magic $^{48}\mathrm{Ca}$ Core in Ca Isotopes beyond $N=28$},
author = {Tanaka, M. and Takechi, M. and Homma, A. and Fukuda, M. and others},
journal = {Phys. Rev. Lett.},
volume = {124},
issue = {10},
pages = {102501},
numpages = {6},
year = {2020},
month = {Mar},
publisher = {American Physical Society},
doi = {10.1103/PhysRevLett.124.102501},
url = {https://link.aps.org/doi/10.1103/PhysRevLett.124.102501}
}

@article{STagami:2022,
title = {Neutron skin in 48Ca determined from p+48Ca and 48Ca+12C scattering},
journal = {Results in Physics},
volume = {33},
pages = {105155},
year = {2022},
issn = {2211-3797},
doi = {https://doi.org/10.1016/j.rinp.2021.105155},
url = {https://www.sciencedirect.com/science/article/pii/S221137972101113X},
author = {Shingo Tagami and Tomotsugu Wakasa and Maya Takechi and Jun Matsui and Masanobu Yahiro}
}

@article{SKarataglidis:2002,
  title = {Discerning the neutron density distribution of ${}^{208}\mathrm{Pb}$ from nucleon elastic scattering},
  author = {Karataglidis, S. and Amos, K. and Brown, B. A. and Deb, P. K.},
  journal = {Phys. Rev. C},
  volume = {65},
  issue = {4},
  pages = {044306},
  numpages = {7},
  year = {2002},
  month = {Mar},
  publisher = {American Physical Society},
  doi = {10.1103/PhysRevC.65.044306},
  url = {https://link.aps.org/doi/10.1103/PhysRevC.65.044306}
}

@article{JZenihiro:2010,
  title = {Neutron density distributions of $^{204,206,208}\mathrm{Pb}$ deduced via proton elastic scattering at ${E}_{p}=295$ MeV},
  author = {Zenihiro, J. and Sakaguchi, H. and Murakami, T. and Yosoi, M. and Yasuda, Y. and Terashima, S. and Iwao, Y. and Takeda, H. and Itoh, M. and Yoshida, H. P. and Uchida, M.},
  journal = {Phys. Rev. C},
  volume = {82},
  issue = {4},
  pages = {044611},
  numpages = {10},
  year = {2010},
  month = {Oct},
  publisher = {American Physical Society},
  doi = {10.1103/PhysRevC.82.044611},
  url = {https://link.aps.org/doi/10.1103/PhysRevC.82.044611}
}

@article{JPiekarewicz:2012,
  title = {Electric dipole polarizability and the neutron skin},
  author = {Piekarewicz, J. and Agrawal, B. K. and Col\`o, G. and Nazarewicz, W. and Paar, N. and Reinhard, P.-G. and Roca-Maza, X. and Vretenar, D.},
  journal = {Phys. Rev. C},
  volume = {85},
  issue = {4},  
  pages = {041302},
  numpages = {5},
  year = {2012},
  month = {Apr},
  publisher = {American Physical Society},
  doi = {10.1103/PhysRevC.85.041302},
  url = {https://link.aps.org/doi/10.1103/PhysRevC.85.041302}
}

@article{VEStarodubsky:1994,
  title = {Extraction of neutron densities from elastic proton scattering by $^{206,207,208}\mathrm{Pb}$ at 650 MeV},
  author = {Starodubsky, V. E. and Hintz, N. M.},
  journal = {Phys. Rev. C},
  volume = {49},
  issue = {4},
  pages = {2118--2135},
  numpages = {0},
  year = {1994},
  month = {Apr},
  publisher = {American Physical Society},
  doi = {10.1103/PhysRevC.49.2118},
  url = {https://link.aps.org/doi/10.1103/PhysRevC.49.2118}
}

@article{STagami:2021,
  title = {Neutron skin thickness of $^{208}\mathrm{Pb}$ determined from the reaction cross section for proton scattering},
  author = {Tagami, Shingo and Wakasa, Tomotsugu and Matsui, Jun and Yahiro, Masanobu and Takechi, Maya},
  journal = {Phys. Rev. C},
  volume = {104},
  issue = {2},
  pages = {024606},
  numpages = {4},
  year = {2021},
  month = {Aug},
  publisher = {American Physical Society},
  doi = {10.1103/PhysRevC.104.024606},
  url = {https://link.aps.org/doi/10.1103/PhysRevC.104.024606}
}

@article{GGiacalone:2023,
  title = {Determination of the Neutron Skin of $^{208}\mathrm{Pb}$ from Ultrarelativistic Nuclear Collisions},
  author = {Giacalone, Giuliano and Nijs, Govert and van der Schee, Wilke},
  journal = {Phys. Rev. Lett.},
  volume = {131},
  issue = {20},
  pages = {202302},
  numpages = {6},
  year = {2023},
  month = {Nov},
  publisher = {American Physical Society},
  doi = {10.1103/PhysRevLett.131.202302},
  url = {https://link.aps.org/doi/10.1103/PhysRevLett.131.202302}
}

@article{BKlos:2007,
title = {Neutron density distributions from antiprotonic $^{208}\mathrm{Pb}$ and $^{209}\mathrm{Bi}$ atoms},
author = {K\l{}os, B. and Trzci\ifmmode \acute{n}\else \'{n}\fi{}ska, A. and Jastrz\ifmmode \mbox{\k{e}}\else \k{e}\fi{}bski, J. and Czosnyka, T. and others},
journal = {Phys. Rev. C},
volume = {76},
issue = {1},
pages = {014311},
numpages = {13},
year = {2007},
month = {Jul},
publisher = {American Physical Society},
doi = {10.1103/PhysRevC.76.014311},
url = {https://link.aps.org/doi/10.1103/PhysRevC.76.014311}
}

@article{SWycech:2007,
  title = {Nuclear surface studies with antiprotonic atom x rays},
  author = {Wycech, S. and Hartmann, F. J. and Jastrz\ifmmode \mbox{\k{e}}\else \k{e}\fi{}bski, J. and K\l{}os, B. and Trzci\ifmmode \acute{n}\else \'{n}\fi{}ska, A. and Egidy, T. von},
  journal = {Phys. Rev. C},
  volume = {76},
  issue = {3},
  pages = {034316},
  numpages = {12},
  year = {2007},
  month = {Sep},
  publisher = {American Physical Society},
  doi = {10.1103/PhysRevC.76.034316},
  url = {https://link.aps.org/doi/10.1103/PhysRevC.76.034316}
}

@article{BABrown:2007,
  title = {Neutron skin deduced from antiprotonic atom data},
  author = {Brown, B. Alex and Shen, G. and Hillhouse, G. C. and Meng, J. and Trzci\ifmmode \acute{n}\else \'{n}\fi{}ska, A.},
  journal = {Phys. Rev. C},
  volume = {76},
  issue = {3},
  pages = {034305},
  numpages = {5},
  year = {2007},
  month = {Sep},
  publisher = {American Physical Society},
  doi = {10.1103/PhysRevC.76.034305},
  url = {https://link.aps.org/doi/10.1103/PhysRevC.76.034305}
}

@article{AKlimkiewicz:2007,
title = {Nuclear symmetry energy and neutron skins derived from pygmy dipole resonances},
author = {Klimkiewicz, A. and Paar, N. and Adrich, P. and Fallot, M. and others},
collaboration = {LAND Collaboration},
journal = {Phys. Rev. C},
volume = {76},
issue = {5},
pages = {051603},
numpages = {4},
year = {2007},
month = {Nov},
publisher = {American Physical Society},
doi = {10.1103/PhysRevC.76.051603},
url = {https://link.aps.org/doi/10.1103/PhysRevC.76.051603}
}

@article{ACarbone:2010,
  title = {Constraints on the symmetry energy and neutron skins from pygmy resonances in $^{68}\mathrm{Ni}$ and $^{132}\mathrm{Sn}$},
  author = {Carbone, Andrea and Col\`o, Gianluca and Bracco, Angela and Cao, Li-Gang and Bortignon, Pier Francesco and Camera, Franco and Wieland, Oliver},
  journal = {Phys. Rev. C},
  volume = {81},
  issue = {4},
  pages = {041301},
  numpages = {5},
  year = {2010},
  month = {Apr},
  publisher = {American Physical Society},
  doi = {10.1103/PhysRevC.81.041301},
  url = {https://link.aps.org/doi/10.1103/PhysRevC.81.041301}
}

@article{ATamii:2011,
title = {Complete Electric Dipole Response and the Neutron Skin in $^{208}\mathrm{Pb}$},
author = {Tamii, A. and Poltoratska, I. and von Neumann-Cosel, P. and Fujita, Y. and others},
journal = {Phys. Rev. Lett.},
volume = {107},
issue = {6},
pages = {062502},
numpages = {5},
year = {2011},
month = {Aug},
publisher = {American Physical Society},
doi = {10.1103/PhysRevLett.107.062502},
url = {https://link.aps.org/doi/10.1103/PhysRevLett.107.062502}
}

@ARTICLE{ATamii:2014,
       author = {{Tamii}, A. and {von Neumann-Cosel}, P. and {Poltoratska}, I.},
        title = "{Electric dipole response of $^{208}$Pb from proton inelastic scattering: Constraints on neutron skin thickness and symmetry energy}",
      journal = {European Physical Journal A},
         year = 2014,
        month = feb,
       volume = {50},
       number = {2},
          eid = {28},
        pages = {28},
          doi = {10.1140/epja/i2014-14028-7},
larchivePrefix = {arXiv},
       leprint = {1307.2706},
 lprimaryClass = {nucl-ex},
       ladsurl = {https://ui.adsabs.harvard.edu/abs/2014EPJA...50...28T},
      ladsnote = {Provided by the SAO/NASA Astrophysics Data System}
}

@article{AKrasznahorkay:1991,
title = {Excitation of the isovector giant dipole resonance by inelastic \ensuremath{\alpha} scattering and the neutron skin of nuclei},
author = {Krasznahorkay, A. and Bacelar, J. and Bordewijk, J. A. and Brandenburg, S. and others},
journal = {Phys. Rev. Lett.},
volume = {66},
issue = {10},
pages = {1287--1290},
numpages = {0},
year = {1991},
month = {Mar},
publisher = {American Physical Society},
doi = {10.1103/PhysRevLett.66.1287},
url = {https://link.aps.org/doi/10.1103/PhysRevLett.66.1287}
}

@article{AKrasznahorkay:2004,
title = {Neutron-skin thickness in neutron-rich isotopes},
journal = {Nuclear Physics A},
volume = {731},
pages = {224-234},
year = {2004},
issn = {0375-9474},
doi = {https://doi.org/10.1016/j.nuclphysa.2003.11.034},
url = {https://www.sciencedirect.com/science/article/pii/S0375947403018712},
author = {A. Krasznahorkay and H. Akimune and A.M. {van den Berg} and N. Blasi and others}
}

@article{AKrasznahorkay:2013,
doi = {10.1088/0031-8949/2013/T154/014018},
url = {https://dx.doi.org/10.1088/0031-8949/2013/T154/014018},
year = {2013},
month = {may},
publisher = {IOP Publishing},
volume = {2013},
number = {T154},
pages = {014018},
author = {A Krasznahorkay and N Paar and D Vretenar and M N Harakeh},
title = {Neutron-skin thickness of 208Pb from the energy of the anti-analogue giant dipole resonance},
journal = {Physica Scripta}
}

@article{JYasuda:2013,
author = {Yasuda, Jumpei and Wakasa, Tomotsugu and Okamoto, Midori and Dozono, Masanori and others},
title = {Extraction of anti-analog giant dipole resonance and neutron skin thickness for 208Pb},
journal = {Progress of Theoretical and Experimental Physics},
volume = {2013},
number = {6},
pages = {063D02},
year = {2013},
month = {06},
issn = {2050-3911},
doi = {10.1093/ptep/ptt038},
url = {https://doi.org/10.1093/ptep/ptt038},
eprint = {https://academic.oup.com/ptep/article-pdf/2013/6/063D02/19300640/ptt038.pdf},
}

@article{CMTarbert:2014,
title = {Neutron Skin of $^{208}\mathrm{Pb}$ from Coherent Pion Photoproduction},
author = {Tarbert, C. M. and Watts, D. P. and Glazier, D. I. and Aguar, P. and others},
collaboration = {Crystal Ball at MAMI and A2 Collaboration},
journal = {Phys. Rev. Lett.},
volume = {112},
issue = {24},
pages = {242502},
numpages = {6},
year = {2014},
month = {Jun},
publisher = {American Physical Society},
doi = {10.1103/PhysRevLett.112.242502},
url = {https://link.aps.org/doi/10.1103/PhysRevLett.112.242502}
}

@article{SAbrahamyan:2012,
title = {Measurement of the Neutron Radius of $^{208}\mathrm{Pb}$ through Parity Violation in Electron Scattering},
author = {Abrahamyan, S. and Ahmed, Z. and Albataineh, H. and Aniol, K. and others},
collaboration = {PREX Collaboration},
journal = {Phys. Rev. Lett.},
volume = {108},
issue = {11},
pages = {112502},
numpages = {6},
year = {2012},
month = {Mar},
publisher = {American Physical Society},
doi = {10.1103/PhysRevLett.108.112502},
url = {https://link.aps.org/doi/10.1103/PhysRevLett.108.112502}
}

@article{FJFattoyev:2018,
  title = {Neutron Skins and Neutron Stars in the Multimessenger Era},
  author = {Fattoyev, F. J. and Piekarewicz, J. and Horowitz, C. J.},
  journal = {Phys. Rev. Lett.},
  volume = {120},
  issue = {17},
  pages = {172702},
  numpages = {6},
  year = {2018},
  month = {Apr},
  publisher = {American Physical Society},
  doi = {10.1103/PhysRevLett.120.172702},
  url = {https://link.aps.org/doi/10.1103/PhysRevLett.120.172702}
}

@article{Bender:2003,
title = {Self-consistent mean-field models for nuclear structure},
author = {Bender, Michael and Heenen, Paul-Henri and Reinhard, Paul-Gerhard},
journal = {Rev. Mod. Phys.},
volume = {75},
issue = {1},
pages = {121--180},
numpages = {0},
year = {2003},
month = {Jan},
publisher = {American Physical Society},
doi = {10.1103/RevModPhys.75.121},
url = {https://link.aps.org/doi/10.1103/RevModPhys.75.121}
}

@article{JRStone:2007,
title = {The Skyrme interaction in finite nuclei and nuclear matter},
journal = {Progress in Particle and Nuclear Physics},
volume = {58},
number = {2},
pages = {587-657},
year = {2007},
issn = {0146-6410},
doi = {https://doi.org/10.1016/j.ppnp.2006.07.001},
url = {https://www.sciencedirect.com/science/article/pii/S0146641006000627},
author = {J.R. Stone and P.-G. Reinhard}
}

@article{MBeiner:1975,
title = {Nuclear ground-state properties and self-consistent calculations with the skyrme interaction: (I). Spherical description},
journal = {Nuclear Physics A},
volume = {238},
number = {1},
pages = {29-69},
year = {1975},
issn = {0375-9474},
doi = {https://doi.org/10.1016/0375-9474(75)90338-3},
url = {https://www.sciencedirect.com/science/article/pii/0375947475903383},
author = {M. Beiner and H. Flocard and Nguyen {Van Giai} and P. Quentin}
}

@article{NVanGiai:1981,
title = {Spin-isospin and pairing properties of modified Skyrme interactions},
journal = {Physics Letters B},
volume = {106},
number = {5},
pages = {379-382},
year = {1981},
issn = {0370-2693},
doi = {https://doi.org/10.1016/0370-2693(81)90646-8},
url = {https://www.sciencedirect.com/science/article/pii/0370269381906468},
author = {Nguyen {Van Giai} and H. Sagawa}
}

@ARTICLE{MRayet:1982,
author = {{Rayet}, M. and {Arnould}, M. and {Paulus}, G. and {Tondeur}, F.},
title = "{Nuclear forces and the properties of matter at high temperature and density}",
journal = {\aap},
year = 1982,
month = dec,
volume = {116},
number = {1},
pages = {183-187},
adsurl = {https://ui.adsabs.harvard.edu/abs/1982A&A...116..183R}
}

@article{JBartel:1982,
title = {Towards a better parametrisation of Skyrme-like effective forces: A critical study of the SkM force},
journal = {Nuclear Physics A},
volume = {386},
number = {1},
pages = {79-100},
year = {1982},
issn = {0375-9474},
doi = {https://doi.org/10.1016/0375-9474(82)90403-1},
url = {https://www.sciencedirect.com/science/article/pii/0375947482904031},
author = {J. Bartel and P. Quentin and M. Brack and C. Guet and H.-B. Håkansson}
}

@article{JDobaczewski:1984,
title = {Hartree-Fock-Bogolyubov description of nuclei near the neutron-drip line},
journal = {Nuclear Physics A},
volume = {422},
number = {1},
pages = {103-139},
year = {1984},
issn = {0375-9474},
doi = {https://doi.org/10.1016/0375-9474(84)90433-0},
url = {https://www.sciencedirect.com/science/article/pii/0375947484904330},
author = {J. Dobaczewski and H. Flocard and J. Treiner}
}

@article{FTondeur:1984,
title = {Static nuclear properties and the parametrisation of Skyrme forces},
journal = {Nuclear Physics A},
volume = {420},
number = {2},
pages = {297-319},
year = {1984},
issn = {0375-9474},
doi = {https://doi.org/10.1016/0375-9474(84)90444-5},
url = {https://www.sciencedirect.com/science/article/pii/0375947484904445},
author = {F. Tondeur and M. Brack and M. Farine and J.M. Pearson}
}

@article{JFriedrich:1986,
title = {Skyrme-force parametrization: Least-squares fit to nuclear ground-state properties},
author = {Friedrich, J. and Reinhard, P.-G.},
journal = {Phys. Rev. C},
volume = {33},
issue = {1},
pages = {335--351},
numpages = {0},
year = {1986},
month = {Jan},
publisher = {American Physical Society},
doi = {10.1103/PhysRevC.33.335},
url = {https://link.aps.org/doi/10.1103/PhysRevC.33.335}
}

@article{LBennour:1989,
title = {Charge distributions of $^{208}\mathrm{Pb}$, $^{206}\mathrm{Pb}$, and $^{205}\mathrm{Tl}$ and the mean-field approximation},
author = {Bennour, L. and Heenen, P-H. and Bonche, P. and Dobaczewski, J. and Flocard, H.},
journal = {Phys. Rev. C},
volume = {40},
issue = {6},
pages = {2834--2839},
numpages = {0},
year = {1989},
month = {Dec},
publisher = {American Physical Society},
doi = {10.1103/PhysRevC.40.2834},
url = {https://link.aps.org/doi/10.1103/PhysRevC.40.2834}
}

@article{PGReinhard:1995,
title = {Nuclear effective forces and isotope shifts},
journal = {Nuclear Physics A},
volume = {584},
number = {3},
pages = {467-488},
year = {1995},
issn = {0375-9474},
doi = {https://doi.org/10.1016/0375-9474(94)00770-N},
url = {https://www.sciencedirect.com/science/article/pii/037594749400770N},
author = {P.-G. Reinhard and H. Flocard}
}

@article{EChabanat:1997,
author = {Chabanat, E. and Bonche, P. and Haensel, P. and Meyer, J. and Schaeffer, R.},
journal = {Nucl. Phys. A},
volume = {627},
pages = {710},
year = {1997}
}

@article{EChabanat:1998,
author = {E. Chabanat and P. Bonche and P. Haensel and J. Meyer and R. Schaeffer},
journal = {Nucl. Phys. A},
volume = {635},
pages = {231},
year = {1998}
}

@article{ABrown:1998,
title = {New Skyrme interaction for normal and exotic nuclei},
author = {Alex Brown, B.},
journal = {Phys. Rev. C},
volume = {58},
issue = {1},
pages = {220--231},
numpages = {0},
year = {1998},
month = {Jul},
publisher = {American Physical Society},
doi = {10.1103/PhysRevC.58.220},
url = {https://link.aps.org/doi/10.1103/PhysRevC.58.220}
}

@article{PGReinhard:1999,
title = {Shape coexistence and the effective nucleon-nucleon interaction},
author = {Reinhard, P.-G. and Dean, D. J. and Nazarewicz, W. and Dobaczewski, J. and Maruhn, J. A. and Strayer, M. R.},
journal = {Phys. Rev. C},
volume = {60},
issue = {1},
pages = {014316},
numpages = {20},
year = {1999},
month = {Jun},
publisher = {American Physical Society},
doi = {10.1103/PhysRevC.60.014316},
url = {https://link.aps.org/doi/10.1103/PhysRevC.60.014316}
}

@article{JMargueron:2002,
title = {Instabilities of infinite matter with effective Skyrme-type interactions},
author = {Margueron, J. and Navarro, J. and Van Giai, Nguyen},
journal = {Phys. Rev. C},
volume = {66},
issue = {1},
pages = {014303},
numpages = {8},
year = {2002},
month = {Jul},
publisher = {American Physical Society},
doi = {10.1103/PhysRevC.66.014303},
url = {https://link.aps.org/doi/10.1103/PhysRevC.66.014303}
}

@article{AWSteiner:2005,
title = {Isospin asymmetry in nuclei and neutron stars},
journal = {Physics Reports},
volume = {411},
number = {6},
pages = {325-375},
year = {2005},
issn = {0370-1573},
doi = {https://doi.org/10.1016/j.physrep.2005.02.004},
url = {https://www.sciencedirect.com/science/article/pii/S0370157305001043},
author = {A.W. Steiner and M. Prakash and J.M. Lattimer and P.J. Ellis}
}

@article{LGCao:2006,
title = {From Brueckner approach to Skyrme-type energy density functional},
author = {Cao, L. G. and Lombardo, U. and Shen, C. W. and Giai, Nguyen Van},
journal = {Phys. Rev. C},
volume = {73},
issue = {1},
pages = {014313},
numpages = {7},
year = {2006},
month = {Jan},
publisher = {American Physical Society},
doi = {10.1103/PhysRevC.73.014313},
url = {https://link.aps.org/doi/10.1103/PhysRevC.73.014313}
}

@article{TLesinski:2006,
title = {Isovector splitting of nucleon effective masses, ab initio benchmarks and extended stability criteria for Skyrme energy functionals},
author = {Lesinski, T. and Bennaceur, K. and Duguet, T. and Meyer, J.},
journal = {Phys. Rev. C},
volume = {74},
issue = {4},
pages = {044315},
numpages = {15},
year = {2006},
month = {Oct},
publisher = {American Physical Society},
doi = {10.1103/PhysRevC.74.044315},
url = {https://link.aps.org/doi/10.1103/PhysRevC.74.044315}
}

@article{TLesinski:2007,
title = {Tensor part of the Skyrme energy density functional: Spherical nuclei},
author = {Lesinski, T. and Bender, M. and Bennaceur, K. and Duguet, T. and Meyer, J.},
journal = {Phys. Rev. C},
volume = {76},
issue = {1},
pages = {014312},
numpages = {34},
year = {2007},
month = {Jul},
publisher = {American Physical Society},
doi = {10.1103/PhysRevC.76.014312},
url = {https://link.aps.org/doi/10.1103/PhysRevC.76.014312}
}

@article{SGoriely:2007,
title = {Further explorations of Skyrme-Hartree-Fock-Bogoliubov mass formulas. VII. Simultaneous fits to masses and fission barriers},
author = {Goriely, S. and Samyn, M. and Pearson, J. M.},
journal = {Phys. Rev. C},
volume = {75},
issue = {6},
pages = {064312},
numpages = {7},
year = {2007},
month = {Jun},
publisher = {American Physical Society},
doi = {10.1103/PhysRevC.75.064312},
url = {https://link.aps.org/doi/10.1103/PhysRevC.75.064312}
}

@article{NChamel:2008,
title = {Further explorations of Skyrme–Hartree–Fock–Bogoliubov mass formulas. IX: Constraint of pairing force to 1S0 neutron-matter gap},
journal = {Nuclear Physics A},
volume = {812},
number = {1},
pages = {72-98},
year = {2008},
issn = {0375-9474},
doi = {https://doi.org/10.1016/j.nuclphysa.2008.08.015},
url = {https://www.sciencedirect.com/science/article/pii/S0375947408006921},
author = {N. Chamel and S. Goriely and J.M. Pearson}
}

@article{SGoriely:2009,
title = {Skyrme-Hartree-Fock-Bogoliubov Nuclear Mass Formulas: Crossing the 0.6 MeV Accuracy Threshold with Microscopically Deduced Pairing},
author = {Goriely, S. and Chamel, N. and Pearson, J. M.},
journal = {Phys. Rev. Lett.},
volume = {102},
issue = {15},
pages = {152503},
numpages = {4},
year = {2009},
month = {Apr},
publisher = {American Physical Society},
doi = {10.1103/PhysRevLett.102.152503},
url = {https://link.aps.org/doi/10.1103/PhysRevLett.102.152503}
}

@article{SGoriely:2013_BSk27,
title = {Hartree-Fock-Bogoliubov nuclear mass model with 0.50 MeV accuracy based on standard forms of Skyrme and pairing functionals},
author = {Goriely, S. and Chamel, N. and Pearson, J. M.},
journal = {Phys. Rev. C},
volume = {88},
issue = {6},
pages = {061302},
numpages = {5},
year = {2013},
month = {Dec},
publisher = {American Physical Society},
doi = {10.1103/PhysRevC.88.061302},
url = {https://link.aps.org/doi/10.1103/PhysRevC.88.061302}
}

@article{MKortelainen:2010,
title = {Nuclear energy density optimization},
author = {Kortelainen, M. and Lesinski, T. and Mor\'e, J. and Nazarewicz, W. and Sarich, J. and Schunck, N. and Stoitsov, M. V. and Wild, S.},
journal = {Phys. Rev. C},
volume = {82},
issue = {2},
pages = {024313},
numpages = {18},
year = {2010},
month = {Aug},
publisher = {American Physical Society},
doi = {10.1103/PhysRevC.82.024313},
url = {https://link.aps.org/doi/10.1103/PhysRevC.82.024313}
}

@article{DGambacurta:2011,
title = {Determination of local energy density functionals from Brueckner-Hartree-Fock calculations},
author = {Gambacurta, D. and Li, L. and Col\`o, G. and Lombardo, U. and Van Giai, N. and Zuo, W.},
journal = {Phys. Rev. C},
volume = {84},
issue = {2},
pages = {024301},
numpages = {9},
year = {2011},
month = {Aug},
publisher = {American Physical Society},
doi = {10.1103/PhysRevC.84.024301},
url = {https://link.aps.org/doi/10.1103/PhysRevC.84.024301}
}

@article{MKortelainen:2012,
title = {Nuclear energy density optimization: Large deformations},
author = {Kortelainen, M. and McDonnell, J. and Nazarewicz, W. and Reinhard, P.-G. and Sarich, J. and Schunck, N. and Stoitsov, M. V. and Wild, S. M.},
journal = {Phys. Rev. C},
volume = {85},
issue = {2},
pages = {024304},
numpages = {15},
year = {2012},
month = {Feb},
publisher = {American Physical Society},
doi = {10.1103/PhysRevC.85.024304},
url = {https://link.aps.org/doi/10.1103/PhysRevC.85.024304}
}

@article{XRocaMaza:2013,
title = {New Skyrme interaction with improved spin-isospin properties},
author = {Roca-Maza, X. and Col\`o, G. and Sagawa, H.},
journal = {Phys. Rev. C},
volume = {86},
issue = {3},
pages = {031306},
numpages = {6},
year = {2012},
month = {Sep},
publisher = {American Physical Society},
doi = {10.1103/PhysRevC.86.031306},
url = {https://link.aps.org/doi/10.1103/PhysRevC.86.031306}
}

@article{BDSerot:1997,
author = {Serot, Brian D. and Walecka, John Dirk},
title = {Recent Progress in Quantum Hadrodynamics},
journal = {International Journal of Modern Physics E},
volume = {06},
number = {04},
pages = {515-631},
year = {1997},
doi = {10.1142/S0218301397000299},
URL = { https://doi.org/10.1142/S0218301397000299 },
eprint = { https://doi.org/10.1142/S0218301397000299}
}

@article{MMSharma:1993,
title = {Rho meson coupling in the relativistic mean field theory and description of exotic nuclei},
journal = {Physics Letters B},
volume = {312},
number = {4},
pages = {377-381},
year = {1993},
issn = {0370-2693},
doi = {https://doi.org/10.1016/0370-2693(93)90970-S},
url = {https://www.sciencedirect.com/science/article/pii/037026939390970S},
author = {M.M. Sharma and M.A. Nagarajan and P. Ring}
}

@article{KSumiyoshi:1995,
title = {Relativistic mean-field theory with non-linear σ and ω terms for neutron stars and supernovae},
journal = {Nuclear Physics A},
volume = {581},
number = {3},
pages = {725-746},
year = {1995},
issn = {0375-9474},
doi = {https://doi.org/10.1016/0375-9474(94)00335-K},
url = {https://www.sciencedirect.com/science/article/pii/037594749400335K},
author = {K. Sumiyoshi and H. Kuwabara and H. Toki}
}

@article{GALalazissis:1997,
  title = {New parametrization for the Lagrangian density of relativistic mean field theory},
  author = {Lalazissis, G. A. and K\"onig, J. and Ring, P.},
  journal = {Phys. Rev. C},
  volume = {55},
  issue = {1},
  pages = {540--543},
  numpages = {0},
  year = {1997},
  month = {Jan},
  publisher = {American Physical Society},
  doi = {10.1103/PhysRevC.55.540},
  url = {https://link.aps.org/doi/10.1103/PhysRevC.55.540}
}

@article{WLong:2004,
title = {New effective interactions in relativistic mean field theory with nonlinear terms and density-dependent meson-nucleon coupling},
author = {Long, Wenhui and Meng, Jie and Giai, Nguyen Van and Zhou, Shan-Gui},
journal = {Phys. Rev. C},
volume = {69},
issue = {3},
pages = {034319},
numpages = {15},
year = {2004},
month = {Mar},
publisher = {American Physical Society},
doi = {10.1103/PhysRevC.69.034319},
url = {https://link.aps.org/doi/10.1103/PhysRevC.69.034319}
}

@article{STypel:1999,
title = {Relativistic mean field calculations with density-dependent meson-nucleon coupling},
journal = {Nuclear Physics A},
volume = {656},
number = {3},
pages = {331-364},
year = {1999},
issn = {0375-9474},
doi = {https://doi.org/10.1016/S0375-9474(99)00310-3},
url = {https://www.sciencedirect.com/science/article/pii/S0375947499003103},
author = {S. Typel and H.H. Wolter}
}

@article{TNiksic:2002,
title = {Relativistic Hartree-Bogoliubov model with density-dependent meson-nucleon couplings},
author = {Nik\ifmmode \check{s}\else \v{s}\fi{}i\ifmmode \acute{c}\else \'{c}\fi{}, T. and Vretenar, D. and Finelli, P. and Ring, P.},
journal = {Phys. Rev. C},
volume = {66},
issue = {2},
pages = {024306},
numpages = {15},
year = {2002},
month = {Aug},
publisher = {American Physical Society},
doi = {10.1103/PhysRevC.66.024306},
url = {https://link.aps.org/doi/10.1103/PhysRevC.66.024306}
}

@article{GALalazissis:2005,
title = {New relativistic mean-field interaction with density-dependent meson-nucleon couplings},
author = {Lalazissis, G. A. and Nik\ifmmode \check{s}\else \v{s}\fi{}i\ifmmode \acute{c}\else \'{c}\fi{}, T. and Vretenar, D. and Ring, P.},
journal = {Phys. Rev. C},
volume = {71},
issue = {2},
pages = {024312},
numpages = {10},
year = {2005},
month = {Feb},
publisher = {American Physical Society},
doi = {10.1103/PhysRevC.71.024312},
url = {https://link.aps.org/doi/10.1103/PhysRevC.71.024312}
}

@article{XRocaMaza:2011,
title = {Relativistic mean-field interaction with density-dependent meson-nucleon vertices based on microscopical calculations},
author = {Roca-Maza, X. and Vi\~nas, X. and Centelles, M. and Ring, P. and Schuck, P.},
journal = {Phys. Rev. C},
volume = {84},
issue = {5},
pages = {054309},
numpages = {16},
year = {2011},
month = {Nov},
publisher = {American Physical Society},
doi = {10.1103/PhysRevC.84.054309},
url = {https://link.aps.org/doi/10.1103/PhysRevC.84.054309}
}

@article{WLong:2006,
title = {Density-dependent relativistic Hartree–Fock approach},
journal = {Physics Letters B},
volume = {640},
number = {4},
pages = {150-154},
year = {2006},
issn = {0370-2693},
doi = {https://doi.org/10.1016/j.physletb.2006.07.064},
url = {https://www.sciencedirect.com/science/article/pii/S0370269306009610},
author = {Wen-Hui Long and Nguyen {Van Giai} and Jie Meng}
}

@article{WLong:2007,
title = {Shell structure and \ensuremath{\rho}-tensor correlations in density dependent relativistic Hartree-Fock theory},
author = {Long, WenHui and Sagawa, Hiroyuki and Giai, Nguyen Van and Meng, Jie},
journal = {Phys. Rev. C},
volume = {76},
issue = {3},
pages = {034314},
numpages = {11},
year = {2007},
month = {Sep},
publisher = {American Physical Society},
doi = {10.1103/PhysRevC.76.034314},
url = {https://link.aps.org/doi/10.1103/PhysRevC.76.034314}
}

@article{WLong:2008,
doi = {10.1209/0295-5075/82/12001},
url = {https://dx.doi.org/10.1209/0295-5075/82/12001},
year = {2008},
month = {mar},
publisher = {},
volume = {82},
number = {1},
pages = {12001},
author = {WenHui Long and Hiroyuki Sagawa and Jie Meng and Nguyen Van Giai},
title = {Evolution of nuclear shell structure due to the pion exchange potential},
journal = {Europhysics Letters}
}

@article{SGoriely:2016,
  title = {Further explorations of Skyrme-Hartree-Fock-Bogoliubov mass formulas. XVI. Inclusion of self-energy effects in pairing},
  author = {Goriely, S. and Chamel, N. and Pearson, J. M.},
  journal = {Phys. Rev. C},
  volume = {93},
  issue = {3},
  pages = {034337},
  numpages = {11},
  year = {2016},
  month = {Mar},
  publisher = {American Physical Society},
  doi = {10.1103/PhysRevC.93.034337},
  url = {https://link.aps.org/doi/10.1103/PhysRevC.93.034337}
}

@article{JMargueron:2018a,
title = {Equation of state for dense nucleonic matter from metamodeling. I. Foundational aspects},
author = {Margueron, J\'er\^ome and Hoffmann Casali, Rudiney and Gulminelli, Francesca},
journal = {Phys. Rev. C},
volume = {97},
issue = {2},
pages = {025805},
numpages = {28},
year = {2018},
month = {Feb},
publisher = {American Physical Society},
doi = {10.1103/PhysRevC.97.025805},
url = {https://link.aps.org/doi/10.1103/PhysRevC.97.025805}
}

@article{JMargueron:2018b,
title = {Equation of state for dense nucleonic matter from metamodeling. II. Predictions for neutron star properties},
author = {Margueron, J\'er\^ome and Hoffmann Casali, Rudiney and Gulminelli, Francesca},
journal = {Phys. Rev. C},
volume = {97},
issue = {2},
pages = {025806},
numpages = {24},
year = {2018},
month = {Feb},
publisher = {American Physical Society},
doi = {10.1103/PhysRevC.97.025806},
url = {https://link.aps.org/doi/10.1103/PhysRevC.97.025806}
}

@ARTICLE{RSomasundaram:2021,
author = {{Somasundaram}, R. and {Drischler}, C. and {Tews}, I. and {Margueron}, J.},
title = "{Constraints on the nuclear symmetry energy from asymmetric-matter calculations with chiral N N and 3 N interactions}",
journal = {\prc},
year = 2021,
month = apr,
volume = {103},
number = {4},
eid = {045803},
pages = {045803},
doi = {10.1103/PhysRevC.103.045803},
archivePrefix = {arXiv},
eprint = {2009.04737},
primaryClass = {nucl-th},
adsurl = {https://ui.adsabs.harvard.edu/abs/2021PhRvC.103d5803S}
}

@misc{nuda:rep,
title = {Public GitHub repository},
year = {2025},
howpublished = {\url{https://github.org/jeromemargueron/nucleardatapy}}
}

@misc{NudaDocumentation2025,
  author    = {J. Margueron and others},
  title     = {nucleardatapy documentation},
  year      = 2025,
  howpublished = {\url{https://nucleardatapy.readthedocs.io/en/latest/}},
  note      = {Accessed: 2025-06-03}
}

@misc{NudaTutorials2025,
  author       = {J. Margueron and others},
  title        = {nucleardatapy tutorials},
  year         = 2025,
  howpublished = {\url{https://jeromemargueron.github.io/nucleardatapy/landing.html}},
  note         = {Accessed: 2025-06-03}
}

@book{Book:BohrMottelson:1969,
author    = "A. Bohr and B. R. Mottelson",
title     = "Nuclear structure vol I",
year      = "1969",
publisher = "Addison-Wesley"
}

@book{Book:Ring:Schuck:1980,
author = {Ring, Peter and Schuck, Peter},
publisher = {Springer-Verlag},
title = {{The Nuclear Many-Body Problem}},
year = {1980}
}

@book{Book:Haensel:Potekhin:Yakovlev:2007,
author = {Pawel Haensel and Alexander Y. Potekhin and Dmitry G. Yakovlev},
title = {Neutron Stars 1: Equation of State and Structure.},
year = {2007},
publisher = "Springer",
url = {https://link.springer.com/book/10.1007/978-0-387-47301-7}
}

@book{Book:Glendenning:2000,
author = {Glendenning, Norman K.},
publisher = {Springer},
title = {{Compact Stars: Nuclear Physics, Particle Physics and General Relativity}},
year = {2000}
}

@book{Book:Rezzolla:2018,
year = 2018,
author = {L. Rezzolla and P. A. M. Pizzochero and D. I. Jones and N. Rea and I. Vidaña},
title = {The Physics and Astrophysics of Neutron Stars},
journal = {Astrophysics and Space Science Library},
volume = {457},
publisher = {Springer International Publishing},
doi = {10.1007/978-3-319-97616-7},
url = {https://www.springer.com/gp/book/9783319976150}
}

@article{JDobaczewski:2014,
doi = {10.1088/0954-3899/41/7/074001},
url = {https://dx.doi.org/10.1088/0954-3899/41/7/074001},
year = {2014},
month = {may},
publisher = {IOP Publishing},
volume = {41},
number = {7},
pages = {074001},
author = {Dobaczewski, J and Nazarewicz, W and Reinhard, P-G},
title = {Error estimates of theoretical models: a guide},
journal = {Journal of Physics G: Nuclear and Particle Physics}
}

@article{LNeufcourt:2018,
title = {Bayesian approach to model-based extrapolation of nuclear observables},
author = {Neufcourt, L\'eo and Cao, Yuchen and Nazarewicz, Witold and Viens, Frederi},
journal = {Phys. Rev. C},
volume = {98},
issue = {3},
pages = {034318},
numpages = {17},
year = {2018},
month = {Sep},
publisher = {American Physical Society},
doi = {10.1103/PhysRevC.98.034318},
url = {https://link.aps.org/doi/10.1103/PhysRevC.98.034318}
}

@misc{CompARE:2023,
title = {Current status of NICER’s measurements of the neutron star masses and radii (online)},
author = {S. Guillot and others},
year = {2023},
howpublished = {\url{https://indico.gsi.de/event/17017/contributions/73365/}}
}

@article{CompOSE:2022,
author = "S. Typel and M. Oertel and T. Klähn and D. Chatterjee and V. Dexheimer and C. Ishizuka and M. Mancini and J. Novak and H. Pais and C. Providência and Ad. R. Raduta and M. Servillat and L. Tolos",
collaboration = "CompOSE Core Team",
title = "{CompOSE Reference Manual}",
eprint = "2203.03209",
archivePrefix = "arXiv",
primaryClass = "astro-ph.HE",
doi = "10.1140/epja/s10050-022-00847-y",
journal = "Eur. Phys. J. A",
volume = "58",
number = "11",
pages = "221",
year = "2022"
}

@misc{StellarCollapse,
title = {Stellar Collapse},
author = "",
howpublished = {\url{https://stellarcollapse.org}},
year = "2025",
}

@misc{O2SCL,
title = {Object Oriented Scientific Computing Library},
author = "",
howpublished = {\url{https://awsteiner.org/code/o2scl}},
year = "2025"
}

@misc{BAND,
title = {Bayesian Analysis of Nuclear Dynamics},
author = "",
howpublished = {\url{https://bandframework.github.io}},
year = "2025"
}

@article{DPhillips:2021,
doi = {10.1088/1361-6471/abf1df},
url = {https://dx.doi.org/10.1088/1361-6471/abf1df},
year = {2021},
month = {may},
publisher = {IOP Publishing},
volume = {48},
number = {7},
pages = {072001},
author = {Phillips, D R and Furnstahl, R J and Heinz, U and Maiti, T and Nazarewicz, W and Nunes, F M and Plumlee, M and Pratola, M T and Pratt, S and Viens, F G and Wild, S M},
title = {Get on the BAND Wagon: a Bayesian framework for quantifying model uncertainties in nuclear dynamics},
journal = {Journal of Physics G: Nuclear and Particle Physics}
}

@article{OBohigas:1979,
title = {Sum rules for nuclear collective excitations},
journal = {Physics Reports},
volume = {51},
number = {5},
pages = {267-316},
year = {1979},
issn = {0370-1573},
doi = {https://doi.org/10.1016/0370-1573(79)90079-6},
url = {https://www.sciencedirect.com/science/article/pii/0370157379900796},
author = {O. Bohigas and A.M. Lane and J. Martorell}
}

\end{document}